\PassOptionsToPackage{dvipsnames}{xcolor}

\documentclass[review]{elsarticle}

\usepackage[colorlinks,citecolor=blue,linktoc=all,linkcolor=cyan]{hyperref}
\usepackage{graphicx}
\usepackage[draft]{changes} % use 'draft' to show all changes, 'final' to accept all
\definechangesauthor[color=blue]{Juan}
\definechangesauthor[color=purple]{Gloria}

\usepackage[T1]{fontenc}
\usepackage{dsfont}               % use mathds instead of mathbb for outline fonts
\usepackage{mathrsfs}             % provides mathscr without overwriting mathcal
\usepackage{slashed}              % For Dirac slash notation.
\usepackage{amsmath}
\usepackage{amssymb}
\usepackage{amsbsy}
\usepackage{amsfonts}
\usepackage{bm}
\usepackage{xspace}
\usepackage{subcaption}
\usepackage{tikz} 
\usepackage{tikz-feynman}
\usepackage[dvipsnames]{xcolor}
\newcommand{\be}{\begin{equation}}
\newcommand{\ee}{\end{equation}}
\newcommand{\ii}{\mathrm{i}}
\newcommand{\mpole}{m_{\textrm{pole}}}
\newcommand{\msc}{m_{\textrm{sc}}}
\newcommand{\mycomment}[1]{}
\newcommand{\re}{\ensuremath{\text{Re}\,}}
\newcommand{\im}{\ensuremath{\text{Im}\,}}
\newcommand{\diff}{\ensuremath{\mathrm{d}}}
\newcommand{\mevnospace}{\ensuremath{{\mathrm{\,Me\kern -0.1em V}}}}
\newcommand{\gevnospace}{\ensuremath{{\mathrm{\,Ge\kern -0.1em V}}}}
\newcommand{\mev}{\mevnospace\xspace}
\newcommand{\gev}{\gevnospace\xspace}

\newcommand{\nn}{\nonumber}
\newcommand{\sutwo}{\ensuremath{\textrm{SU}(2)}}
\newcommand{\suthree}{\ensuremath{\textrm{SU}(3)}}
\newcommand{\sufour}{\ensuremath{\textrm{SU}(4)}}
\newcommand{\sumint}{\mathop{\mathchoice
  {\sum\mkern-24mu\int}{\textstyle\sum\mkern-17.5mu\int}
  {\scriptstyle\sum\mkern-14mu\int}{\scriptscriptstyle\sum\mkern-12mu\int}}\nolimits}
\numberwithin{equation}{section}
\numberwithin{table}{section}
\numberwithin{figure}{section}

\journal{Progress in Particle and Nuclear Physics}

\usepackage{titlesec}
\usepackage{sectsty}
\titleformat{\section}{\normalfont\Large\bfseries}{\thesection}{1em}{}
\titleformat{\subsection}{\normalfont\large\bfseries}{\thesubsection}{1em}{}
\titleformat{\subsubsection}{\normalfont\normalsize\bfseries}{\thesubsubsection}{1em}{}
\usepackage[many]{tcolorbox} % Annotation box

\begin{document}

\begin{frontmatter}

\title{Hadron properties at finite temperature}

%authors, affiliations, corresponding author mention 

\author[mymainaddress,mysecondaryaddress]{Juan M. Torres-Rincon\corref{mycorrespondingauthor}}
\cortext[mycorrespondingauthor]{Corresponding author}
\ead{torres@fqa.ub.edu}

\author[mymainaddress,mysecondaryaddress]{Gl\`oria Monta\~na}
\ead{gmontana@fqa.ub.edu}
\address[mymainaddress]{Departament de F\'isica Qu\`antica i Astrof\'isica, Universitat de Barcelona, Mart\'i i Franqu\`es 1, 08028 Barcelona, Spain}
\address[mysecondaryaddress]{Institut de Ci\`encies del Cosmos, Universitat de Barcelona,  Mart\'i i Franqu\`es 1, 08028 Barcelona, Spain}

\begin{abstract}
This review provides an overview of thermal effects on hadron properties, focusing on the theoretical frameworks used to describe in-medium modifications of masses, decay widths, and spectral functions. We examine the application of finite-temperature quantum field theory---specifically the imaginary-time formalism (ITF)---to analyze both light- and heavy-hadron sectors.
For light hadrons, we discuss the role of chiral symmetry restoration and the different definitions of thermal masses in effective field theories, like chiral perturbation theory. In the heavy-flavor sector, we review recent progress in describing open-heavy mesons and quarkonia using self-consistent unitarized approaches and nonrelativistic effective field theories. All these results are complemented by analyses of recent lattice-QCD calculations using the Euclidean formulation of QCD at finite temperature, relevant to extracting screening masses and reconstructing spectral functions. 
Finally, we discuss the phenomenological impact of the thermal modifications on experimental observables in relativistic heavy-ion collisions, including numerical simulations, dilepton spectra, transport coefficients, and hadron femtoscopy.
By combining phenomenological considerations with robust theoretical tools, this review provides a coherent picture of how thermal effects emerge in the hadronic phase and how they can be systematically studied within controlled frameworks. Ultimately, the discussion serves as a bridge between experimental observations in relativistic heavy-ion collisions and fundamental developments in finite-temperature QCD and effective field theories for hadronic systems.
\end{abstract}


\end{frontmatter}

\newpage

\thispagestyle{empty}
\tableofcontents

%to begin the line numbers: 
%\linenumbers

%beginning of the core of the manuscript
\newpage
\section{Introduction}\label{intro}

The study of strongly interacting matter under extreme conditions of temperature and density is a central and dynamic research area in high-energy nuclear physics. It aims at characterizing the quantum chromodynamics (QCD) phase diagram, which features a rich structure, including a transition from hadronic matter---where quarks and gluons are confined within hadrons---to a deconfined state known as the quark-gluon plasma (QGP) at sufficiently high temperatures and/or baryon densities. A thorough understanding of the nature and properties of these phases, as well as the transition between them, is fundamental to shedding light on a diverse range of phenomena.

In the first microseconds after the Big Bang, it is believed that the universe existed in a QGP state. As the early universe expanded and cooled, it underwent a transition (crossover) to the hadronic matter we observe today. Studying the properties of the QGP and the confining phase transition, including how hadrons are formed and evolve in such a hot and dense medium, is crucial for accurately modeling the evolution of the early universe as well as the systems created in heavy-ion collisions (HICs), where a similar transition from a deconfined QGP to a system of hadrons in a thermal medium takes place for high-energy collision events. Leaving aside the possible applications to the thermal history of the early universe, in this review we mainly focus on the field of HICs. In these systems, medium effects influence hadrons, both light and heavy, for a finite time, so their vacuum properties are expected to be modified by temperature and density. This can potentially have measurable effects on several of the observables that can be addressed in HIC experiments.  

In this review, we focus on the effect of temperature, $T$. Hadrons, as bound states of quarks, serve as invaluable probes of the hot and dense medium created in HICs. Their properties, such as their masses and decay widths, are expected to be modified by interactions with the surrounding medium. The in-medium behavior of hadrons can therefore provide insight into the restoration of chiral symmetry in QCD. Studying how different hadrons, ranging from light pions to heavy quarkonia, are modified in a hot medium offers a tool to gain a better understanding of the QCD phase diagram at vanishing baryon density. This review summarizes the theoretical progress made in understanding the properties of hadrons at finite temperature and their relevance to the study of HICs.

\subsection{Hadrons in a thermal medium}

Relativistic HICs provide controlled experimental settings to study strongly interacting matter under extreme conditions of temperature and energy density. Over the past decades, experiments at high-energy nuclear facilities have established that such collisions can produce a hot and rapidly expanding medium, whose early-time properties are consistent with those of a deconfined QGP. Among the most important observables that probe the thermal nature of the system are electromagnetic signals, such as real photons and dileptons, which are emitted throughout the evolution of the collision and escape the medium largely unaltered due to the small value of the electromagnetic coupling. Their spectra therefore provide direct information on the temperature and spacetime dynamics of the fireball~\cite{Shuryak:1978ij,Feinberg:1976ua,Rapp:1999ej}.

As the system expands and cools, it undergoes a transition into a hadronic phase in which colorless bound states dominate the relevant degrees of freedom. In addition, the medium crosses the chiral transition from a chirally restored to a chirally broken phase. At $\mu_B=0$ (zero baryochemical potential) this transition is known to be a crossover~\cite{Aoki:2006we} and, in this situation, the pseudocritical temperature depends on the particular observable used to define it. Nevertheless, the commonly used value calculated in lattice QCD at finite temperature is $T_c \simeq 156$--$158\mev$ (see Section~\ref{sec:chiralrestoration} for more details). The late stages of the collision are characterized by a hadronic medium that persists for a finite time on the order of a few~fm/$c$ while it still expands until the particles decouple from the medium. During this phase, the whole system is largely inhomogeneous, with different regions causally disconnected from each other. However, locally one still expects the expansion rate to be smaller than the scattering rate, so that local equilibrium can be maintained. In this situation, each homogeneity region can be approximately described as a thermalized hadron gas, until the expansion rate exceeds the scattering rate and the decoupling occurs. In this scenario, the thermal equilibrium assumption can be locally applied. Even within this relatively short-lived phase, finite-temperature effects can substantially alter hadronic properties compared to their vacuum counterparts. In particular, hadron masses, decay widths, and interaction strengths may acquire a nontrivial temperature dependence, reflecting the influence of the surrounding medium. These in-medium modifications are essential ingredients for a consistent interpretation of experimental observables and for understanding the gradual approach to freeze-out.

A phenomenological description of the hadronic stage is commonly achieved using the Statistical Hadronization Model~\cite{Braun-Munzinger:2003pwq,Andronic:2017pug}. In this framework, the hadron yields observed in experiments are well reproduced by assuming an approximate thermal equilibrium at a characteristic temperature, commonly referred to as the chemical freeze-out temperature ($T_\textrm{ch} \simeq 155\mev$), at which inelastic interactions effectively cease. This temperature provides an estimate of the thermal conditions under which the composition of the hadronic system becomes effectively fixed and partial chemical equilibrium can be applied. At later times, elastic and quasi-elastic interactions still drive the system toward local thermal equilibrium. Quasi-elastic processes are those in which number-changing reactions compensate each other on average, such as $\pi\pi \rightarrow \rho \rightarrow \pi\pi$ or $\pi N\rightarrow \Delta \rightarrow \pi N$. All these processes continue until kinetic freeze-out ($T_\textrm{kin} \simeq 110\mev$), when particles decouple entirely from the medium, and their momentum distributions no longer change.

From a theoretical standpoint, the description of hadronic matter at finite temperature requires extending quantum field theory to thermal environments. Thermal field theory provides a consistent framework to incorporate temperature effects into correlation functions, $n$-point Green functions, and decay widths~\cite{Weldon:1983jn,Das:1997gg,Kapusta:2006pm,Bellac:2011kqa,Laine:2016hma}. In particular, the imaginary-time (Euclidean) formulation allows one to treat finite-temperature systems by compactifying the temporal direction, thereby enabling the systematic calculation of thermodynamic and spectral quantities (after analytic continuation) relevant for strongly interacting matter. These approaches are essential for understanding how thermal effects modify hadronic observables and for establishing a link between microscopic dynamics and macroscopic thermodynamics.

Complementary insight is provided by first-principles calculations based on lattice QCD. By formulating the QCD Lagrangian on a Euclidean spacetime lattice, it is possible to access equilibrium properties of strongly interacting matter at finite temperature~\cite{Petreczky:2012rq}, including the equation of state~\cite{Borsanyi:2013bia,HotQCD:2014kol} or screening masses associated with hadronic excitations~\cite{Bazavov:2019www}. Moreover, the Euclidean correlation functions computed on the lattice contain information about the real-time spectral properties of hadrons, which can be extracted through spectral reconstruction techniques. Although lattice-QCD studies at low temperatures, which are relevant for hadron physics, remain technically challenging due to the need for large temporal extents and controlled continuum extrapolations, steady advances in numerical algorithms and computational power continue to improve the reliability and scope of such calculations.

Together, experimental observations from relativistic HICs and theoretical developments in thermal field theory and lattice QCD provide a coherent picture of hadronic matter at finite temperature. This interplay forms the foundation for quantitative studies of thermal modifications of hadrons and sets the stage for more detailed investigations based on effective descriptions and microscopic models discussed in the following sections.

\subsection{Finite-temperature quantum field theory: imaginary-time formalism}

We briefly review the basics of the imaginary-time formalism (ITF), which is useful for computing hadron properties at finite temperature. Although alternative formalisms do exist, such as the real-time formalism, the ITF is the simplest approach that can be applied to equilibrium physics, which is the topic of this review. 

The imaginary-time formalism provides a systematic framework for describing a quantum field theory in thermal equilibrium. It is based on the close connection between quantum statistical mechanics and Euclidean field theory and is particularly well suited for the computation of thermodynamic quantities and equilibrium correlation functions. For a more in-depth treatment of the formalism, the classic references are Refs.~\cite{Weldon:1983jn,Das:1997gg,Kapusta:2006pm,Bellac:2011kqa,Laine:2016hma}.

The starting point is the grand-canonical partition function (we will use natural units where $\hbar=c=k_{\text B}=1$),
\begin{equation}
{\cal Z} = \mathrm{Tr}\,\exp\!\left[-\beta\left(H - \mu N\right)\right] \ ,
\qquad
\beta = \frac{1}{T} \ ,
\end{equation}
where $H$ is the Hamiltonian of the system, $N$ the number operator for a conserved charge, and $\mu$ the associated chemical potential. $T$ is the temperature. By performing a Wick rotation to imaginary time,
\begin{equation}
t \rightarrow - \ii \tau \ ,
\qquad
0 \le \tau \le \beta \ ,
\end{equation}
the partition function can be expressed as a functional integral over fields defined in Euclidean spacetime,
\begin{equation}
{\cal Z} = \int \mathcal{D}\Phi \,
\exp\!\left(-S^{\text E}[\Phi]\right) \ ,
\end{equation}
where $S^{\text E}$ denotes the Euclidean action and $\Phi$ represents the fields of the quantum field theory. The imaginary-time direction is then compactified to a circle of circumference $\beta$, with this parameter corresponding to the inverse temperature of the system.

The compact imaginary-time dimension implies specific boundary conditions on the fields $\Phi$, determined by their quantum statistics. This is a consequence of the Kubo--Martin--Schwinger relation~\cite{Kubo:1957mj,Martin:1959jp}. Bosonic fields satisfy periodic boundary conditions in the imaginary-time variable,
\begin{equation}
\phi(\tau+\beta,\bm{x}) = \phi(\tau,\bm{x}) \ ,
\end{equation}
whereas fermionic fields obey antiperiodic boundary conditions,
\begin{equation}
\psi(\tau+\beta,\bm{x}) = -\psi(\tau,\bm{x}) \ ,
\end{equation}
where the time variable $t=-\ii \tau$ runs along the imaginary axis of the complex time plane, from $0$ to $-\ii \beta$, as depicted in Fig.~\ref{fig:itf}. The ITF takes its name from this definition. 

\begin{figure}[t]
  \centering
\begin{tikzpicture}[scale=1.1]
  % --- Axes (plain lines, no arrowheads) ---
  \draw (-2,0) -- (2,0);
  \draw (0,-2) -- (0,2);

  % Axis labels
  \node[below right] at (1.2,0.6) {\large{$\re t$}};
  \node[above left]  at (1,1.6) {\large{$\im t$}};

  % --- Thick vertical line from origin to -i beta ---
  \coordinate (O) at (0,0);
  \coordinate (B) at (0.0,-1.5); % arbitrary length representing -i\beta

  \draw[line width=2pt] (O) -- (B);

  % Label only at -i beta
  \node[right] at (B) {\large{\ \ $-\ii\beta$}};

  % --- Arrow on the thick vertical line (placed at mid point) ---
  % Draw a short arrow segment along the same line so it looks like an arrow "on" it.
  \draw[-{Triangle[length=3mm,width=2mm]},line width=1.2pt] (0,-0.7) -- (0,-0.8);

\end{tikzpicture}
\caption{Integration path along the imaginary axis of the complex time plane used in the imaginary-time formalism of thermal quantum field theory.}
\label{fig:itf}
\end{figure}

As a consequence, the fields admit a discrete Fourier expansion in the imaginary time,
\begin{equation}
\Phi(\tau,\bm{x}) =
T \sum_{n} \int\!\frac{\diff^3p}{(2\pi)^3}
\, e^{\ii (\omega_n \tau + \bm{p}\cdot\bm{x})}
\, \tilde{\Phi}(\omega_n,\bm{p}) \ ,
\end{equation}
where the so-called Matsubara frequencies are given by
\begin{equation}
\omega_n =
\begin{cases}
2\pi n T \ , & \text{bosons}, \\
(2n+1)\pi T \ , & \text{fermions}.
\end{cases}
\end{equation}
Loop integrals over continuous energies are thus replaced by discrete sums,
\begin{equation}
\int\frac{\diff p_0}{2\pi}
\;\longrightarrow\;
T\sum_n \ ,
\end{equation}
which encode the thermal nature of the system. The following shorthand notation is sometimes used for the four-momentum sum-integral,
\begin{equation}
    \sumint_p \equiv T \sum_n \int \frac{\diff^3p}{(2\pi)^3} \ . \label{eq:sumint}
\end{equation}
It is worth mentioning that Feynman rules in the ITF have the same form as at $T=0$, except that the zeroth component of the four-momenta is replaced by $p^0\rightarrow \ii \omega_n$ and loop integrals by sum-integrals. 

Within the ITF, propagators are defined as Euclidean time-ordered correlation functions. For a free scalar field, the Euclidean thermal free propagator takes the form
\begin{equation}
D^{\text E}_0 (\omega_n,\bm{p}) =
\frac{1}{\omega_n^2 + \bm{p}^2 + m^2} \ , \label{eq:boseprop}
\end{equation}
while for a fermionic field,
\begin{equation}
\Delta^{\text E}_0(\omega_n,\bm{p}) =
\frac{-\ii\gamma_E^0 \omega_n + \ii\boldsymbol{\gamma}_E\cdot\bm{p} + m \mathbb{I}}
{\omega_n^2 + \bm{p}^2 + m^2} \ , \label{eq:fermiprop}
\end{equation}
with $\gamma_E^\mu$ the Euclidean Dirac $4 \times 4$ matrices. These are related to the standard (Minkowski) Dirac matrices as $\gamma_E^0=\gamma^0$ and $\gamma_E^k=-i\gamma^k$ with $k=1,2,3$~\cite{Laine:2016hma}.

After performing the corresponding Matsubara sums, the thermal contributions are 
naturally expressed in terms of statistical distribution functions. 
As an example, a standard application of Matsubara summation is given by the following result for the bosonic free propagator~\cite{Kapusta:2006pm},
\be T \sum_n \frac{1}{\omega_n^2+E_p^2} = \frac{1}{2 E_p} \left[ 1 + 2 n_{\text{B}}(E_p) \right] \ , \ee 
where $E_p^2=\bm{p}^2+m^2$ and
\be n_{\text{B}}(E) = \frac{1}{e^{\beta E}-1}  \label{eq:BE} \ee 
is the Bose--Einstein distribution function. One of the advantages of this formalism is that vacuum and thermal effects are clearly separated.

Thermodynamic observables follow directly from the partition function. In particular, the pressure is given by
\begin{equation}
P = \frac{T}{V}\,\log {\cal Z} \ ,
\end{equation}
while the entropy density, energy density, and susceptibilities can be obtained by taking appropriate derivatives with respect to temperature and chemical potentials.

In interacting theories, thermal effects are conveniently encoded in the finite-temperature effective action or, equivalently, in the effective potential for homogeneous field configurations. Temperature-dependent corrections modify masses, couplings, and order parameters, playing a central role in the description of symmetry restoration and phase transitions. 

For full two-point functions, the modifications to Eqs.~\eqref{eq:boseprop} and~\eqref{eq:fermiprop} are described by the Dyson equation. The complete Euclidean boson propagator ${\cal D}^{\text E}(\omega_n,\bm{p}; T)$ is related to the free propagator through the thermal self-energy $\Pi^{\text E}(\omega_n,\bm{p};T)$, according to
\be ({\cal D}^{\text E})^{-1} (\omega_n,\bm{p};T) = (D^{\text E}_0)^{-1} (\omega_n,\bm{p}) + \Pi^{\text E} (\omega_n,\bm{p};T) \ . \ee
The Euclidean self-energy $\Pi^{\text E}$ encodes the interactions of the boson with the thermal bath, and similarly for fermions. This self-energy is essential for determining the in-medium spectral properties discussed in the following sections.

It should be mentioned that while the ITF is a powerful tool for calculating static equilibrium properties---such as thermodynamic potentials, in-medium masses, and susceptibilities---it possesses inherent limitations regarding real-time dynamics. The ITF cannot directly describe the time evolution of a system out of equilibrium.
Physical real-time observables can be accessed through analytic continuation from discrete Matsubara frequencies to continuous real energies,
\begin{equation}
\ii \omega_n \rightarrow \omega + \ii 0^+ \ .
\end{equation}
In doing so, Euclidean quantities are transformed into retarded ones. For example, the retarded propagator is obtained with the prescription~\cite{Laine:2016hma},
\begin{equation}
    {\cal D}^{\text R}(\omega + \ii 0^+, \bm{p};T) = -{\cal D}^{\text E} ( \omega_n \rightarrow  -\ii (\omega + \ii 0^+), \bm{p};T)\ . 
\end{equation}
This procedure is particularly important for extracting pole masses, spectral functions of thermal states (quasiparticles or collective excitations), and decay widths, although it may become technically involved in interacting systems. For phenomena where the system is far from local thermal equilibrium, or where the time-dependent response is the primary focus, the real-time formalism becomes necessary~\cite{Bellac:2011kqa,Laine:2016hma}. 

In the works presented in the next sections, the focus is on the hadronic phase at local thermal equilibrium, which allows the use of the ITF to describe the modification of meson properties within the thermal bath. For example, after analytic continuation to real energies, the boson spectral function is defined as the imaginary part of the retarded propagator,
\be \label{eq:sfunc}
S (\omega, \bm{p};T) = - \frac{1}{\pi} \im {\cal D}^{\text R} (\omega+\ii 0^+, \bm{p}; T) \ , 
\ee
which contains information on the distribution of states (excitations) in the medium. With the normalization in Eq.~\eqref{eq:sfunc}, the spectral function satisfies the sum rule $\int_{-\infty}^{\infty} \diff\omega\,\omega\, S(\omega,\bm{p};T)=1$ and reduces to $\mathrm{sgn}(\omega)\,\delta(\omega^2-E_p^2)$ for a free particle. This convention is used throughout this review, except for lattice-QCD spectral functions, denoted by $\rho$, which correspond to mesonic currents and are normalized differently (see Section~\ref{sec:D-LQCD}).

The ITF provides a natural framework for incorporating thermal effects into low-energy effective theories of QCD. In effective hadronic models, such as chiral perturbation theory or the linear sigma model, the degrees of freedom are mesonic fields rather than quarks and gluons, although some models also contain quark degrees of freedom. Thermal corrections then arise from mesonic (or quark) loop diagrams evaluated using Matsubara sums with well-established techniques. As a result, the ITF serves as a key tool for studying the thermodynamics of hadronic matter and the restoration of chiral symmetry at finite temperature.

While the continuation to real energies can often be performed explicitly in continuum analytical calculations, it poses a fundamental challenge in numerical approaches such as lattice QCD. Extracting real-time spectral functions from Euclidean lattice-QCD correlation functions represents an intrinsically ill-posed inverse problem, as continuous real-time information must be reconstructed from a discrete, finite, and stochastically noisy set of lattice data points. Stabilizing this reconstruction requires regularized numerical techniques, such as advanced Bayesian methods, which are crucial when interpreting the lattice-QCD results discussed later in this review.

The remainder of this review is structured as follows. In Section~\ref{sec:light}, we focus on the light-flavor sector, discussing different effective field theories (EFTs) and models (Section~\ref{sec:eftslight}) used to extract light-meson thermal properties (Section~\ref{sec:polescreening}). We include a discussion on the restoration of chiral symmetry in Section~\ref{sec:chiralrestoration} and its classification according to the nature of the chiral partners. We comment on the results from different models and lattice QCD in Section~\ref{sec:lqcdlight} and conclude with a discussion on baryons in Section~\ref{sec:lightbaryons}. We then move to the heavy-flavor sector, covering open-heavy hadrons in Section~\ref{sec:openHF} and hidden-heavy states in Section~\ref{sec:hiddenHF}.
The open-heavy sector includes an overview of heavy-quark effective theory in Section~\ref{sec:HQET}, hadronic approaches to heavy mesons at finite temperature in Section~\ref{sec:D-hadronic}, and the application of unitarized hadronic models in Section~\ref{sec:unitarizedEFT}. We further review results from lattice QCD in Section~\ref{sec:D-LQCD}, QCD sum-rule analyses in Section~\ref{sec:D-QCDSR}, and heavy baryons at finite temperature in Section~\ref{sec:heavy-baryons}, followed by a summary of the various frameworks and future perspectives in Section~\ref{sec:openHFsummary}.
For hidden-heavy states, we discuss nonrelativistic effective field theories in Section~\ref{sec:EFT-QQ}, lattice-QCD results in Section~\ref{sec:QQ-LQCD}, complementary approaches based on QCD sum rules and in-medium $T$-matrix formalisms in Section~\ref{sec:QQ-other}, and a summary and overview of future perspectives for quarkonia in Section~\ref{sec:hiddenHFsummary}. We also comment on exotic states and their thermal properties in Section~\ref{sec:exotics}.
Finally, in Section~\ref{sec:applications}, we connect these theoretical developments to experimental observables in relativistic HICs, exploring the impact of in-medium modifications on hadronic simulations in Section~\ref{sec:app-hics}, dilepton production measurements in Section~\ref{sec:app-dileptons}, transport coefficients in Section~\ref{sec:app-transport}, hadron femtoscopy in Section~\ref{sec:app-femtoscopy}, and the model dependence of in-medium observables in Section~\ref{sec:th_uncertainties}. We conclude in Section~\ref{sec:summary} with a summary of the current state of the field and an outlook on future directions.

\section{Light hadrons}\label{sec:light}

In this section, we focus on the thermal effects on the lightest hadrons, both mesons and baryons, composed of up, down, and strange quarks. These are particularly sensitive to the properties of the QCD medium and serve as effective probes of the dynamics of chiral symmetry and confinement. When hadronic matter composed of light hadrons is subjected to finite temperature, as in the final stages of a HIC, its properties deviate from those observed in vacuum. The most prominent changes appear in their masses, decay widths, and spectral functions.

At finite temperature, interactions between hadrons and the surrounding medium give rise to thermal masses, which differ from their vacuum values. These originate from many-body effects such as hadron-hadron scattering, coupling to thermal excitations, and a possible (partial) restoration of chiral symmetry. Regarding the latter, the distinction between pseudoscalar mesons and other states, such as the $\rho$ meson, is essential for understanding the spontaneous breaking of chiral symmetry and its restoration. The fact that the main components of the QCD medium at low temperature---the lightest mesons, or pions---are the pseudo-Goldstone bosons of the broken chiral symmetry has important implications for the description of the medium properties at finite temperature.

\subsection{Effective hadronic frameworks}\label{sec:eftslight}

At low energies, the dynamics of pions dominates the thermodynamics of hadronic matter due to their role as (pseudo-)Goldstone bosons of the spontaneous breaking of chiral symmetry in QCD. Since their interactions are governed primarily by symmetry constraints rather than details of the QCD Lagrangian, pions can be systematically described using effective field theories. Within these approaches, medium effects such as thermal masses, decay constants, and interaction rates can be consistently evaluated, making pions the natural degrees of freedom for exploring the behavior of strongly interacting matter at low temperatures.

The dynamics of pions at very low energies is accurately captured by chiral perturbation theory (ChPT)~\cite{Gasser:1983yg, Gasser:1984gg}, the low-energy effective theory of QCD based on $\sutwo$ chiral symmetry, which can be extended to $N_f=3$ flavors to include kaons and $\eta$ mesons. ChPT is constructed following a bottom-up approach~\cite{Weinberg:1978kz}, where the most general Lagrangian consistent with the symmetries of QCD is written as an expansion in derivatives and quark masses, organized according to a predetermined power counting scheme. 

The leading-order chiral Lagrangian formally corresponds to the so-called nonlinear sigma model (NL$\sigma$M)~\cite{Coleman:1969sm,Callan:1969sn,Weinberg:1968de}. In this formulation, the Goldstone bosons are encoded in a unitary matrix field $U(x)=\exp(\ii\pi^a(x)\tau^a/F)$, which parametrizes the coset space $[\sutwo_\text{L}\times \sutwo_\text{R}]/\sutwo_\text{V}$.
The spontaneous breaking of chiral symmetry $\sutwo_\text{L} \times \sutwo_\text{R} \rightarrow \sutwo_\text{V}$ is realized nonlinearly: the three pions transform nonlinearly under chiral rotations, and no explicit scalar chiral partner appears in the model. The scalar-isoscalar mode commonly associated with the $\sigma$ resonance is effectively integrated out, reflecting the fact that it is significantly heavier than the pions at low energies. As a consequence, the nonlinear realization is particularly suited for describing the Goldstone sector well below the chiral symmetry restoration scale.

A closely related framework is provided by the linear sigma model (L$\sigma$M)~\cite{Gell-Mann:1960mvl}, in which the chiral symmetry is realized linearly and the $\sigma$ scalar field explicitly appears as the chiral partner of the pion. In this model, the spontaneous breaking of the $\text{O}(4)\rightarrow \text{O}(3)$ symmetry leads to a massive scalar and three massless pseudoscalar modes. As in the case of the NL$\sigma$M, the chiral symmetry can be explicitly broken to account for finite pion masses in vacuum. The explicit presence of the scalar mode makes the L$\sigma$M particularly useful for studying the mechanism of spontaneous symmetry breaking and its restoration at finite temperature~\cite{Kapusta:1979fh,Pisarski:1983ms}. Below the critical temperature, the $\sigma$ remains massive, while near chiral restoration it becomes degenerate with the pion, signaling the restoration of chiral symmetry~\cite{Pisarski:1983ms}. A generalization of the L$\sigma$M to more pion fields can be envisaged in the so-called $\text{O}(N)$ models~\cite{Dolan:1973qd}, which, in the large-$N$ limit~\cite{Coleman:1974jh}, provide further analytical control and are widely used to study critical behavior and universality near the chiral phase transition.

The chiral effective framework can be extended to include additional hadronic degrees of freedom, which become more and more important as the temperature is increased. Vector mesons can be incorporated using massive Yang--Mills approaches~\cite{Kaymakcalan:1984bz,Gomm:1984at,Meissner:1987ge} 
or within the hidden gauge formalism~\cite{Bando:1987br,Harada:2003jx}, which has proven particularly useful for studying medium modifications of vector resonances~\cite{Harada:2003jx}. Baryons can also be systematically included through baryon chiral perturbation theory~\cite{Gasser:1987rb,Bernard:1995dp}, allowing for a consistent description of pion-nucleon interactions and baryonic effects in a medium.

At higher temperatures, in particular near the chiral restoration region, the applicability of purely hadronic effective theories becomes increasingly limited. While chiral symmetry restoration can be addressed within chiral models by calculating the quark condensate (order parameter) and chiral susceptibilities, the emergence of deconfined degrees of freedom reduces their quantitative reliability. To overcome this limitation, several effective approaches incorporate both hadronic and quark degrees of freedom. For example, the quark-meson model (QMM)~\cite{Jungnickel:1995fp,Schaefer:2007pw} couples constituent quarks to chiral meson fields within a unified effective Lagrangian, allowing for a simultaneous description of chiral symmetry breaking and its restoration. The effective couplings in these models are typically fixed by vacuum phenomenology and lattice-QCD constraints. Adding more hadronic states has also become possible in this kind of model.

Alternatively, models formulated purely in terms of quark degrees of freedom, such as the Nambu--Jona--Lasinio (NJL) model~\cite{Nambu:1961tp,Klevansky:1992qe} and its extensions like the Polyakov-loop--NJL model (PNJL)~\cite{Fukushima:2003fw,Megias:2004hj,Ratti:2005jh}, can describe mesonic and baryonic excitations as collective modes generated through many-body techniques~\cite{Blanquier:2011zz, Torres-Rincon:2015rma}. Together, these approaches provide a complementary bridge between low-energy hadronic physics and the quark-gluon description relevant at higher temperatures.

In summary, these frameworks should be viewed as complementary rather than competing descriptions, each of them optimized for a specific temperature range:
\begin{itemize}
    \item \textbf{Low-temperature regime ($T < T_c$):} ChPT provides a model-independent, systematically improvable expansion rooted in the symmetry constraints of QCD, and its extensions to accommodate vector mesons (massive Yang--Mills and hidden gauge approaches) extend its reach at the cost of some model dependence. They offer the most reliable description for cold and dilute hadronic matter, but their applicability breaks down near $T_c$ due to the presence of resonance excitations and an eventual onset of a deconfined phase, which cannot be described within these frameworks.
    \item \textbf{Chiral restoration region ($T \sim T_c$):} Linear models (L$\sigma$M, $\text{O}(N)$, and QMM) explicitly include the scalar partner $\sigma$, making them well suited for investigating the order parameter, the transition temperature, and the mechanism of chiral symmetry restoration, at the cost of losing a systematic derivative expansion. In addition, the QMM (and its extensions) considers deconfined matter (quarks) as part of its degrees of freedom.
    \item \textbf{Transition to deconfined matter ($T \gtrsim T_c$):} Quark-based models (NJL, PNJL) bridge the hadronic phase and the deconfined regime by generating mesonic and baryonic excitations as collective modes. They can also contain certain information on gluodynamics when the Polyakov-loop potential is added. They also describe the phase transition region, but they lack real confinement and they are constrained to low-energy processes among quarks.
\end{itemize}
Together, this series of effective approaches provides a continuous conceptual bridge from low-energy hadronic physics to the description of the chiral transition.

In the following, we present some of the most relevant results obtained by several of these models at finite temperature.

\subsection{Light-meson thermal properties}~\label{sec:polescreening}

The properties of light mesons change due to their interaction with particles in the medium. At low temperatures, pions dominate the thermal bath, and more massive degrees of freedom and resonances progressively populate the medium as the temperature increases, according to their increasing excitation energy, or mass.

The mass of a propagating state is related to the pole of the excitation propagator. In vacuum, the propagator has the form
\be 
{\cal D}^{\text R} (p^2) \sim  \frac{1}{p^2-m_0^2-\Pi^{\text R}(p^2)} \ , 
\ee
where $m_0$ is the bare mass, $p^2=p_\mu p^\mu$, and the retarded self-energy arises from loop corrections to the propagator due to the interactions among particles. The mass is defined by the condition,
\be 
\left. p^2 - m_0^2 - \re \Pi^{\text R}(p^2) \right|_{p^2=m^2} =0 \ . \label{eq:mass} 
\ee
The divergent contributions to the self-energy are then absorbed in the bare mass through a convenient renormalization scheme to eventually produce a pole at a finite value of $p^2=m^2$, which is the dressed mass of the hadron.

At finite temperature, the self-energy contains thermal corrections, since in the ITF loops automatically generate Matsubara frequency sums. While the thermal mass is defined in an analogous way, one needs to take into account that the presence of a privileged reference frame (the one of the bath) makes the relation noncovariant. Then, the propagator will depend on $\omega=p^0$ and $\bm{p}$ separately,
\be 
{\cal D}^{\text R}(\omega, \bm{p};T) \sim  \frac{1}{\omega^2-\bm{p}^2-m^2-\Pi^{\text R}(\omega, \bm{p}; T)} \ , \label{eq:propagator} 
\ee
where in $\Pi^{\text R}(\omega,\bm{p};T)$ we have considered only the temperature-dependent terms in the retarded self-energy (the vacuum contributions having been absorbed into the renormalized vacuum mass $m$ at $T=0$).

Then, two standard definitions for the thermal mass can be introduced depending on how the zeros of the denominator in Eq.~\eqref{eq:propagator} are found. The first is the pole mass $\mpole$, which is defined in the limit $\bm{p} \rightarrow 0$ and is usually obtained from the retarded propagator after analytic continuation,
\be 
\left. \omega^2 - m^2 - \re \Pi^{\text R} (\omega+ \ii 0^+,\bm{p} \rightarrow 0 ;T) \right|_{\omega^2 = \mpole^2}= 0 \ . 
\ee
The mass $\mpole$ acts as the effective mass that controls the real-time quasiparticle excitations, like the thermal pions.

On the other hand, one can take the static limit $\omega_n = 0$ and define the screening mass $\msc$ directly from the Euclidean propagator,
\be 
\left. -\bm{p}^2 - m^2 - \re \Pi^{\text{E}}(0,\bm{p}; T) \right|_{\bm{p}^2 = -\msc^2 }= 0 \ ,\label{eq:screeningmass} 
\ee
where $\Pi^{\text E}(\omega_n, \bm{p};T)$ is the Euclidean self-energy (before analytic continuation). Physically, the screening mass $\msc$ characterizes the exponential decay of the spatial correlation function at large distances in the corresponding mesonic channel. It can be interpreted as the inverse of the screening length, $\lambda_\textrm{sc}=1/\msc$, that is, the distance over which spatial correlations in that channel are damped by the thermal bath.

\begin{figure}[!t] 
\centering
\includegraphics[scale=0.42]{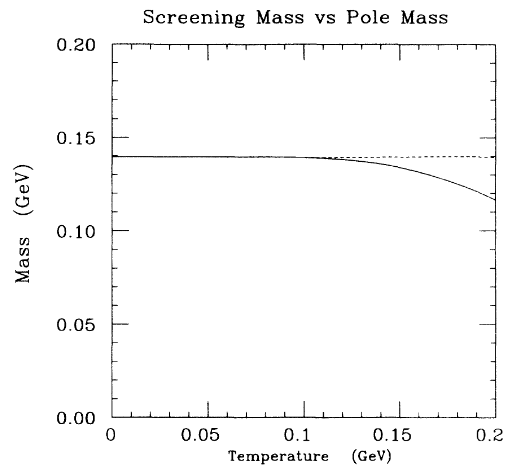}
\caption{Pion screening mass (dashed line) and pion pole mass (solid line) according to the calculation of Ref.~\cite{Song:1993ipa} within ChPT. Figure taken from this reference.}
\label{fig:Song_PolevsScreeningMass}
\end{figure}

The pole and screening masses must coincide at $T=0$, where Lorentz invariance is restored, but they increasingly differ as $T$ increases. As shown in Fig.~\ref{fig:Song_PolevsScreeningMass}, the ChPT calculations of Ref.~\cite{Song:1993ipa} reveal a relative difference of approximately 19\% at $T=200\mev$. Note that in this specific work, the screening mass is defined differently, namely as
\be 
\msc = \sqrt{ m^2+ \re \Pi^{\text E}(\omega_n=0, \bm{p} \rightarrow 0;T)} \ , 
\ee
which, compared to Eq.~\eqref{eq:screeningmass}, also evaluates the self-energy correction at $\bm{p} \rightarrow 0$ instead of at $\bm{p}^2 = - \msc^2$.

As discussed in Section~\ref{sec:eftslight}, at low energies the pion dynamics is governed by ChPT~\cite{Gasser:1983yg}, which in the massless $N_f=2$ case describes the low-energy processes of QCD in terms of three pions acting as Goldstone bosons. The coset construction of the spontaneous breaking of the chiral symmetry $\sutwo_\text{L} \times \sutwo_\text{R} \rightarrow  \sutwo_\text{V}$ generates an effective Lagrangian that can be expanded in powers of the pion momentum. It is also possible to add a finite mass, which explicitly breaks chiral symmetry, by treating it as a perturbation, typically counted as being of the same order as the momentum in the power counting. 

Soon after the introduction of ChPT, thermal effects were added in Ref.~\cite{Gasser:1986vb}, showing a reduction of the light condensate with temperature and a modification of the pion mass. The pion damping coefficient was calculated in Ref.~\cite{Goity:1989gs} using kinetic theory and based on the leading-order (LO) scattering rate of ChPT. An average collision time was estimated as $\tau \simeq 12 F_\pi^4/T^5$ where $F_\pi$ is the pion decay constant at LO and $T$ the temperature. In Ref.~\cite{Schenk:1991xe}, both real and imaginary parts of the pion self-energy were calculated using ChPT at low energy, and phenomenological information on pion phase shifts in the elastic region at higher energies. For the damping coefficient at moderate temperatures ($T=40$--$100\mev$), the inclusion of the phenomenological interaction leads to a correction of about 30\% with respect to LO ChPT. The pion mass decreases with temperature, as opposed to plain ChPT. Similar results were obtained in Refs.~\cite{Song:1993ipa,Toublan:1997rr}. The latter also computed the temporal and spatial pion decay constants in two-loop ChPT. Ref.~\cite{Song:1993ipa} computed the difference between the pole and the screening mass, as discussed above (cf. Fig.~\ref{fig:Song_PolevsScreeningMass}). In Ref.~\cite{Schenk:1993ru}, a one-loop calculation of the pion self-energy was performed in the dilute limit. The retarded pion self-energy reads,
\be 
\Pi^{\text{R}}(\bm{p};T) = -\int \frac{\diff^3q}{(2\pi)^3 2 \omega_q} n_B(\omega_q) \overline{T}_{\pi\pi} (s) \ , 
\ee
which contains the isospin-averaged forward scattering amplitude of the elastic process $\pi \pi \rightarrow \pi \pi$, $\overline{T}_{\pi\pi}$. $n_B(\omega_q)$ is the Bose--Einstein distribution function defined in Eq.~\eqref{eq:BE}. The real part of the self-energy gives a correction to the vacuum dispersion relation $\omega_p = \sqrt{\bm{p}^2+m_\pi^2(T=0)}$, which is obtained from the pole of the boson propagator Eq.~\eqref{eq:propagator},
\be 
\omega^2(\bm{p};T) - \omega_p^2 - \re \Pi^{\textrm{R}} (\bm{p};T)=0 \ , 
\ee
which yields
\be 
\omega(\bm{p};T) \simeq \omega_p +\frac{1}{2\omega_p} \re \Pi^{\textrm{R}}(\bm{p};T)=\omega_p - \frac{1}{2\omega_p} \int \frac{\diff^3q}{(2\pi)^3 2\omega_q} n_B(\omega_q) \re \overline{T}_{\pi\pi} (s) \ , \label{eq:disprel} 
\ee
while the damping rate (or half the thermal width $\gamma(\bm{p};T)= \Gamma(\bm{p};T)/2$) is proportional to the imaginary part of the retarded self-energy,
\be \gamma(\bm{p};T) \simeq -\frac{1}{2\omega_p} \im \Pi^{\text{R}}(\bm{p};T)=\frac{1}{2\omega_p} \int \frac{\diff^3q}{(2\pi)^3 2 \omega_q} n_B(\omega_q) \im \overline{T}_{\pi\pi} (s) \ . \label{eq:damping} \ee
% OURS  gamma  = Gamma/2  = - Im Pi/(2 omega)
% Schenk   ... = gamma/2  = - Im Pi/(2 omega)
% Smilga   xi  = gamma/2  = - Im Pi/(2 omega)
% Teaney       = gamma/2 = - Im Pi/(2 omega)
%
\begin{figure}[!t] 
\centering
\includegraphics[width=0.45\textwidth]{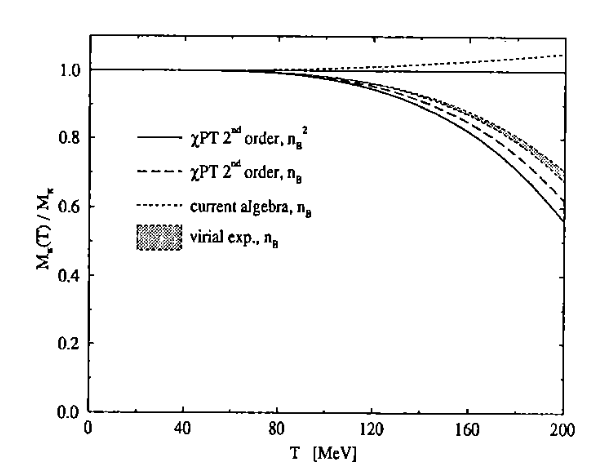}
\includegraphics[width=0.45\textwidth]{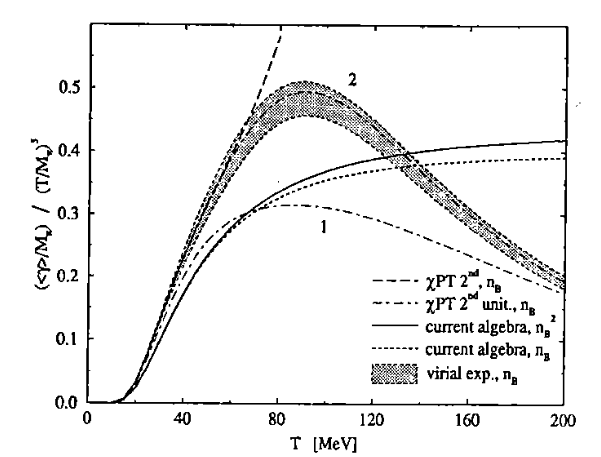}
\caption{Pion thermal mass (left panel) and thermal decay width (right panel) calculated from ChPT in Ref.~\cite{Schenk:1993ru}. See Ref.~\cite{Schenk:1993ru} for an explanation of the different approximations.}
\label{fig:Schenk_massandwidth}
\end{figure}

In Fig.~\ref{fig:Schenk_massandwidth} we reproduce the results of Ref.~\cite{Schenk:1993ru}. In the left panel, we present the thermal pion mass as a function of the temperature. The LO ChPT results in the dilute limit show an increase with $T$. The NLO expansion of ChPT shows a decrease, which is confirmed by the use of phenomenological interactions and the virial expansion. In the right panel, we observe the thermally averaged damping rate, which presents a nonmonotonic behavior when phenomenological amplitudes are used. 

In the chiral limit, the ChPT calculation presents additional problems. In Ref.~\cite{Smilga:1996cm} the damping rate for chiral pions in $\sutwo_f$ was calculated for soft pions $p \ll T$ from a two-loop self-energy diagram, giving
\be 
\gamma( \bm{p};T) \simeq \frac{p^2 T^3}{18\pi F_\pi^4} \log \left(\frac{T}{p} \right) \ ,  
\ee
up to logarithmic accuracy (in fact, the actual coefficient inside the logarithm was computed to be 1.56 in Ref.~\cite{Torres-Rincon:2022ssx}). Also in Ref.~\cite{Smilga:1996cm}, the infrared divergence $p\rightarrow 0$ was argued to be cut off by the pion damping itself (or the scattering rate), providing an estimate (up to logarithmic accuracy) of
\be 
\gamma(\bm{p};T) \simeq \frac{2p^2T^3}{9\pi F_\pi^4} \log \left(2.2 \frac{F_\pi}{T} \right) \ , 
\ee
which is now consistent with the hydrodynamic expectation ($\gamma(\bm{p}) \propto p^2$) in the infrared limit. The complete determination of the factor inside the logarithm requires extending the calculation to the kinetic and hydrodynamic limit. It was obtained much later in Ref.~\cite{Torres-Rincon:2022ssx} for the general $\text{SU}(N)_f$ case as
\be 
\gamma(\bm{p};T) = \frac{N^2 p^2 T^3}{18\pi F_\pi^4} \log \left(16.40 \frac{ F_\pi}{T} \right) \ . 
\ee

\begin{figure}[!t] 
\centering
\includegraphics[width=0.75\textwidth]{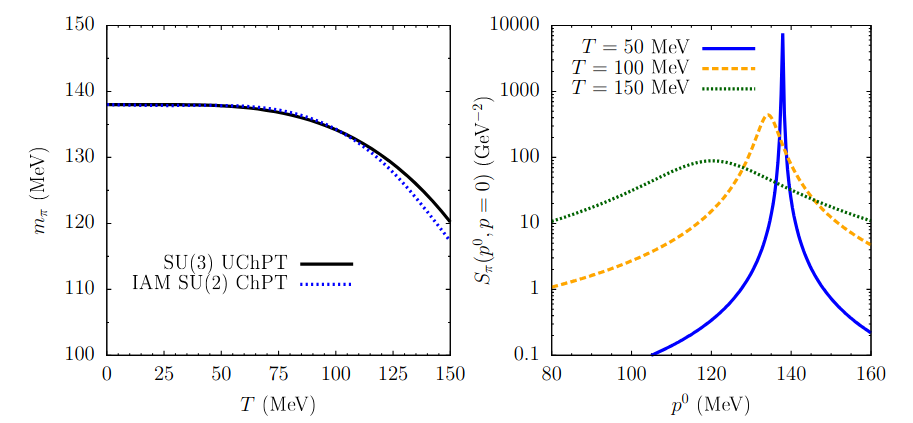}
\caption{Left panel: Pion thermal mass computed from ChPT with the Inverse Amplitude Method and 2 flavors~\cite{Fernandez-Fraile:2009axg} (dotted line) and from ChPT unitarized by the Bethe--Salpeter equation for the $T$ matrix with 3 flavors~\cite{Oller:1998hw} (solid line). Right panel: Pion spectral function for pions at rest using ChPT unitarized by the Bethe--Salpeter equation with 3 flavors. Figures taken from Ref.~\cite{Torres-Rincon:2021wyd}.}
\label{fig:pionmass_unit}
\end{figure}
One way to extend the perturbative amplitudes obtained from an EFT is to apply unitarization methods~\cite{Oller:2020guq}. Essentially, they consist in imposing exact unitarity on the scattering amplitudes (as opposed to a perturbative or approximate unitarity), though at the expense of introducing some model dependence. Various methods exist to unitarize partial-wave amplitudes, such as the Bethe--Salpeter approach for the $T$ matrix (to be described later in Sec.~\ref{sec:unitarizedEFT} in the context of heavy mesons), the $N/D$ method, the $K$-matrix approach, and the Inverse Amplitude Method (IAM). For more detailed information on unitarization methods, we refer the reader to Refs.~\cite{Oller:2020guq,Salas-Bernardez:2026gda}.

The use of unitarized amplitudes allows one to extend the validity of the EFT to higher energies and, therefore, temperatures. 
The reduction of the pion mass with temperature is common in this class of models. In the left panel of Fig.~\ref{fig:pionmass_unit} we present two such calculations. This figure is taken from Ref.~\cite{Torres-Rincon:2021wyd}.
In the left panel, the result labeled $\mathrm{SU}(3)$ UChPT corresponds to a determination of the thermal pion mass $m_\pi(T)=\omega(\bm{p}=0;T)$ employing Eq.~\eqref{eq:disprel} and
incorporating the interaction from $\suthree_f$ unitarized ChPT from Ref.~\cite{Oller:1998hw}, which obtains the two-body $T$-matrix amplitudes by solving the Bethe--Salpeter equation. The unitarization procedure is performed in coupled channels so that the $\pi \pi$ interaction is also affected by the coupled channel $K\bar{K}$ when the energy reaches this mass threshold. The result denoted as IAM $\sutwo_f$ ChPT is taken from Ref.~\cite{Fernandez-Fraile:2009axg}. It uses two light flavors (therefore, not coupled to the $K\bar{K}$ channel) and employs the IAM unitarization method~\cite{Dobado:1989qm,Dobado:1992ha,Dobado:1996ps}.

Both calculations show a decrease of the thermal mass with temperature. Although both unitarization methods introduce different model dependencies, the results agree well up to $T\simeq 110\mev$. Beyond this temperature, the effect of the $K\bar{K}$ channel and other systematic differences cause them to slightly diverge from each other. 

In the right panel of Fig.~\ref{fig:pionmass_unit}, the spectral function of a pion at rest ($\bm{p}=0$) is shown as a function of its energy and temperature. The calculation utilizes the interactions of $\mathrm{SU}(3)_f$ unitarized ChPT from Ref.~\cite{Oller:1998hw}, as well as the dispersion relation and damping coefficient calculated using Eqs.~\eqref{eq:disprel} and~\eqref{eq:damping}. Near the quasiparticle pole ($p^0$), the spectral function is computed as
\be
S_\pi(p^0,\bm{p})=\frac{1}{2\pi \omega (\bm{p})} \frac{\gamma(\bm{p})}{[p^0-\omega(\bm{p})]^2+[\gamma(\bm{p})]^2} \ . 
\ee
In the figure, the peak of the spectral function shifts to lower energies when the temperature increases. This is a reflection of the thermal mass behavior shown in the left panel. In addition, the width of the spectral function gets broader with temperature, signaling an increasing collisional broadening of the pion in the medium. This is a reflection of the increase of the coefficient $\gamma(\bm{p} =0)$ with temperature.

\begin{figure}[!t] 
\centering
\includegraphics[width=0.45\textwidth]{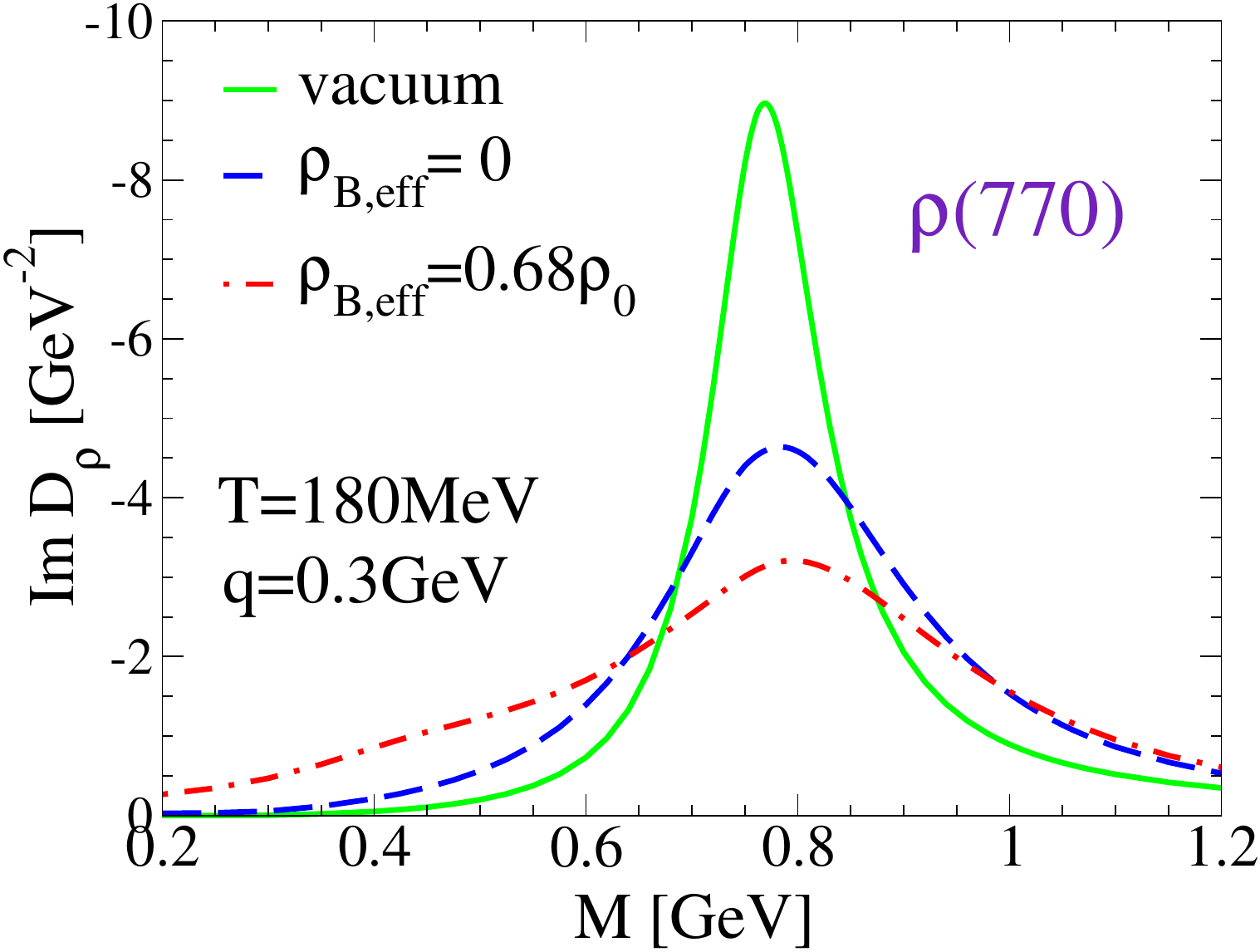}
\includegraphics[width=0.42\textwidth]{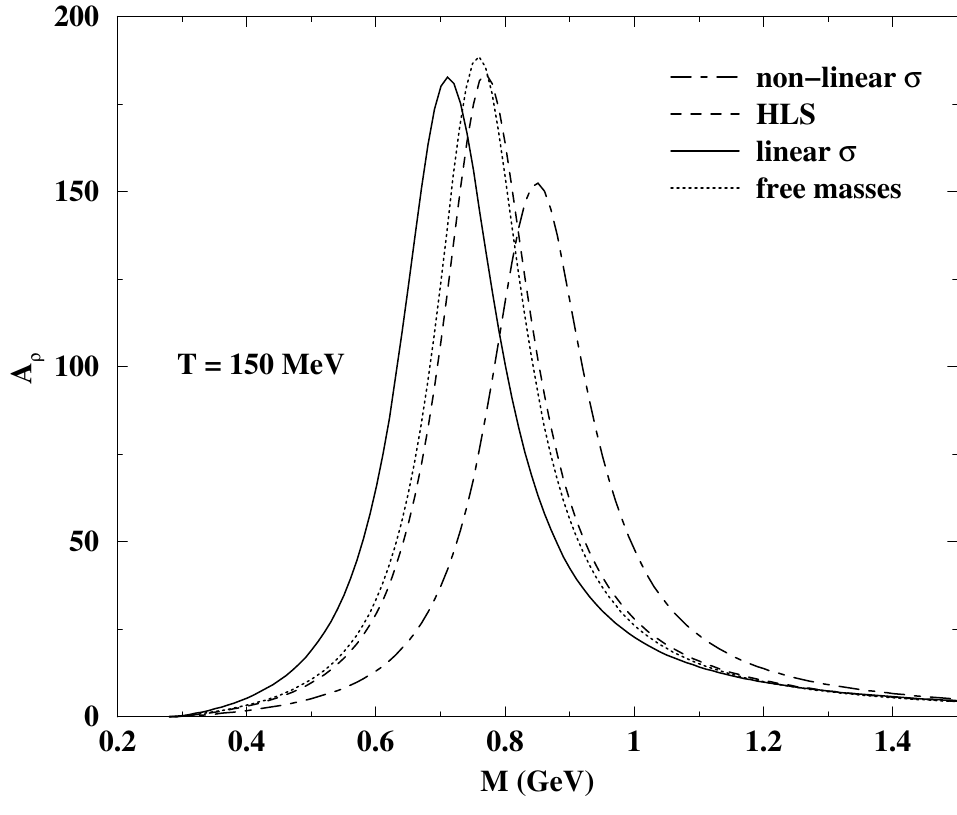}
\caption{Left panel: $\rho$-meson in-medium spectral function from the low-energy effective Lagrangian of Refs.~\cite{Rapp:2000pe,Rapp:2004zh}. Right panel: $\rho$-meson in-medium spectral function from different low-energy models (NL$\sigma$M, hidden-local symmetry model, and L$\sigma$M), from Ref.~\cite{Alam:1999sc}.}
\label{fig:rhoRappAlam}
\end{figure}

The $\rho$ meson is a vector meson that decays to $\pi\pi$. A model for the in-medium spectral function of this state was studied in Refs.~\cite{Rapp:1999us,Rapp:2000pe}. There, its connection to the electromagnetic spectral function makes it possible to use experimentally observed dilepton rates to test the medium modification and the effect of chiral symmetry restoration~\cite{Gale:1990pn}. In Ref.~\cite{Rapp:1999us} density effects were also included since, even at high collision energies, the total baryon density, as opposed to the net-baryon density, is nonnegligible. In the left panel of Fig.~\ref{fig:rhoRappAlam}, we reproduce the results from Ref.~\cite{Rapp:2004zh} (see also the original publication~\cite{Rapp:2000pe}) where the purely thermal modification to the spectral function of the $\rho$ meson can be seen at $T=180\mev$. No mass shift can be observed at this temperature, but the broadening of the state is evident. 

Different effective models at finite temperature have been reviewed in Ref.~\cite{Alam:1999sc}. In the right panel of Fig.~\ref{fig:rhoRappAlam}, we show the $\rho$-meson spectral function at $T=150\mev$ presented in this reference for the (gauged) L$\sigma$M, the NL$\sigma$M, and also the hidden-local symmetry formalisms. The latter presents the smallest shift in the mass with respect to the vacuum one ($\Delta m_\rho \simeq + 10\mev$), the L$\sigma$M presents a downward shift $\Delta m_\rho \simeq - 45\mev$, while the NL$\sigma$M has the largest shift $\Delta m_\rho \simeq + 90\mev$, which is accompanied by a larger broadening due to the increased phase space for the decay into two pions. A more recent calculation by the group is given in Ref.~\cite{Ghosh:2010hap}. 

The vector meson properties can also be accessed through quark-antiquark rescattering in quark models like the (P)NJL model. These will be discussed in the next section in the context of chiral symmetry restoration at finite temperature.

\begin{figure}[!ht] 
\centering
\includegraphics[width=0.75\textwidth]{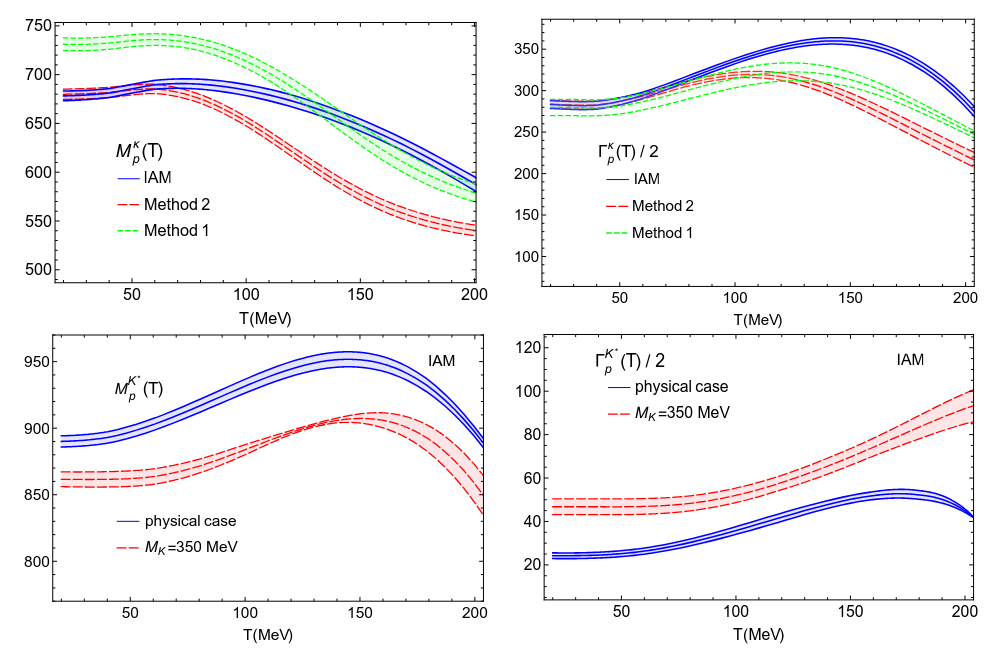}
\caption{Thermal masses (left panels) and half-widths (right panels) of the $K_0^*(700)/\kappa$ (top panels) and the vector $K^*(892)$ (bottom panels) using unitarized $\mathrm{SU}(3)_f$ ChPT. Figures taken from Ref.~\cite{GomezNicola:2023rqi}.}
\label{fig:kaoniam}
\end{figure}

In the nonzero strangeness sector, for $N_f=3$, recent calculations using unitarized ChPT exist~\cite{GomezNicola:2023rqi}. In that case, the poles of the scalar $K_0^*(700)$ (also known as the $\kappa$ meson) and the vector $K^*(892)$ were calculated using the IAM and compared with different methods of computing the kaon-pion loop function (see Ref.~\cite{GomezNicola:2023rqi} for details). The results are shown in Fig.~\ref{fig:kaoniam}.
The $K_0^*(700)$ is shown in the upper panels of Fig.~\ref{fig:kaoniam}. It reveals a nearly constant mass, with a mild increase until $T\simeq 80\mev$
and a subsequent decrease with temperature until the transition temperature. The half-width presents a smooth increase until $T=150\mev$ and then a decrease up to $T=200\mev$. One of the results used for comparison (``Method 1'') is that of Ref.~\cite{Gao:2019idb}, where the authors used unitarized ChPT to account for the masses of scalar mesons, namely $\sigma$, $f_0(980)$, $K_0^*(700)/\kappa$ and the $a_0(980)$. They found that the masses and widths of both the $\sigma$ and the $\kappa$ decrease moderately up to $T=200\mev$, while the masses and widths of the $f_0(980)$ and the $a_0(980)$ do not have much dependence on temperature. 
In the bottom panels we show the thermal modification of the mass and width of the vector $K^*(892)$. The mass increases with temperature until $T=150\mev$, at which point it decreases rapidly, while the decay width increases by a factor of 2 at a temperature close to $T_c$.

An alternative powerful technique that allows one to compute nonperturbative information from hadronic correlation functions is provided by QCD sum rules. In this approach, the analytic properties of the meson correlation function are exploited to connect its real and imaginary parts through a dispersion relation (up to subtraction terms)~\cite{Shifman:1978bx,Shifman:1978by}:
\begin{subequations}
    \begin{align} \label{eq:qcdSR}
    \Pi_\Gamma(q^2)&= \ii\int \diff ^4x\,e^{\ii q \cdot x}\langle T[\mathcal{O}_\Gamma(x)\mathcal{O}_\Gamma^\dagger(0)]\rangle\\ \label{eq:qcdSR-DR}
    &=\frac{1}{\pi}\int_0^\infty \diff s\,\frac{\im \Pi_\Gamma(s)}{s-q^2-\ii\epsilon} \ ,
    \end{align}
\end{subequations}
where $\mathcal{O}_\Gamma=\bar{q}\Gamma q$ is the interpolating operator of the meson, and the Dirac matrix $\Gamma$ selects the channel.
The real part of the correlator is evaluated in Euclidean space using the operator product expansion (OPE),
\begin{equation}
    \ii\int \diff^4x\,e^{\ii q\cdot x}\langle T[\mathcal{O}_\Gamma(x)\mathcal{O}_\Gamma^\dagger(0)]\rangle = C_I(q^2) \mathbb{I} +\sum_nC_n (q^2)\langle O_n \rangle_T  \ ,
\end{equation}
where the first term and the Wilson coefficients $C_n$ can be calculated perturbatively, and nonperturbative corrections are encoded in QCD condensates $\langle O_n \rangle_T$, which carry the thermal corrections.
The imaginary part is expressed as a meson spectral function, incorporating contributions from all physical states with appropriate quantum numbers, typically the ground state, excited states, and the continuum of scattering states. By matching the OPE representation to a suitable spectral function parametrization, one can extract spectral properties such as temperature-dependent masses, decay constants, and widths. In practice, a Borel transformation is often applied to improve the convergence of the OPE series and suppress contributions from higher excited states and the continuum. The QCD sum rules have also been extended to finite temperature~\cite{Bochkarev:1985ex}, where the correlator depends on $q_0$ and $|\bm{q}|$ separately, additional non-scalar condensates appear in the OPE, and the spectral side acquires a scattering (Landau-damping) contribution. For detailed reviews of QCD sum rules and their extension to finite temperature, see Refs.~\cite{Gubler:2018ctz,Ayala:2016vnt}.

\begin{figure}[!t] 
\centering
\includegraphics[width=0.45\textwidth]{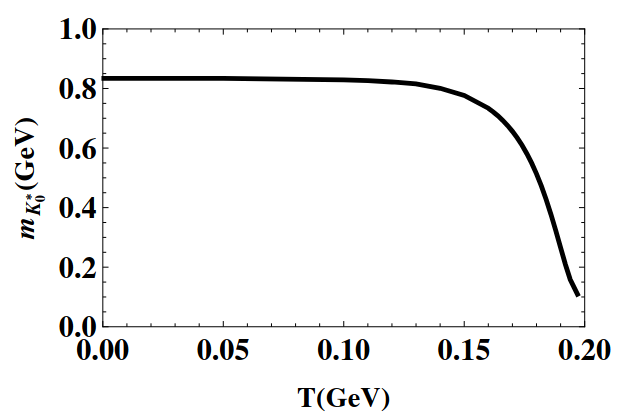}
\includegraphics[width=0.45\textwidth]{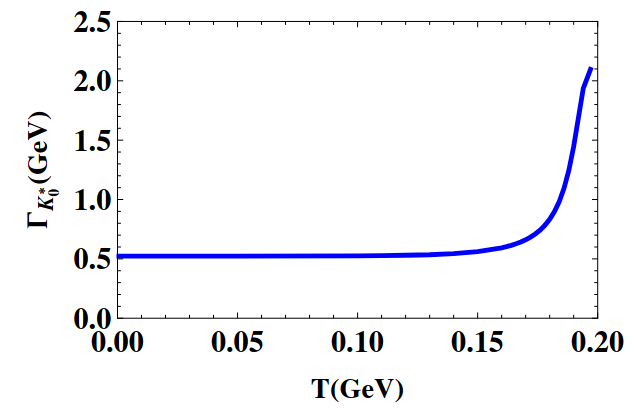}
\caption{Thermal masses (left panel) and decay widths (right panel) of the scalar $K_0^* (700)/\kappa$ meson from thermal QCD sum rules. Figures adapted from Ref.~\cite{Azizi:2019kzj}.}
\label{fig:azizikaon}
\end{figure}

The strange scalar meson $K^*_0(700)$ in the medium has been considered in Ref.~\cite{Azizi:2019kzj} using thermal QCD sum rules. The results for its mass and decay width are shown in Fig.~\ref{fig:azizikaon}, where a flat mass and width can be observed up to $T \simeq 150\mev$. Then, a sudden decrease (increase) of the mass (width) occurs at higher temperatures.
Light-meson properties have also been studied in other models, like the L$\sigma$M, the quark meson model and the (P)NJL model. The results within these models will be discussed in Section~\ref{sec:chiralrestoration} in the context of chiral symmetry restoration.

\begin{figure}[!t] 
\centering
\includegraphics[width=0.5\textwidth]{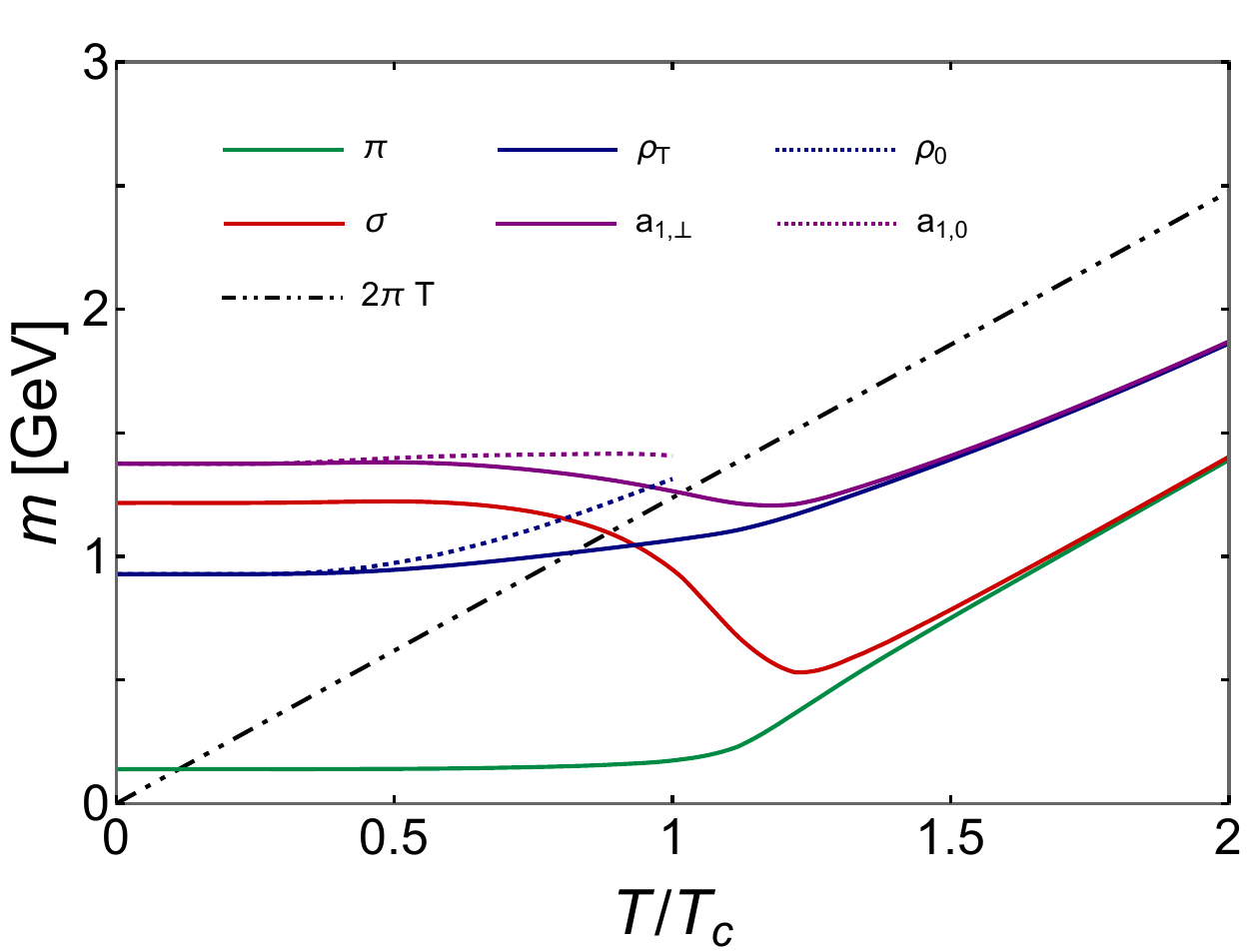}
\caption{Meson screening masses computed within the vector $\times$ vector contact interaction model of Ref.~\cite{Chen:2024emt}.}
\label{fig:BS4light}
\end{figure}

Another approach is based on the Dyson--Schwinger equations, which were used to determine the masses of spin-zero mesons (scalar and pseudoscalar) using a symmetry-preserving approach of a vector $\times$ vector contact interaction model~\cite{Ramirez-Garrido:2025rsu}. The light, heavy-light, and heavy-heavy meson masses were calculated up to $T=500\mev$. In general, the results show that the pseudoscalar meson masses have an almost constant behavior up to $T_c$ and then an increasing trend with temperature. Moreover, the lighter the meson, the steeper the increase. For scalar mesons, a decreasing trend at low temperature manifests up to a temperature above $T_c$, and then a similar increase occurs. The difference may be related to the repulsion of the spin-orbit term in the scalar sector. Within a similar model, screening masses of mesons with quantum numbers $0^\pm$ and $1^\pm$ were computed in Ref.~\cite{Chen:2024emt}. They used the Bethe--Salpeter equation for the quark-antiquark scattering at finite temperature. In Fig.~\ref{fig:BS4light} we reproduce the results from this work in the strangeness $S=0$ sector, showing the screening masses for scalar, pseudoscalar, vector, and axial-vector states, up to $T=2 T_c$ (with $T_c=197\mev$ in this model). One observes a monotonic increase in the pseudoscalar ($\pi$) and vector ($\rho_{\text{T}}$) cases, and an initial decrease at low temperatures for the scalar ($\sigma$) and axial-vector cases ($a_{1,\perp}$). The dashed-dotted-dotted line marks the asymptotic noninteracting limit $2\pi T$.

From the different results one observes that the pion mass presents different temperature dependencies (increase, decrease, or nearly constant). Rather than signaling different realizations of the chiral symmetry (since all of them treat the pion as the Goldstone boson of its spontaneous breaking), the distinctions stem from the different dynamical mechanisms and from the temperature range in which each approach is reliable.
In hadronic EFTs, the mass shift is driven by the thermal average of the forward $\pi \pi$ amplitude. At leading order, ChPT gives a slight increase, but as the temperature increases and the typical energies probe the resonant $\sigma$ and $\rho$ interactions, the real part of the self-energy produces a decrease in the pion mass. This is clearly seen in the unitarized versions of the chiral interactions, extending the validity up to $T \simeq 120$--$140\mev$. Beyond that range, the dilute approximation, the omission of heavier states and baryons, and the lack of the crossover limit their reliability.
In quark-based approaches, such as QCD sum rules, Dyson--Schwinger and NJL-type models (to be seen later), the pion mass is dominated by the melting of the quark condensate. The pion mass stays nearly flat up to the vicinity of $T_c$, since the decrease of the chiral condensate mostly translates into a decrease of the pion decay constant. Then, around $T_c$ the pion mass increases when approaching the free quark-antiquark limit. Part of the differences is also a matter of definition: pole and screening masses separate with $T$ (about 19\% at $T = 200\mev$ in Fig.~\ref{fig:Song_PolevsScreeningMass}), with the latter growing toward $2 \pi T$ at high temperature. Comparisons among frameworks must therefore use consistent mass definitions. 

Taking the light-meson results presented in this section together, some conclusions appear robust across the different theoretical approaches, while others remain model dependent. Among the robust features, all hadronic approaches agree that, within their range of applicability ($T \lesssim 120$--$140\mev$), the pion and the $\rho$ meson remain well-defined quasiparticles that acquire thermal widths generally increasing with temperature, as collisions with the particles of the pion gas become more frequent (see Figs.~\ref{fig:Schenk_massandwidth} and~\ref{fig:pionmass_unit}). In the same temperature range, the thermal mass shift of the pion remains moderate, as expected from its Goldstone-boson nature, and different unitarization methods agree on its decreasing trend up to $T\simeq 110\mev$. For the $\rho$ meson, the dominant medium effect is a strong broadening of its spectral function, a conclusion that is also supported by experimental dilepton data (see Section~\ref{sec:app-dileptons}). 
On the other hand, several results remain model dependent. The sign and magnitude of the $\rho$-meson mass shift depend on the effective Lagrangian, ranging from $-45\mev$ to $+90\mev$ at $T=150\mev$ among the models of Ref.~\cite{Alam:1999sc}. The thermal evolution of the broad scalar states, such as the $\sigma$ and the $K_0^*(700)$, depends on the unitarization method and on the treatment of the loop functions~\cite{Gao:2019idb,GomezNicola:2023rqi}, and differs from the QCD sum-rule result~\cite{Azizi:2019kzj}, in which the mass and width remain flat up to $T\simeq 150\mev$ and change abruptly beyond. More generally, hadronic approaches lose their validity for $T\gtrsim 150\mev$, and the properties of light mesons close to and above the crossover can only be addressed with models that incorporate the chiral transition and with lattice-QCD calculations, which are discussed in Sections~\ref{sec:chiralrestoration} and~\ref{sec:lqcdlight}, respectively.

\subsection{Chiral symmetry breaking and its thermal restoration}~\label{sec:chiralrestoration}

In the limit of vanishing quark masses, chiral symmetry of the QCD Lagrangian is one of the main guiding principles for describing the low-energy hadronic physics. The QCD vacuum spontaneously breaks the chiral symmetry according to the pattern,
\[ \text{SU}(N_f)_\text{L} \times \text{SU}(N_f)_\text{R} \rightarrow \text{SU}(N_f)_\text{V}\ , \]
for $N_f$ massless flavors. The $N_f^2-1$ Goldstone bosons emerging from this breaking correspond to a massless multiplet of pseudoscalar mesons.

A finite quark mass already breaks the chiral symmetry explicitly, but if this mass is still small, it can be used as a perturbative parameter, and it can be incorporated into the symmetry-breaking scheme. In this case, the pions become massive and are therefore known as pseudo-Goldstone bosons. If each flavor carries a different quark mass, the remaining $\text{SU}(N_f)_\text{V}$ flavor symmetry is also explicitly broken, as happens in real QCD, leading to the light meson spectrum. The vast topic of chiral symmetry has been reviewed in many previous works~\cite{Pagels:1974se,Coleman:1985rnk,Hatsuda:1994pi,Leutwyler:1994fi,Bernard:1995dp,Alkofer:1995mv,Nowak:1996aj,Koch:1997ei,Cassing:1999es,Brown:2001nh,Shuryak:2004pry}.

Beyond the restoration of the $\text{SU}(N_f)_\text{L} \times \text{SU}(N_f)_\text{R}$ chiral symmetry, the fate of the $\text{U}(1)_\text{A}$ axial symmetry at finite temperature remains a subject of intense investigation. In vacuum, this symmetry is broken by the Adler--Bell--Jackiw anomaly~\cite{Adler:1969gk,Bell:1969ts}, which explains the large mass of the $\eta'$ meson compared to those of the Goldstone octet. As the temperature increases, the topological susceptibility of the QCD vacuum is expected to decrease~\cite{Gavai:2024mcj}, potentially leading to a partial restoration of the $\text{U}(1)_\text{A}$ symmetry. The degree to which this occurs near the chiral transition temperature $T_c$ has profound implications for the meson spectrum; specifically, a restored $\text{U}(1)_\text{A}$ symmetry would lead to the degeneracy of partners that are otherwise split by the anomaly, such as the isovector $(\pi, a_0)$ and isoscalar $(\sigma, \eta)$ pairs. Lattice-QCD analyses and effective models suggest that while $\text{U}(1)_\text{A}$ breaking effects persist at $T_c$~\cite{Buchoff:2013nra}, they are significantly suppressed, leading to a decrease of the $\eta'$ mass and a narrowing of the mass splitting between parity partners~\cite{Buchoff:2013nra,Aoki:2025mue}. In the context of ChPT at low temperatures, this sector has been studied, e.g., in Refs.~\cite{GomezNicola:2019myi,GomezNicola:2020qxo}.

\begin{figure}[!t] 
\centering
\includegraphics[width=0.4\textwidth]{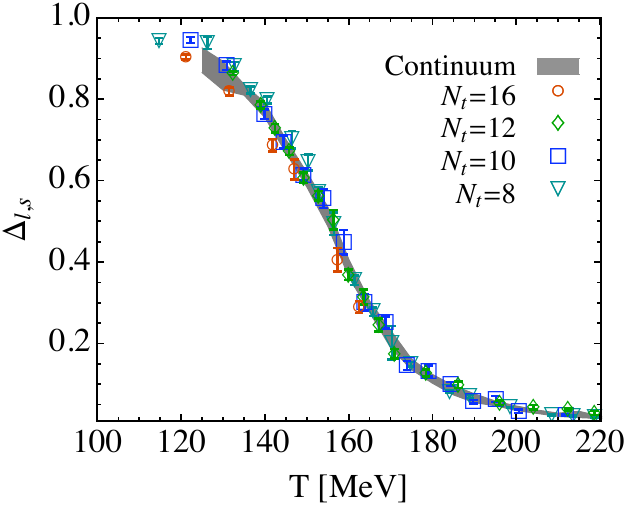}
\caption{Subtracted chiral condensate Eq.~\eqref{eq:condensate} as a function of the temperature, as obtained in the lattice-QCD calculation of Ref.~\cite{Borsanyi:2010bp}.}
\label{fig:crossover}
\end{figure}

Increasing temperature or density can lead to partial restoration of chiral symmetry, potentially through a phase transition. The chiral phase transition separates the Nambu--Goldstone (or broken) phase from the Wigner--Weyl (or symmetric) phase. An order parameter for the restoration of chiral symmetry can be defined through the quark condensate $\langle \bar{q} q \rangle$, which takes a finite value in the Nambu--Goldstone phase and vanishes in the Wigner--Weyl phase in the chiral limit. In real QCD, with physical quark masses, the chiral transition at finite temperature and zero baryochemical potential is known to be a crossover~\cite{Aoki:2006we,Borsanyi:2010bp}. In Fig.~\ref{fig:crossover}, we reproduce the order parameter defined in Ref.~\cite{Borsanyi:2010bp} as a function of the temperature. The chosen order parameter is the subtracted chiral condensate $\Delta_{l,s}$, which combines the light and strange quark condensates,
\be 
\Delta_{l,s} \equiv \frac{ \langle \bar{\psi} \psi \rangle_{l} (T) - \frac{m_l}{m_s} \langle \bar{\psi} \psi \rangle_{s} (T)}{\langle \bar{\psi} \psi \rangle_{l} (0) - \frac{m_l}{m_s} \langle \bar{\psi} \psi \rangle_{s} (0)} \ . \label{eq:condensate}
\ee
The results are shown for several temporal lattice extents and also in the continuum limit as a band.

The pseudocritical temperature of the chiral crossover, which we denote as $T_c$ throughout this review, has been determined in lattice QCD to be $T_c=(158.0 \pm 0.6)\mev$ in Ref.~\cite{Borsanyi:2020fev}
and $T_c=(156.5 \pm 1.5)\mev$ by the HotQCD collaboration~\cite{HotQCD:2018pds}. At finite baryochemical potential, the nature of the transition remains a crossover up to the highest densities currently accessible by lattice-QCD calculations, $\mu_B/T<3.5$~\cite{Borsanyi:2021sxv,Borsanyi:2022qlh}. At higher density, the existence of a first-order transition and a critical endpoint is not precluded, and some recent theoretical calculations applied to QCD suggest that a critical point might exist around $T \sim 100$--$115\mev$ and $\mu_B \sim 430$--$650\mev$~\cite{Fu:2019hdw,Gao:2020fbl,Gunkel:2021oya,Basar:2023nkp,Clarke:2024ugt}. In the chiral limit for light quarks (with a physical strange quark mass), QCD presents a phase transition at $\mu_B=0$ and $T=132_{-6}^{+3} \mev$~\cite{HotQCD:2019xnw}. Should the critical endpoint exist, its temperature is expected to be lower than the chiral transition temperature~\cite{HotQCD:2019xnw,MUSES:2023hyz}. When discussing specific models or lattice-QCD setups, we also use $T_c$ for the (pseudo)critical temperature of the corresponding calculation, which may differ from the physical value.

According to the thermal sum rules~\cite{Kapusta:1993hq}, one of the scenarios for the chiral symmetry restoration is a degeneracy between chiral partners. This idea has been realized in a plethora of effective models for hadrons applied to different flavor sectors. We review here several of these results according to some of the models presented before.

Following the classification in Ref.~\cite{Torres-Rincon:2021wyd}, we comment on different models of chiral symmetry restoration according to the nature of the chiral partners in the effective Lagrangian, namely,
\begin{enumerate}
    \item The chiral partners are fundamental degrees of freedom in the effective Lagrangian. The prototypical example is the $\pi$--$\sigma$ doublet in the L$\sigma$M and its extensions.
    \item The chiral partners are dynamically generated states out of more fundamental degrees of freedom (quarks) via few-body equations. For example, the $\pi$--$\sigma$ or the $\rho$--$a_1$ partners in the NJL or PNJL models, and their extensions.
    \item One partner is a fundamental degree of freedom, while the other is dynamically generated. This is the case in standard ChPT, where pions and kaons are active degrees of freedom, while the chiral partners, the $\sigma$ and the $\kappa$, can be dynamically generated from unitarized pion-pion or pion-kaon two-body interactions, as we have already addressed.
    \item The chiral partners comprise three different states. This appears to be the case in the open charm (and beauty) sector, where the chiral partner of the pseudoscalar $D$ meson is a scalar $D_0^*(2300)$ resonance, which is known to have a double-pole structure~\cite{Meissner:2020khl}, generated from the coupled-channel dynamics of $D\pi$, $D\eta$, and $D_s\bar{K}$.
\end{enumerate}

We will comment on the first two cases, which are rather well studied in the literature of thermal effective theories. The mixed case, represented by ChPT, does not have such a broad literature except for a few studies of the $\sigma/f_0(500)$ behavior at finite temperature by the Madrid group~\cite{Dobado:2002xf,GomezNicola:2002an,Ferreres-Sole:2018djq} (thermal pions in ChPT have a much broader literature as discussed above). This case is discussed in the review of Ref.~\cite{Torres-Rincon:2021wyd}. In addition, the last case concerning three states is also studied within EFTs with dynamically generated states by the Barcelona group~\cite{Montana:2020lfi,Montana:2020vjg}. Since it is a more exceptional case, we do not discuss it here in the context of chiral symmetry restoration but refer the reader to the review of Ref.~\cite{Torres-Rincon:2021wyd}. The thermal properties of these heavy states will be discussed in Section~\ref{sec:openHF}.

We start with the first scenario in which the chiral partners are part of the fundamental degrees of freedom of the model. The simplest case is the standard L$\sigma$M~\cite{Baym:1977qb}, in which both pions and the isoscalar $\sigma$ state are part of an $\text{O}(4)$ multiplet, and after spontaneous symmetry breaking to $\text{O}(3)$ the vacuum expectation value of the field is chosen to be in the direction of the $\sigma$ mode, making it massive, while the pions remain massless (the Goldstone bosons). Extended to the $\text{O}(N+1) \rightarrow \text{O}(N)$ symmetry-breaking pattern, the effective Euclidean Lagrangian reads
\be {\cal L}_{\text E} = \frac12 \partial_\mu \Phi_i \partial^\mu \Phi_i - \bar{\mu}^2 \Phi_i \Phi_i + \frac{\lambda}{N} (\Phi_i \Phi_i)^2 - \epsilon \Phi_{N+1} \ , \ee
where $\Phi_i = (\pi_a,\sigma)$ is a multiplet of $N+1$ scalar fields ($a= 1,...,N$, $i=1,...,N+1$). The quartic coupling $\lambda$ is positive, and $\bar{\mu}^2$ is a positive parameter that forces the spontaneous symmetry breaking in vacuum. The last term explicitly breaks the chiral symmetry through the small parameter $\epsilon$.

At finite temperature, both the $\pi$ and $\sigma$ receive thermal corrections which can be computed from the meson Green function within the ITF (see Ref.~\cite{Torres-Rincon:2021wyd} and references therein). 

\begin{figure}[!t] 
\centering
\includegraphics[width=0.38\linewidth]{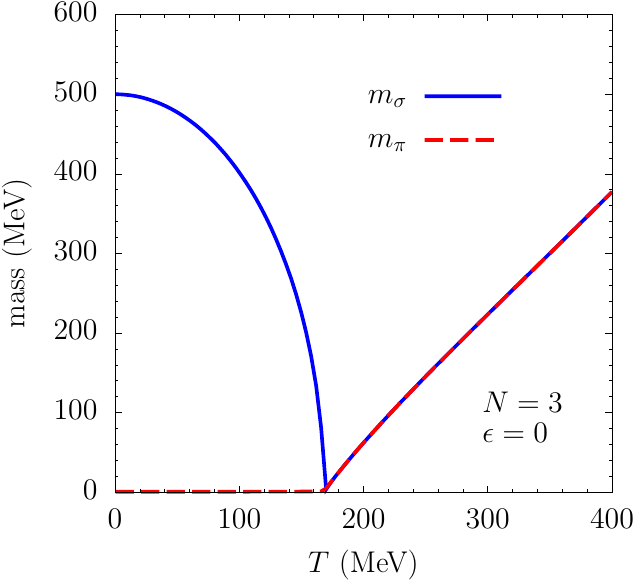}
\includegraphics[width=0.38\linewidth]{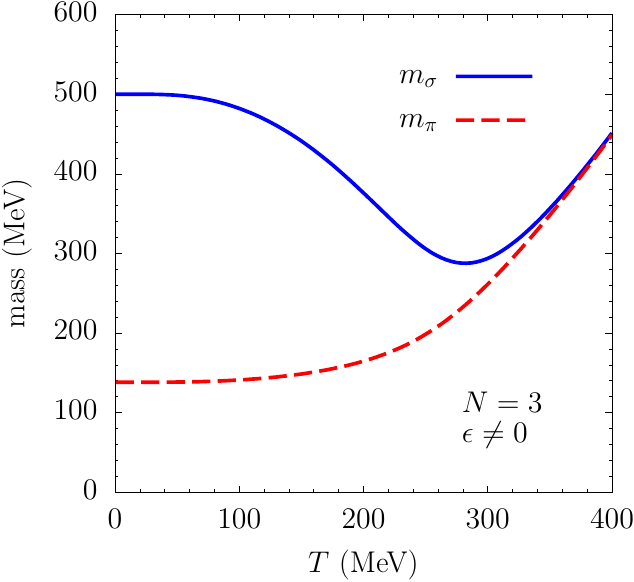}
\caption{Pion and $\sigma$ thermal masses in the $\text{O}(N+1)$ model for $N=3$ as shown in Ref.~\cite{Torres-Rincon:2021wyd}. We plot the case without an explicit symmetry-breaking term $\epsilon=0$ (left panel), and the case with an explicit symmetry-breaking term $\epsilon \neq 0$ (right panel).}
\label{fig:LSM}
\end{figure}

In Fig.~\ref{fig:LSM} we present the results of the $\pi$ and $\sigma$ meson masses as functions of temperature, below and above the chiral transition temperature, from Ref.~\cite{Torres-Rincon:2021wyd}. In the left panel, a value $\epsilon=0$ has been used, removing the explicit chiral-symmetry-breaking term. Therefore, the pions remain massless below the chiral transition temperature and become degenerate with the scalar partner above it. In the right panel, a nonzero value of $\epsilon$ is chosen to match the physical vacuum pion mass. Therefore, the pions are realized as pseudo-Goldstone bosons and become degenerate with the $\sigma$ mode above the chiral transition. In both cases, the vacuum mass of the scalar mode has been chosen to be $m_\sigma(T=0)=500\mev$. 

\begin{figure}[!t] 
\centering
\includegraphics[width=0.44\linewidth]{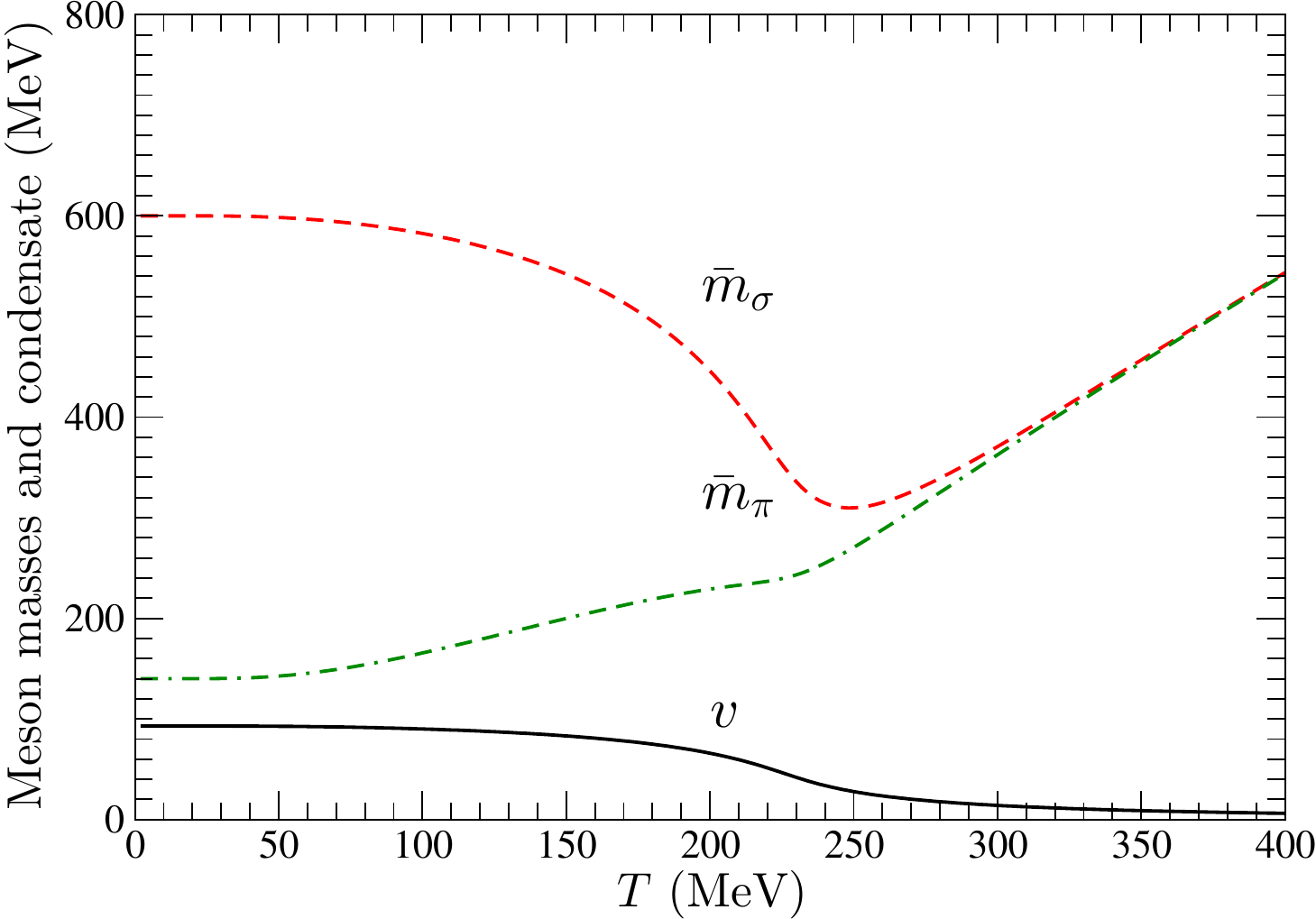}
\includegraphics[width=0.36\linewidth]{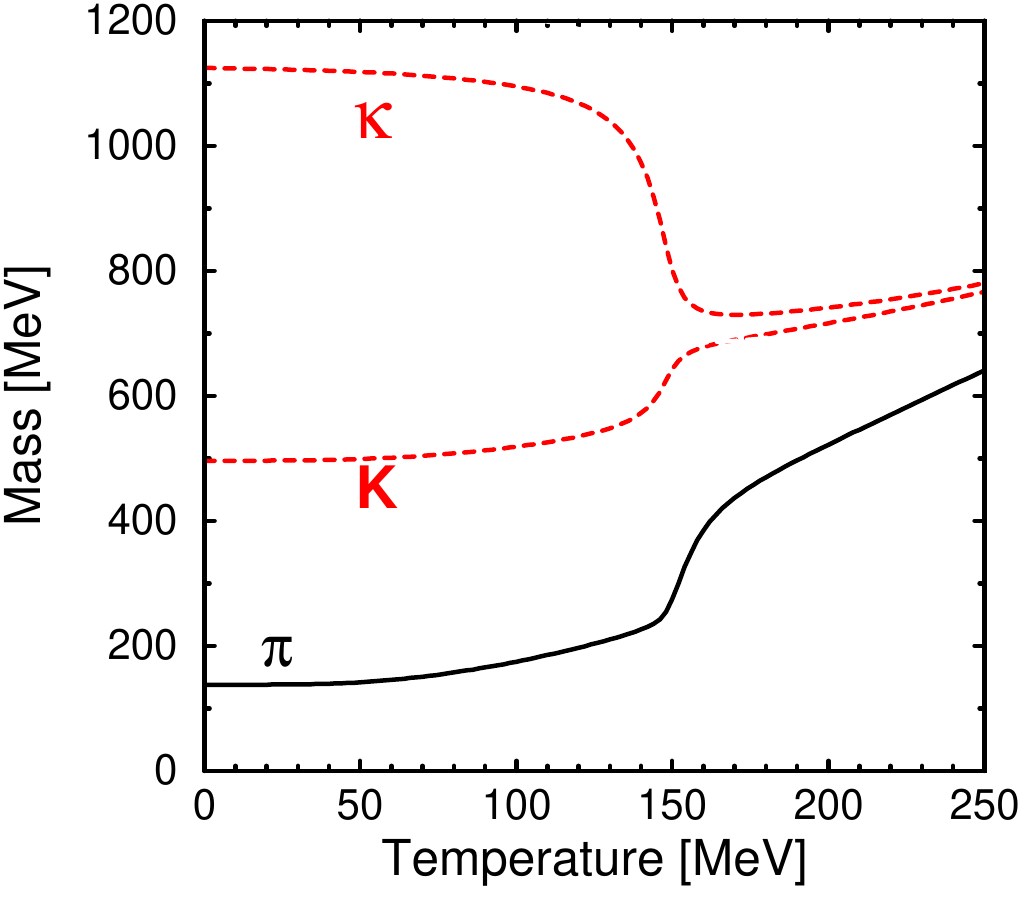}
\caption{Left panel: Pion and $\sigma$ thermal masses from the L$\sigma$M of Ref.~\cite{Chakraborty:2010fr} showing the parity partner degeneracy at high temperatures. Right panel: Kaon and $\kappa$ thermal masses in the 3-flavor L$\sigma$M of Ref.~\cite{Schaffner-Bielich:1999cux}, also presenting a degeneracy above the chiral transition.}
\label{fig:LSM2}
\end{figure}

In Fig.~\ref{fig:LSM2} we reproduce previous results in the L$\sigma$M for 2 and 3 flavors. In the left panel, we show the results of Ref.~\cite{Chakraborty:2010fr} where the masses present a gap in vacuum (with a vacuum $\sigma$ mass fixed to $600\mev$), while at high temperature the two become degenerate. Similar results appear in Refs.~\cite{Petropoulos:2004bt, Dobado:2012zf} under different approximations of the same model, and in Ref.~\cite{Scavenius:2000qd} where the original $\text{O}(4)$ L$\sigma$M is supplemented with a quark sector (QMM) to account for a finite chemical potential together with finite temperature. For three flavors, one can consult Ref.~\cite{Lenaghan:2000ey} for the complete pseudoscalar-scalar multiplets with and without the $\text{U}(1)_\text{A}$ anomaly. In the right panel of Fig.~\ref{fig:LSM2}, we show the results for the pseudoscalar $K$ and scalar $\kappa$ masses as functions of temperature within the $N_f=3$ L$\sigma$M of Ref.~\cite{Schaffner-Bielich:1999cux}. Therefore, a similar degeneracy pattern is present in the nonzero strangeness sector as well.

The degeneracy between chiral partners can also be obtained in models where the hadrons are composite states, like in the NJL~\cite{Nambu:1961tp,Klimt:1989pm,Vogl:1989ea,Vogl:1991qt,Klevansky:1992qe,Hatsuda:1994pi,Alkofer:1995mv, Buballa:2003qv} or the PNJL~\cite{Fukushima:2003fw,Megias:2004hj,Ratti:2005jh,Hansen:2006ee,Fukushima:2008wg,Torres-Rincon:2015rma} models. The fundamental degrees of freedom are quarks interacting at low energies, where the gluons have been integrated out. The fundamental interactions are contact terms among quarks---while other nonlocal versions also exist~\cite{Bowler:1994ir,Plant:1997jr,Hell:2008cc}---plus the expectation value of the Polyakov loop in the PNJL version (there exist some caveats about using the Polyakov loop as a classical variable in this kind of chiral model at finite temperature~\cite{Megias:2004hj}).

The Lagrangian of the model can be obtained from the color-color vector QCD interaction~\cite{Buballa:2003qv} upon Fierz transformations of the quark fields. For three flavors, a standard version of the PNJL Lagrangian is,
\begin{align} 
{\cal L}_{\textrm{PNJL}} &= \sum\limits_i \bar{\psi}_i (\ii \slashed{D}-m_{0i}) \psi_i  \nn\\
&+ G \sum\limits_{a} \sum\limits_{ijkl} \left[ (\bar{\psi}_i \ \ii \gamma_5 \tau^{a}_{ij} \psi_j) \ 
(\bar{\psi}_k \ \ii \gamma_5 \tau^{a}_{kl} \psi_l)
+ (\bar{\psi}_i \tau^{a}_{ij} \psi_j) \ 
(\bar{\psi}_k  \tau^{a}_{kl} \psi_l) \right] \nn \\
&+ G_V \sum\limits_{a} \sum\limits_{ijkl} \left[ (\bar{\psi}_i \ \gamma_5 \gamma_\mu \tau^{a}_{ij} \psi_j) \ 
(\bar{\psi}_k \  \gamma_5 \gamma^\mu \tau^{a}_{kl} \psi_l)
+ (\bar{\psi}_i \tau^{a}_{ij} \gamma_\mu  \psi_j) \ 
(\bar{\psi}_k  \tau^{a}_{kl} \gamma^\mu \psi_l) \right]  \nn\\
% & -    H \det_{ij} \left[ \bar{\psi}_i \ ( \mathbb{I} - \gamma_5 ) \psi_j \right] - H \det_{ij} \left[ \bar{\psi}_i \ ( \mathbb{I} + \gamma_5 ) \psi_j \right]  \nn \\ 
&- {\cal U} (T;\Phi,\bar{\Phi})\ , \label{eq:lagPNJL} 
\end{align}
where we include explicit interactions in the (pseudo)scalar and (axial-)vector channels. The indices $i,j,k,l$ represent the quark flavors (up, down and strange), and $a=0,1,...,8$, where $\tau^{1,...,8}$ are the Gell-Mann matrices and $\tau^0=\sqrt{2/3} \ \mathbb{I}$. The first line corresponds to the kinetic term, including bare quark masses and the coupling with the temporal gauge field ($D^\mu=\partial^\mu-\ii\delta^{\mu 0}A^0$). In the second line, the coupling $G$ controls the pseudoscalar and scalar channels. This term can be used to generate mesons in these sectors.
In the third line, $G_V$ controls the vector and axial-vector couplings, generating the corresponding mesons. In the final line,  ${\cal U}$ is the Polyakov-loop effective potential, parametrized in terms of the Polyakov loop $\Phi$ and the temperature. The particular choice of parameters can be found in Ref.~\cite{Torres-Rincon:2015rma}. For simplicity, here we suppress the 6-point 't Hooft interaction.

In Eq.~\eqref{eq:lagPNJL}, the bare quark masses $m_{0i}$ can be set to zero if the exact chiral limit needs to be achieved. Then, chiral symmetry becomes spontaneously broken in the vacuum due to the formation of a quark condensate, leading to the emergence of composite Goldstone modes. If $m_{0i} \neq 0$, chiral symmetry is explicitly broken, and after spontaneous symmetry breaking, the physical masses of the resulting pseudo-Goldstone bosons can be obtained.

The generation of hadron masses in this model proceeds in several steps.
First, quarks of flavor $i$ propagating in a medium acquire masses $m_i(T)$ due to interactions. At the mean-field level, the dressed quark mass is~\cite{Buballa:2003qv,Hansen:2006ee,Torres-Rincon:2015rma}
\be \label{eq:gap}
m_i (T) = m_{0i} - 4 G \langle \bar{\psi}_i \psi_i \rangle  \ ,
\ee
where the thermal quark condensate is given by
\be
\langle \bar{\psi}_i \psi_i \rangle = \textrm{Tr} \sumint_q \frac{1}{\slashed{q}-m_i} \ ,
\ee
where the trace is taken in color and Dirac spaces, and $\sumint_q$ is defined in Eq.~\eqref{eq:sumint}.

\begin{figure}[!ht] 
\centering
\includegraphics[scale=1.05]{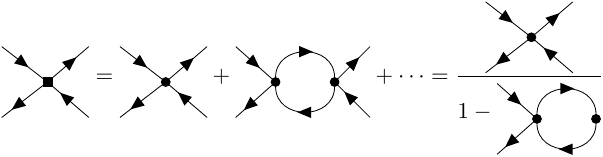}
\caption{Diagrammatic representation of the $T$-matrix Eqs.~\eqref{eq:BSPNJL} and \eqref{eq:meson} in the meson sector of the (P)NJL model.}
\label{fig:NJL_mesondiag}
\end{figure}

Mesonic excitations can be generated dynamically by solving the Bethe–Salpeter equation for the $q\bar{q}$ scattering. Consider the scattering process of a quark–antiquark pair $i+\bar{j} \to k+\bar{l}$. In the random-phase approximation and within the ITF, the $T$-matrix elements satisfy~\cite{Vogl:1991qt,Klevansky:1992qe,Torres-Rincon:2015rma}
\be 
T^{ab}_{i \bar{j},k \bar{l}} (\ii \nu_m,{\bm p};T) = {\cal K}^{ab}_{i \bar{j},k \bar{l}} - \sumint_q
{\cal K}^{ac}_{i \bar{j}, r \bar{s}} \  \Delta_{r} \left( \ii \omega_n, {\bm q};T \right) \ \Delta_{ \bar{s}} \left(\ii \omega_n - \ii \nu_m, {\bm q}-{\bm p} ;T\right)
\ T^{cb}_{r \bar{s},k \bar{l}} (\ii \nu_m,{\bm p};T) \ ,
\label{eq:BSPNJL}
\ee
where $a,b$ label the flavor channel of the collective excitation, and $\Delta_r$ are in-medium quark propagators. The interaction kernel ${\cal K}$ contains essentially the coupling constant $G$, plus all possible flavor and Dirac structures (in the color-singlet channel), which generate the different meson channels. A diagrammatic representation of the $T$-matrix equation for quark-antiquark scattering is given in Fig.~\ref{fig:NJL_mesondiag}.

Given that the interaction kernel is constant, one can obtain a factorization of the equation, and the corresponding $T$-matrix equation (suppressing flavor and Dirac structures) reduces to
\be \label{eq:meson}
t^{ab} = \left[ \frac{2G}{1- 2 G \Pi} \right]^{ab} \ , \ee
where $t^{ab}$ are the $T$-matrix elements. The polarization function $\Pi^{ab}$ at finite temperature is given by 
\be \label{eq:polmeson} \Pi^{ab} (\ii\nu_m,{\bm p}) = - \sumint_k \textrm{ Tr}_{\gamma}
 \left[\bar{\Omega}^a_{\bar{j}i} \ \Delta_i \left(\ii\omega_n , {\bm k} \right) \ \Omega^b_{i\bar{j}} \ \Delta_{\bar j} \left( \ii\omega_n-\ii\nu_m ,{\bm k} - {\bm p} \right)  \right] \ , 
 \ee
where the trace is taken in color and Dirac space and the matrix $\Omega^{a}_{i \bar{j}} = \left( \mathbb{I}_\text{color} \otimes \tau^a_{i \bar{j}} \otimes \{ 1, \ii \gamma_5, \gamma^\mu,\gamma_5 \gamma^\mu \} \right)$ selects the appropriate flavor-spin channel.
After analytic continuation to real energies, the poles of $t^{ab}(p_0+\ii\epsilon,{\bm p})$ correspond to dynamically generated meson states. Expanding $(t^{-1})^{ab}(p_0,0)$ around a pole at $p_0=m_M$ yields
\be
t^{ab} (p_0,\bm{p}=0) \simeq \frac{-g^2_{M\rightarrow q\bar{q}}}{p_0^2-m_M^2} \ ,
\ee
with effective coupling
\be
g^2_{M\rightarrow q\bar{q}} \equiv \frac{2m_M}{ \left. \frac{\partial \Pi^{ab} (p_0,\bm{p}=0)}{\partial p_0} \right|_{p_0=m_M} } \ .
\ee

Thus, $t^{ab}(p_0,\bm{p})$ can be interpreted as the meson propagator in the corresponding channel, and the pole mass is determined by (with $G$ replaced by $G_V$ in the vector and axial-vector channels)
\be \label{eq:mesonmass}
1- 2 G \Pi^{ab} (p_0=m_M,{\bm p}=0) = 0 \ .
\ee

If $m_M$ exceeds the sum of constituent quark masses, the meson can decay into a $q\bar{q}$ pair, reflecting the absence of confinement in the model. In such cases, Eq.~\eqref{eq:mesonmass} yields a complex solution, whose real and imaginary parts correspond to the mass and half-width of the state.

\begin{figure}[!t] 
\centering
\includegraphics[width=0.4\linewidth]{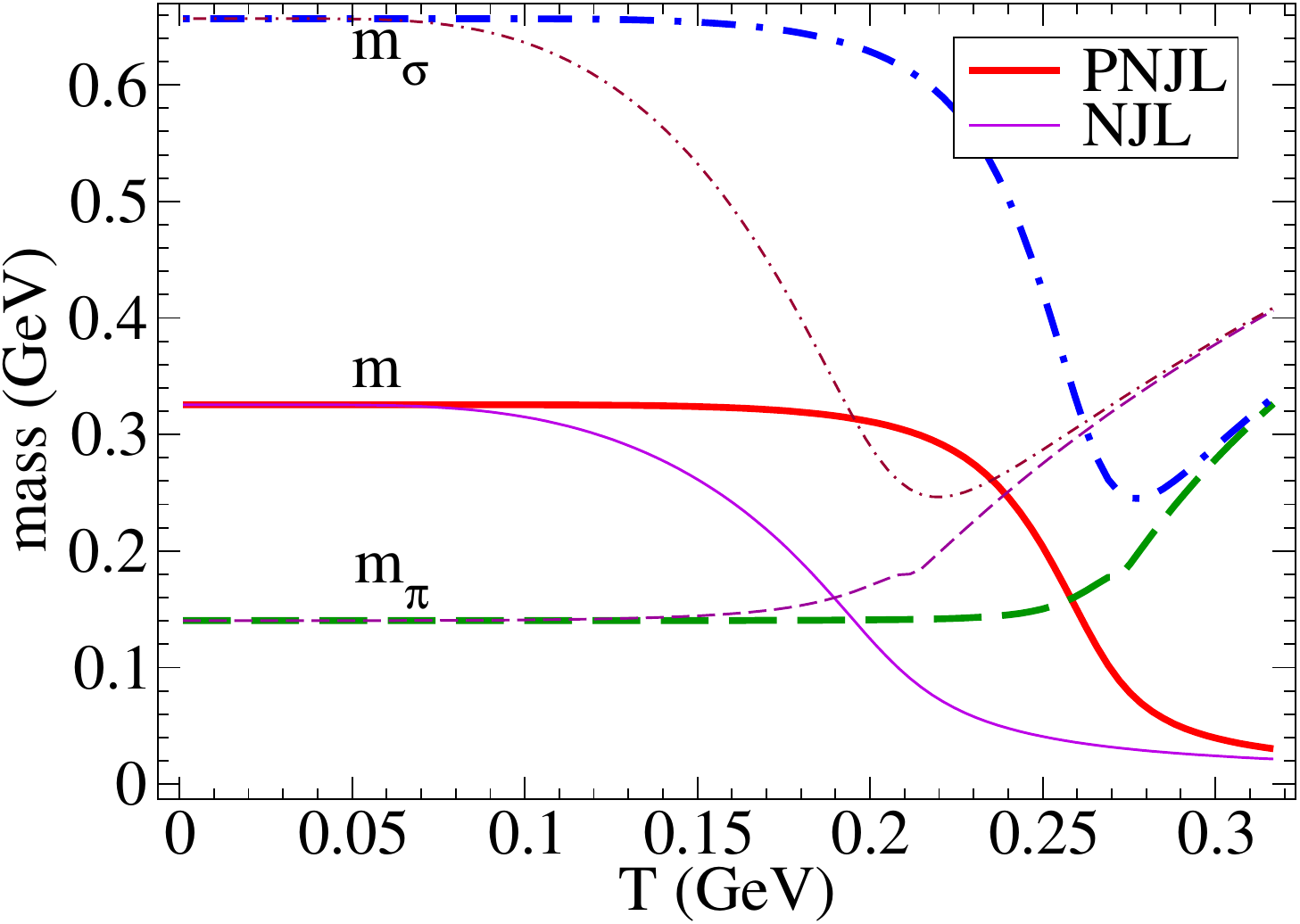}
\includegraphics[width=0.3\linewidth]{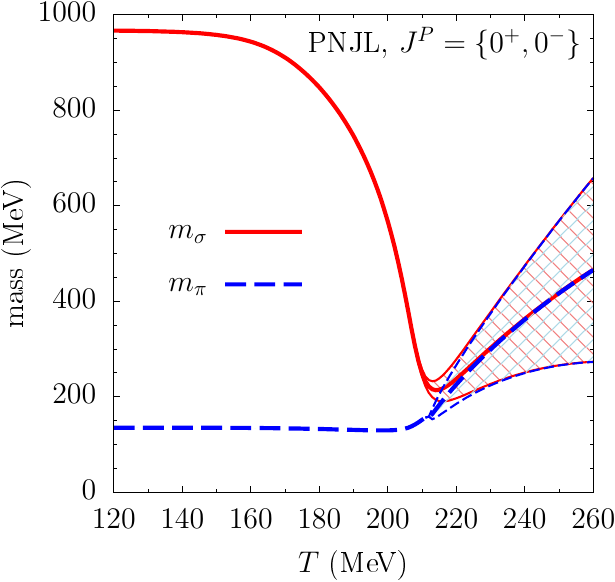}
\caption{Results for the pion and $\sigma$ masses in the (P)NJL models at finite temperature in the calculation with two~\cite{Hansen:2006ee} and three flavors~\cite{Torres-Rincon:2015rma}.}
\label{fig:PNJL1}
\end{figure}

Examples of meson thermal masses are shown in Fig.~\ref{fig:PNJL1}. In the left panel, we show the masses of the lightest mesons with opposite parity, $J^P=0^+$ vs.\ $0^-$ as functions of temperature. The results are from Ref.~\cite{Hansen:2006ee} where the 2-flavor NJL and PNJL models are used. The panel shows the difference between the NJL and PNJL results when their vacuum masses are equal. Notice that for 2 degenerate flavors, the $\sigma$ mass is close to what was at that time known as the $f_0(600)$~\cite{ParticleDataGroup:2006fqo} (now with a pole mass of $400$--$550\mev$~\cite{ParticleDataGroup:2024cfk}). In the right panel, we show the results of the 3-flavor PNJL model with parameters and approximations from Ref.~\cite{Torres-Rincon:2015rma} but without ultraviolet cutoff in the thermal integrals.
In both cases, the resulting behavior is qualitatively similar to that of the L$\sigma$M with explicit symmetry breaking ($\epsilon\neq 0$), shown in Figs.~\ref{fig:LSM} and \ref{fig:LSM2}. However, within the $\suthree_f$ PNJL model with parameters from Ref.~\cite{Torres-Rincon:2015rma}, the scalar–isoscalar state corresponds to the $f_0(980)$, not to the $f_0(500)$ (in Ref.~\cite{Torres-Rincon:2015rma} no scalar state close to $500\mev$ is found with $N_f=3$, which disfavors the $q\bar{q}$ interpretation~\cite{Pelaez:2015qba}).

At high temperatures, both states develop a thermal decay width, represented by the bands around the corresponding masses. However, notice that this decay width is due to the $q\bar{q}$ continuum, allowed in the (P)NJL model, reflecting its lack of true confinement. This happens around $T\simeq 220$--$270\mev$ (depending on the version of the model), where it is more favorable to have two isolated quarks than the bound state.

\begin{figure}[!t] 
\centering
\includegraphics[width=0.4\linewidth]{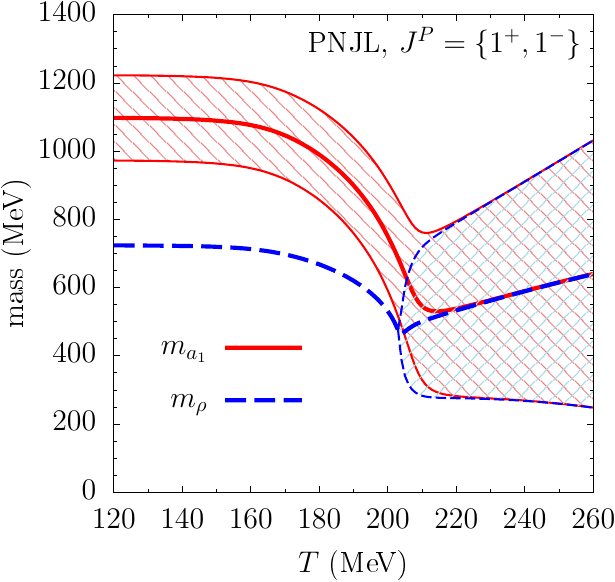}
\includegraphics[width=0.4\linewidth]{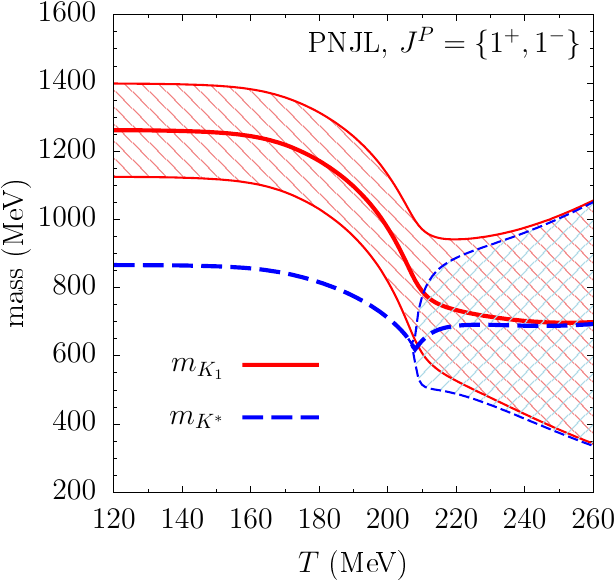}
\caption{Vector and axial-vector meson thermal masses as calculated in the 3-flavor PNJL model. We show the strangeness $S=0$ case (left panel) and the $S=1$ case (right panel) taken from Ref.~\cite{Torres-Rincon:2021wyd}.}
\label{fig:PNJL2}
\end{figure}

In Fig.~\ref{fig:PNJL2} we present the results of the 3-flavor PNJL model for vector/axial-vector mesons~\cite{Torres-Rincon:2021wyd}. In the left panel, we show the situation in the light $J=1$ sector, $\rho$ and $a_1$ mesons, both controlled by the same vector coupling $G_V$. This coupling is fixed to reproduce reasonable values of the vacuum masses, and the thermal behavior is then obtained. Again, the decay width at high temperature is due to the $q\bar{q}$ continuum, which for the $a_1$ is already nonzero at $T=0$. Masses and widths become degenerate above $T_c$.

In the right panel of Fig.~\ref{fig:PNJL2}, we show the vector and axial-vector channels with strangeness. We compare the $K^* (892)$ with the $K_1(1270)$. No extra parameters (beyond those already mentioned) are needed to obtain these states for $T \ge 0$. The $K^*$ does not exhibit a decay width at $T=0$, while the $K_1(1270)$ can decay into a $q\bar{q}$ pair. Chiral degeneracy happens in this channel around $T=250\mev$, both in the mass and in the decay width.

It should be stressed again that the decay width to quark-antiquark pairs present in the random-phase approximation of the (P)NJL model is unphysical, since it reflects the lack of true confinement of the model. Even if one could associate such a width above the melting (Mott) temperature with a sort of deconfinement process, its nonzero value at $T=0$ in some channels makes it difficult to interpret. On the other hand, the expected vacuum width of some states, like the $\rho$ meson, due to decay to other hadron channels, is missing in this approximation. While technically more involved, it can, however, be incorporated if the hadronic final state is included as a coupled channel in the $T$-matrix approach. One of the few calculations in which this is achieved at finite temperature is that of Ref.~\cite{He:1997gn} in the NJL model, where the $\rho$ meson pole mass was calculated, and its vacuum decay width was found to be consistent with experiment. A more general $1/N_c$ expansion to account for meson loops was reported in Ref.~\cite{Oertel:2000jp}, finding good agreement with ChPT in the description of $\pi\pi$ scattering. 

In summary, the theoretical description of chiral symmetry restoration reflects a clear distinction between robust nonperturbative consensus and framework-dependent modeling. First-principles lattice QCD unambiguously establishes that the chiral transition at $\mu_B = 0$ for physical quark masses is a smooth crossover occurring at $T_c \approx 156\text{--}158\mev$, a feature that persists at least up to $\mu_B /T<3.5$. Across theoretical approaches such as linear sigma models and (P)NJL dynamics, the main signature of this restoration is the emergent mass and width degeneracy of chiral parity partners (such as $\pi\text{--}\sigma$, $\rho\text{--}a_1$, and $K\text{--}\kappa$) above $T_c$. In unitarized chiral perturbation theory, as discussed in the previous section, one cannot directly capture the phase transition itself, but the thermal dependence of the chiral partner masses ($\sigma$ and $\kappa$) exhibits a decreasing trend toward the pion and kaon masses. Although the effective models that can address higher temperatures are not first-principles calculations in their own right, we will now see that finite-temperature lattice-QCD screening masses show, at least qualitatively, a similar degeneracy pattern of chiral partners beyond $T_c$.

\subsection{Lattice QCD for light systems}~\label{sec:lqcdlight}

In this section, we briefly review the lattice-QCD calculations that have addressed the hadron thermal masses in the context of chiral symmetry restoration. For more general references on lattice QCD in the medium, we refer to Refs.~\cite{Karsch:2001cy,Karsch:2003jg}.

Calculations of thermal masses on the lattice require the analysis of thermal two-point correlation functions of mesonic operators. Such operators are typically constructed as bilinears of the form ${\cal O}_\Gamma = \bar{\psi} \Gamma \tau^a \psi$, where $\Gamma$ denotes a generic Dirac matrix and $\tau^a$ a flavor generator. Different choices of $\Gamma$ and $\tau^a$ probe mesonic channels with different spin-parity quantum numbers, namely, scalar, pseudoscalar, vector, axial-vector, tensor, and axial tensor. In a thermal Euclidean setup, spatial (screening) correlation functions are defined as
\be
{\cal C}_\Gamma (z;T) = \int_0^\beta \diff\tau \int \diff^2r_\perp 
\left\langle {\cal O}_\Gamma (\bm{r}_\perp, z,\tau ) {\cal O}^\dag_\Gamma (\bm{0},0,0)\right\rangle \ .
\ee
At large spatial separations, these correlators exhibit an exponential decay
\be 
{\cal C}_\Gamma (z;T) \underset{z \rightarrow \infty}{\propto} \exp \left( -\msc (T) \ z \right) \ , 
\ee
or a $\text{cosh}$ behavior on a finite lattice with periodic boundary conditions, allowing one to fit the screening mass $\msc(T)$. 
At high temperature, the natural scale for screening masses is $2\pi T$, which corresponds to two noninteracting quarks with lowest Matsubara frequency $\omega_{n=0}= \pi T$.
However, the relation of the fitted screening masses to the phenomenologically relevant pole masses is not straightforward.

Alternatively, temporal (effective) masses can be extracted from Euclidean time correlation functions
\be
    \mathcal{C}(\tau;T)=\int \diff ^3x \left\langle\mathcal{O}(\bm{x},\tau;T)\mathcal{O}^\dagger(\bm{0},0;T)\right\rangle \ ,  
\ee
as
\be 
m_\text{eff}(\tau;T)=\frac{1}{a_\tau}\log\left[\frac{\mathcal{C}(\tau;T)}{\mathcal{C}(\tau+a_\tau;T)}\right] \ .
\ee
While $m_\text{eff}(\tau;T=0)$ reaches a plateau at large time slices, yielding the physical pole mass, this extraction becomes more challenging at finite temperature because the temporal extent is determined by the inverse of the temperature, $\beta=1/T$, which requires very fine lattices in the temporal direction. More fundamentally, reconstructing physical thermal spectral functions from discrete Euclidean temporal correlators represents an intrinsically ill-posed inverse problem, which we discuss in more detail in Section~\ref{sec:D-LQCD} for heavy hadrons. While spectral reconstruction is essential in the heavy-quark sector to study the survival and sequential melting of bound states near $T_c$, finite-temperature lattice-QCD calculations for light mesons rely predominantly on spatial screening masses, which provide a robust indicator of chiral symmetry restoration without requiring analytic continuation.

One should note that the lattice-QCD calculations typically do not present results much below $T_c$. Therefore, a direct comparison with results from effective field theories is usually difficult, apart from the fact that these low-energy results lack the physics of a phase transition at $T_c$. One should also be aware that the screening masses calculated in lattice QCD need not necessarily coincide with the predicted pole masses. Some of the mentioned models (L$\sigma$M, NJL model) can describe a chiral transition and can even give access to screening-type masses, which allows a closer comparison.

\begin{figure}[!t] 
\centering
\includegraphics[width=0.45\linewidth]{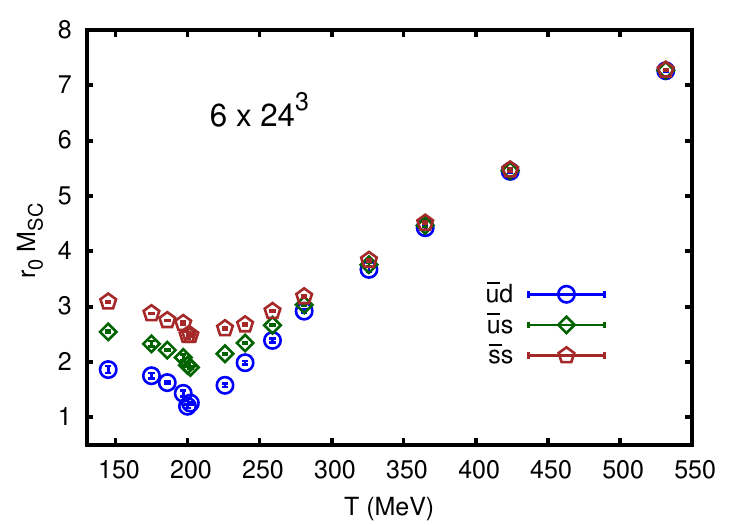}
\includegraphics[width=0.45\linewidth]{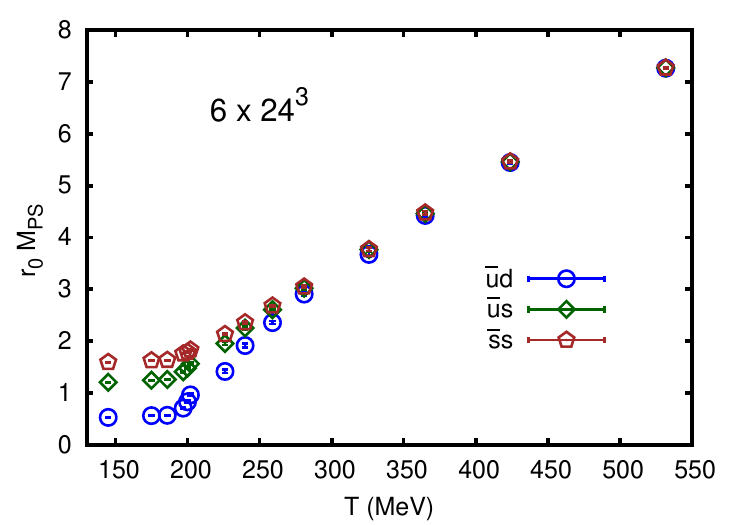}
\caption{Scalar (left panel) and pseudoscalar (right panel) screening meson masses from lattice QCD at finite temperature. Figures taken from Ref.~\cite{Cheng:2010fe}.}
\label{fig:ChengLattice}
\end{figure}

In Ref.~\cite{Cheng:2010fe} screening masses were calculated between $T=140\mev$ and $T=800\mev$. They used $N_f=2+1$ flavors of improved (p4) staggered fermions. The quark masses were chosen so that the pion mass is $200\mev$ and the kaon mass, $500\mev$. The temporal extent of the lattice was $N_\tau=4,6,8$, and the gauge configurations were taken from the RBC-Bielefeld and HotQCD collaborations. We provide the results for $N_\tau=6$ in the case of scalar (left panel) and pseudoscalar (right panel) mesons in Fig.~\ref{fig:ChengLattice}.
We observe that the members of the scalar channel present a decrease with temperature up to $T=200\mev$, respecting the ordering given by their quark masses. Then the three states merge into a single line that increases linearly with temperature. The pseudoscalar states do not present the decrease at low temperature  (at least above $T=140\mev$), and the masses remain stable until $T=200\mev$ and then increase with $T$, with the three states becoming degenerate at high temperatures. Notice that the masses are multiplied by the Sommer parameter $r_0$, whose value is $r_0 = 0.469$ fm in the continuum limit for physical quark masses according to Ref.~\cite{Cheng:2010fe}.

\begin{figure}[!t] 
\centering
\includegraphics[width=0.3\linewidth]{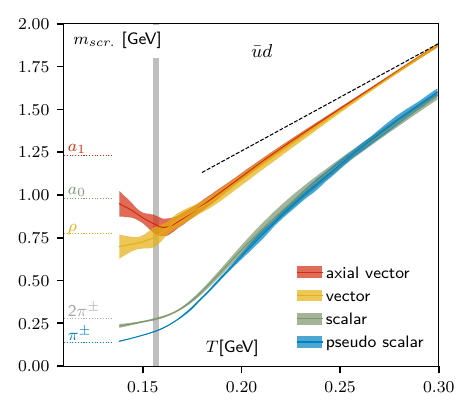}
\includegraphics[width=0.3\linewidth]{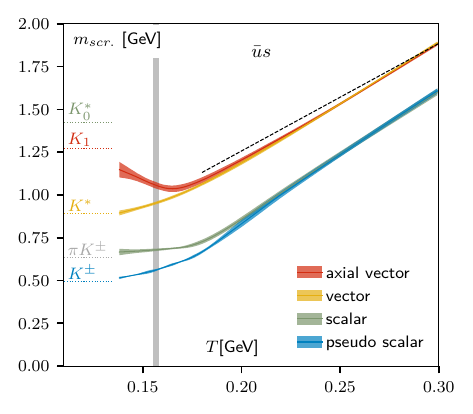}
\includegraphics[width=0.3\linewidth]{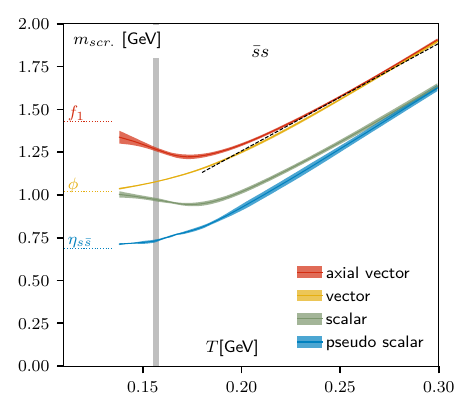}
\caption{Light (left panel), open-strangeness (middle panel), and hidden-strangeness (right panel) screening meson masses from lattice QCD at finite temperature. Figures taken from Ref.~\cite{Bazavov:2019www} by the HotQCD Collaboration.}
\label{fig:HotQCD_ThermalMasses}
\end{figure}

The HotQCD collaboration has presented in Ref.~\cite{Bazavov:2019www} results for the screening masses of mesons at finite temperature in a broad range between $T=140\mev$ and $T=2500\mev$. They worked with $N_f=2+1$ flavors, and they used the action for highly improved staggered quarks (HISQ). The strange quark mass is set to its physical value, and light-quark masses are taken in such a way that the pion mass is $140\mev$ or $160\mev$, that is, very close to the physical point. We present in Fig.~\ref{fig:HotQCD_ThermalMasses} their final results with continuum-extrapolated masses.
%.
The left, middle, and right panels show the results for light mesons, open-strangeness mesons, and hidden-strangeness mesons, respectively. Starting with the first panel, one finds
\begin{itemize}
    \item Pions and their scalar counterparts have monotonically increasing masses and become degenerate above $T=200\mev$. Below $T=300\mev$, these masses increase more rapidly than $2\pi T$. The scalar state in this calculation is a nonphysical $\pi\pi$ state.
    \item The vector and axial-vector states ($\rho$ and $a_1$) begin with a splitting of around $250\mev$. While the $\rho$ mass increases with temperature, the $a_1$ mass decreases until passing $T_c$, reaching a minimum and becoming degenerate with that of the $\rho$ meson. At higher temperatures, both masses increase linearly with temperature with a slope approaching $2\pi$.
\end{itemize}
For the middle panel, the situation is analogous, with the strange chiral-partner masses split at low temperature and becoming degenerate above $T=200\mev$. The splitting of the $K^*$ and $K_1$ is now closer to the expectations at $T=0$ than in the case with light mesons. The slopes at high temperature are now close to $2\pi$.
Finally, the right panel of Fig.~\ref{fig:HotQCD_ThermalMasses} shows the hidden-strangeness case. Again, similar conclusions can be drawn, with the difference that the chiral degeneracy seems to occur at higher temperatures, above $T=250\mev$. The slope at higher temperature is even closer to $2\pi$ for this case, as compared to the lighter cases.

\begin{figure}[!t] 
\centering
\includegraphics[scale=0.9]{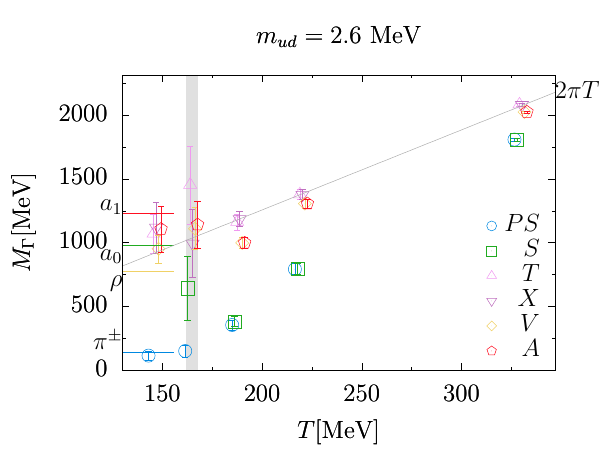}
\caption{Light-meson screening masses in different spin-parity channels from lattice QCD at finite temperature. Figure taken from Ref.~\cite{Aoki:2025mue} by the JLQCD collaboration.}
\label{fig:JLQCD_ThermalMasses}
\end{figure}

The JLQCD collaboration has published results for screening masses across the transition temperature (between $T=147\mev$ and $T=330\mev$) of $N_f=2$ mesons~\cite{Aoki:2025mue}. In that work, they used M\"obius domain-wall fermions with fixed lattice spacing ($a \sim 0.075$ fm) and the lightest bare quark mass below the physical value. All quark bilinear operators were used, spanning scalar, pseudoscalar, vector, axial-vector, tensor, and axial tensor mesons. The results are shown in Fig.~\ref{fig:JLQCD_ThermalMasses}.
Given the small number of data points for each state, one can only extract hints of the following:
\begin{itemize}
    \item The pion mass remains rather flat up to the transition temperature ($T_c \simeq 165\mev$ in that study), and then increases linearly with temperature with a slope larger than $2\pi$.
    \item The scalar state (identified with the $a_0$) exhibits a strong decrease up to $T \simeq 185\mev$, above which it becomes degenerate with its pseudoscalar partner. 
    \item The $\rho$ only shows a slight increase at low temperature, and later, it increases with a slope close to $2\pi$. The axial state shows a similar behavior, with a possible minimum below $T=200\mev$, and then becomes degenerate with the vector meson.
    \item The tensor and axial-tensor states very quickly degenerate in mass as a function of $T$.
\end{itemize}

Also relevant is the lattice-QCD work of Ref.~\cite{DallaBrida:2021ddx}, where the meson screening masses in the vector and axial-vector channels, as well as the scalar and pseudoscalar channels, were computed for three massless quark flavors. This calculation focused on the degeneracy of states expected from the restoration of chiral symmetry at very large temperatures, in the range $T=1$--$160\gev$. 

In a similar range of temperatures $T=0.8$--$164\gev$,  Ref.~\cite{Bala:2025ilf} calculated the spin splitting between pseudoscalar and vector screening masses. To do this, they solved the Schr\"odinger equation for a quark-antiquark pair, each with effective mass $\pi T$ (lowest fermionic Matsubara frequency) with a spin-dependent potential that matches the short-distance part described by electrostatic QCD to the long-distance part described by the nonperturbative spatial string tension.

It is also interesting to mention the efforts to construct nonperturbative thermal spectral functions by imposing constraints from field locality at finite temperature (see Refs.~\cite{Lowdon:2022keu,Nair:2025jgl} and references therein). This approach has been used to construct the spectral functions of light
meson states at temperatures above and below $T_c$, and to use them to predict temporal correlators in Euclidean space, which are compared with the corresponding temporal lattice-QCD data. For example, the pion case has been treated in Ref.~\cite{Lowdon:2022xcl}, and the kaon and hidden-strangeness mesons in Ref.~\cite{Bala:2023iqu}.

\subsection{Baryons at finite temperature}\label{sec:lightbaryons}

Baryon masses at finite temperature have also been studied with different models and in lattice QCD, but the literature is scarce when compared to the light-meson sector.

In Ref.~\cite{Leutwyler:1990uq} by Leutwyler and Smilga, and also reviewed in Ref.~\cite{Smilga:1996cm} by Smilga himself, a system of nucleons moving in a thermal pion gas was studied. They considered the modification of the nucleon pole position and its residue within ChPT at low temperatures. The imaginary part, corresponding to the nucleon damping rate, increases with temperature, but the real part, the thermal mass, is first suppressed by a few MeV up to $T=100\mev$ and then increases with temperature, reaching a shift of the order of $25\mev$ around $T=200\mev$. This result uses a one-loop pion correction for the nucleon propagation at first order in the density (virial expansion).

Within these approximations, the damping is calculated as
\be 
\gamma_N (T)= \frac{1}{4\pi^2} \int_{m_\pi}^\infty 
\diff E \ (E^2-m^2_\pi) \ n_B(E) \left[ \sigma_{\pi^+ p} (E) + \sigma_{\pi^0 p} (E)+\sigma_{\pi^- p} (E) \right] \ , \label{eq:gammaN}
\ee
where $E=\sqrt{m_\pi^2+\bm{p}^2}$ and the Bose--Einstein distribution function $n_B(E)$ is defined in Eq.~\eqref{eq:BE}. The pion-proton cross sections are assumed to be dominated by the $\Delta$ resonance up to temperatures of $T=150\mev$. In the narrow-resonance limit, the expression reads 
\be 
\gamma_N (T)=\frac{1}{8\pi} \Gamma_\Delta \sigma_{\pi N} (E_\Delta^2-m_\pi^2) \ n_B(E_\Delta) \ , \label{eq:gammaNnarrow} 
\ee
where $E_\Delta$ is the pion laboratory energy at the position of the $\Delta$ peak, $\Gamma_\Delta$ the resonance width and $\sigma_{\pi N} \simeq 380$ mb, considering the three pion-proton channels.

\begin{figure}[!t] 
\centering
\includegraphics[scale=0.38]{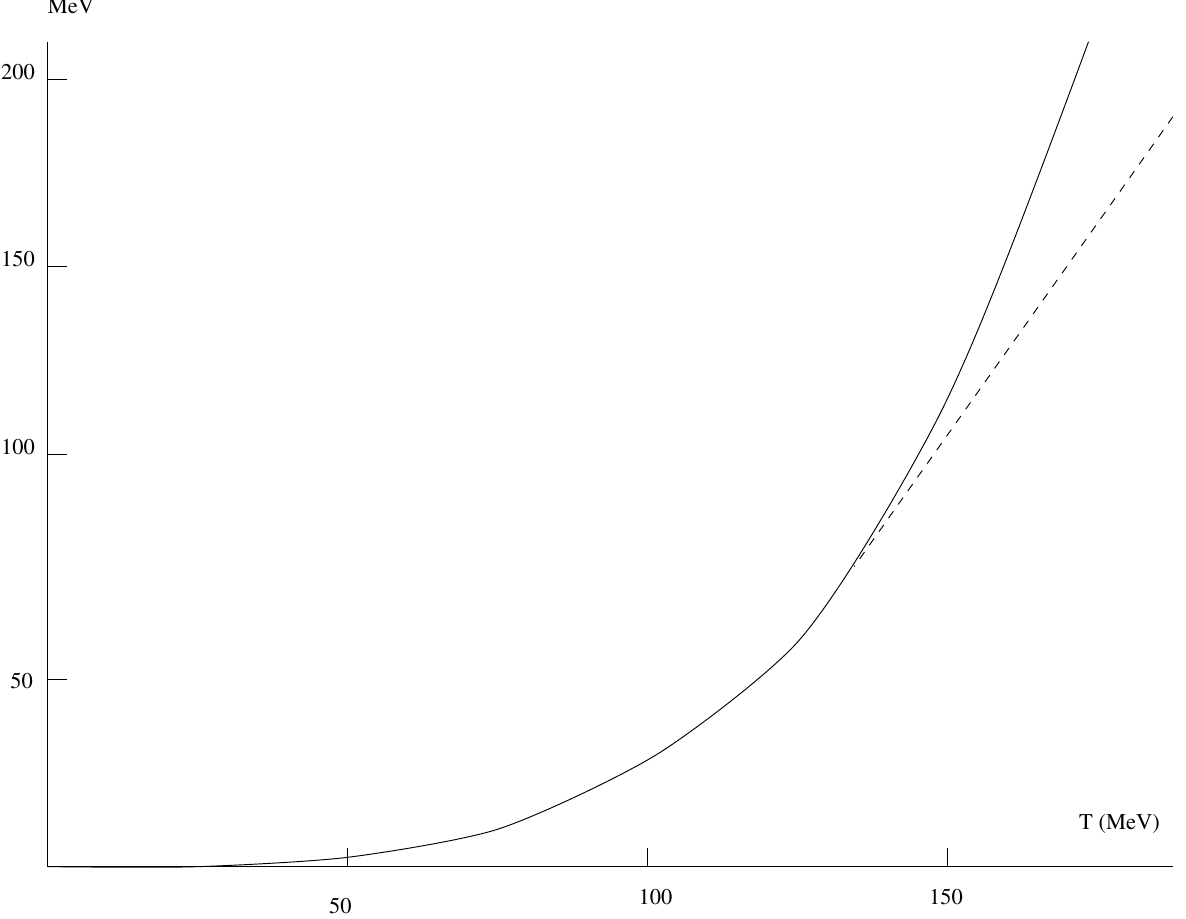}
\includegraphics[scale=0.28]{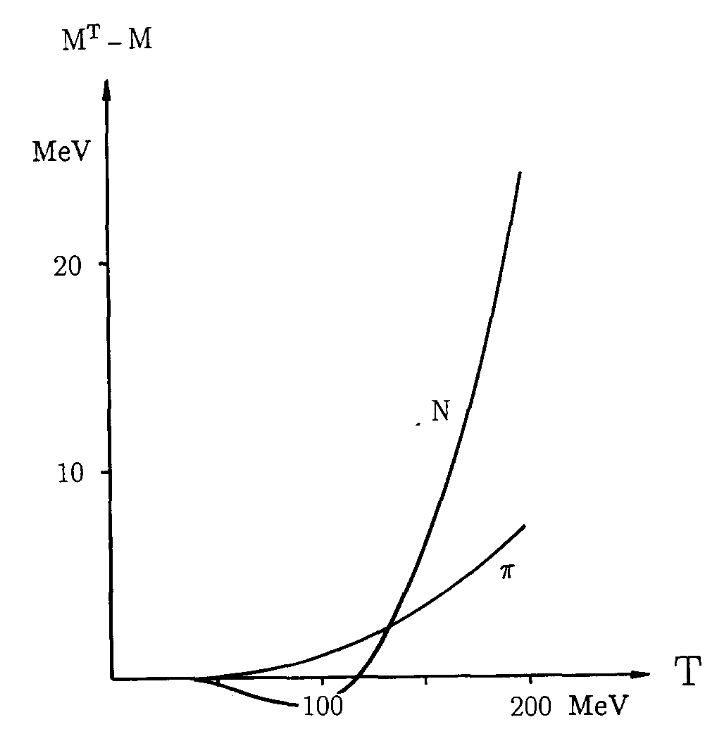}
\caption{Left panel: Damping coefficient of nucleons in a dilute gas of pions, from Ref.~\cite{Smilga:1996cm}. Right panel: Thermal mass of nucleons in a dilute gas of pions, from Ref.~\cite{Leutwyler:1990uq}.}
\label{fig:smilganucleon}
\end{figure}

The damping rate as a function of temperature is shown in the left panel of Fig.~\ref{fig:smilganucleon}, which is taken from Ref.~\cite{Smilga:1996cm}. The solid line is the narrow-resonance limit of Eq.~\eqref{eq:gammaNnarrow},
while the dashed line is the result from Eq.~\eqref{eq:gammaN} using phenomenological scattering data.

The thermal mass shift of the nucleon is shown in the right panel of Fig.~\ref{fig:smilganucleon} together with the pion mass shift from Ref.~\cite{Leutwyler:1990uq}. According to Ref.~\cite{Smilga:1996cm}, the formula for the mass shifts reads
\be 
M(T) = M(T=0)-\sum_{i=1}^3\int \frac{\diff^3p}{(2\pi)^3 2E}  \frac{\re \ T_{\pi^i N} (E)}{2M (T=0)} n_{\text{B}}(E) \ , 
\ee
where $i$ runs over the three pion states ($\pi^-,\pi^0,\pi^+$), and $M(T=0)$ is the vacuum nucleon mass.

\begin{figure}[!t] 
\centering
\includegraphics[scale=0.55]{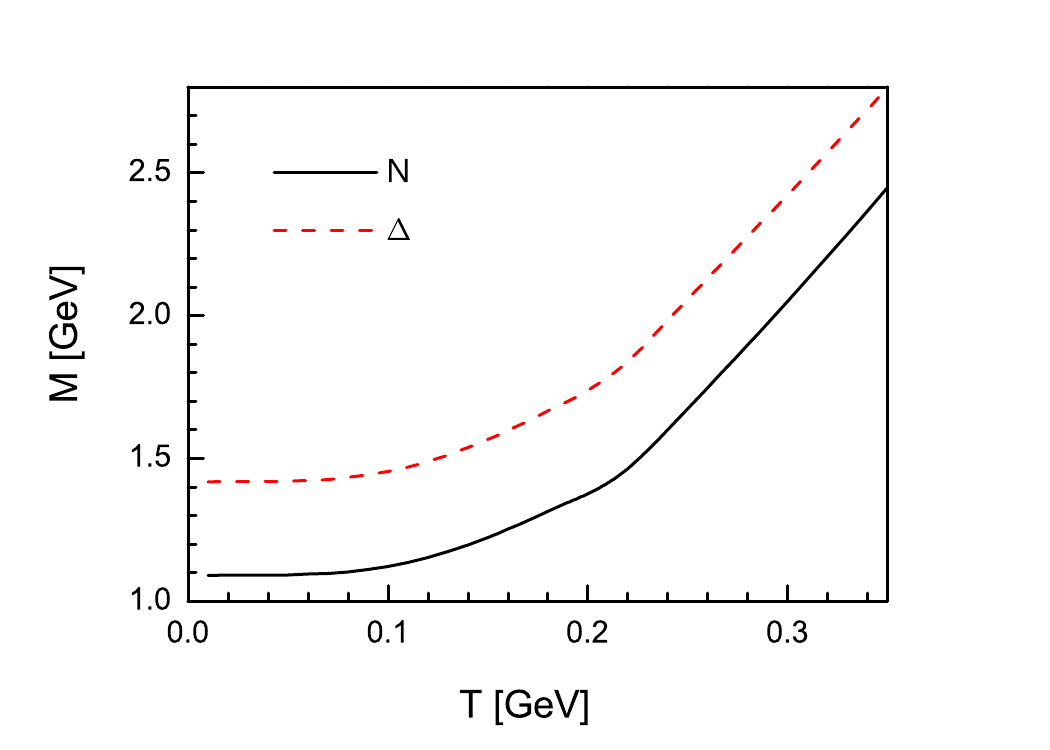}
\caption{Nucleon and $\Delta$ baryon screening masses from the QCD Dyson--Schwinger equations performed in Ref.~\cite{Wang:2013wk}.}
\label{fig:DS_nucleonDelta}
\end{figure}

Alternatively, the nucleon and $\Delta$ baryon screening masses have been investigated in Ref.~\cite{Wang:2013wk} using a symmetry-preserving truncation scheme of the Dyson--Schwinger functional method, based on a Faddeev-like equation for the three-body interaction. The result is shown in Fig.~\ref{fig:DS_nucleonDelta}. In this work, the increase of the nucleon screening mass with temperature is much stronger than the increase of the pole mass found in the pioneering calculations of Refs.~\cite{Leutwyler:1990uq,Smilga:1996cm}, although the two quantities are not directly comparable. In Ref.~\cite{Wang:2013wk}, a strong diquark correlation inside the baryon was inferred from the numerical results. 

\begin{figure}[!t] 
\centering
\includegraphics[scale=0.4]{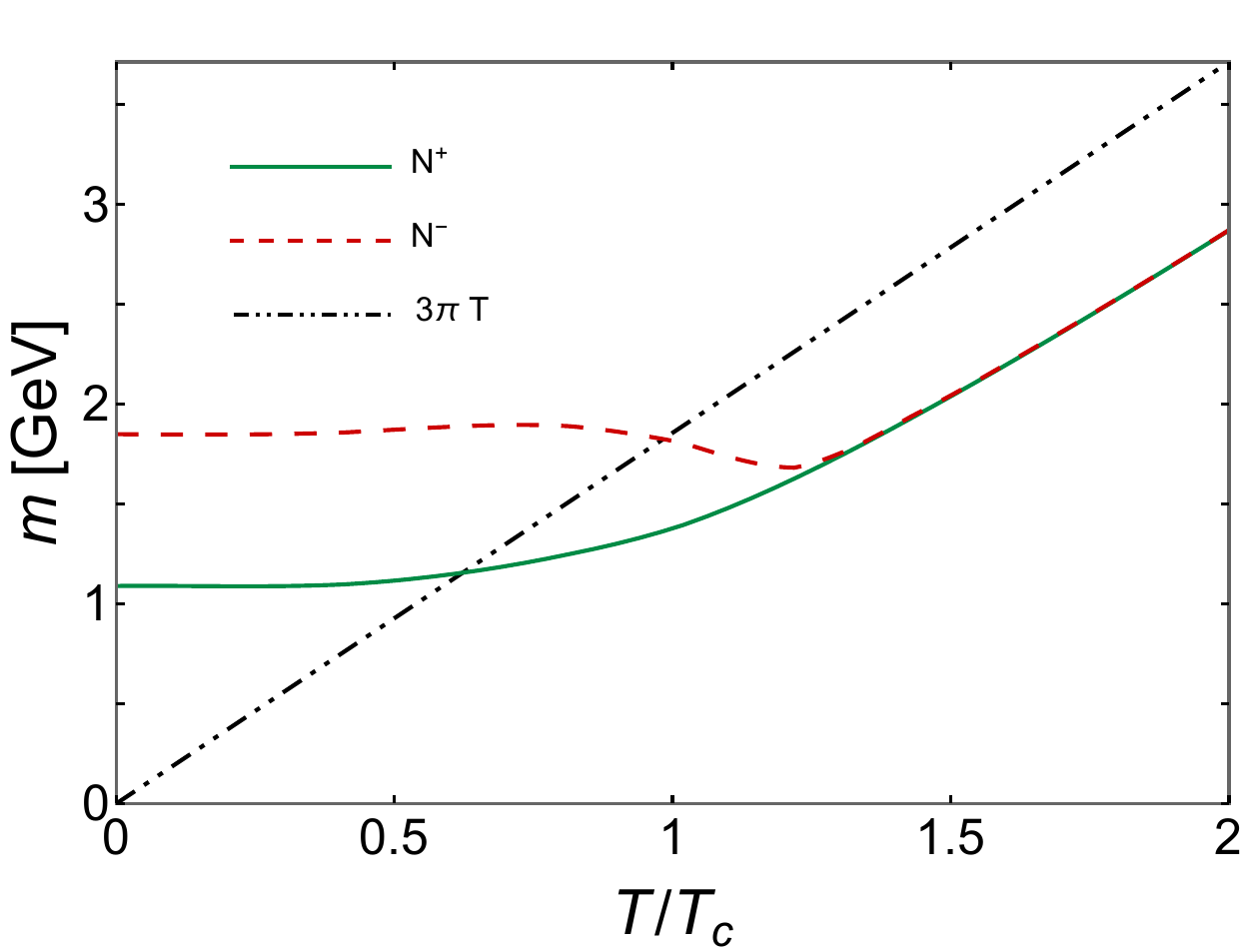}
\includegraphics[scale=0.4]{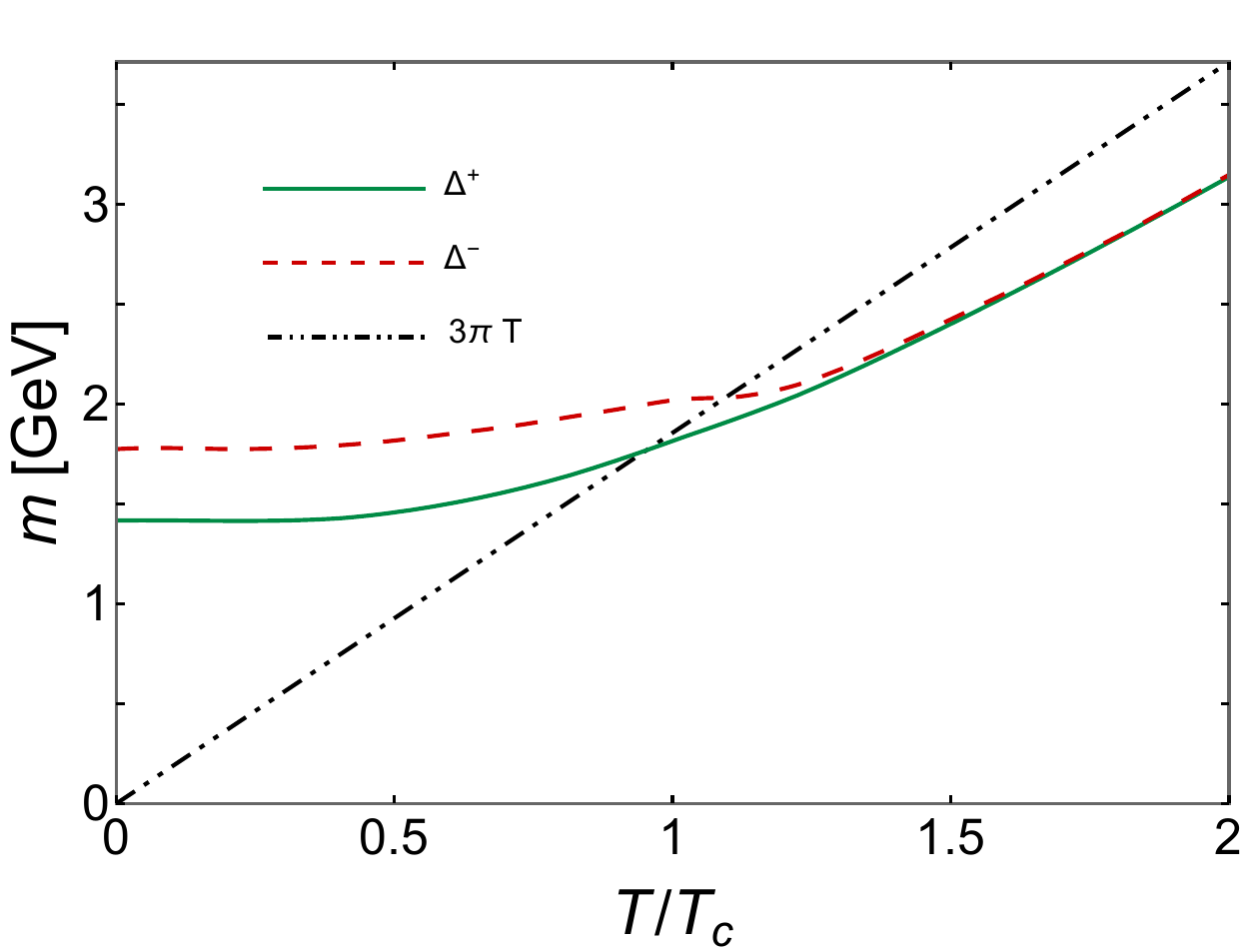}
\caption{Nucleon (left panel) and $\Delta$ baryon (right panel) screening masses from the vector $\times$ vector contact interaction model of Ref.~\cite{Chen:2024emt}. Both positive (solid lines) and negative (dashed lines) parity states have been included.}
\label{fig:ChenBaryons}
\end{figure}

Baryons were also addressed in the vector $\times$ vector contact interaction model of Ref.~\cite{Chen:2024emt}, where quark-diquark scattering was solved in a Faddeev-like equation. We show their results for the nucleon (left panel) and $\Delta$ (right panel) sectors in Fig.~\ref{fig:ChenBaryons}. Both the positive- and negative-parity baryon screening masses are shown.
The results in the left panel show that the positive-parity $N$ increases monotonically after a rather constant trend at low temperatures, while the negative-parity $N$ shows a slight decrease when crossing $T_c$. Immediately afterward, the two masses become degenerate and proportional to the temperature. In the right panel, we observe the $\Delta$ baryons, with both parity states presenting a slightly increasing behavior with a mass gap between them, until passing $T_c$, where the two become degenerate and follow the free-theory expectations for the screening mass. 

\begin{figure}[!t] 
\centering
\includegraphics[scale=0.6]{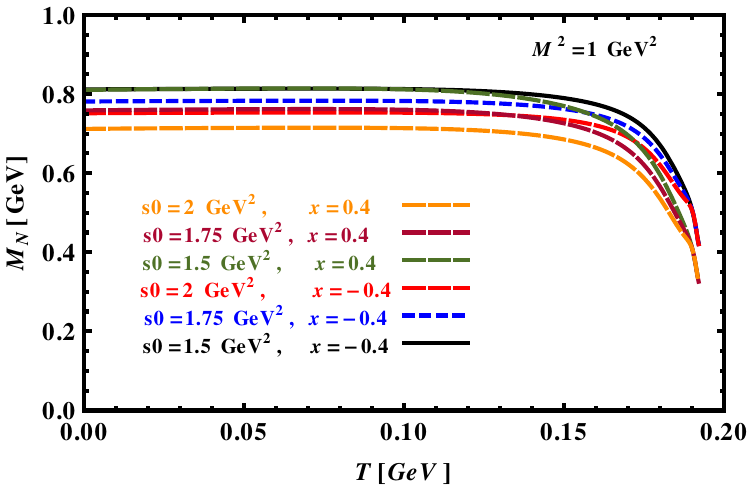}
\includegraphics[scale=0.45]{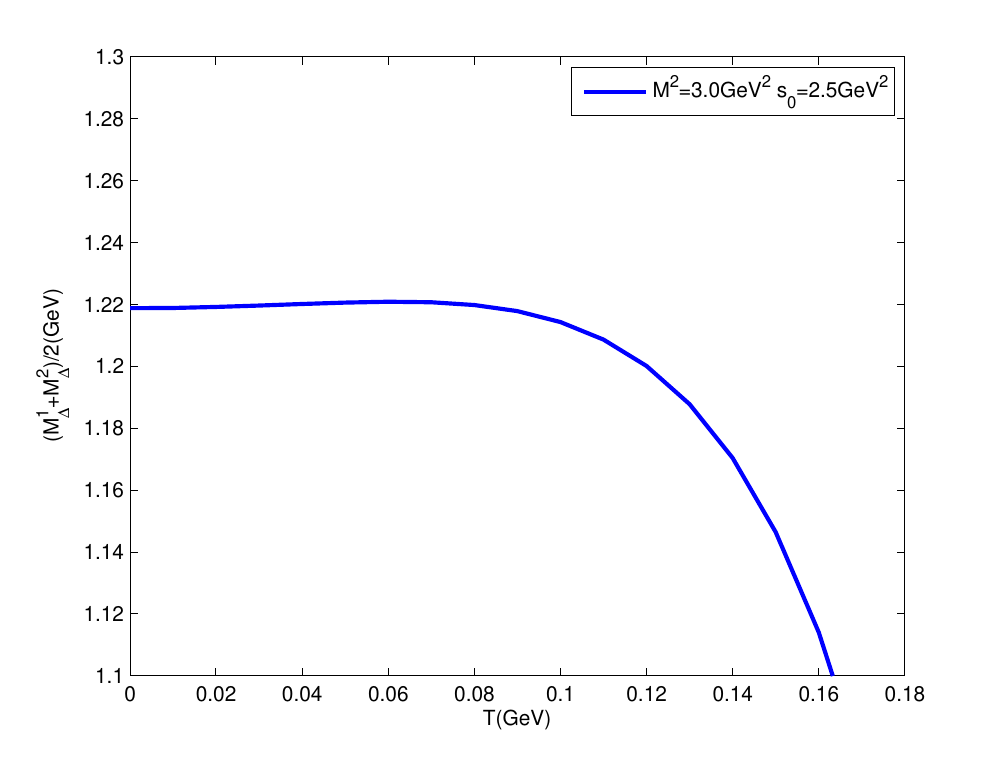}
\caption{Nucleon mass from the calculation of Ref.~\cite{Azizi:2015ona} and $\Delta$ baryon mass from Ref.~\cite{Xu:2015jxa}. Both works use thermal QCD sum-rule techniques.}
\label{fig:nucleonsumrules}
\end{figure}

Light baryon masses have also been addressed using thermal QCD sum rules in Refs.~\cite{Azizi:2015ona} (nucleon), \cite{Azizi:2015oxa} (hyperons), and \cite{Xu:2015jxa,Azizi:2016ddw} (decuplet baryons). In the left panel of Fig.~\ref{fig:nucleonsumrules}, we present the result for the nucleon mass given by the thermal QCD sum rules of Ref.~\cite{Azizi:2015ona} for different values of the parameters used. The masses of hyperons given in Ref.~\cite{Azizi:2015oxa} are qualitatively analogous to these, with corresponding modifications of vacuum masses. No noticeable modifications can be seen until $T \simeq 150\mev$.
Similarly, all masses of the members of the decuplet ($\Delta, \Sigma^*, \Xi^*, \Omega$) remain unaltered with temperature until they drop considerably around $T=110\mev$ according to Ref.~\cite{Xu:2015jxa}, and $T=150\mev$ according to Ref.~\cite{Azizi:2016ddw}. We show the $\Delta$ baryon thermal mass from Ref.~\cite{Xu:2015jxa} in the right panel of Fig.~\ref{fig:nucleonsumrules}. The result shown is the average of the masses obtained in the two independent ways of extracting the baryon mass from the thermal sum rule.

A different approach to access nucleon masses that relies on diquark--quark excitations is the (P)NJL model~\cite{Reinhardt:1989rw,Vogl:1991qt,Buck:1992wz,Ishii:1993rt, Ishii:1995bu,Ebert:1996ab, Oettel:2000ig, Wang:2010iu,Blanquier:2011zz,Torres-Rincon:2015rma,Pfaff:2022sfv}. In this model, the baryon is modeled as a quark-diquark bound state in the different flavor-spin channels in a two-step process. First, two quarks bind to form a diquark in any of the possible channels, and then a third quark is added to the bound system in order to generate a baryon-like excitation. Typically, the Faddeev partitions are reduced to a two-body equation where one of the particles is itself a bound (or resonant) state~\cite{Reinhardt:1989rw,Vogl:1991qt,Buck:1992wz}.

\begin{figure}[!ht] 
\centering
\includegraphics[scale=1.05]{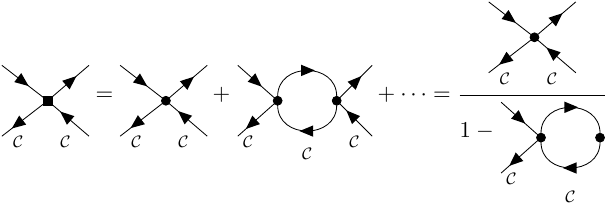}
\caption{Diagrammatic representation of the $T$-matrix Eqs.~\eqref{eq:BSPNJL2} and \eqref{eq:BSPNJL3} in the diquark sector of the (P)NJL model.}
\label{fig:NJL_diquarkdiag}
\end{figure}

\begin{figure}[!ht] 
\centering
\includegraphics[scale=1.0]{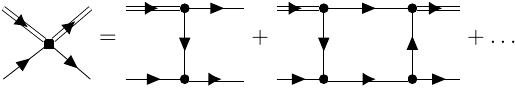}
\hspace{12mm}
\includegraphics[scale=1.0]{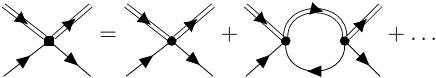}
\caption{Diagrammatic representation of the quark-diquark rescattering in the (P)NJL model for the generation of baryon-like states. Left panel: First two diagrams of the expansion with diquark-quark effective couplings. Right panel: Reduction of the previous diagram in the {\it static limit} where the ``$u$-channel quark'' is taken to be at rest.}
\label{fig:NJL_baryondiag}
\end{figure}

\begin{figure}[!ht] 
\centering
\includegraphics[scale=0.4]{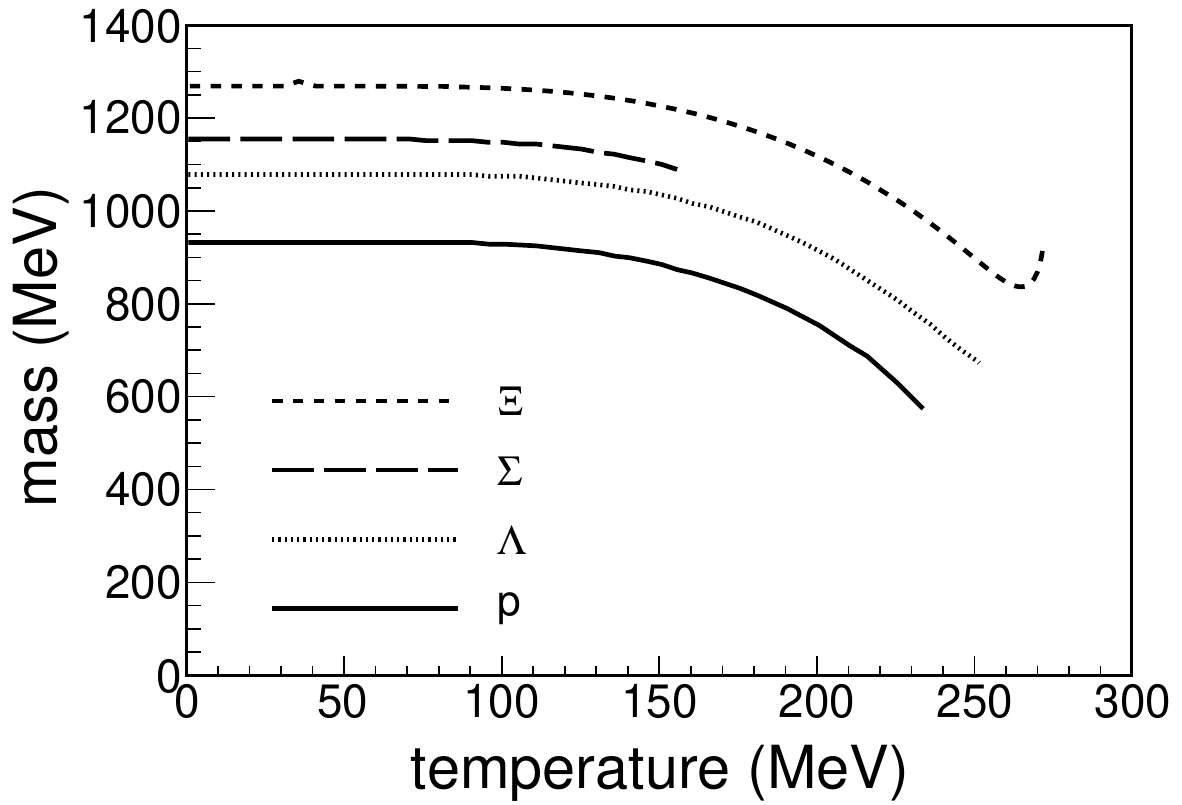}
\includegraphics[scale=0.4]{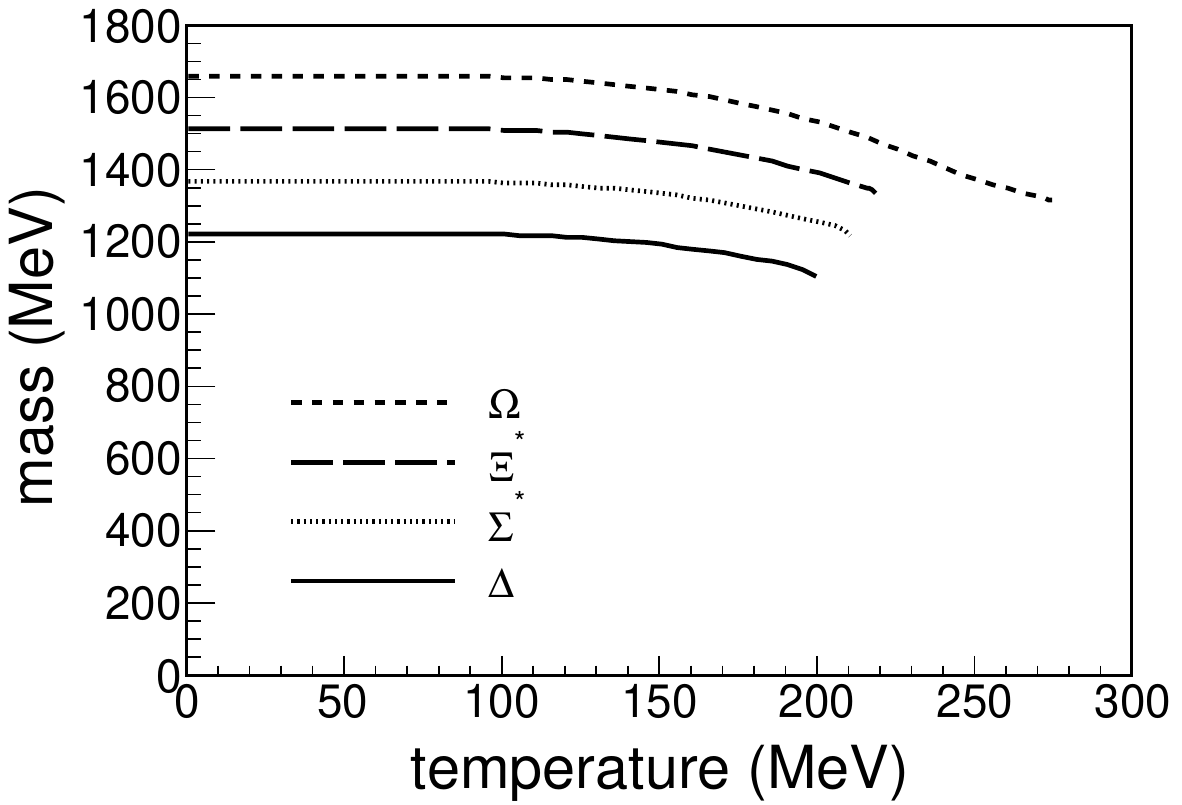}
\caption{Baryon thermal masses in the NJL model of Ref.~\cite{Torres-Rincon:2015rma} for the members of the octet (left) and decuplet (right) $\suthree_f$ sectors.}
\label{fig:NJL_baryonresult}
\end{figure}

\begin{figure}[!ht] 
\centering
\includegraphics[scale=0.28]{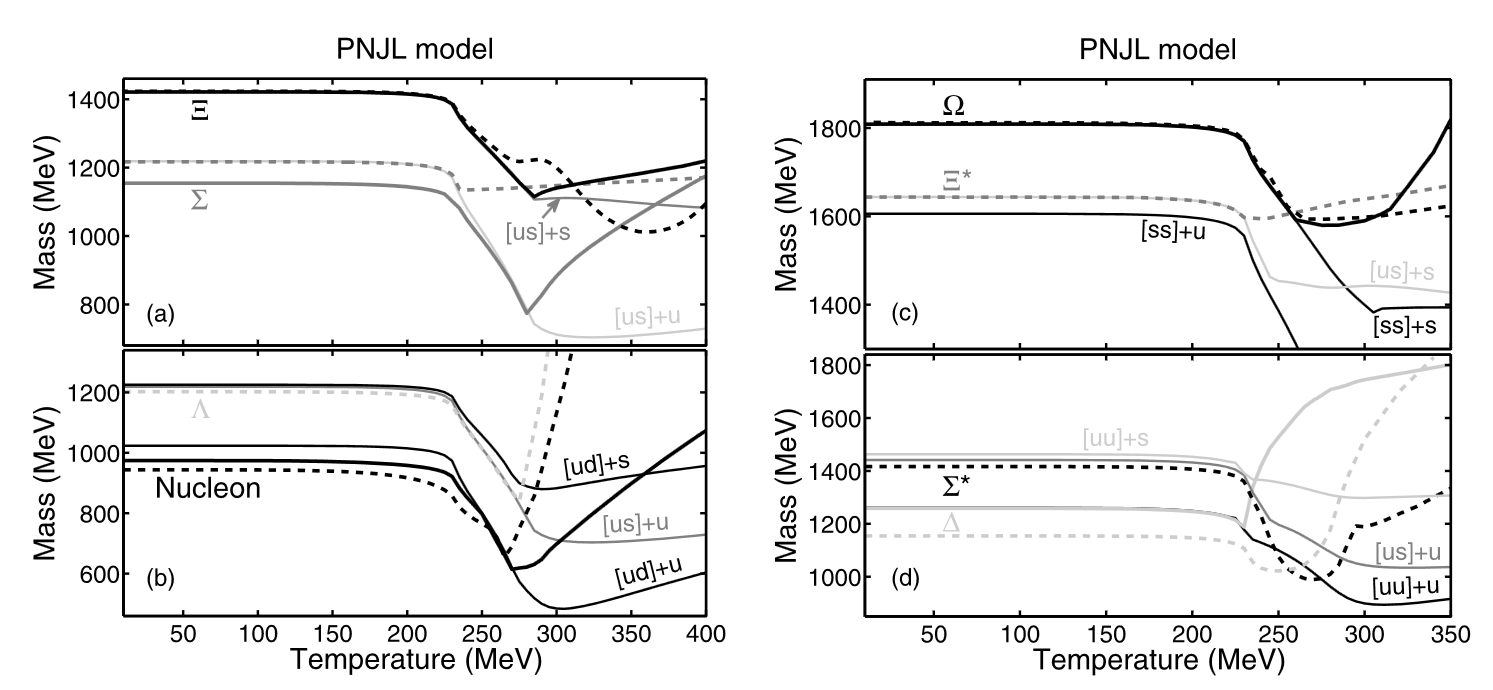}
\caption{Baryon thermal masses in the PNJL model of Ref.~\cite{Blanquier:2025ysh} for the members of the octet (left) and decuplet (right) $\suthree_f$ sectors.}
\label{fig:PNJL_baryonresult}
\end{figure}

The diquark system is obtained in a similar manner to mesons, as described in Section~\ref{sec:chiralrestoration}. However, the kernel ${\cal K}^{ab}_{i j,m n}$ of the corresponding $T$-matrix equation (cf. Eq.~\eqref{eq:BSPNJL}) contains the quark-quark interaction, obtained by Fierz transformation of the interaction terms of the Lagrangian in Eq.~\eqref{eq:lagPNJL}. In addition, the equivalent $T$-matrix equation, 
\be 
T^{ab}_{i j,k l} (\ii\nu_m,{\bm p}) = {\cal K}^{ab}_{i j,k l} - \sumint_q
{\cal K}^{ac}_{i j, rs} \  \Delta_{r} \left( \ii\omega_n, {\bm k} \right) \ \Delta^c_s \left( \ii\omega_n - \ii\nu_m, {\bm k}-{\bm p} \right)
\ T^{cb}_{r s,k l} (\ii\nu_m,{\bm p}) \ , \label{eq:BSPNJL2}
\ee
contains two quark propagators, one of which is the charge-conjugated propagator, $\Delta_q^c (p)=C^{-1}\Delta^T_q(-p)C$. After factorization, the resulting $T$-matrix element reads
\be 
t^{ab} = \left[ \frac{2G_{\textrm{DIQ}}}{1-2 G_{\textrm{DIQ}} \Pi} \right]^{ab} \ , \label{eq:BSPNJL3}
\ee
where 
\be \Pi^{ab}(\ii \nu_m, \bm{p}) = - \sumint_k \textrm{Tr } \left[ 
    \bar{\Omega}^{a}_{ji} \Delta_i (\ii \omega_n, \bm{k}) \ \Omega^{b}_{ij} \Delta^c_j (\ii \omega_n- \ii \nu_m, \bm{k}-\bm{p}) \right] \ .
\ee
The diagrammatic representation of the $T$-matrix equation is shown in Fig.~\ref{fig:NJL_diquarkdiag}, where the round vertex denotes the original diquark coupling and the fermion line with a $\mathcal{C}$ denotes the charge-conjugated quark propagator. The resummed diagrams provide the $T$-matrix amplitude (denoted with a square vertex on the left-hand side) and  $\Pi^{ab}$ is the two-quark propagator.
The poles of the $T$-matrix give rise to the diquark state in an analogous manner as for mesons in quark-antiquark rescattering, with the key difference that the diquarks are not color singlets, but color antitriplets (the members of the color sextet representations cannot be part of baryons and are therefore discarded).
Once the diquarks have been obtained, they are combined in appropriate spin-flavor channels with an additional quark to generate baryons. For $N_f=3$ the possibilities are discussed in Refs.~\cite{Blanquier:2011zz,Torres-Rincon:2015rma}. In diagrammatic terms, the idea is to iterate diquark-quark scattering using the diquark mass and effective coupling to the quarks obtained in the previous $T$-matrix calculation.
The two-body scheme is shown in the left panel of Fig.~\ref{fig:NJL_baryondiag}.
The resummation of diagrams becomes very complicated due to the box structure of the kernel. Therefore, usually the calculation relies on the {\it static approximation} for the intermediate $u$-channel quark propagation (vertical lines), in which the mass of the quark is assumed to be much greater than the momentum exchange so that the propagator becomes local and generates an effective  2-diquark--2-quark contact interaction. In the right panel of Fig.~\ref{fig:NJL_baryondiag}, the resulting scheme is shown for the first two terms. In this limit, only the diquark-quark propagator presents a nontrivial analytical structure with potential poles in the final quark-diquark scattering amplitude. At $T=0$, all members of the $N_f=3$ baryon octet and baryon decuplet were obtained with very reasonable masses using only two diquark couplings from the NJL Lagrangian as free parameters. Despite the many approximations made along the way, the results were very satisfactory and support a picture of baryons with strong quark-diquark correlations~\cite{Pfaff:2022sfv}. At finite temperature, the calculation is easily extended, with the caveat that the static approximation worsens as the temperature increases, since the quark masses decrease with temperature. In Fig.~\ref{fig:NJL_baryonresult} we present the results for the members of the octet (left panel) and the decuplet (right panel) within the NJL model obtained from Ref.~\cite{Torres-Rincon:2015rma}.
In this plot, the baryon masses are shown in their stability range, that is, only for the temperatures for which the masses are real. Above a certain temperature, this is no longer the case for two possible reasons: either the diquark forming the baryon becomes unstable, generating a decay width to two quarks, or the baryon itself becomes unstable by the generation of an imaginary part in its mass due to the possible decay to a quark--diquark final state.
In Fig.~\ref{fig:PNJL_baryonresult}, we present results for the same baryons in the PNJL model from a recent calculation~\cite{Blanquier:2025ysh}, where several improvements have been incorporated with respect to previous works. In particular, the real parts of the baryon masses are plotted even in the region where these baryons become unstable.

\begin{figure}[!ht] 
\centering
\includegraphics[scale=0.45]{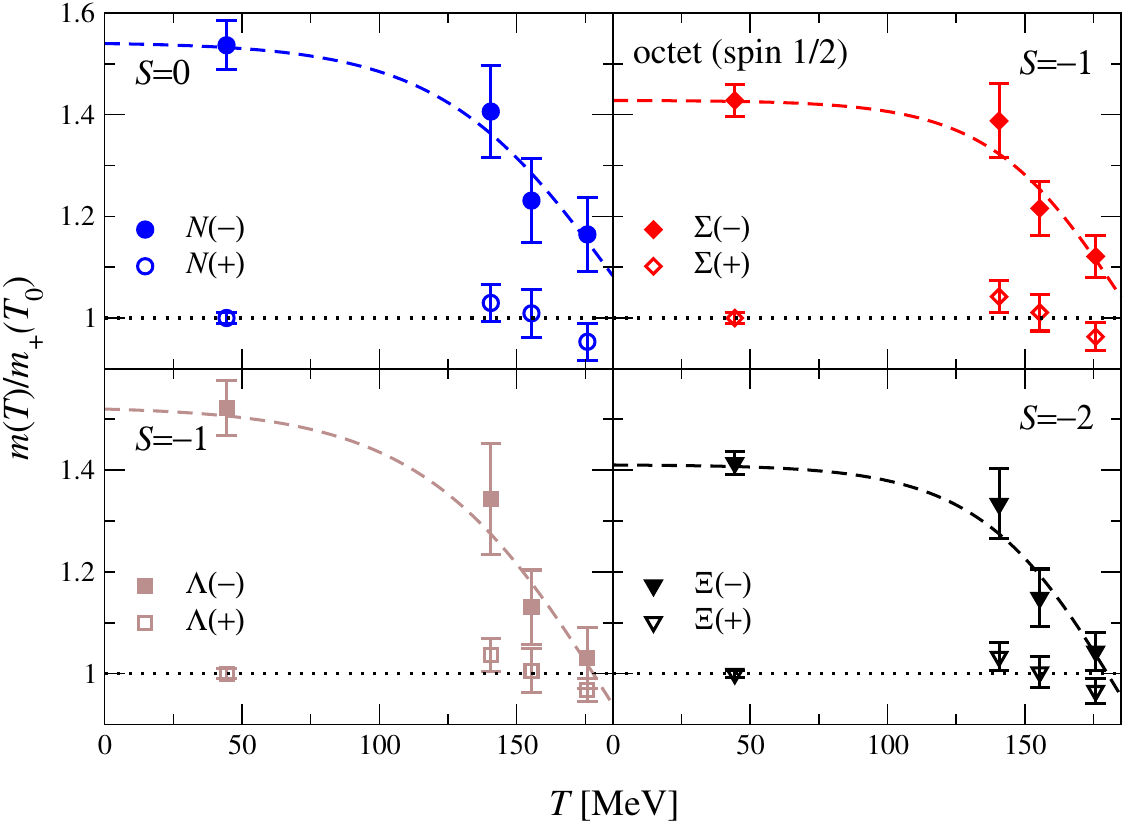}
\includegraphics[scale=0.45]{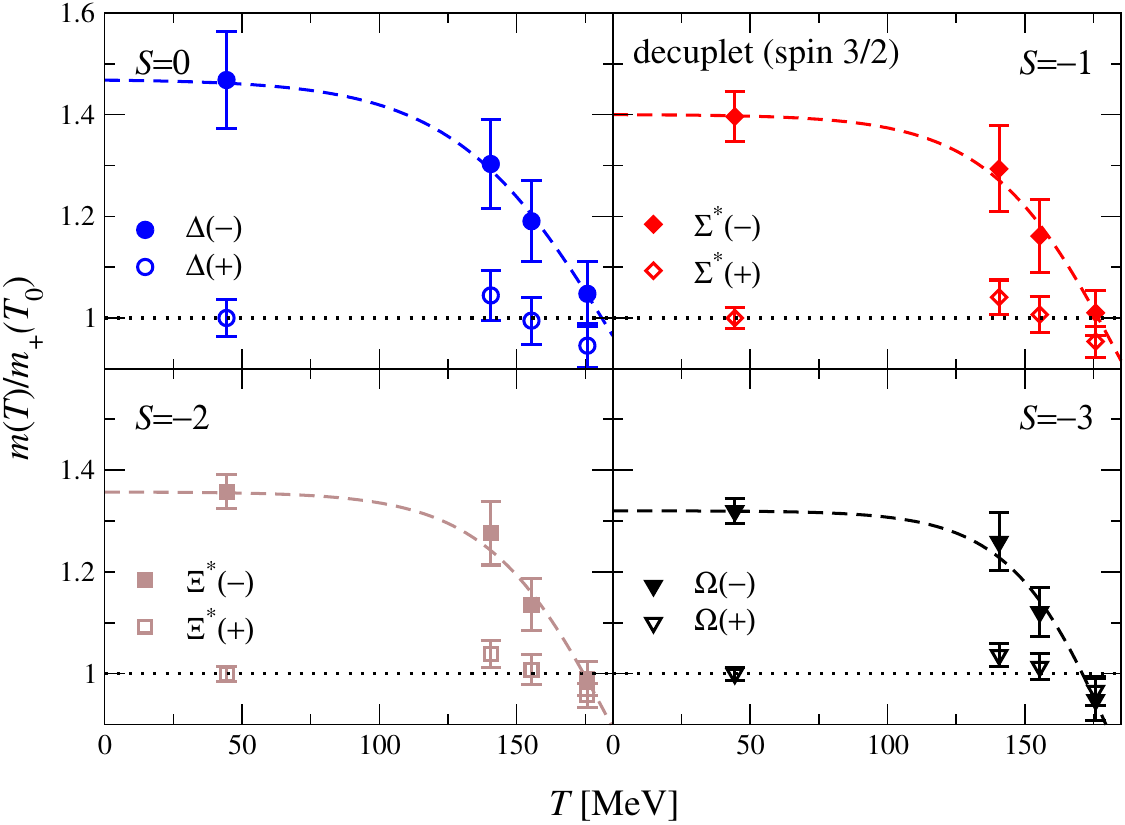}
\caption{Temporal baryon masses calculated by the FASTSUM collaboration in Ref.~\cite{Aarts:2018glk}. The left panel shows the members of the baryon octet ($N, \Sigma, \Lambda, \Xi$) and the right panel the members of the decuplet ($\Delta,\Sigma^*,\Xi^*,\Omega$). All of them are shown with both negative and positive parity ($\mp$).}
\label{fig:baryonlattice}
\end{figure}

From the lattice-QCD perspective, the thermal properties of baryons have not been extensively investigated. 
The nucleon (neutron) thermal screening mass has been considered by the QCD-TARO collaboration in Ref.~\cite{Pushkina:2004wa}. On the other hand, temporal baryon masses have been calculated in more recent works~\cite{Datta:2012fz,Aarts:2015mma,Aarts:2017rrl,Aarts:2018glk}. In Ref.~\cite{Aarts:2018glk}, for example, the FASTSUM collaboration used anisotropic lattices, with $N_f=2+1$ flavors to also account for hyperons. Both octet and decuplet sets of baryons were investigated, paying special attention to the positive- and negative-parity ground states. The physical temperature range spans from $T=44\mev$ to $T=352\mev$. The deconfinement transition for the used higher-than-physical pion mass ($m_\pi = 384(4)\mev$) is estimated to be at $T_c=185(4)\mev$. We present their results in Fig.~\ref{fig:baryonlattice}, showing the members of the octet (decuplet) in the left (right) panel.
The results clearly show that the positive-parity ground states do not present medium modifications, remaining constant within uncertainties. In contrast, the negative-parity baryons show a clear reduction of their masses up to the point of becoming degenerate with the positive-parity states around $T_c$ of the system.
In relative terms, the behavior of the baryon masses looks qualitatively compatible with the (P)NJL results, although in the PNJL model the masses of the negative-parity states remain much more stable (due to the interplay with the Polyakov loop sector) than those seen in lattice QCD.

In summary, while finite-temperature studies of light baryons reveal key phenomena such as thermal damping and parity-partner degeneracy near $T_c$, quantitative theoretical progress remains limited when compared to the meson sector. From the lattice-QCD perspective, baryonic correlators have been restricted to simulations using unphysically heavy pion masses (although the use of new lattice ensembles is lowering this mass~\cite{Aarts:2020vyb}). In continuum effective approaches and phenomenological models, calculations frequently rely on drastic approximations---such as the static limit in the quark-diquark Faddeev equations, or unphysical unbinding thresholds in nonconfining quark models. Looking ahead, future progress relies on high-precision lattice simulations with physical quark masses leveraging advanced spectral reconstruction techniques, as well as the development of nonstatic three-body bound-state kernels, or unitarized in-medium chiral effective theories. These combined advancements will be essential for establishing a quantitative understanding of thermal baryonic excitations and parity restoration across the chiral crossover transition.

\section{Open heavy-flavor hadrons}
\label{sec:openHF}
Following the discussion of light hadrons in the previous section, we now turn to mesons carrying heavy flavor. First, in this section, we concentrate on mesons that contain a single heavy quark ($Q$), while heavy quark-antiquark (hidden-flavor) mesons will be addressed separately in Section~\ref{sec:hiddenHF}. 

In the hot environment created in HICs, and in the limit of vanishing baryonic density, these heavy states propagate through a thermal medium composed of light partons in the QGP phase or light mesons in the hadronic phase. Because heavy quarks are much heavier than the typical temperatures reached in HICs ($m_c\approx 1.3\gev$, $m_b\approx 4.2\gev$ $\gg T_\text{QGP}$), they are produced almost exclusively in the initial stage of the collision. Their relaxation time is larger than the thermalization time of light partons by a factor $\sim m_Q/T$, and they are thus not expected to fully thermalize within the lifetime of the fireball. As a result, heavy hadrons retain valuable information about the history of heavy-flavor interactions with the thermal medium and therefore serve as valuable probes of hot QCD matter.

The theoretical description of heavy hadrons at finite temperature benefits greatly from the hierarchy $m_Q\gg \Lambda_\text{QCD},\, T$. Heavy-light systems can be described using the heavy-quark effective theory (HQET)~\cite{Eichten:1989zv,Georgi:1990um,Neubert:1993mb}, often combined with chiral symmetry via heavy-meson effective theory (HMET)~\cite{Burdman:1992gh,Wise:1992hn}. 
Systems composed of a heavy quark and a light quark are particularly interesting because their interactions with the medium are simultaneously governed by two symmetry regimes. The light-quark dynamics is governed by the (approximate) chiral symmetry, while the heavy quark obeys heavy-quark symmetries in the limit $m_Q \to \infty$. 
In this limit, the interactions become independent of the heavy-quark spin and largely insensitive to its mass, leading to degeneracies between vector and pseudoscalar mesons, and to the same light degree-of-freedom dynamics in the charm and bottom sectors (up to $1/m_Q$ corrections). Furthermore, at finite temperature, the expected chiral symmetry restoration near the critical temperature leads to parity doubling, where the masses of chiral partners (e.g., the $D$ and the scalar $D_{0}^*$) are expected to become degenerate.
The interplay between these symmetries constrains the interactions between open heavy-flavor mesons and the thermal medium.

Although open heavy flavor has long been studied in dense nuclear matter (see, e.g., Refs.~\cite{Rapp:2011zz,Tolos:2013gta,Hosaka:2016ypm,Aarts:2016hap,Das:2024vac} for reviews), comparatively fewer works have focused on its behavior at finite temperature. Much of the early interest in heavy flavor at finite temperature has been driven by the study of quarkonium states, particularly motivated by the observation of $J/\psi$ suppression in HICs~\cite{Gonin:1996wn}. which had been proposed as a signature of deconfinement arising from color screening in the QGP~\cite{Matsui:1986dk}. However, interactions with the medium constituents can strongly modify quarkonium yields and survival probabilities. In this context, open-charm mesons play an important role, as their thermal modifications may affect charmonium absorption through scattering with comoving hadrons~\cite{Gerschel:1998zi,Vogt:1999cu,Capella:2000zp}. Moreover, the heavy-quark potential, which is central to the theoretical description of charmonium suppression, is closely connected to the finite-temperature properties of open-charm mesons, since the energy of an infinitely separated charm-anticharm quark pair can be identified with twice the $D$-meson mass at finite temperature~\cite{Gubler:2020hft}. Therefore, beyond their intrinsic interest, a quantitative understanding of open heavy flavor at finite temperature is essential. 

At finite temperature, open heavy-flavor mesons experience many soft collisions with the light constituents of the medium and can be treated as Brownian particles propagating through a thermal bath. In what follows, we focus primarily on the thermal properties of the ground-state open-charm ($D,~D^*,~D_s,~D_s^*$) and open-bottom ($B,~B^*,~B_s,~B_s^*$) mesons, for which most studies have been performed, while also including results for excited states when available. A wide range of theoretical approaches has been employed to investigate the thermal behavior of these hadrons, including QCD sum rules, lattice QCD, and various effective hadronic models. More recently, a self-consistent finite-temperature approach based on unitarized EFTs has been developed~\cite{Cleven:2017fun,Montana:2020lfi,Montana:2020vjg,MontanaFaiget:2022cog}. Below, we review these studies and discuss their main findings regarding the properties of open heavy-flavor mesons in a hot medium.

\subsection{Heavy-quark effective theory}\label{sec:HQET}
Heavy-quark effective theory (HQET) provides a natural framework for describing singly heavy mesons. This EFT benefits from the simplification that arises from the large mass hierarchy between the heavy-quark mass and the nonperturbative QCD scale, $m_Q\gg \Lambda_{\text{QCD}}$. In this regime, the heavy quark moves almost nonrelativistically within the hadron and, at leading order, behaves as a static color source for the surrounding light quarks and gluons. 
In the limit $m_Q\to\infty$, the dynamics of the light degrees of freedom becomes insensitive to the spin and flavor of the heavy quark. This leads to two approximate symmetries: heavy-quark spin symmetry (HQSS), which arises from the decoupling of the heavy-quark spin, and heavy-quark flavor symmetry (HQFS), which reflects the insensitivity of the interaction to whether the heavy-quark flavor is charm or bottom.  Together, these form a larger approximate heavy-quark spin-flavor symmetry (HQSFS), which strongly constrains the structure of heavy-light interactions both in vacuum and in a thermal medium. 

In this context, HQET provides a systematic tool to separate the physics associated with the two scales~\cite{Eichten:1989zv,Georgi:1990um}. Because the typical momentum exchange between the heavy quark and the light degrees of freedom is of the order of $\Lambda_\text{QCD}$, the heavy quark remains close to its
mass shell ($p_Q^2 = m_Q^2$) and its momentum can be decomposed as $p_Q^\mu=m_Qv^\mu+k^\mu$, where $v^\mu$ is the four-velocity of the hadron ($v^2 = 1$) and the residual momentum $k^\mu \sim \mathcal{O}(\Lambda_{\text{QCD}})$ encodes the soft interactions with the light fields. In this formulation, the dependence on the heavy-quark mass is factored out explicitly. 
In addition, the heavy-quark field can be decomposed into ``large'' and ``small'' components, \mbox{$\psi_Q(x)=e^{- \ii m_Qv\cdot x}\left[h_v(x)+H_v(x)\right]$}, where $h_v$ describes the low-energy degrees of freedom associated with the heavy quark, while $H_v$ represents fluctuations suppressed by powers of $1/m_Q$. Integrating out the small component $H_v$ leads to an effective theory expressed solely in terms of $h_v$.
The resulting HQET Lagrangian is organized as an expansion in powers of $1/m_Q$, where the leading-order term manifestly respects HQSS and HQFS.

At low energies, it is convenient to formulate the effective theory directly in terms of hadronic degrees of freedom rather than quark fields. For systems containing a single heavy quark, this leads to heavy-meson effective theory (HMET), which combines the heavy-quark symmetries of HQET with the approximate chiral symmetry~\cite{Wise:1992hn,Burdman:1992gh,Casalbuoni:1996pg}. In HMET, the ground-state heavy-light mesons are organized into spin doublets (pseudoscalar and vector) that form antitriplets of light flavor, such as $(D^{(*)0},D^{(*)+},D_s^{(*)+})$ or their bottom counterparts. To explicitly preserve HQSS, pseudoscalar and vector mesons are treated on equal footing, and combined into a single superfield, $H_a = \frac{1+\slashed{v}}{2}\left[P_{a,\mu}^* \gamma^\mu - P_a \gamma_5\right]$, where $a$ is the light-flavor $\mathrm{SU}(3)$ index. HQFS implies analogous structures for charm and bottom systems. The interactions with the light pseudo-Goldstone bosons are implemented through a nonlinear realization of chiral symmetry that preserves HQSS. As a result, the HMET Lagrangian is fully constrained by HQSFS and chiral symmetry and depends on a small number of low-energy constants.

\subsection{Hadronic approaches for open heavy-flavor mesons}\label{sec:D-hadronic}

Some of the earliest studies of open-charm mesons at finite temperature were carried out within phenomenological hadronic models, motivated by the role of open heavy flavor in charmonium suppression and transport properties of heavy flavor in relativistic HICs. For a recent review we also refer the reader to Ref.~\cite{Das:2024vac}. 

\begin{figure}[htbp!] 
\centering
    \includegraphics[width=0.4\textwidth]{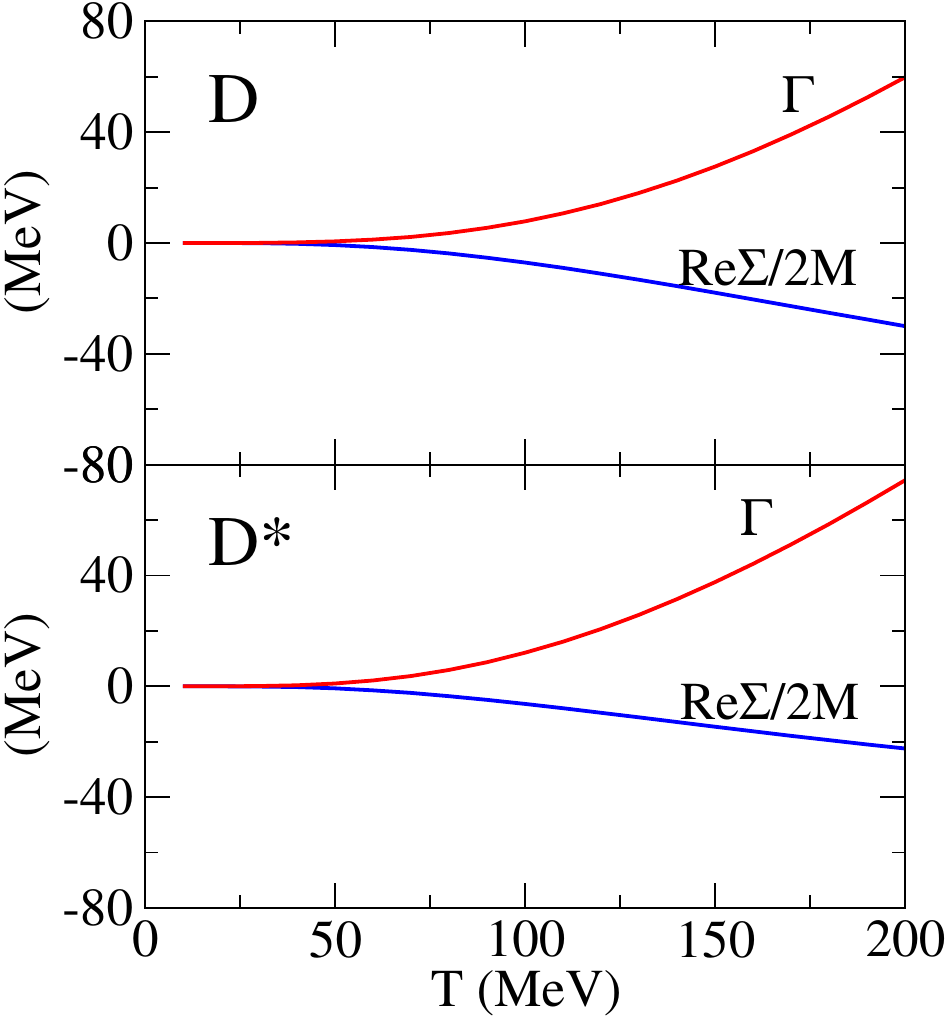}
    \includegraphics[width=0.385\textwidth]{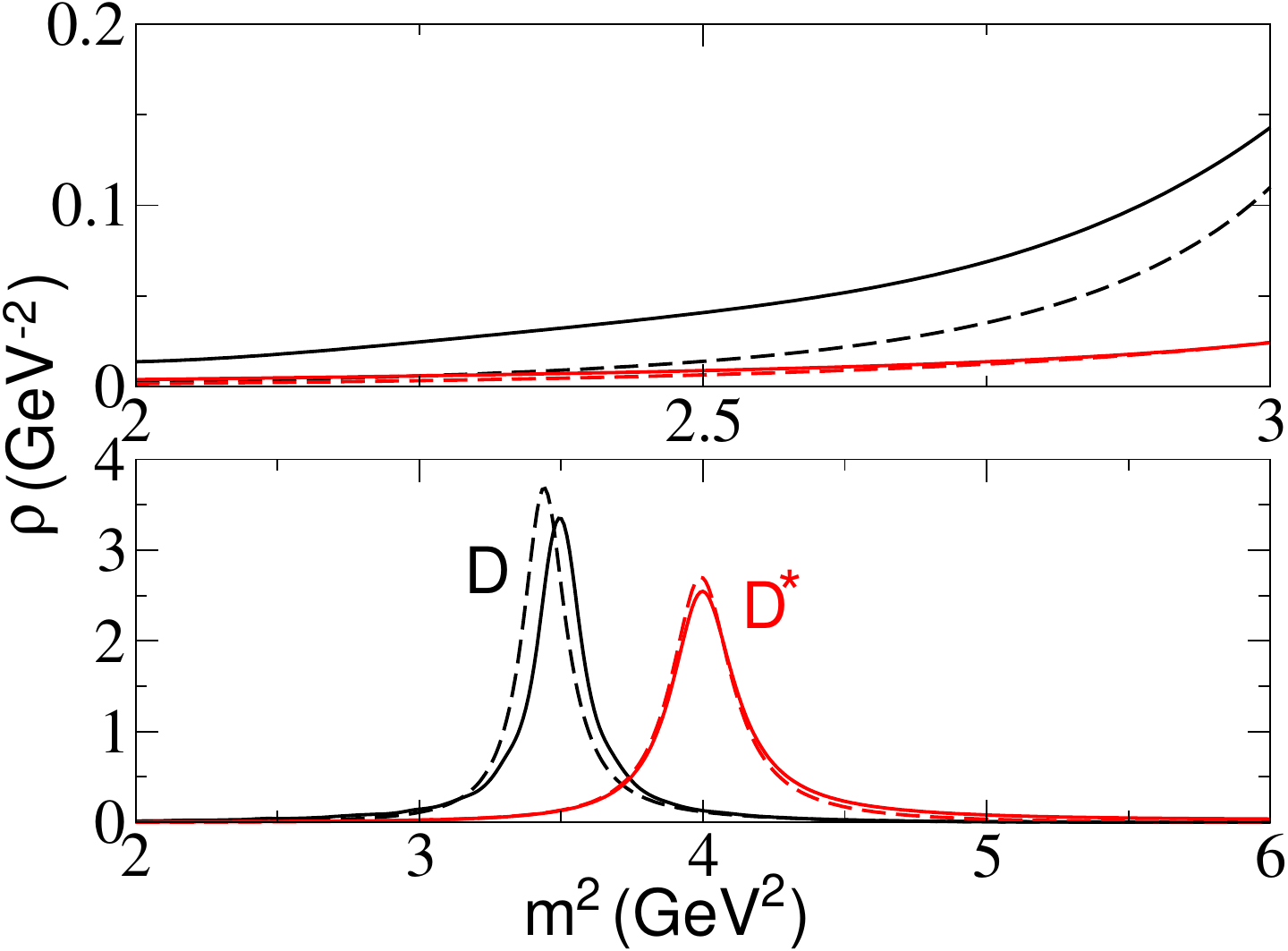}
\caption{Thermal properties of the $D$ and $D^*$ mesons at rest in a pion gas obtained in Ref.~\cite{Fuchs:2004fh}. Left: Collisional width $\Gamma$ and mass shift $\re\Sigma/2M$ as a function of the temperature. Right: Spectral functions at a temperature $T=200\mev$, where dashed and solid curves correspond to the lowest-order and the first iteration of the self-consistent calculation. Figures taken from Ref.~\cite{Fuchs:2004fh}.}
\label{fig:Fuchs}
\end{figure}

In Ref.~\cite{Fuchs:2004fh}, the thermal behavior of the $D$ and $D^*$ mesons in a hot pion gas was explored using relativistic Breit--Wigner parametrizations of the $D\pi$ scattering amplitudes in the isospin $I=1/2$ channel, together with Bose--Einstein distributions for the thermal meson densities. The authors computed the $D^{(*)}$-meson self-energy $\Sigma$ by integrating the scattering amplitude over the thermal distribution of pions, and extracted the temperature dependence of the mass shift ($\re \Sigma^{\text R}/2M$) and collisional width ($\Gamma=-\im\Sigma^{\text R}/M$) from its real and imaginary parts, respectively. The resulting mass shifts and widths are shown in the left panels of Fig.~\ref{fig:Fuchs}. The corresponding spectral functions computed at zeroth and first order in the self-consistent equations that relate the self-energy, the $D$-meson propagator, and the scattering amplitude are displayed in the right panel for $T=200\mev$. At this temperature, a substantial broadening of the spectral peaks is observed, with widths of about $60$--$70\mev$, and a downward mass shift of approximately $30\mev$ for the pseudoscalar and $20\mev$ for the vector meson.
Their results demonstrate that, even in the hadronic phase, open-charm mesons can experience substantial modifications due to interactions with the surrounding thermal bath.

\begin{figure}[htbp!] 
\centering
    \includegraphics[width=0.45\textwidth]{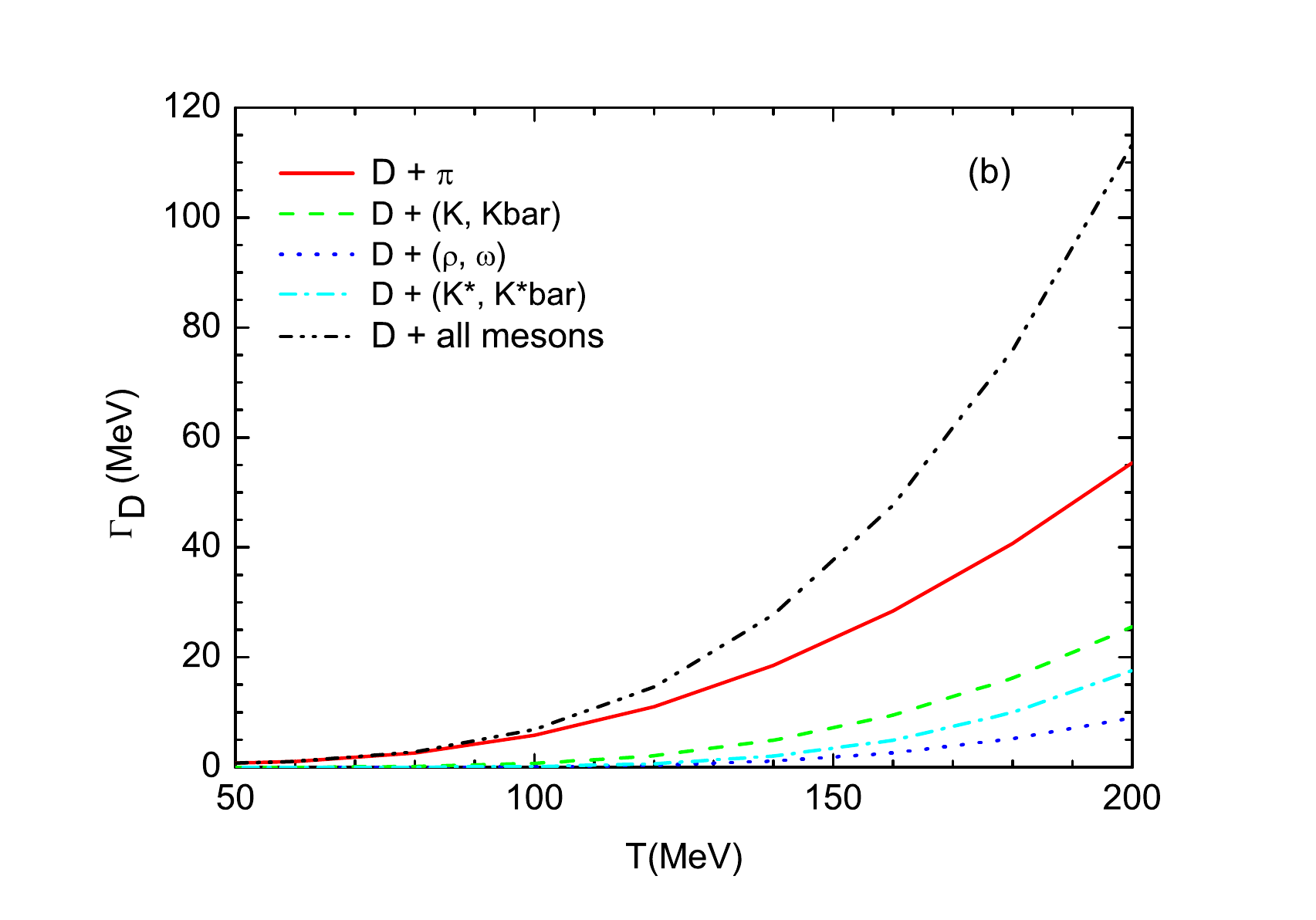}
\caption{$D$-meson collisional width in a hot meson gas as a function of the temperature computed using the Boltzmann equation. The contributions to the total width from individual meson species are also displayed. Figure taken from Ref.~\cite{He:2011yi}.}
\label{fig:He}
\end{figure}

Following the work of Ref.~\cite{Fuchs:2004fh}, the collisional rate of the $D$ meson in a hot meson gas was computed in Ref.~\cite{He:2011yi} using the Boltzmann equation rather than extracting it directly from the imaginary part of the self-energy. The Boltzmann equation describes how the distribution of $D$ mesons evolves due to interactions with the thermal bath. When applied to a dilute gas, it gives the collision rate as an integral over the momentum distributions of the light mesons in the medium, weighted by the corresponding $D$-meson scattering amplitudes. In Ref.~\cite{He:2011yi}, empirical elastic scattering amplitudes were employed, allowing the authors to quantify how pions, kaons, and vector mesons contribute to the total thermal broadening of the $D$ meson. We show the results in Fig.~\ref{fig:He}, which displays the temperature dependence of the total $D$-meson width, together with the individual contributions from each meson species. The contribution arising from the interaction with pions is consistent with the results of Ref.~\cite{Fuchs:2004fh}, which is not surprising given that both studies rely, at lowest order, on vacuum $D\pi$ scattering amplitudes folded with thermal pion distributions. The interactions with heavier mesons become increasingly important as the temperature increases. At $T = 150\mev$, the inclusion of these channels increases the total $D$-meson width to $\Gamma_D \simeq 40\mev$, corresponding to an increase of roughly 50\% compared to the pion-gas contribution alone, with an even steeper rise at higher temperatures.

\begin{figure}[htbp!] 
\centering
    \includegraphics[width=0.32\textwidth]{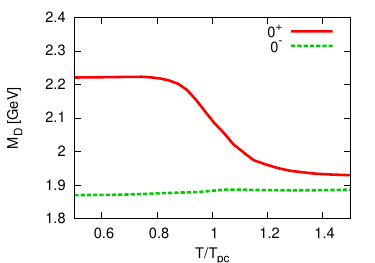}
    \includegraphics[width=0.32\textwidth]{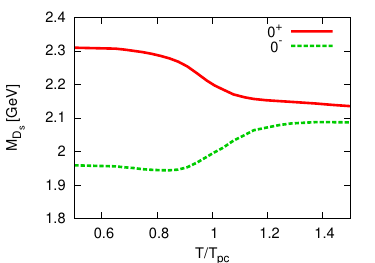}
    \includegraphics[width=0.32\textwidth]{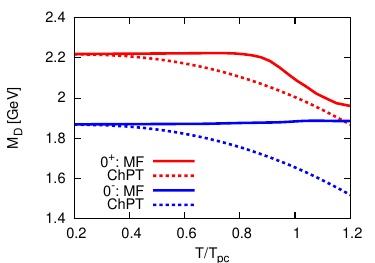}
\caption{Temperature dependence of the $0^\pm$ nonstrange (left panel) and strange (middle panel) charm mesons obtained within the L$\sigma$M incorporating heavy-quark symmetry in the mean-field approximation in Ref.~\cite{Sasaki:2014asa}. The results for the nonstrange states are compared with nonunitarized chiral-EFT (right panel) calculations from the same reference. Figures taken from Ref.~\cite{Sasaki:2014asa}.}
\label{fig:Sasaki}
\end{figure}

The temperature dependence of charmed-meson masses has also been investigated in the context of chiral symmetry restoration in Ref.~\cite{Sasaki:2014asa}, using an extension of the L$\sigma$M for heavy-light mesons that incorporates heavy-quark symmetry, in the mean-field approximation. As shown in Fig.~\ref{fig:Sasaki}, the masses of both strange and nonstrange $0^\pm$ states remain nearly unchanged up to temperatures of order $0.9\,T_c$. Above this point, the mass of the negative-parity ($0^-$) states begins to increase, more markedly for the strange $D_s$ meson, which shifts upward by more than $100\mev$, while the nonstrange $D$ meson shows only a small effect.  In contrast, the positive-parity ($0^+$) states experience a substantial decrease in mass as the temperature goes above $T_c$, with drops of more than $200\mev$ for the nonstrange state and over $100\mev$ for the strange partner. 
This leads to a significant reduction of the mass splitting between parity partners.

For comparison, Ref.~\cite{Sasaki:2014asa} also evaluates the masses of the nonstrange mesons using a finite-temperature framework based on a lowest-order one-loop self-energy calculation, methodologically similar to that of Refs.~\cite{Fuchs:2004fh,He:2011yi}, but using nonunitarized ChPT to model the $D\pi$ interaction. As seen in the right panel of Fig.~\ref{fig:Sasaki}, this leads to a pronounced decrease of more than $300\mev$ in the mass of both the scalar and pseudoscalar states at $1.2\,T_c$.

\subsection{Unitarized thermal EFT approach}\label{sec:unitarizedEFT}

A powerful framework to study the in-medium properties of open heavy-flavor mesons in the hadronic phase is the self-consistent unitarized effective field theory approach developed in Refs.~\cite{Cleven:2017fun,Montana:2020lfi,Montana:2020vjg,MontanaFaiget:2022cog}. 
In the earlier work of Ref.~\cite{Cleven:2017fun}, the interaction was constructed from an $\sufour$ extension of the chiral Lagrangian, which provides a phenomenological framework to incorporate heavy flavors into the symmetry structure~\cite{Gamermann:2006nm,Gamermann:2007fi}. For the pseudoscalar-pseudoscalar and vector-pseudoscalar interactions, the interaction Lagrangians are given, respectively, by
\be
\mathcal{L}_{PPPP}=\frac{1}{12f^2}\text{Tr}\left(J^\mu J_\mu+P^4M\right) \ , \quad \mathcal{L}_{VPVP}=-\frac{1}{4f^2}\text{Tr}\left(J^\mu\mathcal{J}_\mu\right) \ ,
\ee
where $f$ is the meson decay constant ($f_\pi$ for the light mesons and $f_D$ for the charm mesons), and $M={\rm diag}(m_\pi^2,m_\pi^2,2m_K^2-m_\pi^2,2m_D^2-m_\pi^2)$ is the mass matrix accounting for $\suthree$ and $\sufour$ breaking. The currents are defined in terms of the $\sufour$ 15-plets of pseudoscalar mesons, $P$, and vector mesons, $V$,
\be
J_\mu=(\partial_\mu P)P-P(\partial_\mu P) \ ,\quad \mathcal{J}_\mu=(\partial_\mu V_\nu)V^\nu-V_\nu(\partial_\mu V^\nu) \ .
\ee

The later works of Refs.~\cite{Montana:2020lfi,Montana:2020vjg} employ HMET at NLO in the chiral expansion and at LO in the heavy-quark mass expansion~\cite{Guo:2009ct,Geng:2010vw,Abreu:2011ic}. In this approach, the interaction Lagrangian between the heavy mesons ($H,H^*_\mu$) and the light pseudoscalar mesons of the thermal bath ($\Phi$) is given by
\begin{align}
 \mathcal{L}_{\rm LO}&=\langle\nabla^\mu H\nabla_\mu H^\dagger\rangle-m_H^2\langle HH^\dagger\rangle-\langle\nabla^\mu H^{*\nu}\nabla_\mu H^{*\dagger}_{\nu}\rangle+m_{H^*}^2\langle H^{*\nu}H^{*\dagger}_{\nu}\rangle \nonumber \\
 & +\ii g\langle H^{*\mu}u_\mu H^\dagger-Hu^\mu H^{*\dagger}_\mu\rangle+\frac{g}{2m_H}\langle H^*_\mu u_\alpha\nabla_\beta H^{*\dagger}_\nu-\nabla_\beta H^*_\mu u_\alpha H^{*\dagger}_\nu\rangle\epsilon^{\mu\nu\alpha\beta} \ ,
\end{align}
and
\begin{align}\nonumber\label{eq:lagrangianNLO}
 \mathcal{L}_{\rm NLO}=&-h_0\langle HH^\dagger\rangle\langle\chi_+\rangle+h_1\langle H\chi_+H^\dagger\rangle+h_2\langle HH^\dagger\rangle\langle u^\mu u_\mu\rangle \\ \nonumber
 &+h_3\langle Hu^\mu u_\mu H^\dagger\rangle+h_4\langle\nabla_\mu H\nabla_\nu H^\dagger\rangle\langle u^\mu u^\nu\rangle+h_5\langle\nabla_\mu H\{u^\mu,u^\nu\}\nabla_\nu H^\dagger \rangle \\ \nonumber
 &+\tilde{h}_0\langle H^{*\mu}H^{*\dagger}_\mu\rangle\langle\chi_+\rangle-\tilde{h}_1\langle H^{*\mu}\chi_+H^{*\dagger}_\mu\rangle-\tilde{h}_2\langle H^{*\mu}H^{*\dagger}_\mu\rangle\langle u^\nu u_\nu\rangle \\ 
 &-\tilde{h}_3\langle H^{*\mu}u^\nu u_\nu H^{*\dagger}_\mu\rangle-\tilde{h}_4\langle\nabla_\mu H^{*\alpha}\nabla_\nu H^{*\dagger}_\alpha\rangle\langle u^\mu u^\nu\rangle-\tilde{h}_5\langle\nabla_\mu H^{*\alpha}\{u^\mu,u^\nu\}\nabla_\nu H^{*\dagger}_\alpha\rangle \ ,
\end{align}
where $H,H^*_\mu$ are the relativistic fields of the pseudoscalar and vector heavy-meson antitriplets, respectively, i.e., the same states as $P_a$ and $P^*_{a,\mu}$ above, but with relativistic normalization. The light mesons are encoded in $u_\mu=\ii (u^\dagger\partial_\mu u-u\partial_\mu u^\dagger)$ and $\chi_+=u^\dagger \chi u^\dagger+u\chi u$, with $u$ the unitary matrix of Goldstone bosons in the exponential representation, and the mass matrix $\chi={\rm diag}(m_\pi^2,m_\pi^2,2m_K^2-m_\pi^2)$. The low-energy constants $h_i$, $\tilde{h}_i$ that appear at NLO are constrained by lattice-QCD data.

In both cases, the resulting $S$-wave projection of the tree-level interaction, $V$, is unitarized in coupled channels by solving the Bethe--Salpeter equation~\cite{Oller:1997ng,Oset:1997it},
\begin{equation}\label{eq:bs}
T = V + VGT \ ,
\end{equation}
where $G$ is the loop function describing the intermediate propagation of a heavy and a light meson. The unitarized amplitude $T$ is then embedded into a finite-temperature, self-consistent scheme, allowing the ground states and the excited states that are dynamically generated within the EFT to acquire thermal masses and widths.

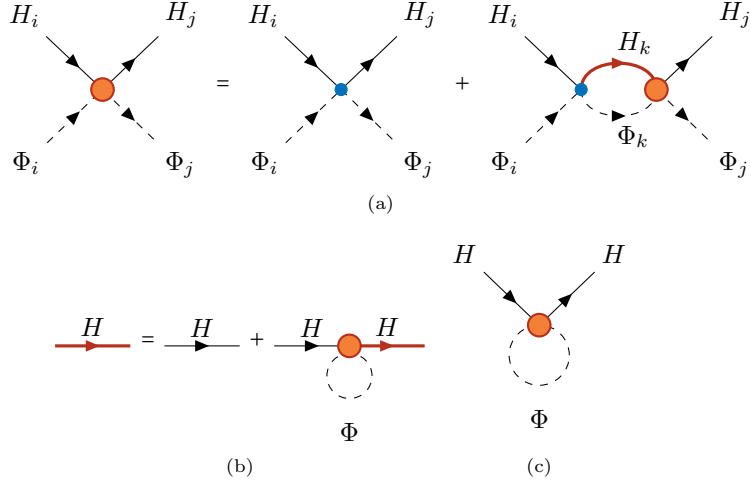
\begin{figure}[htb!]
\centering 
\begin{subfigure}[b]{\textwidth}\centering 
\captionsetup{skip=0pt}
  \begin{tikzpicture}[baseline=(i.base)]
     \begin{feynman}[small]
       \vertex (i) ;
       \vertex [above left = of i] (a) {\(H_i\)};
       \vertex [above right = of i] (b) {\(H_j\)};
       \vertex [below right = of i] (d) {\(\Phi_j\)};
       \vertex [below left = of i] (c) {\(\Phi_i\)};
       \diagram* {
         (a) -- [fermion] (i), 
         (i) -- [fermion] (b),
         (c) -- [charged scalar] (i),
         (i) -- [charged scalar] (d),
        };
      \draw[dot,minimum size=4mm,thick,BrickRed,fill=Orange] (i) circle(1.5mm);
     \end{feynman}
   \end{tikzpicture}
   $=$
   \begin{tikzpicture}[baseline=(i.base)]
     \begin{feynman}[small]
       \vertex (i) ;
       \vertex [above left = of i] (a) {\(H_i\)};
       \vertex [above right = of i] (b) {\(H_j\)};
       \vertex [below right = of i] (d) {\(\Phi_j\)};
       \vertex [below left=of i] (c) {\(\Phi_i\)};
       \diagram*{
         (a) -- [fermion] (i), 
         (i) -- [fermion] (b),
         (c) -- [charged scalar] (i),
         (i) -- [charged scalar] (d),
        } ;    
      \draw[dot,RoyalBlue,fill=RoyalBlue] (i) circle(.8mm);
     \end{feynman}
   \end{tikzpicture}
   $+$
   \begin{tikzpicture}[baseline=(i.base)]
     \begin{feynman}[small]
       \vertex (i) ;
       \vertex [above left = of i] (a) {\(H_i\)};
       \vertex [right = of i] (j);
       \vertex [above right = of j] (b) {\(H_j\)};
       \vertex [below right = of j] (d) {\(\Phi_j\)};
       \vertex [below left=of i] (c) {\(\Phi_i\)};
       \vertex [above right =0.5cm of i] (b1) {\(H_k\)};
       \vertex [below right =0.5cm of i] (d1) {\(\Phi_k\)};
       \diagram*{
         (a) -- [fermion] (i), 
         (j) -- [fermion] (b),
         (i) -- [BrickRed, fermion, very thick, half left, looseness=1.2] (j),
         (i) -- [charged scalar, half right, looseness=1.2] (j),
         (c) -- [charged scalar] (i),
         (j) -- [charged scalar] (d),
        } ;    
      \draw[dot,RoyalBlue,fill=RoyalBlue] (i) circle(0.8mm);
      \draw[dot,minimum size=4mm,thick,BrickRed,fill=Orange] (j) circle(1.5mm);
     \end{feynman}
   \end{tikzpicture}
\caption{}
\label{fig:selfcons-a}
\end{subfigure}
\\[0.2cm]
\hspace{-3cm}\begin{subfigure}[b]{0.6\textwidth}\centering 
\captionsetup{skip=0pt}
  \begin{tikzpicture}[baseline=(a.base)]
     \begin{feynman}[small]
       \vertex (a);
       \vertex [right = of a] (b);
       \diagram* {
         (a) -- [BrickRed, fermion, very thick,edge label=\(\textcolor{black}{H}\)] (b), 
        };
     \end{feynman}
   \end{tikzpicture}
   $=$
  \begin{tikzpicture}[baseline=(a.base)]
     \begin{feynman}[small]
       \vertex (a);
       \vertex [right = of a] (b);
       \diagram* {
         (a) -- [fermion, edge label=\(H\)] (b), 
        };
     \end{feynman}
   \end{tikzpicture}
   $+$
   \begin{tikzpicture}[baseline=(a.base)]
     \begin{feynman}[small, inline=(a)]
       \vertex (i) ;
       \vertex [left = of i] (a);
       \vertex [right = of i] (b);
       \vertex [below = 0.4cm of i] (d);
       \vertex [below = 0.9cm of i] (e) {\(\Phi\)};
       \diagram* {
         (a) -- [fermion,edge label=\(H\)] (i), 
         (i) -- [BrickRed, fermion, very thick,edge label=\(\textcolor{black}{H}\)] (b),
        } ;   
      \draw[dashed] (d) circle(0.3cm);
      \draw[dot,minimum size=4mm,thick,BrickRed,fill=Orange] (i) circle(1.5mm);
     \end{feynman}
   \end{tikzpicture}
\caption{}
\label{fig:selfcons-b}
\end{subfigure}\hspace{-4cm}
\begin{subfigure}[b]{0.25\textwidth}\centering
\captionsetup{skip=5pt}
   \begin{tikzpicture}
     \begin{feynman}[small]
       \vertex (i) ;
       \vertex [above left = of i] (a) {\(H\)};
       \vertex [above right = of i] (b) {\(H\)};
       \vertex [below = 0.4cm of i] (d);
       \vertex [below = 1cm of i] (e) {\(\Phi\)};
       \diagram* {
         (a) -- [fermion] (i), 
         (i) -- [fermion] (b),
        } ;   
      \draw[dashed] (d) circle(0.4cm);
      \draw[dot,minimum size=4mm,thick,BrickRed,fill=Orange] (i) circle(1.5mm);
     \end{feynman}
   \end{tikzpicture}
\caption{}
\label{fig:selfcons-c}
\end{subfigure}
\caption{(a) Coupled-channel Bethe--Salpeter equation. At finite temperature, the $T$ matrix (large circle) follows from unitarizing the interaction kernel (small circle) with a two-meson loop including a dressed heavy-meson propagator (thick red lines).  (b) Dyson equation for the dressed heavy-meson propagator. (c) Heavy-meson self-energy.}
\label{fig:selfcons}
\end{figure}

In particular, finite-temperature effects are incorporated through the ITF. After summing over Matsubara frequencies and performing the analytic continuation to real energies, the thermal two-meson loop function takes the form:
\begin{align}\label{eq:thermalloop}
G_{H\Phi}(E,\bm{p}\, ;T)=\int \frac{\diff^3q}{(2\pi)^3}\int \diff\omega\int \diff\omega' \ \frac{S_H(\omega,\bm{q}\, ;T)S_\Phi(\omega',\bm{p}-\bm{q}\, ;T)}{E-\omega-\omega'+\ii\epsilon}\left[1+n_{\text{B}}(\omega,T)+n_{\text{B}}(\omega',T)\right] \ ,
\end{align}
where the integrals over energy run from $-\infty$ to $+\infty$, and $n_{\text{B}}(\omega,T)$ is the Bose--Einstein distribution function  defined in Eq.~\eqref{eq:BE}. $\Phi$ generically denotes any of the light mesons. The ultraviolet divergences present in the vacuum loop contribution must be regularized. Ref.~\cite{Cleven:2017fun} uses dimensional regularization in vacuum and defines the thermal correction by subtracting the corresponding $T=0$ contribution (regularized with a sharp three-momentum cutoff). Alternatively, Refs.~\cite{Montana:2020lfi,Montana:2020vjg} apply the same momentum cutoff at all temperatures, ensuring a uniform treatment.

The heavy-meson spectral function is given by \begin{equation}
    S_H(\omega,\bm{q}\, ;T)=-\frac{1}{\pi}\im{\mathcal{D}^{\text R}}(\omega,\bm{q}\, ;T)=-\frac{1}{\pi}\im\left(\frac{1}{\omega^2-\bm{q}\,^2-m^2-\Pi^{\text R} (\omega,\bm{q},T)}\right) \ ,
\end{equation}
where the in-medium self-energy $\Pi^{\text R}(\omega,\bm{q}\, ;T)$ encodes thermal corrections to the heavy-meson propagator (Fig.~\ref{fig:selfcons-b}). 
The dressing of the light mesons is neglected, since their thermal modifications are mild (see Section~\ref{sec:light}) and would only provide subleading corrections.

The heavy-meson self-energy $\Pi^{\text R}(\omega,\bm{p}\, ;T)$ is obtained self-consistently through a loop integral over the in-medium scattering amplitude:
\be
\label{eq:selfenergy}
    \Pi^{\text R}(\omega,\bm{p}\,;T)=-\frac{1}{\pi} \int\frac{\diff^3q}{(2\pi)^3}\int \diff\Omega \frac{\omega}{\omega_\Phi}
    \frac{n_{\text{B}}(\Omega,T)-n_{\text{B}}(\omega_\Phi,T)}{\omega^2-(\omega_\Phi-\Omega)^2+\ii\epsilon}\,\im T_{H\Phi}(\Omega,\bm{p}+\bm{q}\,;T) \ .
\ee
Here $\omega_\Phi=\sqrt{\bm{q}^2+m_\Phi^2}$ is the on-shell energy of the light meson. This expression includes only thermal corrections in $\Pi^{\text R}$~\cite{Montana:2020lfi,Montana:2020vjg}. Alternatively, Ref.~\cite{Cleven:2017fun} employed the full self-energy, which also included a vacuum contribution, and subsequently subtracted the $T=0$ self-energy to isolate the genuine thermal effects. 

While technical differences in the implementation of the thermal medium result in quantitative differences, the self-consistent approach, diagrammatically depicted in Fig.~\ref{fig:selfcons}, yields a common qualitative picture for the temperature-dependent spectral functions and scattering amplitudes.

\begin{figure}[htbp!] 
\centering
    \includegraphics[width=0.9\textwidth]{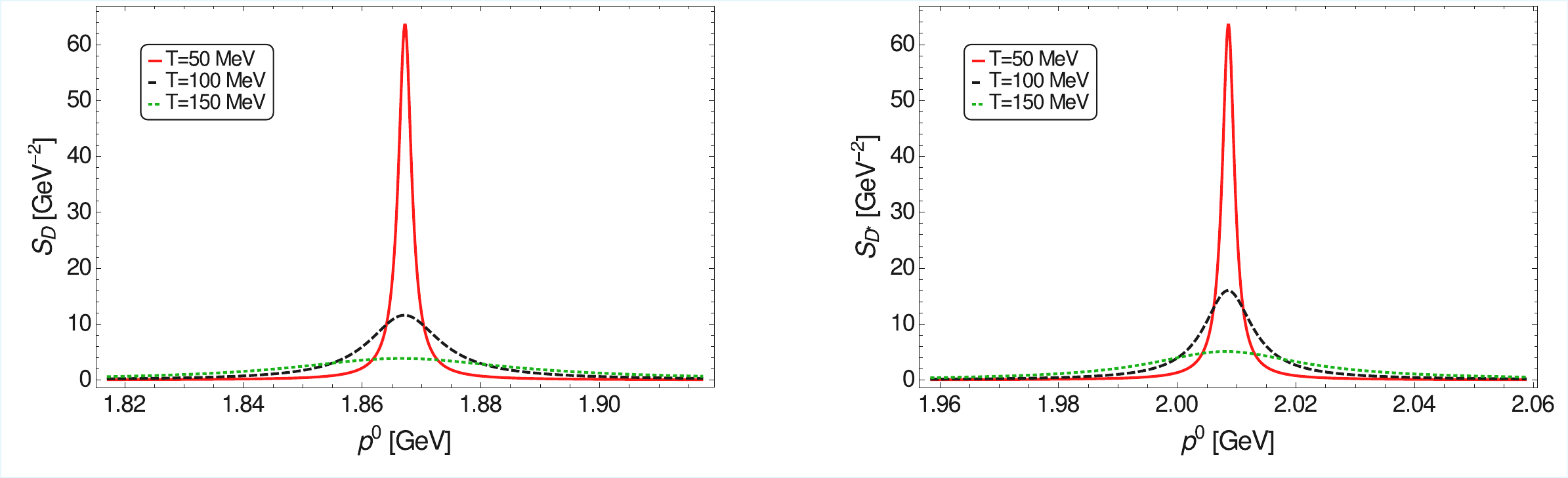}
    \includegraphics[width=0.8\textwidth]{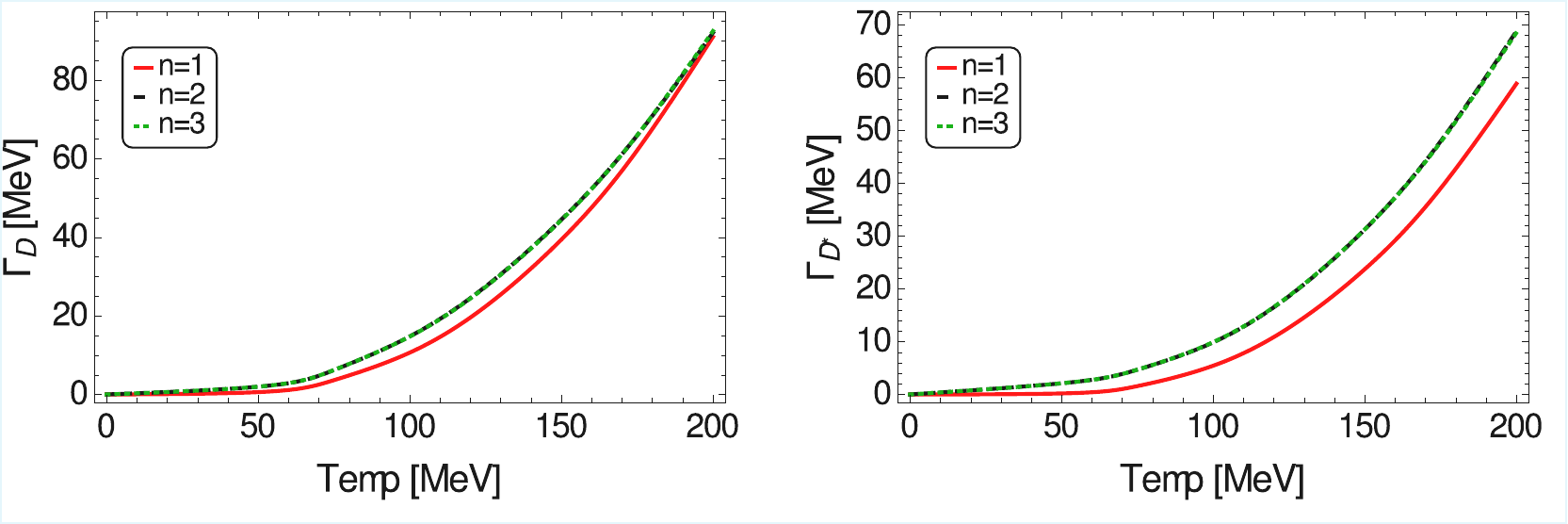}
\caption{Thermal properties of the $D$ (left) and $D^*$ mesons (right) obtained within the self-consistent unitarized $\sufour$ EFT framework of Ref.~\cite{Cleven:2017fun}. Top panels: Spectral functions at rest ($\bm p=0$) as a function of the energy $p^0$ for three different temperatures. Bottom panels: Thermal decay width as a function of the temperature for three subsequent iterations of the self-consistent approach (see details in the text). Figures taken from Ref.~\cite{Cleven:2017fun}.}
\label{fig:Cleven-Dspectralfunctions}
\end{figure}

Figure~\ref{fig:Cleven-Dspectralfunctions} shows the spectral functions of the $D$ and $D^*$ mesons at various temperatures (top panels) and the corresponding temperature dependence of the thermal decay widths (bottom panels), as obtained in Ref.~\cite{Cleven:2017fun}. A clear thermal broadening is observed. In these calculations, the authors set the real parts of the corresponding self-energies to zero, thereby neglecting any temperature-induced mass shifts.

\begin{figure}[htbp!] 
\centering
    \includegraphics[width=0.9\textwidth]{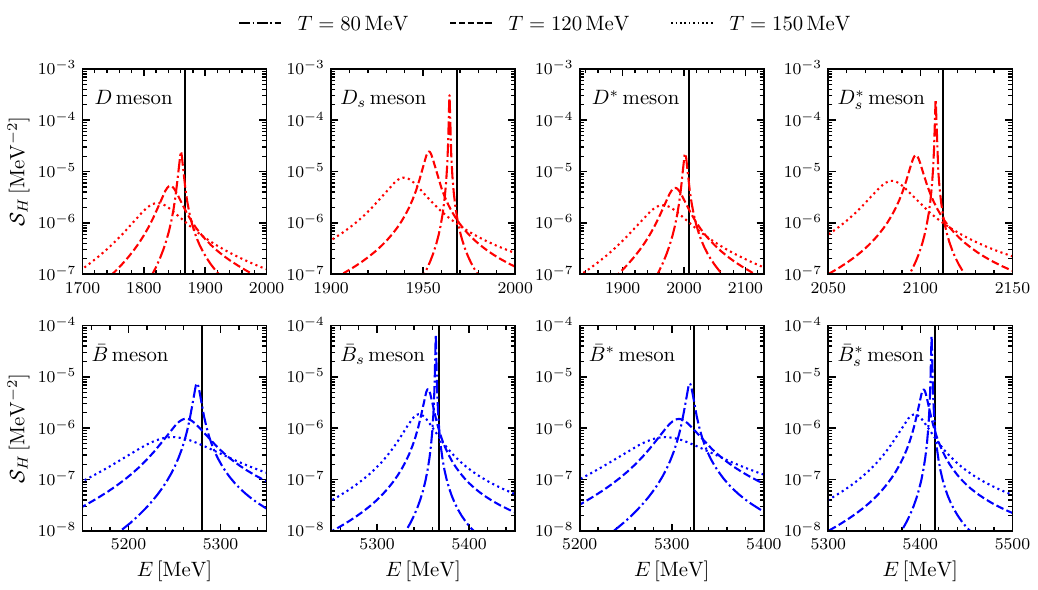}
\caption{Thermal spectral functions of the ground-state open heavy-flavor mesons ($D$, $D^{*}$, $D_{s}$, $D_{s}^{*}$, and their bottom counterparts $B$, $B^{*}$, $B_{s}$, $B_{s}^{*}$), computed using the self-consistent unitarized HMET approach of Refs.~\cite{Montana:2020lfi,Montana:2020vjg,MontanaFaiget:2022cog}. Figure taken from Ref.~\cite{Montana:2023sft}.}
\label{fig:MontanaGSSpectralFunctions}
\end{figure}

The same framework has been applied in Refs.~\cite{Montana:2020lfi,Montana:2020vjg} to study the thermal properties of heavy-light pseudoscalar and vector ground-state mesons in the charm sector, and subsequently extended to the bottom sector in Ref.~\cite{Montana:2023sft}. The resulting spectral functions are displayed in Fig.~\ref{fig:MontanaGSSpectralFunctions}. Besides the expected thermal broadening, a moderate downward mass shift is also observed. Although small compared to the values of the meson masses in vacuum, the mass shift is well defined within the self-consistent EFT framework. 

\begin{figure}[htbp!] 
\centering
    \includegraphics[width=0.6\textwidth]{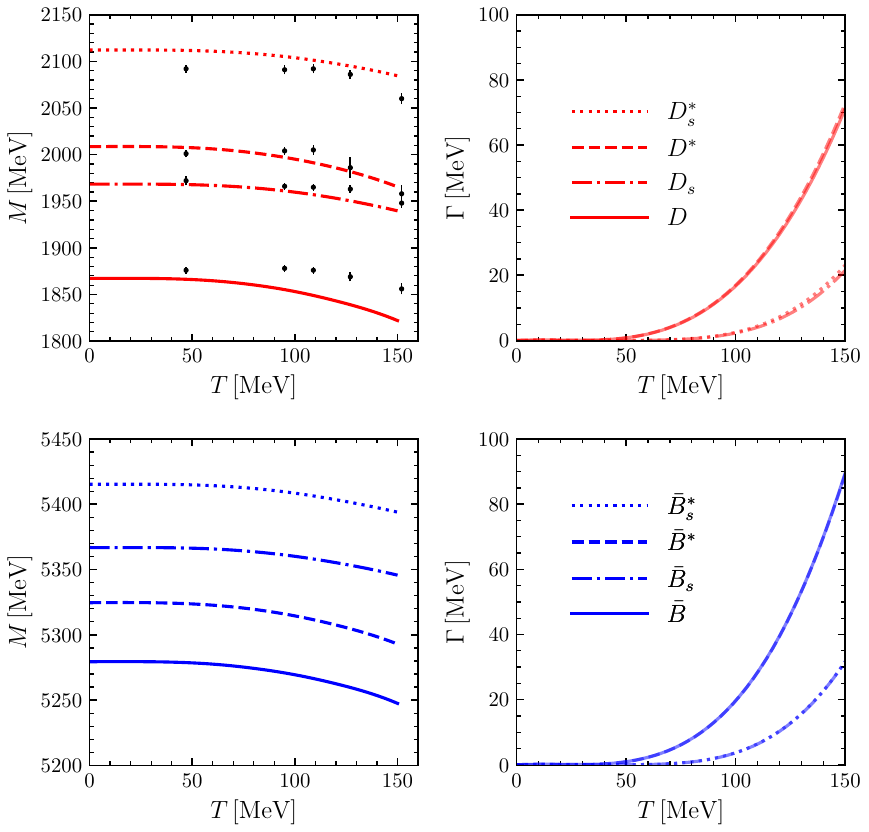}
\caption{Temperature dependence of the masses (left) and widths (right) of the ground-state heavy-light mesons extracted from the spectral functions in Fig.~\ref{fig:MontanaGSSpectralFunctions}. Lattice-QCD results of Ref.~\cite{Aarts:2022krz} ($m_\pi=239\mev$) are included in the upper left panel. Figure taken from Ref.~\cite{Montana:2023sft}.}
\label{fig:MontanaMassWidth}
\end{figure}

The temperature dependence of the masses and widths extracted from these spectral functions is shown in Fig.~\ref{fig:MontanaMassWidth}. As the temperature approaches the applicability limit of the EFT, $T=150\mev$, the ground-state masses decrease by several tens of MeV. Lattice QCD results in the charm sector~\cite{Aarts:2022krz} are also shown for comparison (upper left panel), where a smaller mass shift is found, depending on the channel. This difference can be attributed to the heavier pion mass used in the lattice simulations (see Section~\ref{sec:D-LQCD}), which suppresses thermal medium effects and thus reduces the in-medium modification of the heavy mesons. The right panels of Fig.~\ref{fig:MontanaMassWidth} display the corresponding thermal widths, which grow rapidly with temperature. For nonstrange mesons, the widths reach $70$–$90\mev$ at $T=150\mev$, whereas strange partners broaden more moderately, with widths up to $20$–$30\mev$.

\begin{figure}[htbp!] 
\centering
    \includegraphics[width=0.9\textwidth]{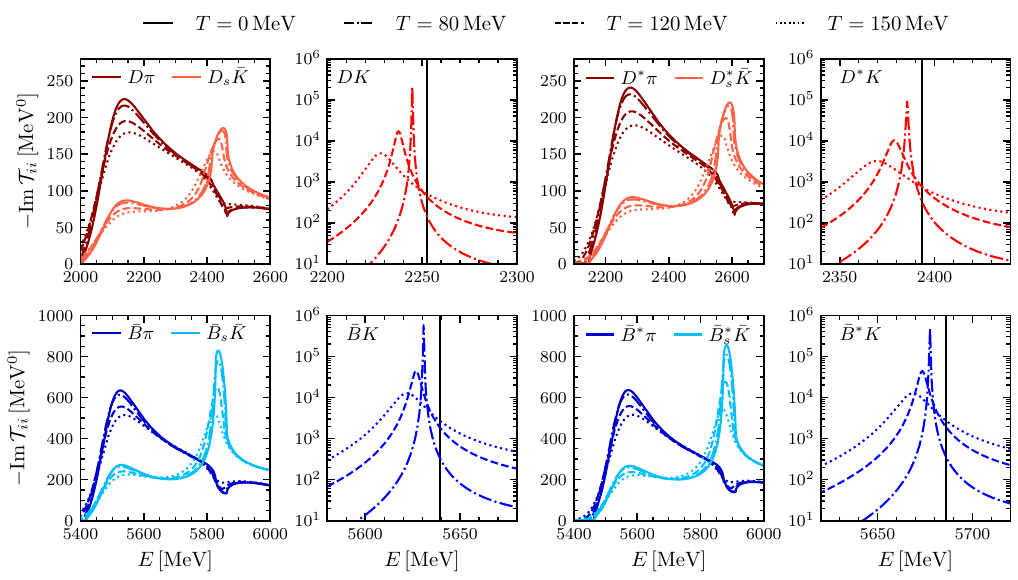}
\caption{Imaginary part of the $T$-matrix diagonal elements showing the lineshapes of the dynamically generated excited heavy-light mesons in the charm and bottom sectors. Figure taken from Ref.~\cite{Montana:2023sft}.}
\label{fig:MontanaExcitedSpectralFunctions}
\end{figure}

Refs.~\cite{Montana:2020lfi,Montana:2020vjg,Montana:2023sft} also reported results for the lowest-lying excited states that are dynamically generated within this unitarized framework. These include the double-pole structure associated with the $D_{0}^*(2300)$ in the $J^P=0^+$ sector, as well as the $D_{1}(2430)$ in the $J^P=1^+$ sector, and the strange $D_{s0}^*(2317)$ and $D_{s1}(2460)$ bound states and their bottom counterparts. The imaginary part of the $T$ matrix, as a proxy for their thermal spectral functions~\cite{Montana:2020vjg}, is shown in Fig.~\ref{fig:MontanaExcitedSpectralFunctions}. The thermal evolution of the bound states follows the same trend as that of the ground states, with moderate downward mass shifts and increasing widths with temperature. For the broad resonant states, the thermal broadening is visible directly in the lineshapes, although the precise determination of the temperature evolution of the pole positions characterizing the thermal masses and widths would require extending the self-consistent framework to the complex-energy plane.

These $(0^+, 1^+)$ excited states may be interpreted as the chiral partners of the ground-state $(0^-,1^-)$ mesons. The implications of their thermal modifications for the restoration of chiral symmetry in the charm sector were examined in detail in Ref.~\cite{Torres-Rincon:2021wyd}.

\subsection{Lattice QCD for open heavy-flavor mesons}\label{sec:D-LQCD}

The spectral properties of open heavy-flavor mesons at finite temperature have also been studied using lattice QCD. The meson spectral function is related to the relativistic meson correlator in Euclidean time by an integral transform of the form
\begin{equation}
    \mathcal{C}(\tau,\bm{p};\, T)=\int_0^\infty \diff \omega\, K(\tau,\omega;\, T)\,\rho(\omega,\bm{p};\, T) \ ,
\end{equation}
where $\mathcal{C}$ is the Euclidean correlator, $\rho$ is the (unknown) spectral function, and $K$ is a known temperature-dependent integration kernel.
Extracting the spectral function from a finite set of noisy correlator data constitutes an ill-posed inverse problem, further aggravated at finite temperature by the limited extent in Euclidean time. In this context, anisotropic lattices (with lattice spacings $a_\tau\ll a_s$) are often employed to enhance temporal resolution and improve sensitivity to spectral features. To address the inverse problem, several approaches have been developed, including fitting sophisticated physically motivated Ansätze for the spectral function, applying Bayesian inference methods such as the \textit{Maximum Entropy Method} (MEM)~\cite{Asakawa:2000tr} and \textit{Bayesian Reconstruction} (BR)~\cite{Burnier:2013nla} to determine the most probable spectral function, and reconstructing a smeared version of the spectral function based on the Backus--Gilbert method~\cite{Backus:1968svk}. For a comprehensive overview of these techniques, see Ref.~\cite{Rothkopf:2022fyo}.

Unlike hidden heavy-flavor mesons (see Section~\ref{sec:hiddenHF}), for which additional simplifications can often be exploited, open-charm mesons require a relativistic treatment of the light quark, making simulations more demanding. Consequently, only a few studies have addressed their thermal behavior.

\begin{figure}[!ht] 
\centering
\begin{subfigure}{0.48\textwidth}
    \includegraphics[width=0.49\textwidth]{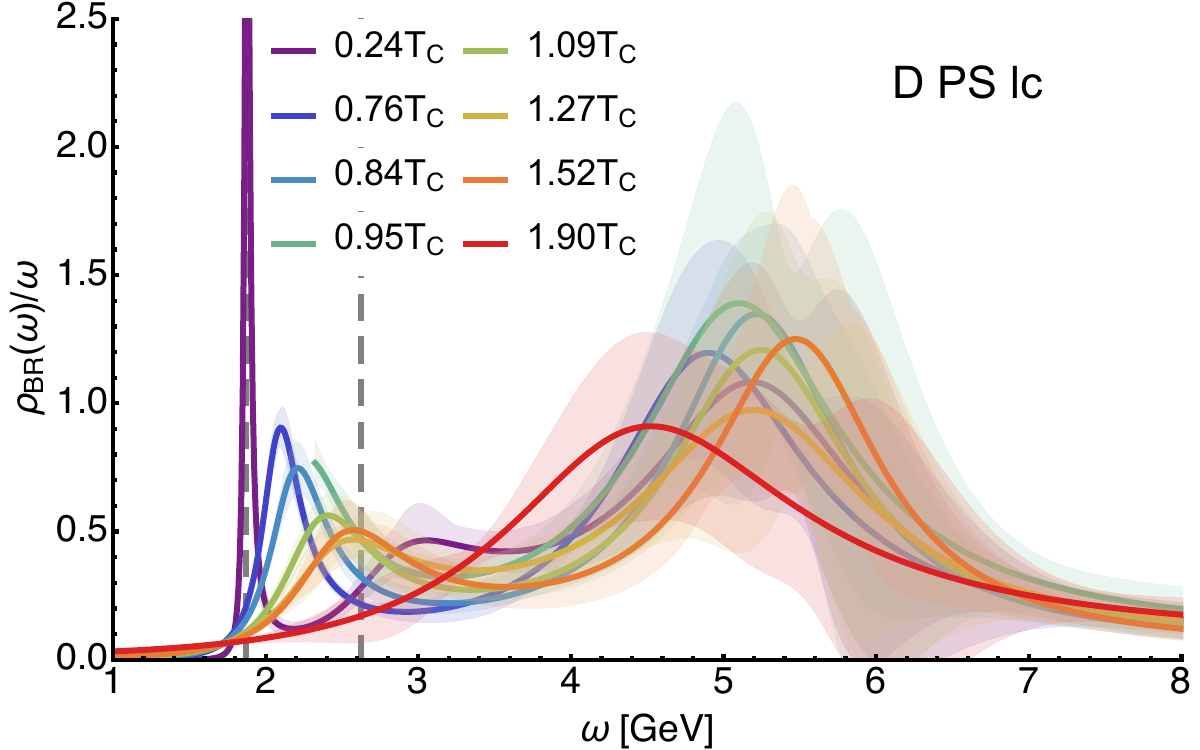}
    \includegraphics[width=0.49\textwidth]{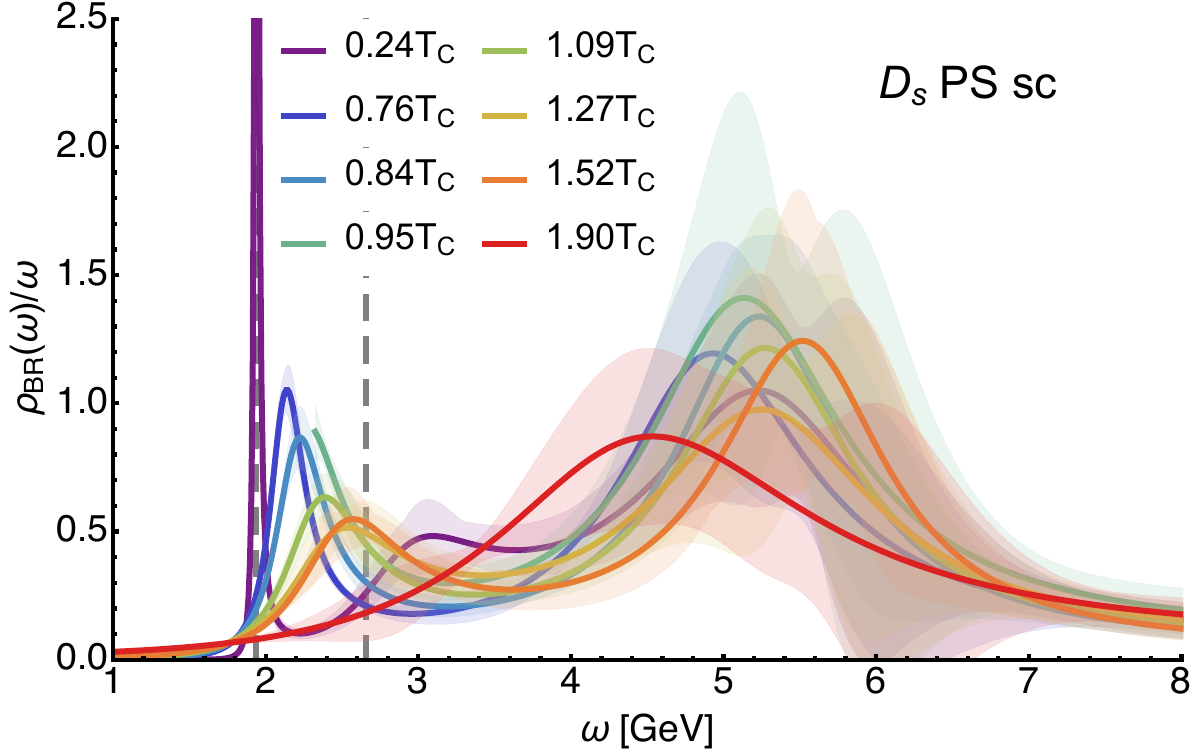} \\
    \includegraphics[width=0.49\textwidth]{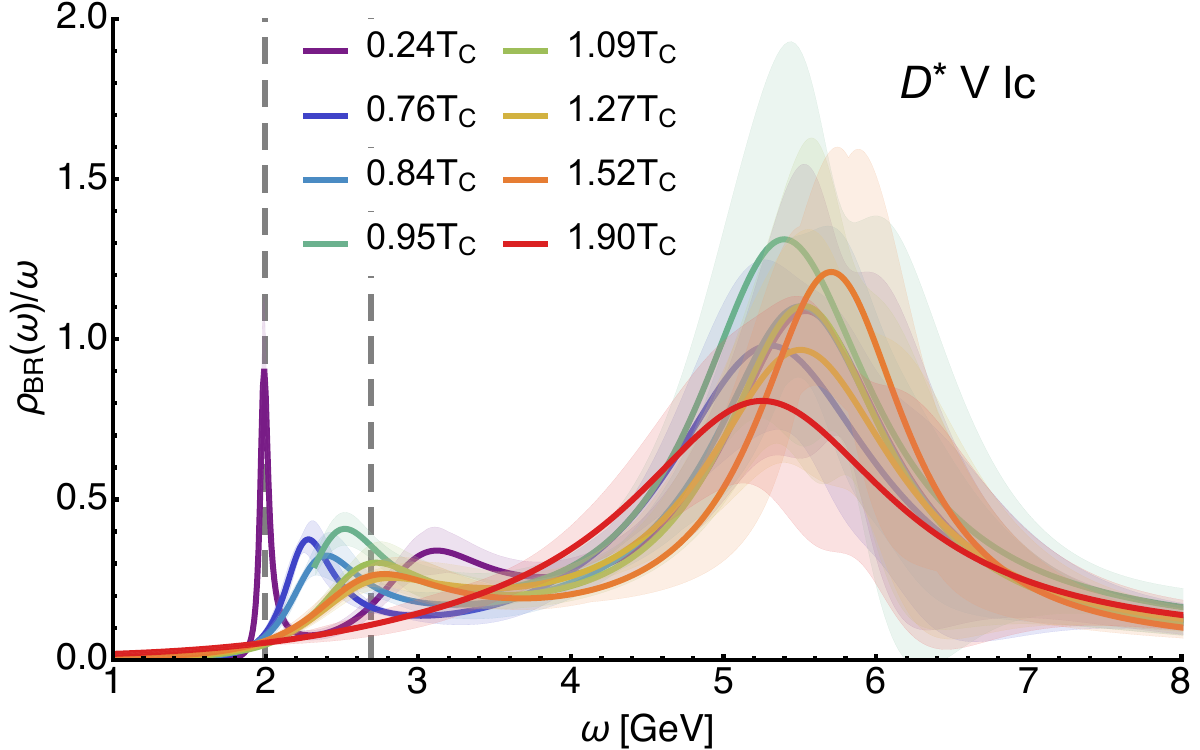}
    \includegraphics[width=0.49\textwidth]{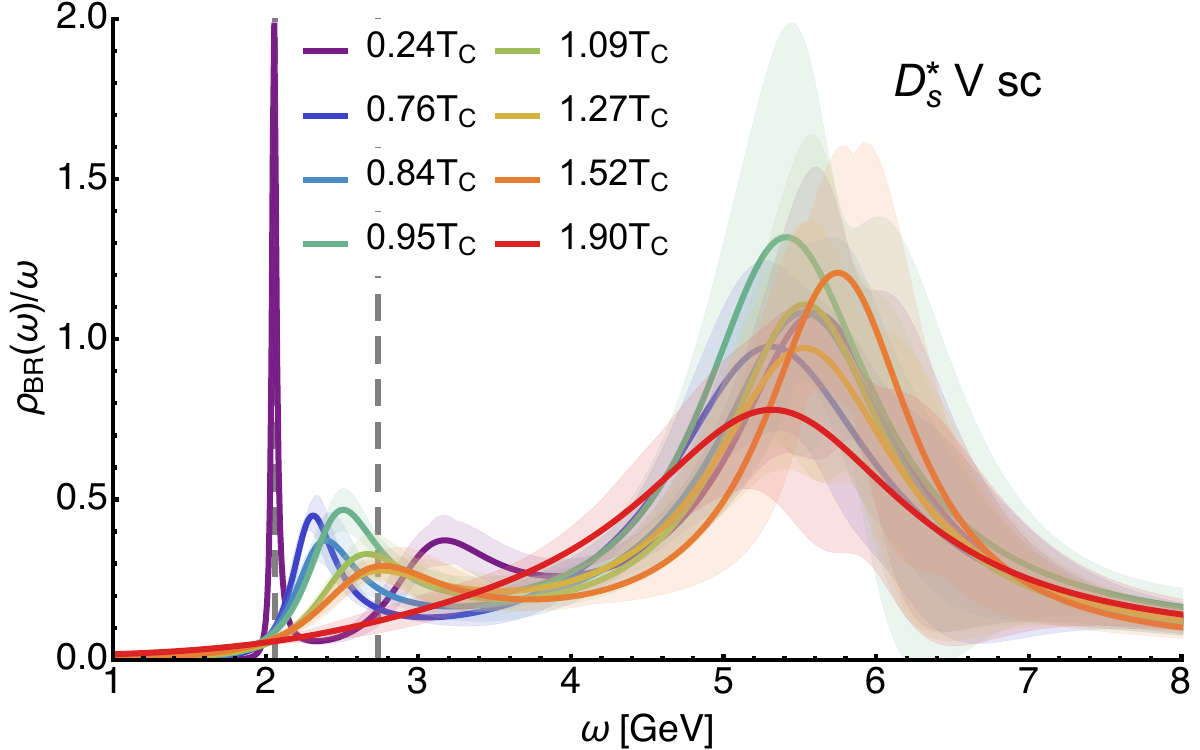}
    \caption{BR reconstruction}
\end{subfigure}
\begin{subfigure}{0.48\textwidth}
    \includegraphics[width=0.49\textwidth]{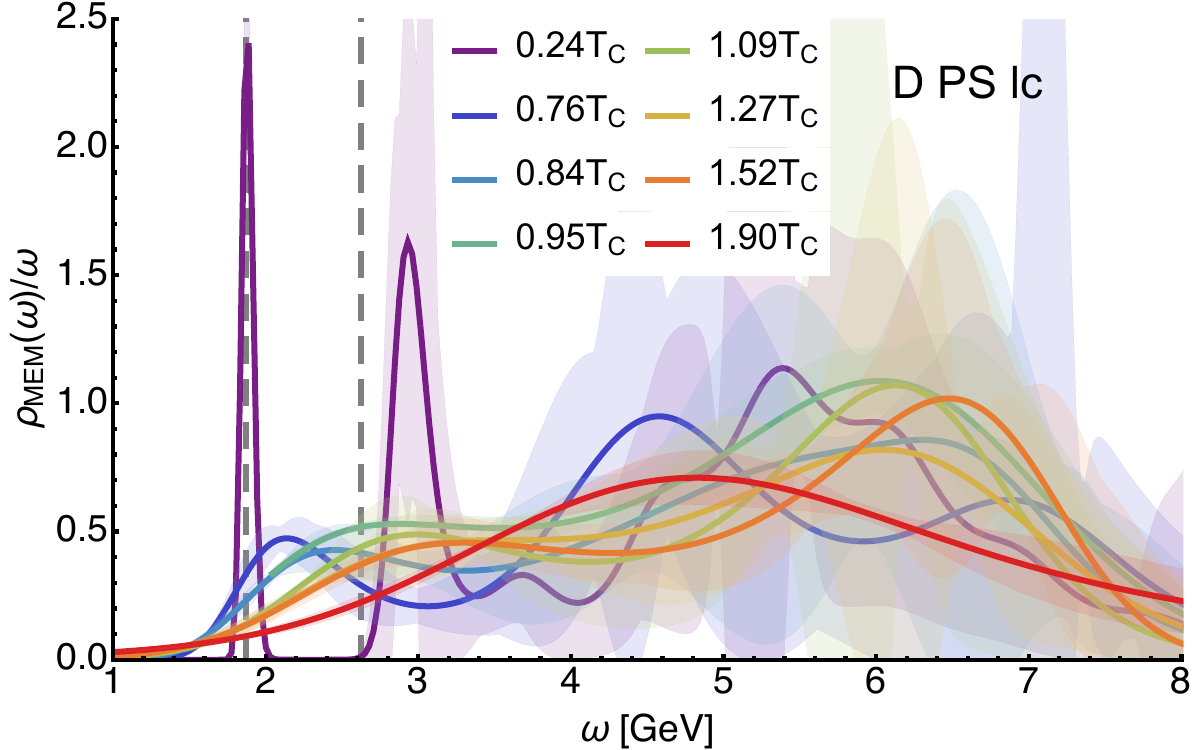}
    \includegraphics[width=0.49\textwidth]{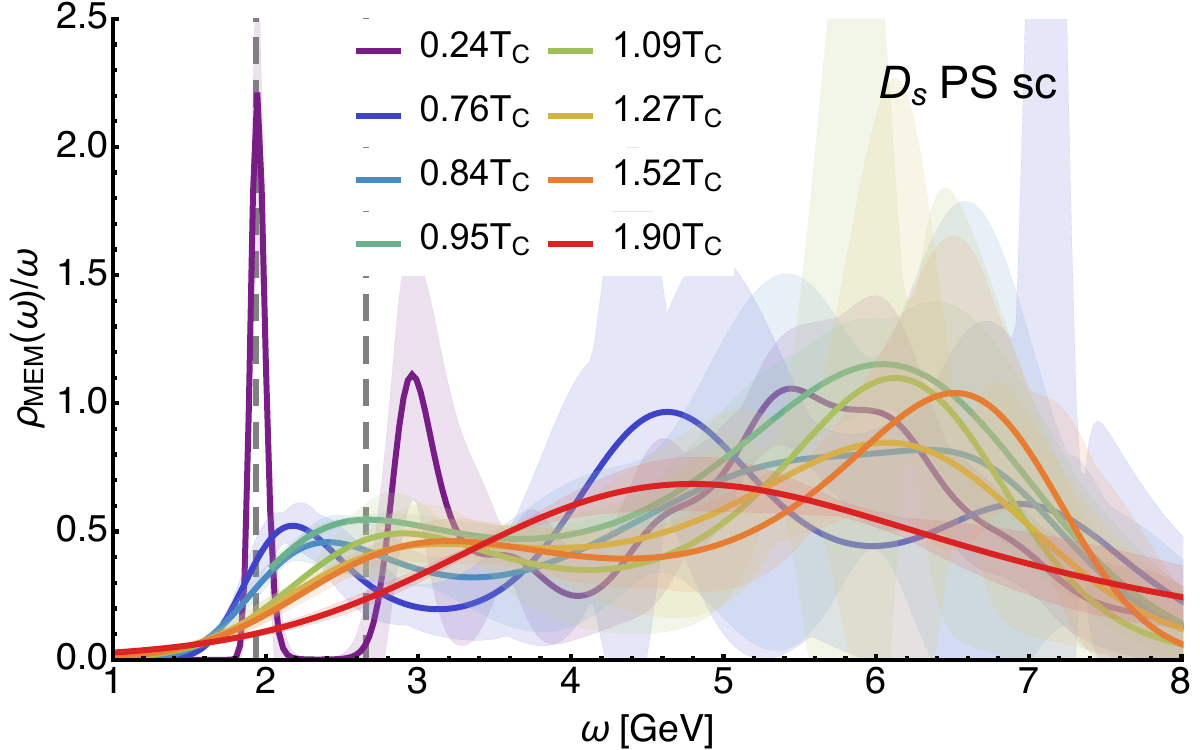} \\
    \includegraphics[width=0.49\textwidth]{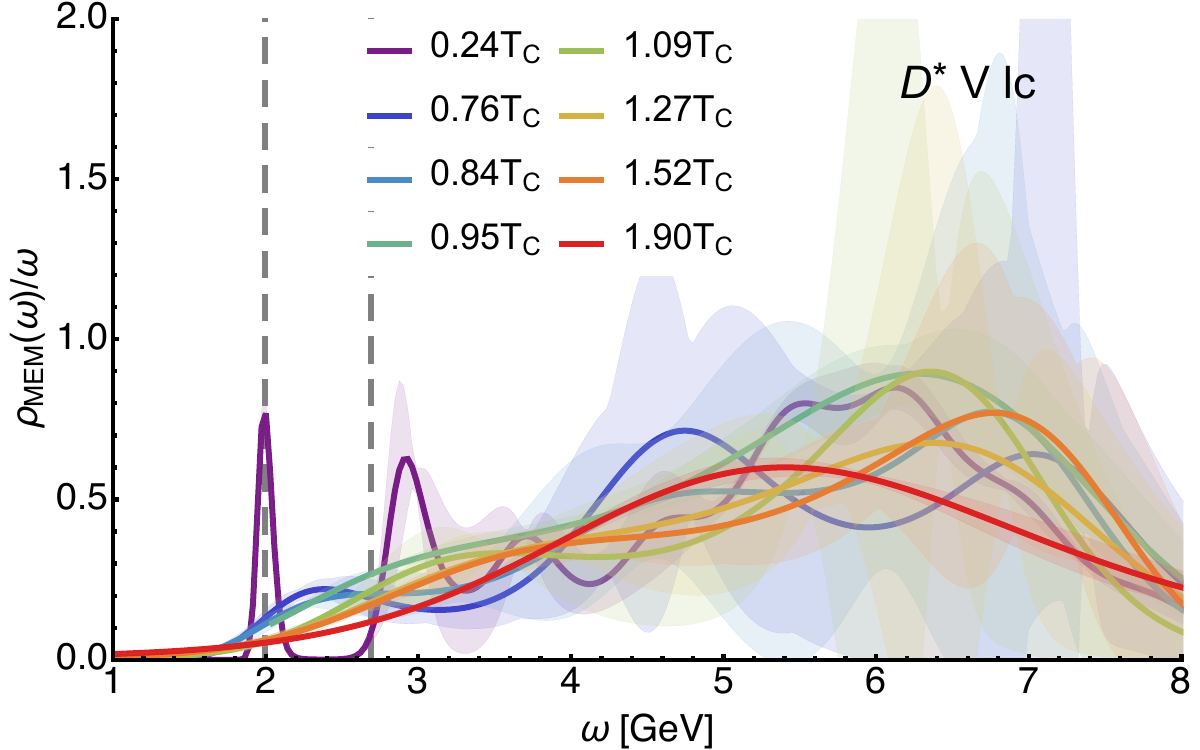}
    \includegraphics[width=0.49\textwidth]{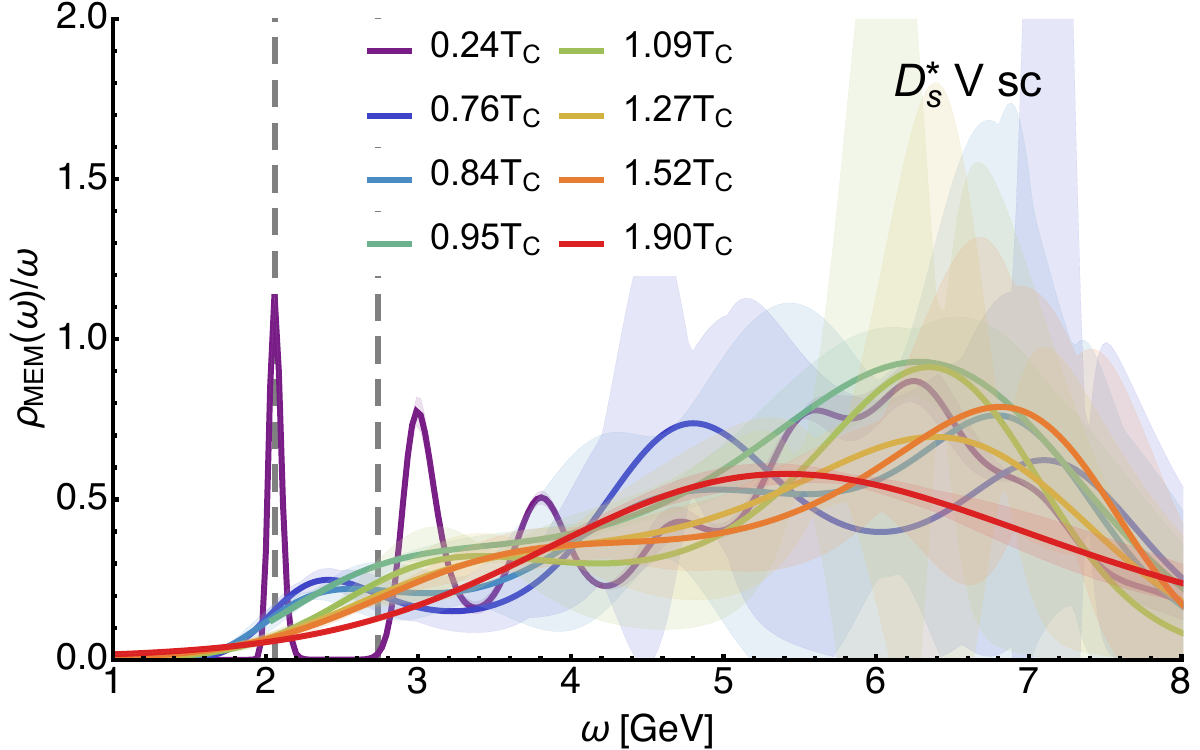}
    \caption{MEM reconstruction}
\end{subfigure}\caption{Spectral functions of the open-charm states reconstructed from lattice-QCD Euclidean correlators at various temperatures. Figures obtained from Ref.~\cite{Kelly:2018hsi}.}
\label{fig:KellyDmesons}
\end{figure}

The first detailed lattice-QCD study of open-charm mesons at temperatures around and above the transition temperature was presented in Ref.~\cite{Kelly:2018hsi}. Using relativistic anisotropic lattices with $N_f=2+1$ flavors of clover fermions from the FASTSUM collaboration and a pion mass of $m_\pi\approx 380\mev$, the authors analyzed Euclidean correlators and reconstructed spectral functions for $D$, $D_s$, $D^*$, $D_s^*$ channels (as well as for quarkonium states) using both BR and MEM. As shown in Fig.~\ref{fig:KellyDmesons}, both methods revealed broadening of the ground-state peak with increasing temperature, with vector channels being consistently slightly more affected than pseudoscalar ones. While no significant thermal modification was observed below $T_c$, the ground-state peaks disappeared by $1.9\,T_c$ in all channels. However, large systematic differences between MEM and BR prevented precise conclusions and highlighted the need for improved reconstruction techniques and higher-precision lattice correlator data, in order to enable a robust determination of the spectral properties of open-charm mesons from lattice QCD. 

Complementary insights into the challenging task of spectral function reconstruction from temporal Euclidean correlators can be obtained from the study of spatial correlators of mesons. Although their relation to the spectral function is more involved,
\begin{equation}
    \mathcal{C}(z;\,T)=\int_0^\infty  \frac{2\, \diff \omega}{\omega}\int_{-\infty}^\infty \diff p_z\,\text{e}^{\ii p_zz}\rho(\omega,p_z;\,T) \ ,
\end{equation}
spatial correlators have the advantage that, at large spatial separations, they allow the extraction of screening masses, as described in Section~\ref{sec:lqcdlight}. 

\begin{figure}[!ht] 
\centering
\includegraphics[width=0.4\textwidth]{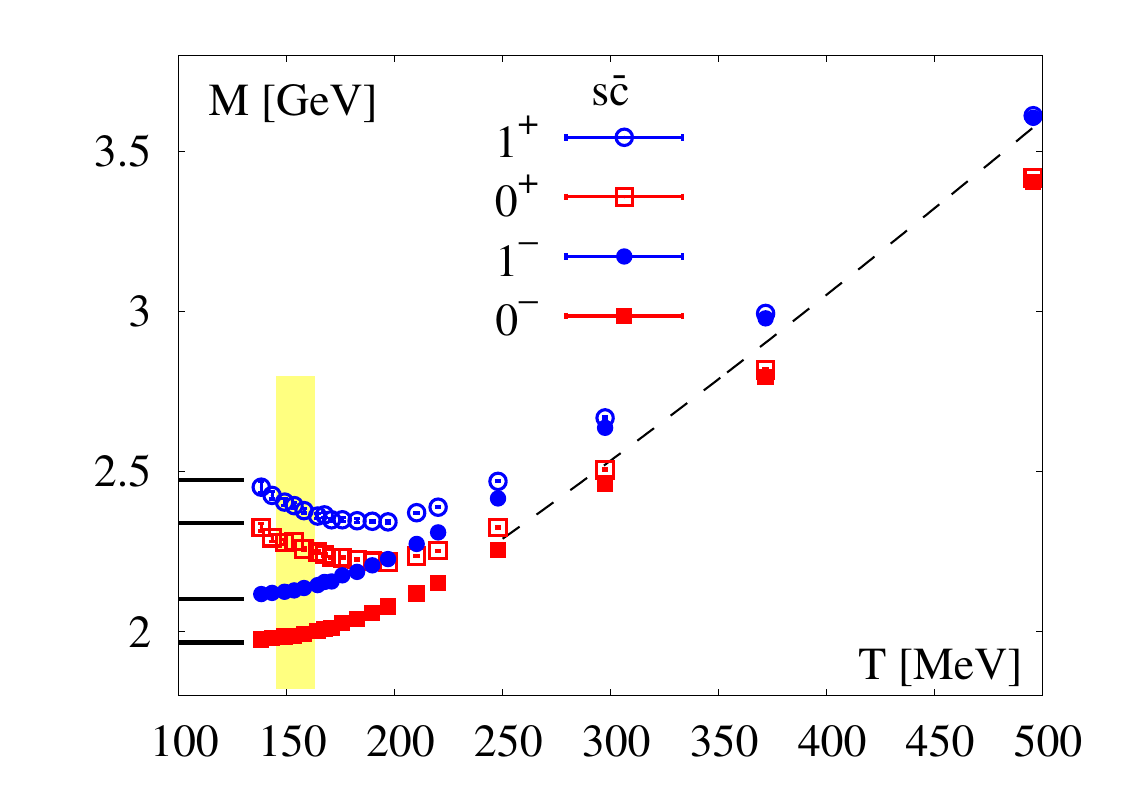}
\includegraphics[width=0.4\textwidth]{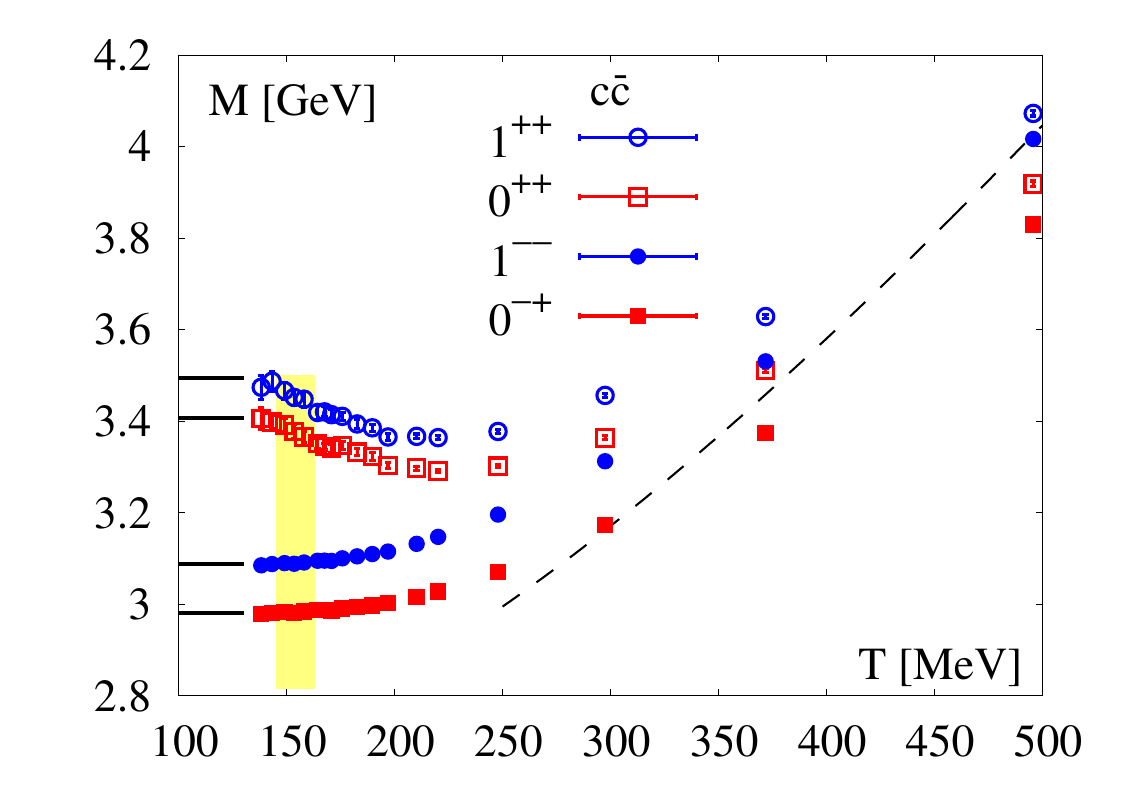}
\caption{Screening mass for $s\bar{c}$ (left) and $c\bar{c}$ states (right) as a function of the temperature extracted from lattice QCD spatial correlators. Figure obtained from Ref.~\cite{Bazavov:2014cta}.}
\label{fig:Bazavov-sc-mass}
\end{figure}

Bazavov et al.~\cite{Bazavov:2014cta} computed screening masses for $s\bar{c}$ states (as well as $s\bar{s}$ and $c\bar{c}$ states) using $N_f=2+1$ flavors of HISQ ensembles with $m_\pi\simeq 160\mev$ over a wide range of temperatures from below to well above $T_c$. The results in the left panel of Fig.~\ref{fig:Bazavov-sc-mass} show that the screening masses of open-charm mesons begin to deviate from their zero-temperature values already near $T_c$, and the splitting between parity partners was observed to decrease rapidly above $T_c$, signaling the approximate restoration of chiral symmetry in the charm sector. At sufficiently high temperatures, the screening masses approach the free-theory limit, consistent with a gradual transition toward quasi-free propagation of charm quarks in the medium.

\begin{figure}[!ht] 
\centering
\includegraphics[width=0.4\textwidth]{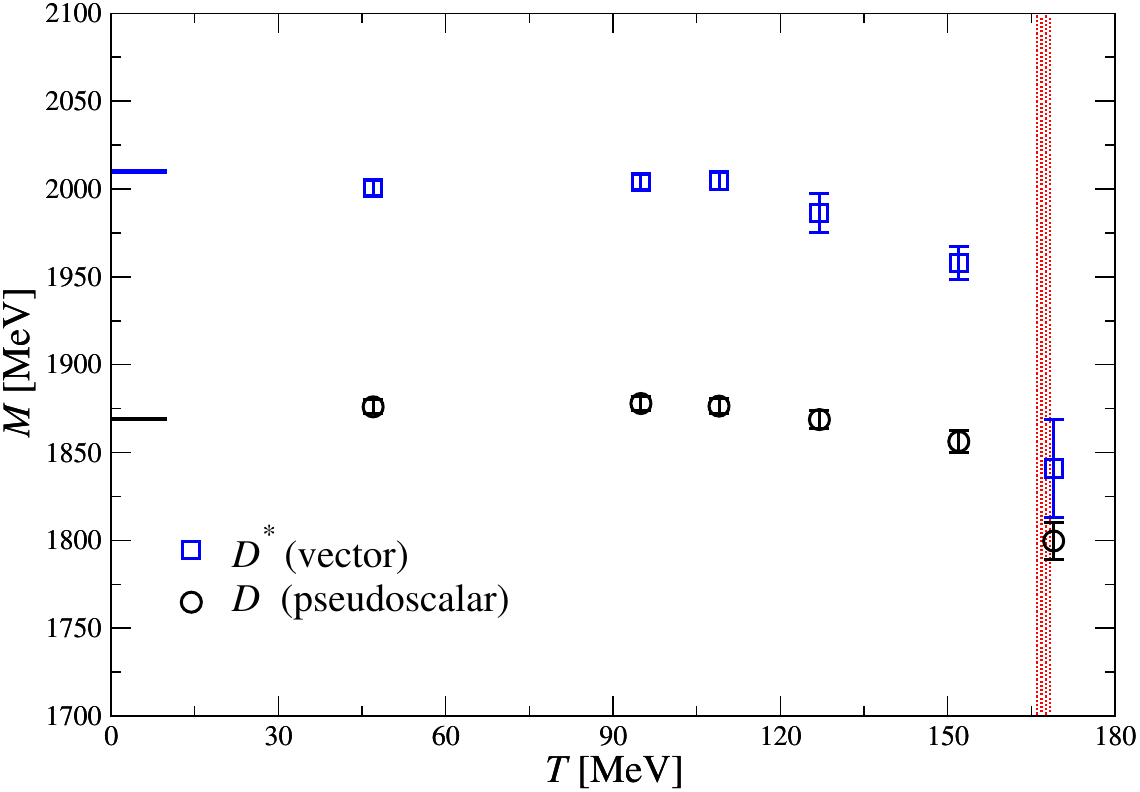}
\includegraphics[width=0.4\textwidth]{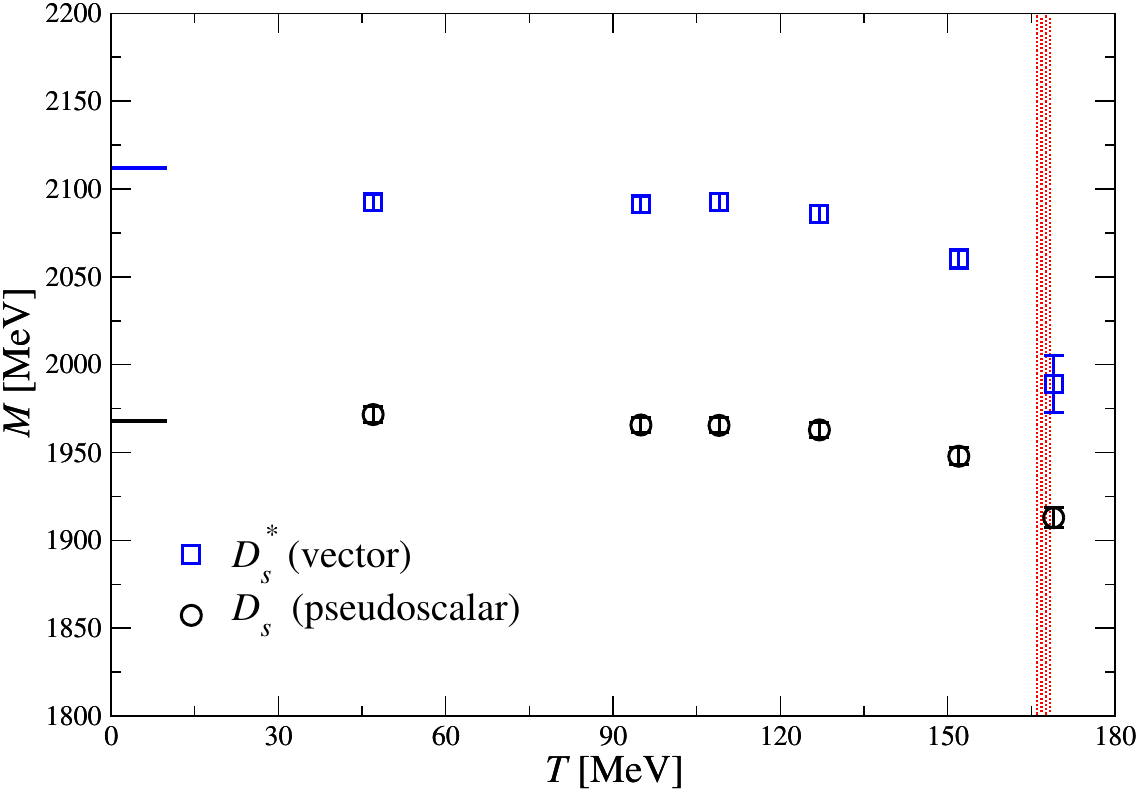}
\caption{Temperature dependence of the mass of the $D$ mesons (left: $D$, $D^*$; right: $D_s$, $D_s^*$), obtained via the double-ratio method. Figure obtained from Ref.~\cite{Aarts:2022krz}.}
\label{fig:Aarts-D-mass}
\end{figure}

More recently, Aarts et al.~\cite{Aarts:2022krz} introduced a double-ratio method to isolate the genuine temperature dependence of the ground-state masses in the hadronic phase from the temporal correlators, without full spectral reconstruction. The approach compares Euclidean correlators at temperature $T$ to those at a reference temperature $T_0$ while correcting for the temperature dependence of the kernel:
\begin{equation}
    R(\tau;\,T,T_0)=\frac{\mathcal{C}(\tau;\,T)/\mathcal{C}(\tau;\,T_0)}{\mathcal{C}_\text{model}(\tau;\,T,T_0)/\mathcal{C}_\text{model}(\tau;\,T_0,T_0)} \ ,
\end{equation}
where $\mathcal{C}_\text{model}$ is constructed from the ground-state peak determined at $T_0$. Given the thermal EFT results for these states~\cite{Montana:2020lfi,Montana:2020vjg}, the fitting procedure employed in Ref.~\cite{Aarts:2022krz} assumed that these states remain very narrow in the hadronic phase, $\Gamma\ll M$, and therefore the width was neglected in order to simplify the modeling. Using this method on the FASTSUM anisotropic lattice ensembles with $N_f=2+1$ flavors of dynamical quarks and $m_\pi=239\mev$, Ref.~\cite{Aarts:2022krz} found that pseudoscalar and vector channels exhibit only small thermal modifications (mass reductions of $\sim 1$--$3\%$) as $T$ approaches $T_c$, as seen in Fig.~\ref{fig:Aarts-D-mass}. The authors also attempted to study thermal effects on the scalar ($D_0^*$, $D_{s0}^*$) and axial-vector ($D_1$, $D_{s1}$) channels. However, in this case, the assumption of a narrow spectral function is not justified, as the $D_0^*$ and $D_1$ are expected to be broad even in vacuum, and the nearby threshold dynamics plays a nontrivial role already at zero temperature. Consequently, the strong thermal dependence observed in the double ratios for these channels largely originates from the temperature dependence of the Euclidean kernel and correlators themselves, rather than genuine changes in the spectral function. This exposes the limitations of the method when applied to channels with intrinsically large widths and in the proximity of two-particle thresholds.

\begin{figure}[!hb] 
\centering
\includegraphics[width=0.45\textwidth]{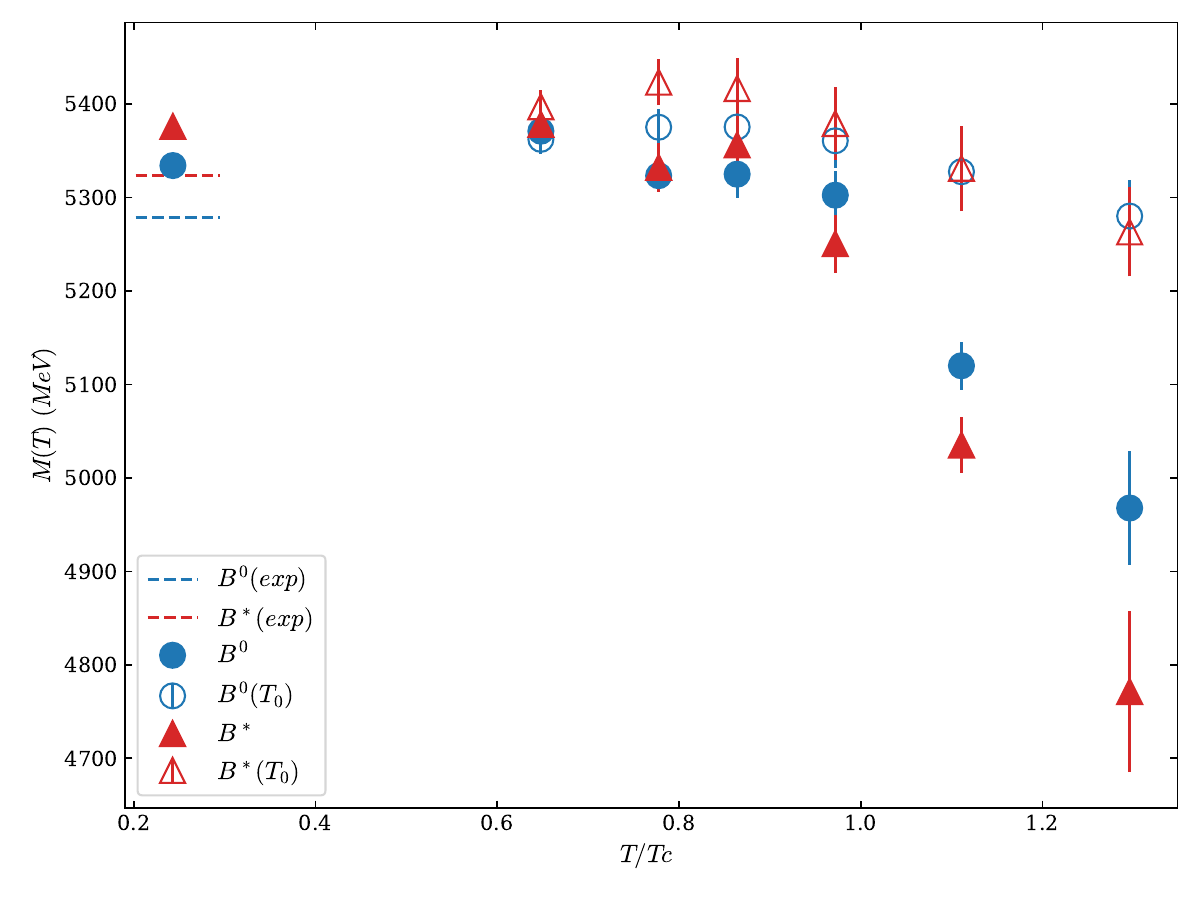}
\includegraphics[width=0.45\textwidth]{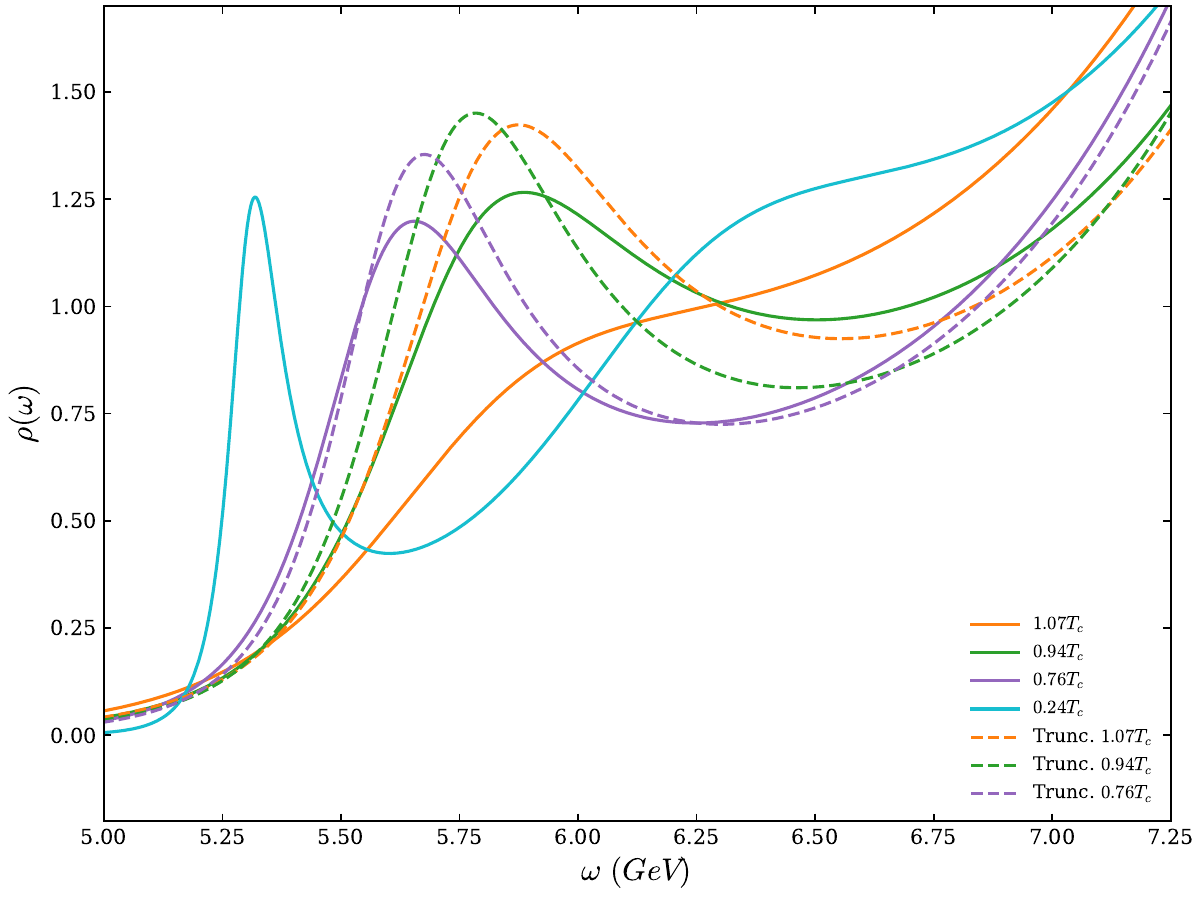}
\caption{Temperature dependence of the mass of the $B$ and $B^*$ mesons (left) and $B$-meson spectral functions (right), compared to those extracted from $T=0$ correlators truncated at the same temporal extent. Figures obtained from Ref.~\cite{Skullerud:2026sek}.}
\label{fig:Skullerud-B}
\end{figure}

The first results from lattice QCD for masses and spectral functions of $B$ mesons at finite temperature were recently presented in Ref.~\cite{Skullerud:2026sek}, using anisotropic lattices from the Hadron Spectrum collaboration with $N_f=2+1$ active quark flavors and $m_\pi\approx 380\mev$. The correlators were constructed by combining nonrelativistic QCD propagators for the heavy $b$ (anti)quark and relativistic propagators for the light (anti)quark. The left panel in Fig.~\ref{fig:Skullerud-B} shows the thermal masses of the $B$ and $B^*$ mesons obtained from standard exponential fits to the Euclidean correlators at various temperatures, compared with the results from the $T=0$ ensembles truncated to the temporal extent of the corresponding finite-temperature correlator. After subtracting this truncation effect, negative mass shifts remain, which are of the same magnitude as those expected from the thermal unitarized EFT approach in the hadronic phase~\cite{Montana:2023sft} and increase substantially above $T_c$. In the right panel, we show the corresponding $B$-meson spectral functions reconstructed with the BR method, again contrasted with the extraction from truncated $T=0$ correlators. The ground-state peak of the spectral function disappears around $T_c$, suggesting the melting of the $B$ meson at larger temperatures.

\subsection{QCD sum rules for open heavy-flavor mesons}\label{sec:D-QCDSR}

\begin{figure}[htbp!] 
\centering
    \includegraphics[width=0.45\textwidth]{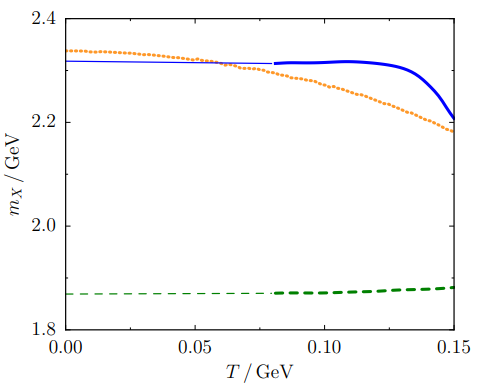}\hspace{0.5cm}
    \includegraphics[width=0.45\textwidth]{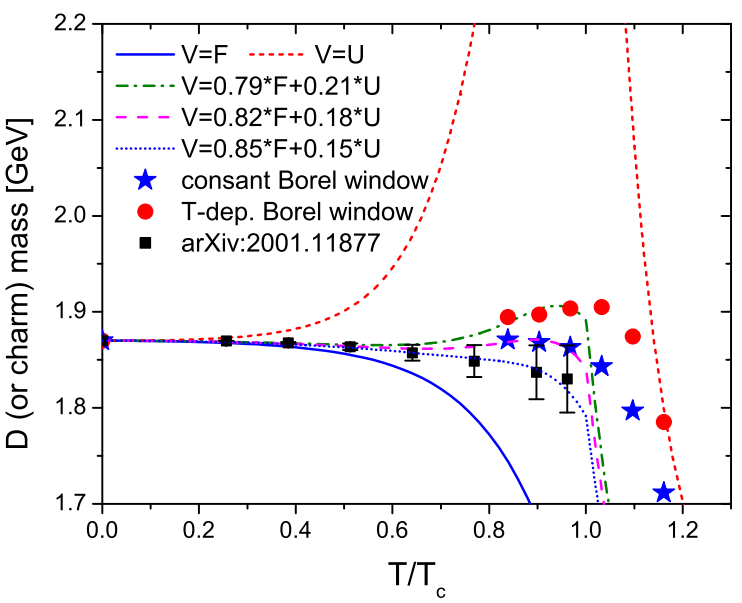}
\caption{Temperature dependence of the $D$-meson mass from QCD sum-rule analyses. Left panel: Scalar ($0^+$) channel from Ref.~\cite{Buchheim:2018kss} (orange dotted curve), compared with the EFT results of Ref.~\cite{Sasaki:2014asa} for both $0^+$ and $0^-$ states. Right panel: Pseudoscalar ($0^-$) channel from Ref.~\cite{Gubler:2020hft} (blue stars and red circles), alongside various heavy-quark potentials and the EFT results of Ref.~\cite{Montana:2020lfi}. Figures obtained from Refs.~\cite{Buchheim:2018kss,Gubler:2020hft}.}
\label{fig:qcdSR-Dmesons}
\end{figure}

The formalism of QCD sum rules, introduced for light mesons in Section~\ref{sec:light}, has been applied to open heavy flavor in several recent works. As in the light sector, the sum rules are Borel transformed, and the analysis is restricted to a Borel window in which both the truncation of the OPE and the pole-plus-continuum Ansatz for the spectral function remain under control. In Ref.~\cite{Buchheim:2018kss}, the temperature dependence of the spectral properties of both the scalar ($0^+$) and pseudoscalar ($0^-$) $D$ mesons was discussed in the context of dynamical chiral symmetry breaking. The authors found that the scalar $D$-meson mass decreases significantly with temperature, dropping from $2.334\gev$ at $T=0$ to $2.182\gev$ at the upper limit of applicability of their approach, $T=150\mev$. In contrast, the pseudoscalar channel is rather insensitive to thermal modifications, though the mass itself could not be reliably extracted at finite temperature. Nevertheless, the results suggest that the two chiral partners move closer in mass as $T$ increases, which would provide a signal of partial chiral symmetry restoration. This behavior is in qualitative agreement with the hadronic approach of Ref.~\cite{Sasaki:2014asa}, although the detailed temperature dependence differs; in particular, the temperature evolution is smoother in the QCD sum-rule approach, as illustrated in the left panel of Fig.~\ref{fig:qcdSR-Dmesons}.

Ref.~\cite{Gubler:2020hft} also investigated the pseudoscalar channel using an alternative QCD sum-rule setup that incorporates constraints from the heavy-quark potential at finite temperature. Their strategy is based on the observation that the value of the heavy-quark potential at asymptotically large separations obtained from lattice QCD can be related to twice the thermal $D$-meson mass. Their analysis shows that the pseudoscalar mass remains relatively stable up to temperatures close to the crossover temperature, where it begins to decrease. This can be seen in the right panel of Fig.~\ref{fig:qcdSR-Dmesons}, where the sum-rule results lie close to the thermal EFT predictions of Ref.~\cite{Montana:2020lfi}, although in the EFT case the mass starts to decrease already at lower temperatures.

\subsection{Heavy-flavor baryons}\label{sec:heavy-baryons}

\begin{figure}[htbp!] 
\centering
\includegraphics[width=0.35\linewidth]{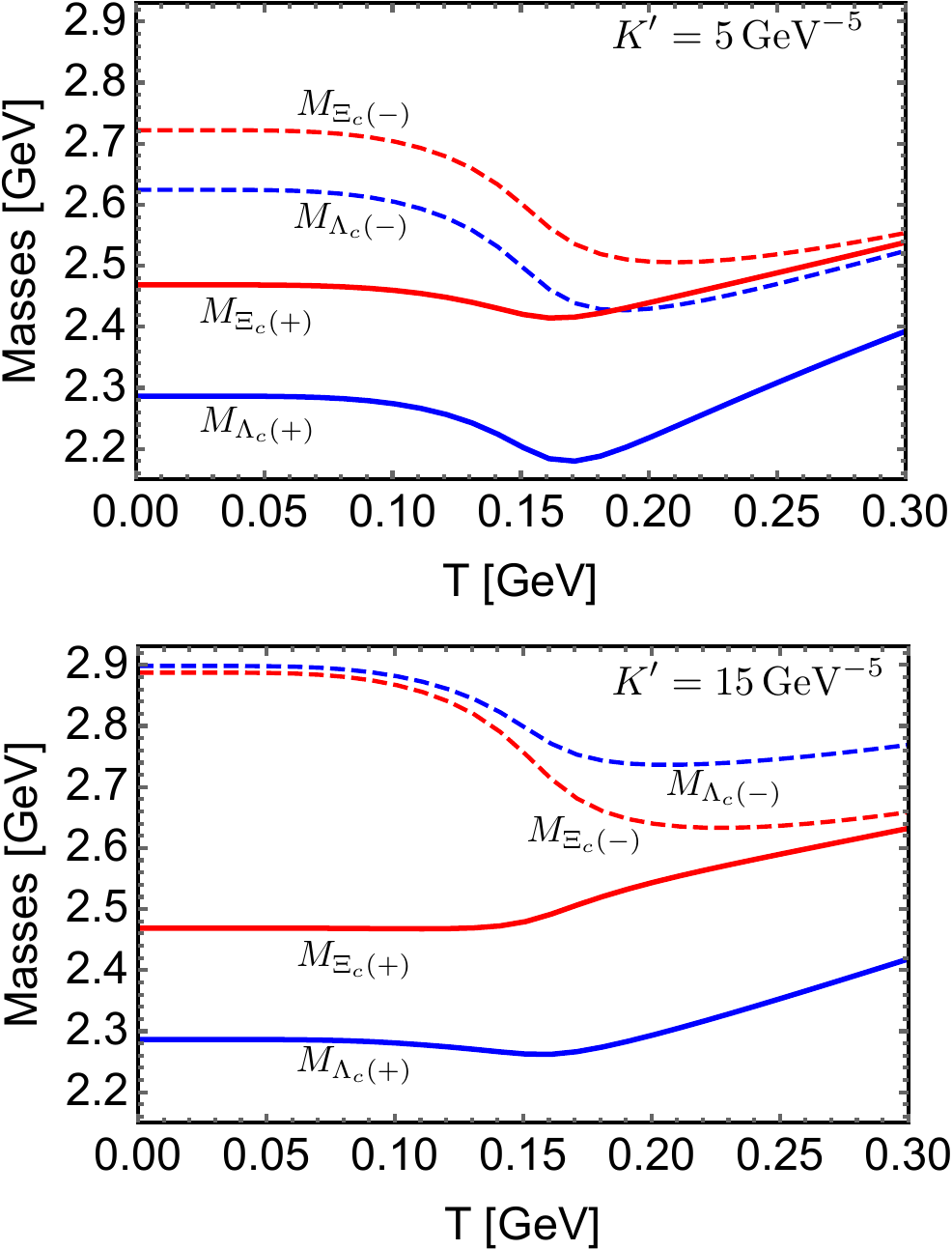}
\includegraphics[width=0.35\linewidth]{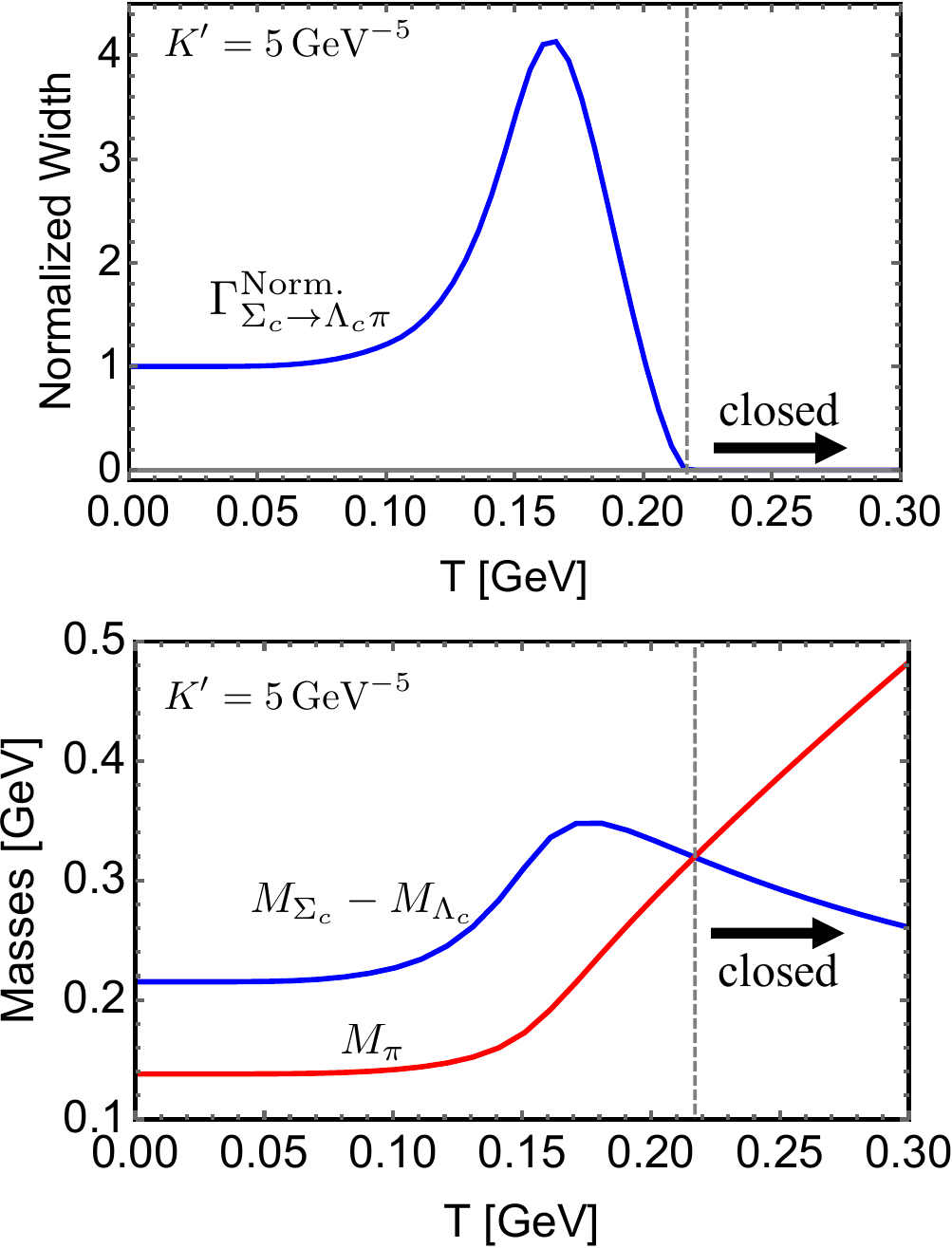}
\caption{Left panels: Heavy-baryon masses of opposite parity for different values of the anomalous coupling constant $K'$ as functions of the temperature. Upper right panel: Two-body decay width of the $\Sigma_c \rightarrow \Lambda_c \pi$ process, normalized to the vacuum value as a function of the temperature. Lower right panel: Comparison of thermal masses of $\Sigma_c, \Lambda_c$ and $\pi$, showing when the two-body $\Sigma_c \rightarrow \Lambda_c \pi$ decay is kinematically closed. Figures taken from Ref.~\cite{Suenaga:2024vwr}.}
\label{fig:SuenagaHeavyBaryons}
\end{figure}

Although the heavy-baryon sector in medium has not received as much attention as its light-baryon counterpart, the thermal behavior of heavy baryons has been investigated within various theoretical approaches. For instance, thermal QCD sum rules were applied in Ref.~\cite{Azizi:2019cmj} to spin-$3/2$ baryons ($\Sigma_b^*, \Xi_b^*, \Omega_b^*$) in the infinite heavy-quark mass limit. Potential models at finite temperature have also been utilized: Ref.~\cite{Shi:2019tji} calculated the binding energies of singly, doubly and triply charmed baryons ($\Lambda_c,\Xi_c,\Omega_c,\Xi_{cc},\Omega_{cc}, \Omega_{ccc}$) at temperatures above $T_c=155\mev$, while Ref.~\cite{Zhao:2023qww} examined triply heavy baryons ($\Omega_{ccc},\Omega_{bcc},\Omega_{bbc},\Omega_{bbb}$) by solving the Schr\"odinger equation with a thermal potential. Furthermore, recent NJL model calculations within the diquark-quark picture~\cite{Suenaga:2024vwr} have explored the behavior of $\Lambda_c,\Xi_c,\Sigma_c,\Xi_c',\Lambda_c'$ and $\Omega_c$ baryons from the perspective of chiral symmetry restoration.
In Fig.~\ref{fig:SuenagaHeavyBaryons} we present a couple of results from Ref.~\cite{Suenaga:2024vwr}.
In the left panels, the masses of the $\Lambda_c$ and $\Xi_c$ baryons (of both parities) are shown as functions of temperature. The upper and lower left panels show the effect of using different values of the coupling constant $K'$ that regulates the effect of the axial anomaly on the diquarks. The degeneracy of the $\Xi_c$ states can be observed around $T=0.3\gev$, but not for the $\Lambda_c$ states. In the right panels, we observe the thermal dependence of the two-body decay $\Sigma_c \rightarrow \Lambda_c \pi$. The upper right panel shows the decay width as a function of temperature, where an increase close to $T_c=0.15\gev$ can be seen, followed by a rapid drop to zero at $T=0.21\gev$. At this temperature, the $\Sigma_c$ mass becomes less than $M_{\Lambda_c} + M_\pi$ and the phase space of the final state vanishes, as can be seen in the lower right panel. This kind of sensitivity to the mass thresholds and phase space can have important phenomenological implications in relativistic HICs.

\begin{figure}[htbp!] 
\centering
\includegraphics[width=0.45\linewidth]{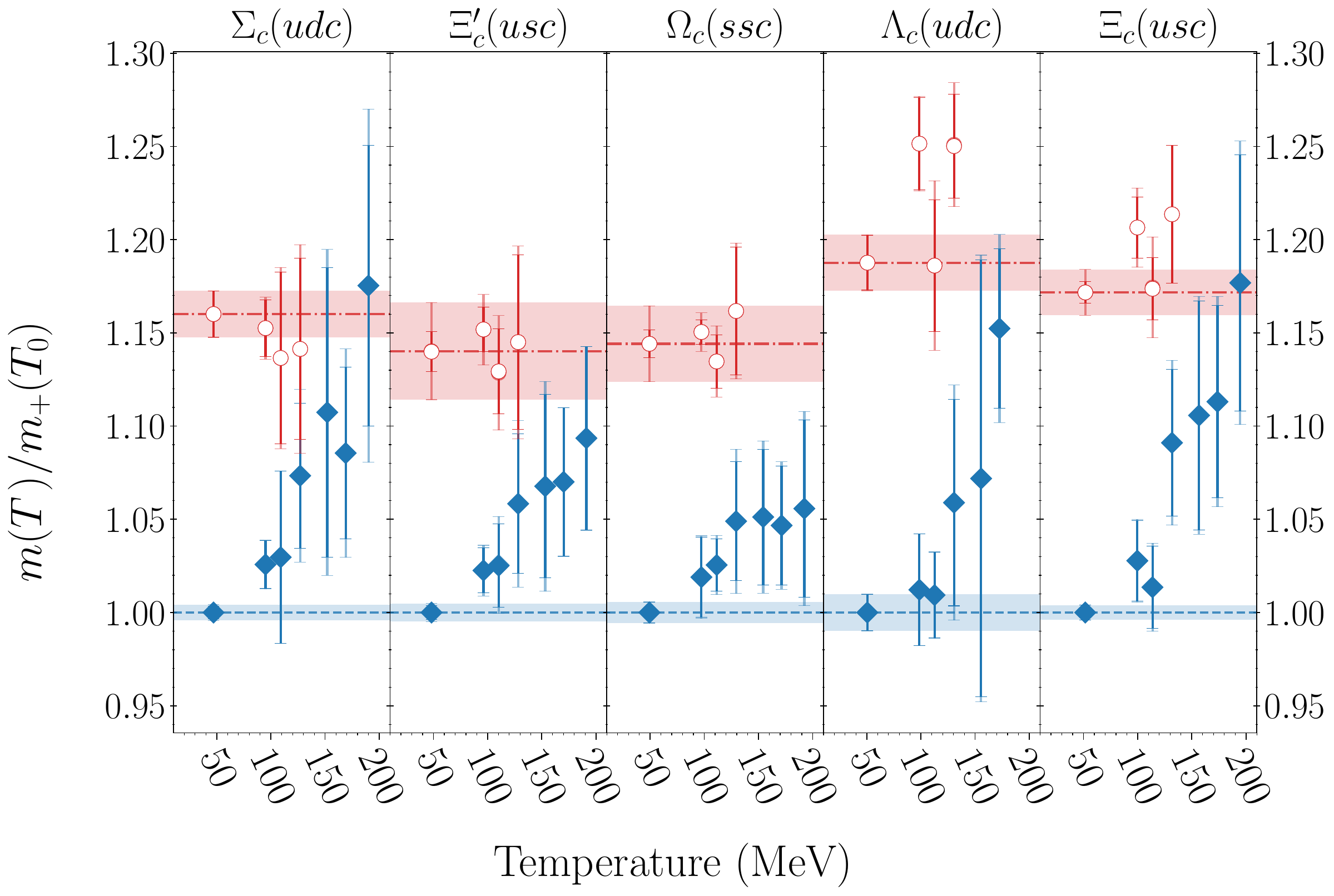}
\includegraphics[width=0.45\linewidth]{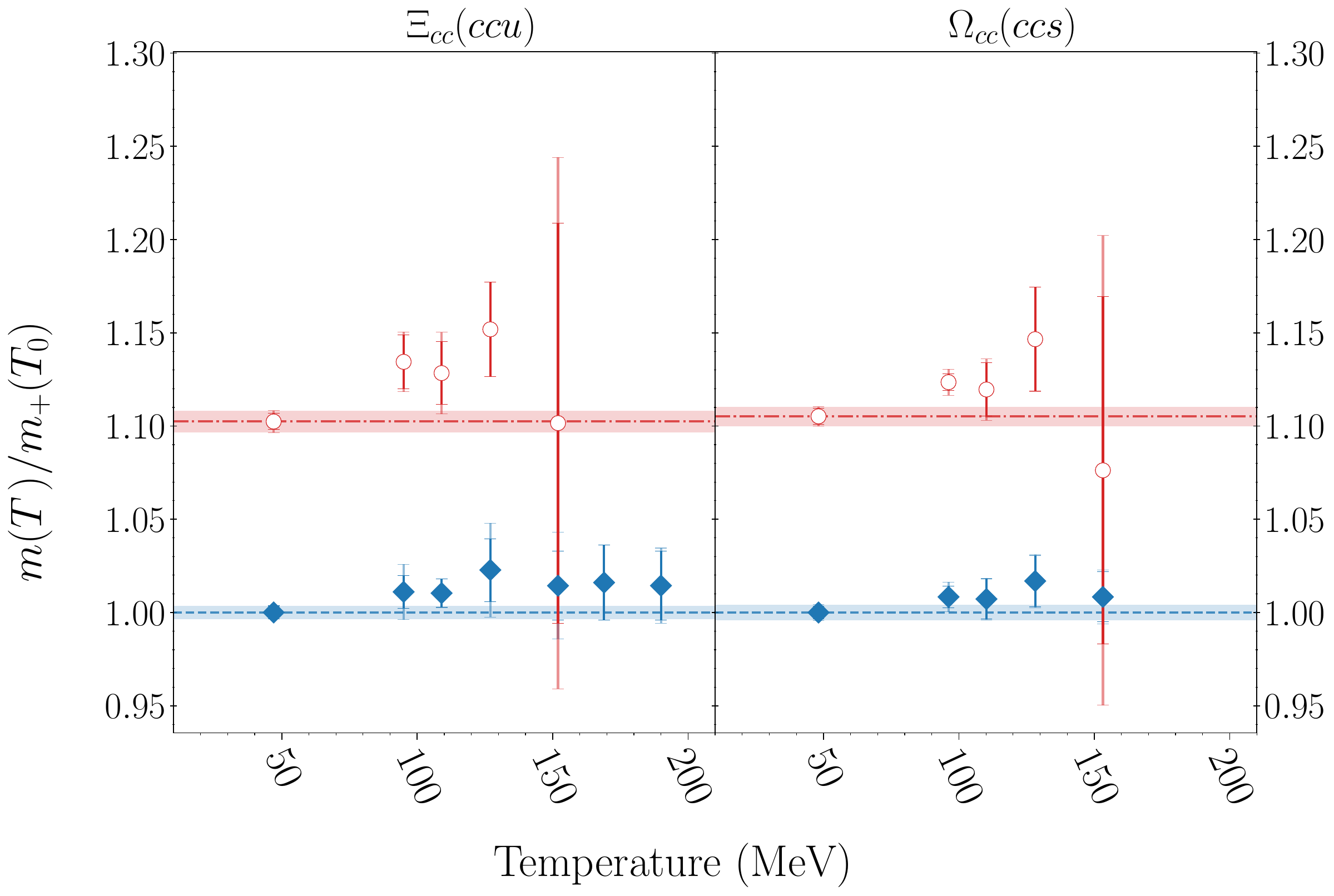}
\caption{Temperature dependence of the ground-state masses of the spin-$1/2$ singly charmed (left panel) and doubly charmed baryons (right panel) extracted from lattice QCD simulations using the method based on the double ratio of correlators. Blue, filled symbols correspond to positive-parity states, and red, open symbols to negative-parity states. The horizontal dashed lines depict the results at the lowest temperature. Figures taken from Ref.~\cite{Aarts:2023nax}.}
\label{fig:AartsHeavyBaryons}
\end{figure}

In lattice QCD, the method based on the double ratio of Euclidean correlators developed for open-charm mesons in Ref.~\cite{Aarts:2022krz} was extended to the sector of spin-$1/2$ charmed baryons in Ref.~\cite{Aarts:2023nax}. Using lattice-QCD simulations with $N_f=2+1$ flavors on anisotropic lattices, with $m_\pi=239\mev$, temperature effects on the mass of the positive- and negative-parity ground states were investigated using ratios of thermal correlators. This avoids the complications associated with fitting or reconstructing the full spectral functions. In Fig.~\ref{fig:AartsHeavyBaryons}, we show the masses of the singly charmed baryons ($\Sigma_c$, $\Xi_c'$, $\Omega_c$, $\Lambda_c$, $\Xi_c$) in the left panel, and the doubly charmed baryons ($\Xi_{cc}$, $\Omega_{cc}$) on the right, normalized to the positive-parity ground-state mass at the lowest temperature available in the lattice analysis. 
The results suggest that in the positive-parity sector, the masses of the singly charmed baryons increase with temperature already at temperatures below $T_c$. For the negative-parity states, as well as the doubly charmed baryons with both parities, it is difficult to conclude whether the masses change in a systematic way or remain relatively unchanged by the thermal medium.
This effort, together with the mesonic studies discussed above, underlines the leading role of the FASTSUM collaboration in advancing finite-temperature lattice QCD for open heavy-flavor systems~\cite{Allton:2024frr}.

\subsection{Summary and future perspectives}\label{sec:openHFsummary}
The approaches reviewed in this section rely on different degrees of freedom, apply in different temperature ranges, and give access to different observables. Phenomenological hadronic models and EFTs describe the hadronic phase below the transition temperature, $T\lesssim T_c\simeq 156$--$158\mev$~\cite{HotQCD:2018pds,Borsanyi:2020fev}, where the thermal bath is dominated by pions and the interactions of the heavy mesons with the bath are described either by empirical or model amplitudes or, in EFTs, through a combined chiral and heavy-quark expansion. In the imaginary-time formalism, thermal effects enter through loop integrals weighted by Bose--Einstein distributions. At the one-loop level with vacuum amplitudes and propagators~\cite{Fuchs:2004fh,He:2011yi} (see also the one-loop ChPT comparison in Ref.~\cite{Sasaki:2014asa}), these approaches give a transparent estimate of collisional widths and mass shifts. Unitarization adds the nonperturbative resummation of the heavy-light scattering amplitudes, and with it, the excited states that are generated dynamically in coupled channels. The self-consistent treatment of Refs.~\cite{Cleven:2017fun,Montana:2020lfi,Montana:2020vjg,Montana:2023sft} then yields full spectral functions together with the in-medium scattering amplitudes. Since these frameworks are built on hadronic degrees of freedom, they are not reliable above $T_c$.

Lattice QCD is the only first-principles method that covers both the hadronic and the deconfined phases, with the temperature set by the temporal extent of the lattice, $1/T = N_\tau a_\tau$. Its main limitations for open heavy flavor are the ill-posed reconstruction of spectral functions from Euclidean correlators, the small number of temporal points available at high temperature, the computational cost of large lattices required at low temperatures, and the use of heavier-than-physical pion masses in most of the anisotropic ensembles employed for spectral studies~\cite{Kelly:2018hsi,Aarts:2022krz,Skullerud:2026sek}. Different lattice observables also probe different quantities: spatial correlators give screening masses~\cite{Bazavov:2014cta}, which are not the pole masses of the spectral function, while double ratios of temporal correlators~\cite{Aarts:2022krz} give pole-mass shifts only under the assumption of a narrow ground state.

QCD sum rules provide an analytic bridge between the hadronic spectrum and the QCD vacuum structure. They relate thermal modifications of heavy-light mesons to the temperature dependence of quark and gluon condensates through the OPE and a dispersion relation. Their predictive power lies mainly in the sign and qualitative temperature dependence of mass shifts, and in the evolution of the splitting between chiral partners~\cite{Buchheim:2018kss,Gubler:2020hft}. Their quantitative precision is limited by the truncation of the OPE (partly mitigated by the Borel transformation), by the pole-plus-continuum Ansatz for the spectral function, by the input required for the thermal condensates, and by the narrowing of the Borel window as the temperature increases.

Despite their methodological differences, several features of the in-medium modification of open heavy-flavor mesons emerge consistently across the various approaches:
\begin{itemize}
    \item \textbf{Ground states survive in the hadronic phase.} All approaches that address this question agree that the ground-state $D^{(*)}$ and $B^{(*)}$ mesons remain well-defined quasiparticles below $T_c$. Hadronic and EFT calculations find thermal widths that grow with temperature, reaching several tens of MeV at $T\simeq 150\mev$~\cite{Fuchs:2004fh,He:2011yi,Cleven:2017fun,Montana:2020lfi,Montana:2020vjg,Montana:2023sft}. Within the EFT, strange mesons broaden less than nonstrange ones, reflecting the weaker interaction of $D_s^{(*)}$ and $B_s^{(*)}$ mesons with the pion bath~\cite{Montana:2020vjg,Montana:2023sft}.
    Lattice reconstructions show no significant modification of the ground-state peaks below $T_c$ within their present resolution, which is compatible with widths of a few tens of MeV. They do show substantial broadening above $T_c$, with the peaks having disappeared by $1.9\,T_c$ for open-charm mesons~\cite{Kelly:2018hsi} and already around $T_c$ for the $B$ meson~\cite{Skullerud:2026sek}.

    \item \textbf{Moderate downward mass shifts.} Unitarized EFT calculations predict a reduction of the ground-state masses by several tens of MeV as $T$ approaches $150\mev$~\cite{Montana:2020lfi,Montana:2020vjg,Montana:2023sft}. The lattice double-ratio analysis of Ref.~\cite{Aarts:2022krz} finds reductions of $\sim1$--$3\%$ for $D^{(*)}_{(s)}$ mesons, depending on the channel. The somewhat smaller effect can be largely attributed to the heavier-than-physical pion mass, which suppresses the thermal pion population. For $B^{(*)}$ mesons, the negative mass shifts found on the lattice in the hadronic phase are of the same magnitude as the EFT predictions~\cite{Skullerud:2026sek,Montana:2023sft}. The QCD sum-rule analysis of the pseudoscalar channel~\cite{Gubler:2020hft}, which is constrained by the lattice heavy-quark potential, also finds a decreasing $D$-meson mass.

    \item \textbf{Partial chiral symmetry restoration.} All approaches that are sensitive to the chiral structure of the spectrum find that the mass splitting between parity partners remains in the hadronic phase and shrinks as the temperature approaches and exceeds $T_c$. In the L$\sigma$M with heavy-quark symmetry~\cite{Sasaki:2014asa}, this happens mainly because the positive-parity state becomes lighter, while the nonstrange negative-parity ground state is comparatively stable (the strange $D_s$ instead shifts upward by more than $100\mev$). Lattice screening masses show that the splitting between opposite-parity $s\bar c$ channels closes rapidly above $T_c$, with the screening masses approaching the free-quark limit at high temperatures~\cite{Bazavov:2014cta}.
    In unitarized EFT frameworks, the excited open heavy-flavor states emerge as dynamically generated states, with the nonstrange sector exhibiting a double-pole structure, and are strongly affected by nearby thresholds~\cite{Montana:2020lfi,Montana:2020vjg,Montana:2023sft}. The possible scenarios for chiral restoration in this context were discussed in Ref.~\cite{Torres-Rincon:2021wyd}.
\end{itemize}

Several points nevertheless remain unsettled or are not yet directly comparable across frameworks. The magnitude and onset of the mass shifts depend strongly on how the heavy-light interaction is treated. The discrepancies are largest for the positive-parity partners, which some frameworks treat as genuine states or degrees of freedom, whereas unitarized EFTs generate them dynamically from heavy-light meson scattering. The size of the thermal broadening for the various channels is also not yet well established.

Several developments would improve the theoretical understanding of the properties of heavy-flavor hadrons at finite temperature and reduce the tensions between different approaches.

On the EFT side, the unitarized framework of Refs.~\cite{Montana:2020lfi,Montana:2020vjg,Montana:2023sft} treats the light mesons of the thermal bath with vacuum propagators, an approximation justified by their mild thermal modifications in the hadronic phase (see Section~\ref{sec:light}). Dressing them self-consistently would improve the consistency of the calculation as $T$ approaches $T_c$. Although pions dominate the heavy-meson self-energy~\cite{Montana:2020vjg}, including kaons, $\eta$ mesons, and vector mesons, which contribute significantly at $T\gtrsim150\mev$~\cite{He:2011yi}, would give a more complete description of the medium. Further steps include tracking the thermal pole positions of the broad dynamically generated states in the complex-energy plane and adding subleading heavy-quark mass corrections.

On the lattice side, progress requires finer temporal resolution, larger $N_\tau$, and ensembles closer to the physical pion mass. It also requires further development of reconstruction techniques with controlled systematic uncertainties~\cite{Rothkopf:2022fyo}. The extension of these studies to the bottom sector~\cite{Skullerud:2026sek} and to heavy baryons~\cite{Aarts:2023nax} is particularly promising.
A direct confrontation with EFTs, bypassing the inverse problem, was initiated in Ref.~\cite{Montana:2020var} by computing Euclidean correlators from the EFT spectral functions. Extending this strategy systematically, for instance by evaluating the unitarized EFT at the lattice pion masses or by computing the double ratios of Ref.~\cite{Aarts:2022krz}, would make the comparison quantitative.

Finally, the thermal masses, widths, and in-medium amplitudes of open heavy-flavor hadrons should be more closely integrated with heavy-ion phenomenology, through their implementation in transport simulations of the hadronic stage, the computation of transport coefficients~\cite{Torres-Rincon:2021yga}, and the interpretation of femtoscopic correlations, along the lines of the works discussed in Section~\ref{sec:applications}.

\section{Hidden heavy-flavor hadrons}\label{sec:hiddenHF}

Heavy quarkonium states (charmonium and bottomonium) are classic probes of the initial states of HICs and the hot QCD medium, and have therefore been the focus of extensive theoretical and experimental investigation. Even in vacuum, quarkonia provide a rich playground for studying the strong interaction. For a comprehensive review covering the spectroscopy and decay of quarkonia, their production, and in-medium behavior, see Ref.~\cite{Brambilla:2010cs}.

In vacuum, the quarkonium spectrum is well described by the heavy-quark potential, defined as a static color-singlet quark--antiquark potential consisting of a short-range Coulomb part and a long-range confining term, typically parametrized via the Cornell potential~\cite{Eichten:1978tg},
\be
V(r) = -\frac{4}{3}\frac{\alpha_s}{r} + \sigma r + C \ ,
\ee
where $r$ is the distance between the quark and antiquark,  the coefficient $4/3$ is the Casimir of the fundamental $\suthree$ representation, and the coupling $\alpha_s$ is the strong coupling of QCD, while $\sigma$ is known as the string tension, and $C$ is a normalization constant.

At finite temperature, this interaction is modified by the presence of the thermal medium, which led Matsui and Satz to conjecture the suppression of quarkonium production in HICs as a signal of QGP formation~\cite{Matsui:1986dk}. The basic idea was that in a deconfined medium, color screening weakens the binding between the $Q\bar Q$ pair, reducing the formation probability of bound states. This idea was later developed into a characteristic, sequential pattern of suppression depending on the binding energy of each state.
A more comprehensive understanding has been achieved over the years thanks to a combination of nonperturbative lattice QCD and EFT calculations, as well as transport models and simulations of HICs. 

The thermal behavior of quarkonia is encoded in their spectral functions, which determine their in-medium binding energies, thermal widths, and melting temperatures. Extracting these spectral functions from lattice-QCD correlators, however, entails solving an intrinsically ill-posed inverse problem, subject to the same fundamental limitations discussed earlier for open heavy flavor. In practice, quarkonium correlators often achieve higher statistical precision than their open heavy-flavor counterparts, which can somewhat alleviate these difficulties. 
Quarkonia also benefit from EFT formulations that provide simplified descriptions of heavy-quark bound states in the medium compared to fully relativistic QCD. 

The resulting in-medium spectral properties can then be implemented in transport models and simulations of HICs (see Section~\ref{sec:applications}).

\subsection{EFTs for quarkonia}\label{sec:EFT-QQ}

Successful effective theoretical descriptions of $Q\bar Q$ quarkonia are provided by nonrelativistic QCD (NRQCD)~\cite{Caswell:1985ui,Thacker:1990bm,Lepage:1992tx}, and potential NRQCD (pNRQCD)~\cite{Pineda:1997bj,Brambilla:1999xf,Brambilla:2004jw}. These frameworks exploit the hierarchy of energy scales inherent to heavy-quark systems, $m_Q\gg m_Q v\gg m_Q v^2$ together with $m_Q\gg\Lambda_\text{QCD}$. By sequentially integrating out higher-energy physics, the complicated, fully relativistic Dirac dynamics are replaced by a more tractable effective nonrelativistic description.

In NRQCD, the hard scale $m_Q$, where $Q\bar Q$ creation and annihilation occur, is integrated out, yielding an effective theory of Pauli spinors for the heavy quark and antiquark interacting with soft gluons and light quarks. NRQCD has been extensively used in lattice QCD, where it provides an efficient discretization framework for heavy quarks and is widely employed for quarkonium spectroscopy.

The next step addresses the soft scale $m_Qv$, which governs the relative momentum of the bound state. Integrating this scale out yields pNRQCD. If the resulting ultrasoft scale satisfies $m_Qv^2 \gtrsim \Lambda_\text{QCD}$, the degrees of freedom are color-singlet and color-octet heavy-quark wave functions coupled to ultrasoft gluons via potentials derived directly from QCD. If instead $m_Qv^2 \lesssim \Lambda_\text{QCD}$, the system enters the strong-coupling regime; octet states become nondynamical, and the theory reduces to a color-singlet quarkonium field scattering with ultrasoft Goldstone bosons (e.g., pions). Detailed derivations of these limits can be found in Refs.~\cite{Grinstein:1998xb,Brambilla:2004jw}.

When extending these frameworks to a thermal medium, an additional hierarchy involving the temperature $T$ and the QCD coupling $g$ emerges: $T\gg gT\gg g^2T$~\cite{Brambilla:2008cx,Brambilla:2010vq,Brambilla:2011sg}. Here, the scale $gT\sim m_D$ corresponds to the Debye mass that governs the screening of chromoelectric interactions, while chromomagnetic screening occurs at $g^2T$. The construction of a thermal EFT depends crucially on how the heavy-quark vacuum scales compare to these thermal scales:
\begin{itemize}
    \item $T < m_Qv^2$: The vacuum pNRQCD potential remains unchanged, although thermal effects generate corrections to the binding energies and decay widths.
    \item $m_Qv > T > m_Qv^2$: The thermal scale $T$ is integrated out. The potential receives thermal corrections, acquiring both real and imaginary parts.
    \item $T > m_Qv$: The temperature scale is integrated out before the soft scale $m_Qv$. The gluon and light-quark sectors of NRQCD are replaced by the Hard Thermal Loop (HTL) effective Lagrangian~\cite{Pisarski:1988vd}, yielding NRQCD$_\text{HTL}$. Integrating out the soft scale then leads to pNRQCD$_\text{HTL}$.
\end{itemize}

The emergence of a complex heavy-quark potential represents an important change in the understanding of quarkonium suppression. Its imaginary part is related to the thermal decay width and comes from processes such as Landau damping and thermal gluon dissociation (singlet-to-octet transitions). This suggests that quarkonia can melt dynamically due to thermal scattering long before static color screening destroys the bound-state potential.

\subsection{Lattice QCD for quarkonia}\label{sec:QQ-LQCD}

Heavy quarkonium systems have been extensively studied in lattice QCD. While fully relativistic lattice QCD studies of charmonium states at finite temperature have been performed for over two decades, several challenges have limited quantitative insights into their thermal properties. In particular, in addition to the difficulties associated with the inverse problem discussed in Section~\ref{sec:D-LQCD}, spatial lattice spacings much smaller than the heavy-quark Compton wavelength ($a_s\ll 1/m_Q$) are required to avoid large discretization effects; and relativistic spectral functions contain a low-frequency transport peak associated with heavy-quark diffusion, which complicates the extraction of the bound-state signal. Lattice formulations of NRQCD avoid some of these issues and make the extraction of spectral properties less demanding. For instance, by integrating out the heavy-quark mass scale, the zero-mode transport contribution is separated from the bound-state dynamics. This comes, however, at the cost of simulating an EFT of QCD truncated at a given order. Additionally, approximations to quarkonium spectral functions can be obtained on the lattice using the in-medium potential derived from pNRQCD. In the following, we summarize some of the lattice QCD results for quarkonia over the years, and refer the reader to other reviews focusing specifically on quarkonia for a more comprehensive discussion~\cite{Bazavov:2009us,Mocsy:2013syh,Rothkopf:2019ipj}.

\begin{figure}[htbp!] 
\centering
\includegraphics[width=0.35\linewidth]{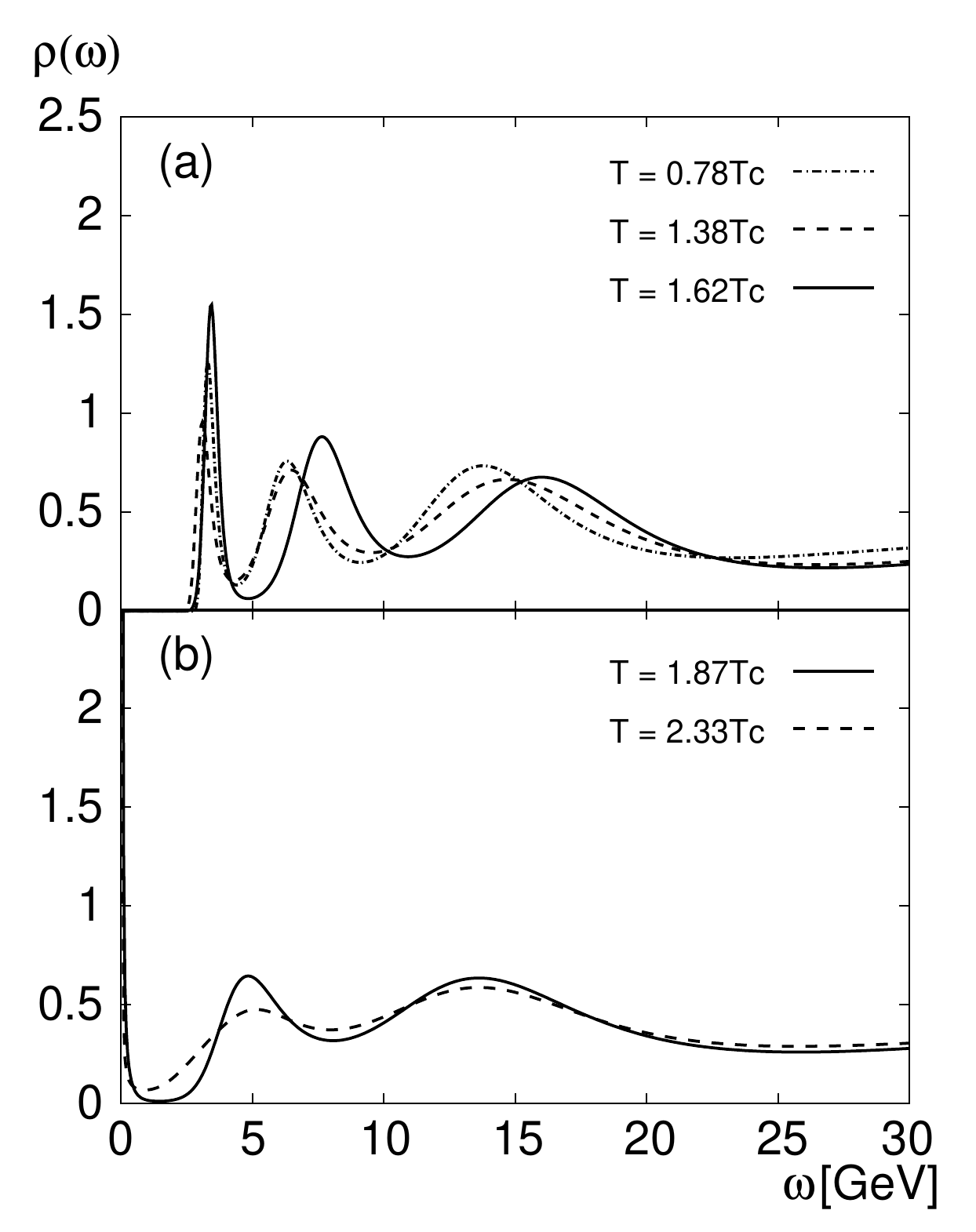}
\includegraphics[width=0.35\linewidth]{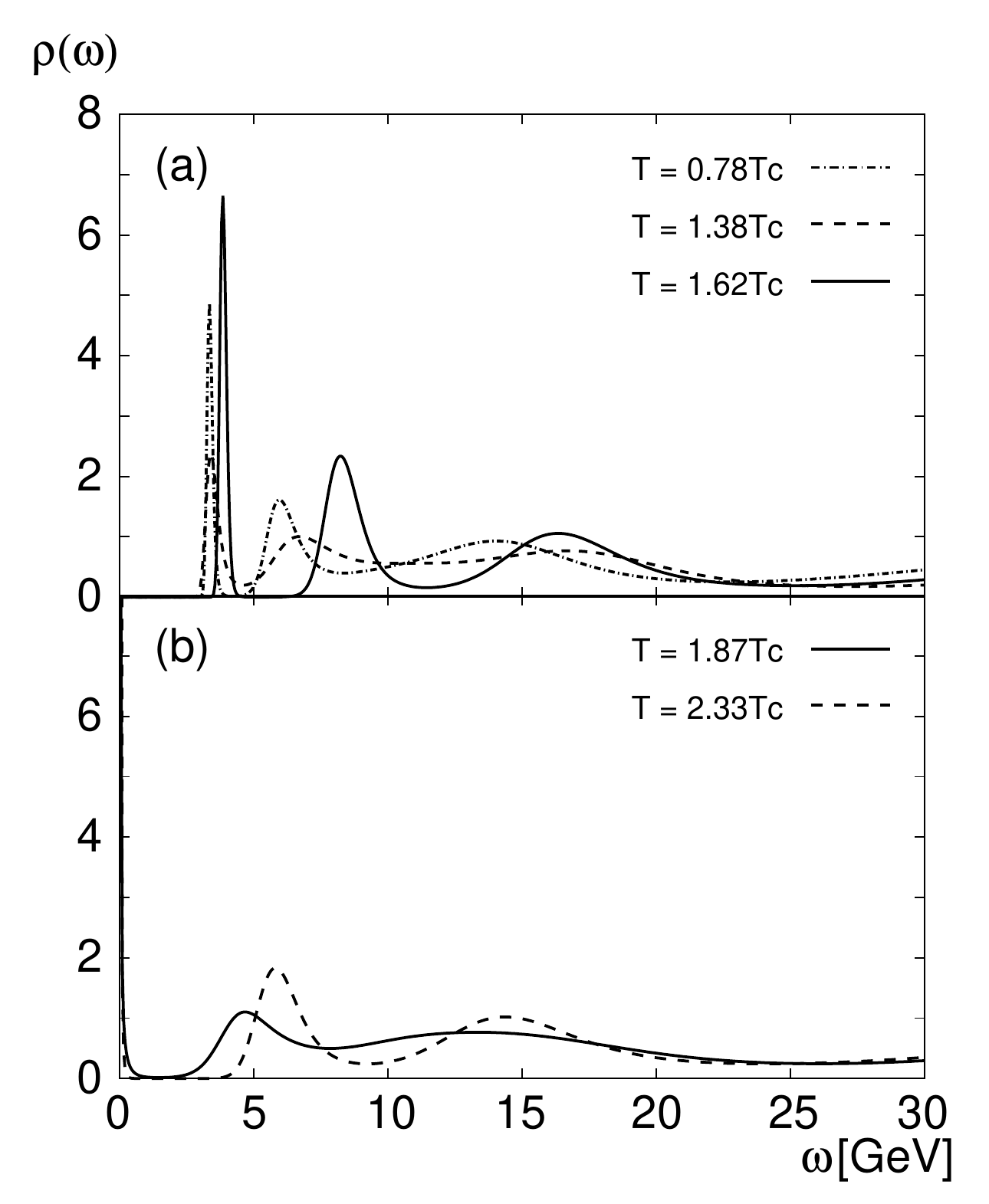}
\caption{Spectral function of $J/\psi$ (left) and $\eta_c$ (right) from quenched lattice QCD at finite temperature, reconstructed using MEM. Figure taken from Ref.~\cite{Asakawa:2003re}.}
\label{fig:Asakawa_Sfunc}
\end{figure}

\begin{figure}[htbp!] 
\centering
\includegraphics[angle=270,width=0.35\linewidth]{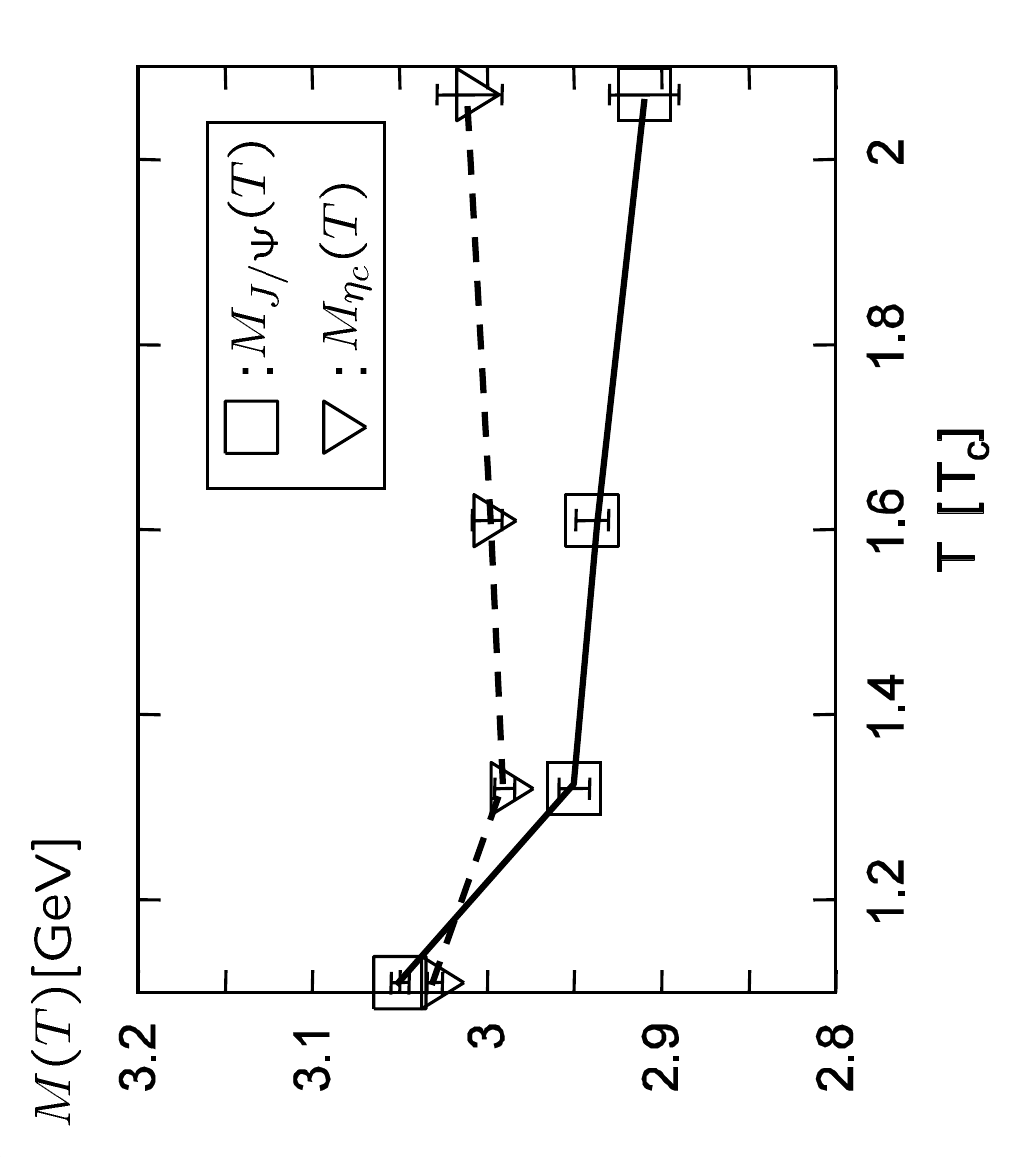}
\caption{Temperature dependence of the pole mass of $J/\psi$ and $\eta_c$ extracted from the effective mass plot in quenched lattice QCD. Figure taken from Ref.~\cite{Iida:2006mv}.}
\label{fig:Iida_mass}
\end{figure}

The first relativistic lattice QCD studies of charmonium spectral functions at finite temperature were performed in the quenched approximation (i.e., ignoring dynamical quark loops) using anisotropic lattices. Using MEM for spectral function reconstruction, Ref.~\cite{Asakawa:2003re} found that the $J/\psi$ and $\eta_c$ can survive well above the transition temperature, at least up to temperatures of about $1.6\,T_c$, as seen in Fig.~\ref{fig:Asakawa_Sfunc}. These results were confirmed by other quenched studies~\cite{Datta:2003ww,Jakovac:2006sf}, which also found that the $\chi_{c0}$ and $\chi_{c1}$ states are more strongly affected already near $T_c$. Pole masses derived in Ref.~\cite{Iida:2006mv} from the quenched quarkonium correlator for $J/\psi$ and $\eta_c$ through the effective mass plot are shown in Fig.~\ref{fig:Iida_mass}.


\begin{figure}[htbp!] 
\centering
\includegraphics[width=0.4\linewidth]{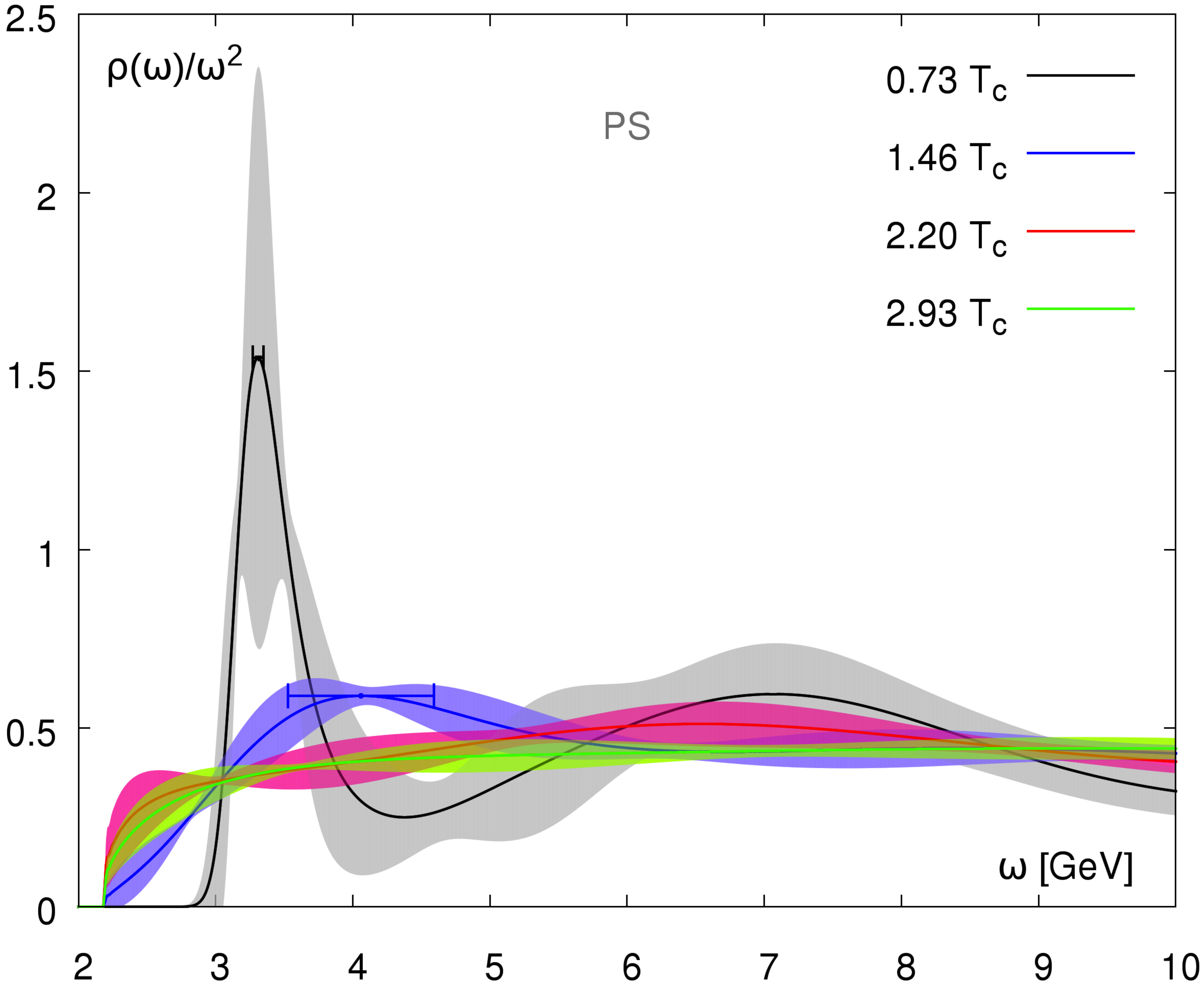}
\includegraphics[width=0.35\linewidth]{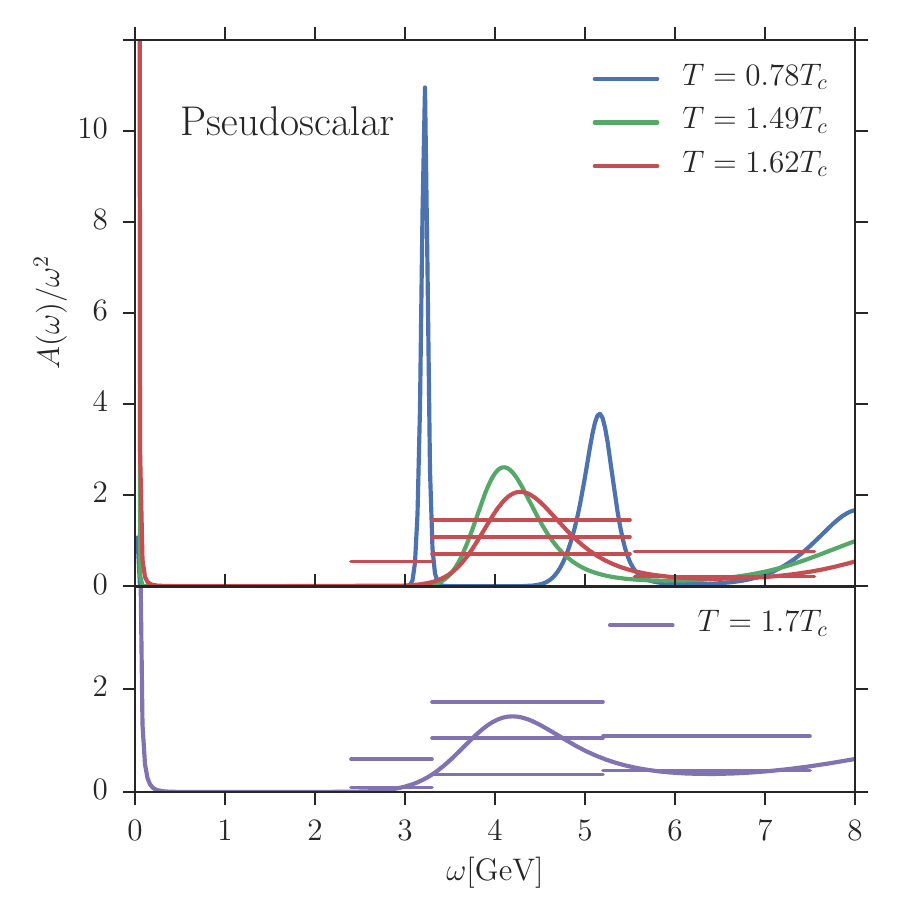}
\caption{Pseudoscalar spectral functions from quenched lattice QCD reconstructed using MEM. Figures taken from Ref.~\cite{Ding:2012sp} (left) and Ref.~\cite{Ikeda:2016czj} (right).}
\label{fig:PS_spectral_quenched}
\end{figure}

More recent extractions of the charmonium spectral functions from quenched lattice QCD include those of Refs.~\cite{Ding:2012sp,Ikeda:2016czj}. Both studies used MEM to reconstruct the spectral functions. Ref.~\cite{Ikeda:2016czj} used anisotropic lattices following the strategy of earlier studies, which enhanced temporal resolution at the cost of larger spatial cutoff effects. In contrast, the authors of Ref.~\cite{Ding:2012sp} employed large isotropic lattices to reduce cutoff and discretization effects. The two studies also reached significantly different conclusions, as shown in Fig.~\ref{fig:PS_spectral_quenched} for the pseudoscalar channel. While Ikeda et al. observed clear charmonium peaks at $T=1.62\,T_c$ (right panel), Ding et al. found that the peaks had already melted at $T=1.46\,T_c$ (left panel). This highlights how strongly the extracted spectral features depend on the lattice setup and reconstruction method.

\begin{figure}[!ht] 
\centering
\begin{subfigure}{0.48\textwidth}
    \includegraphics[width=0.49\textwidth]{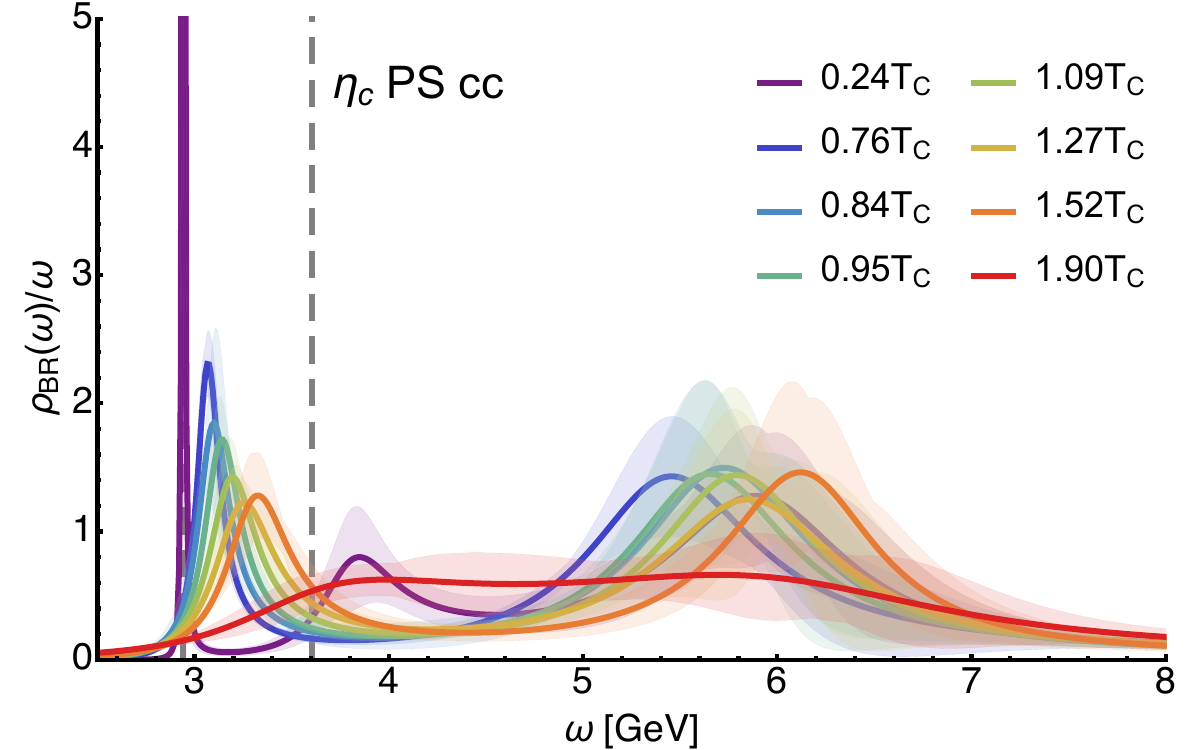}
    \includegraphics[width=0.49\textwidth]{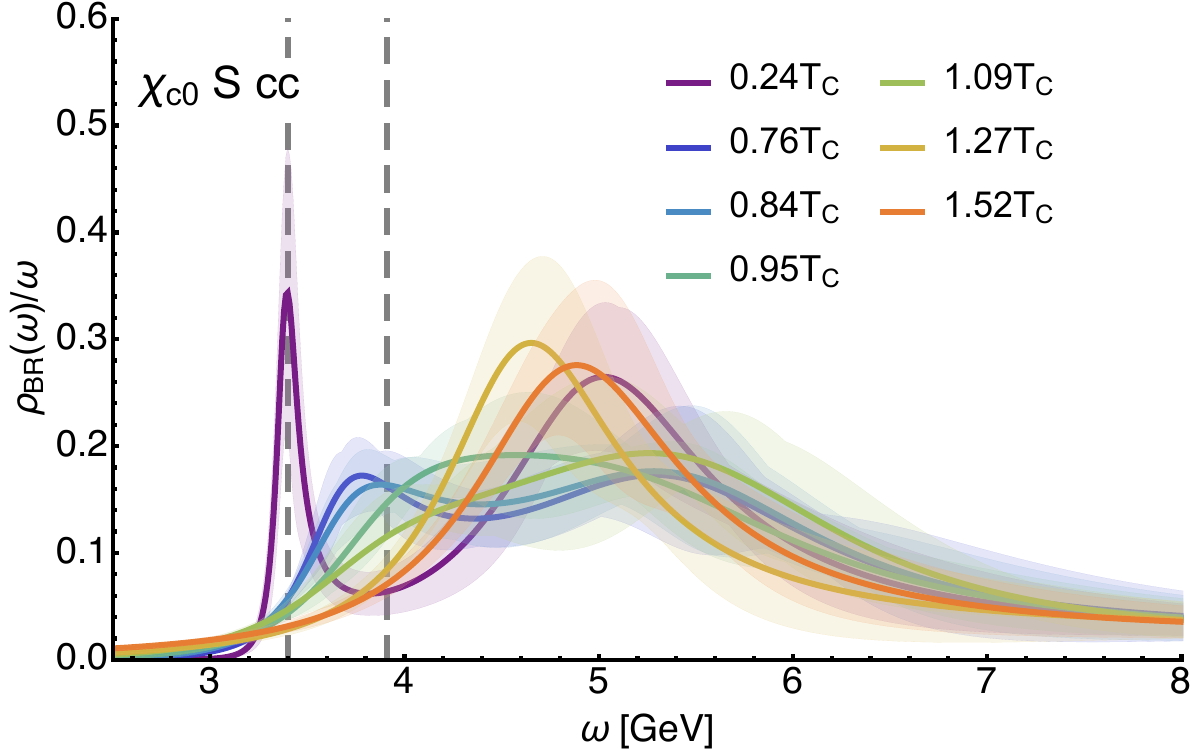} \\
    \includegraphics[width=0.49\textwidth]{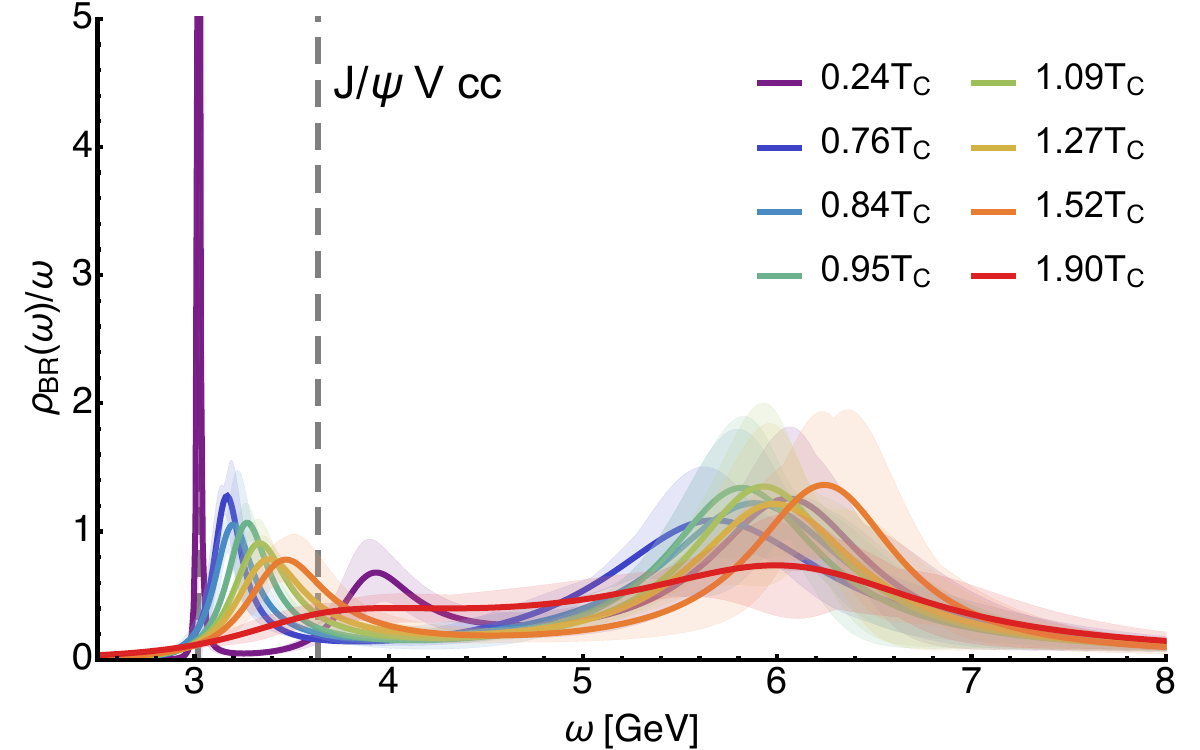}
    \includegraphics[width=0.49\textwidth]{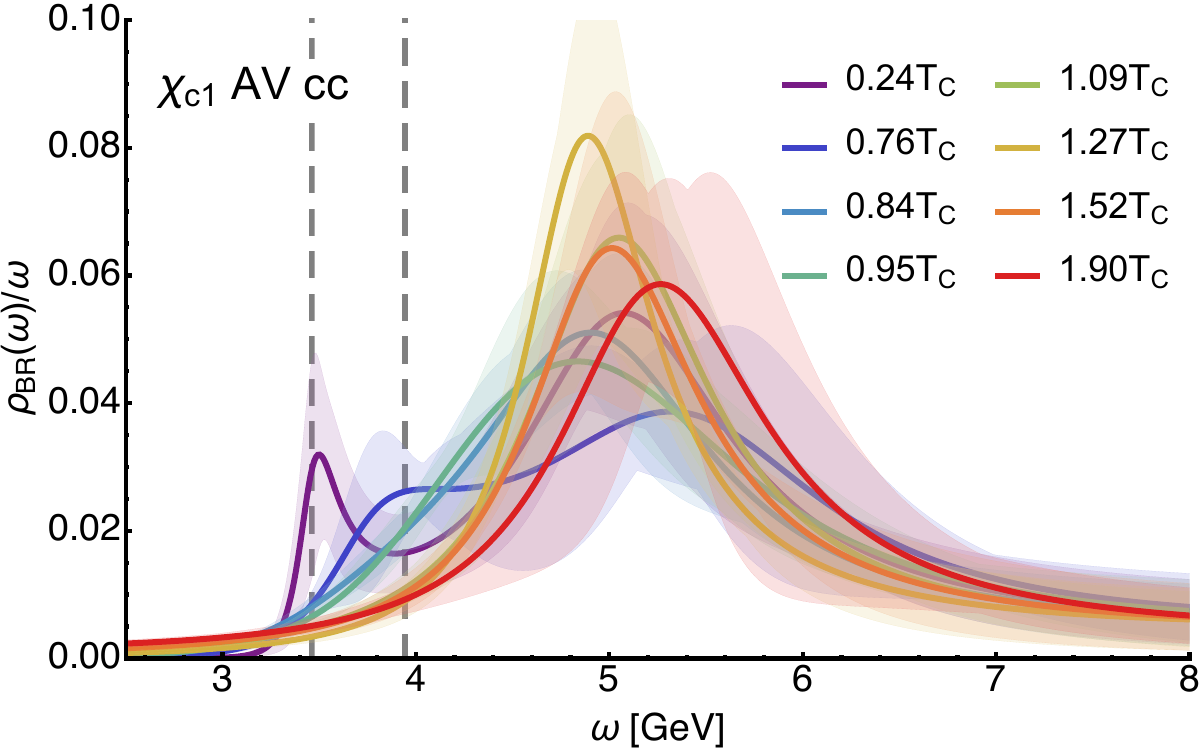}
    \caption{BR reconstruction}
\end{subfigure}
\begin{subfigure}{0.48\textwidth}
    \includegraphics[width=0.49\textwidth]{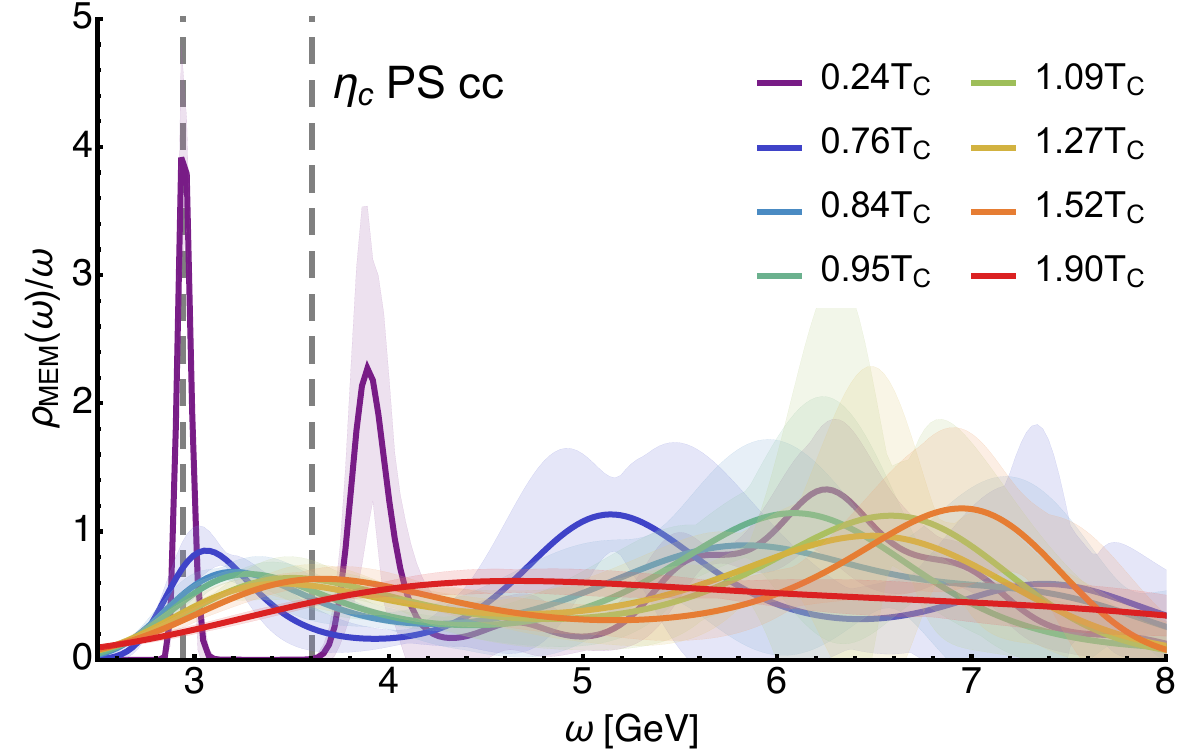}
    \includegraphics[width=0.49\textwidth]{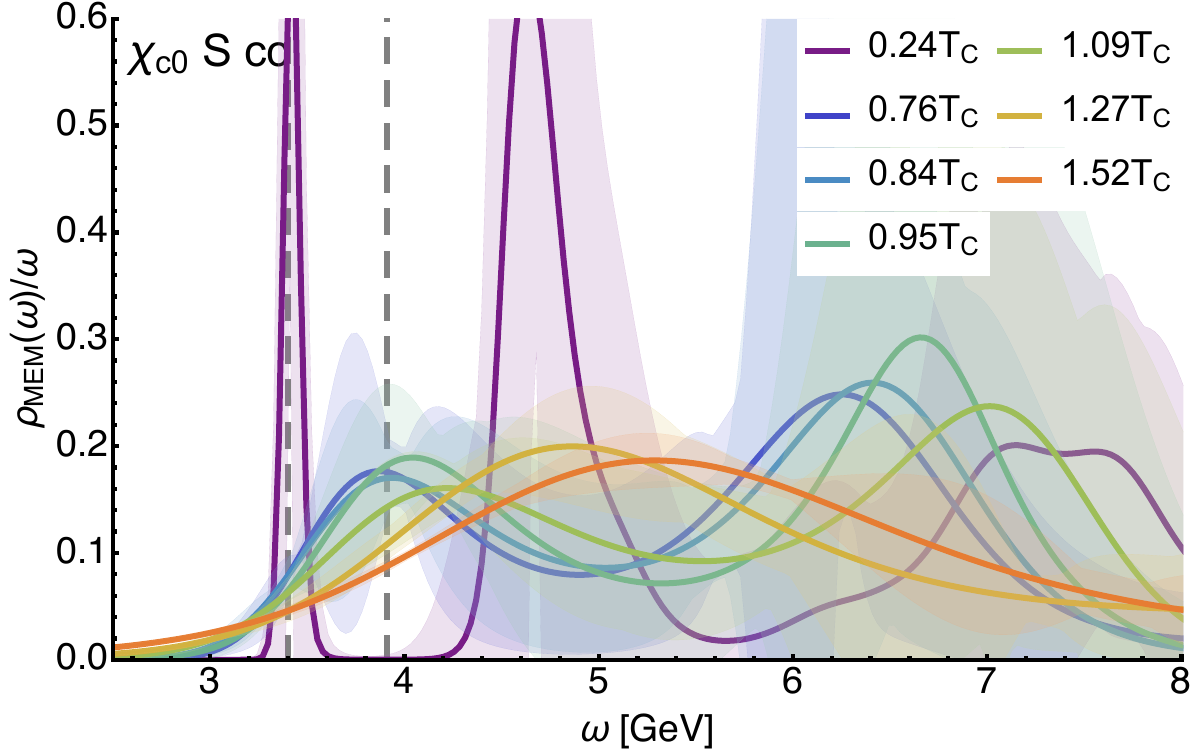} \\
    \includegraphics[width=0.49\textwidth]{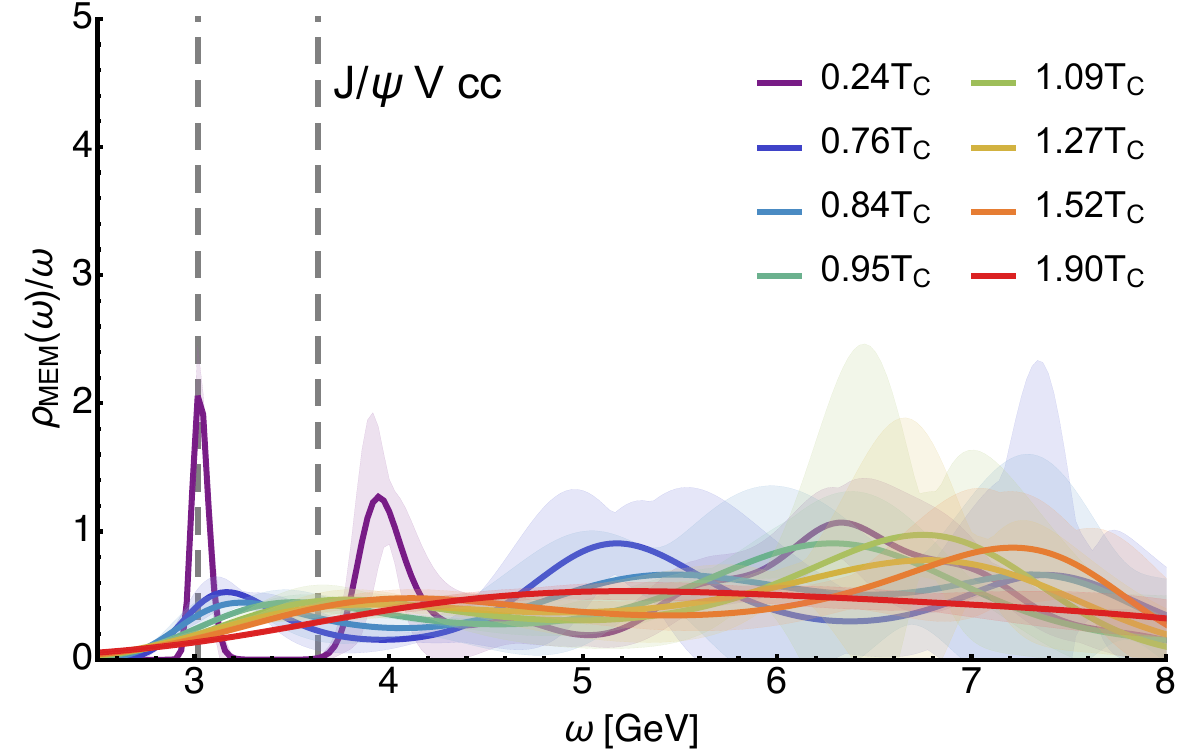}
    \includegraphics[width=0.49\textwidth]{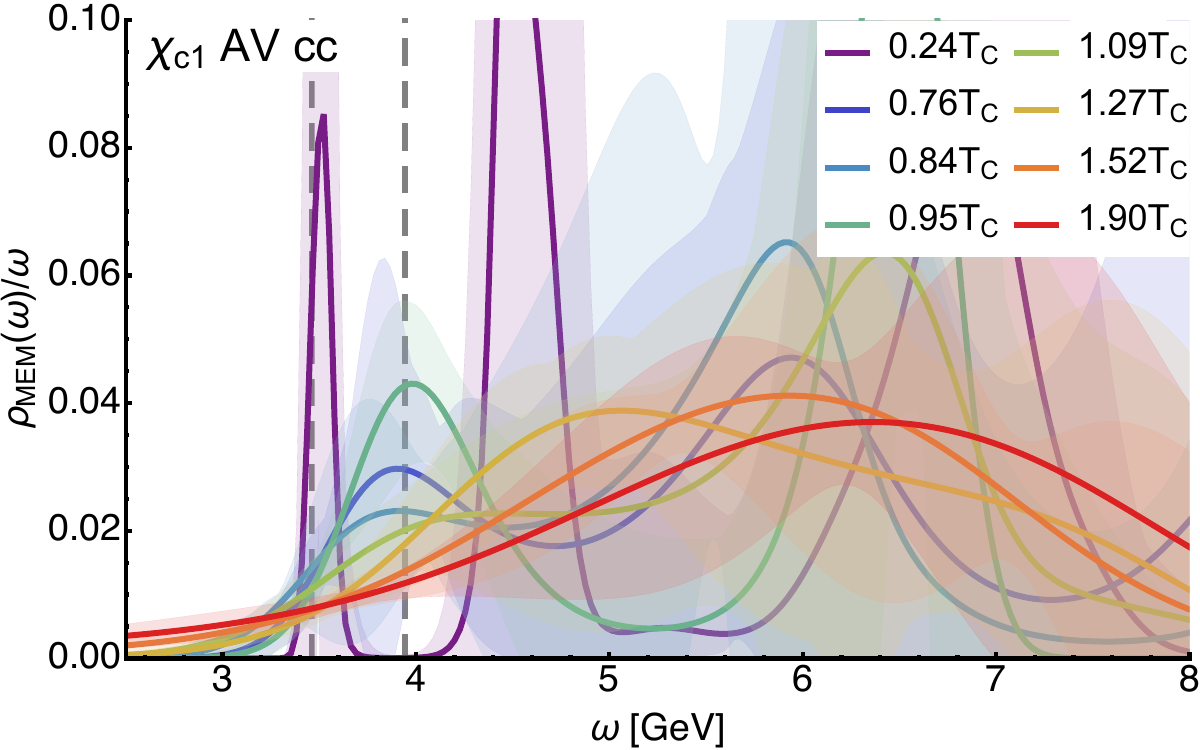}
    \caption{MEM reconstruction}
\end{subfigure}\caption{Spectral functions of the charmonium states reconstructed from $N_f=2+1$ lattice QCD Euclidean correlators at various temperatures. Figures obtained from Ref.~\cite{Kelly:2018hsi}.}
\label{fig:KellyCharmonia}
\end{figure}

Studies incorporating dynamical quarks in fully relativistic lattice QCD have also been performed. Results broadly consistent with those obtained in the quenched approximation have been reported in two-flavor QCD simulations on anisotropic lattices using MEM~\cite{Aarts:2007pk}, as well as in calculations with $N_f=2+1$ dynamical quarks on fine isotropic lattices with $m_\pi\approx 545\mev$~\cite{Borsanyi:2014vka}. More recently, Kelly et al. reconstructed spectral functions of both charmonium and open charm states from $N_f=2+1$ FASTSUM anisotropic ensembles~\cite{Kelly:2018hsi}. Figure~\ref{fig:KellyCharmonia} compares their results across the four charmonium channels, reconstructed via BR (left) and MEM (right). Although the differences between the two techniques illustrate the large systematic uncertainties inherent in spectral reconstruction, the authors concluded that there was little modification of the $J/\psi$ and $\eta_c$ spectral functions around $T_c$, whereas the $\chi_{c0}$ and $\chi_{c1}$ are much more strongly affected by the thermal medium. 

Screening masses for $c\bar{c}$ states have also been computed in $N_f=2+1$ lattice QCD from spatial correlators~\cite{Bazavov:2014cta} (see right panel of Fig.~\ref{fig:Bazavov-sc-mass}), also finding larger modifications at $T\gtrsim T_c$ for the P-wave states ($\chi_{c0}$ and $\chi_{c1}$) than for the S-wave states ($\eta_c$ and $J/\psi$).
The calculation of screening masses has also been extended to bottomonium at high temperatures.

Reference~\cite{Petreczky:2021zmz} used 2+1 flavors of quarks discretized with the relativistic HISQ action to calculate spatial screening correlators of bottomonium in the $0^{-+}$, $1^{--}$, $0^{++}$, and $1^{+-}$ channels for temperatures in the range $350\mev\lesssim T<1\gev$. They employed $N_\tau=8,10,12$ lattices to control discretization effects. The extracted screening masses show sequential melting: the $1P$ $\chi_{b0}$ and $h_b$ states dissolve already at $T\gtrsim350\mev$, while the $1S$ $\eta_b$ and $\Upsilon$ states remain identifiable to substantially higher temperatures, with melting indicated at $T\gtrsim500\mev$.

At lower temperature, however, relativistic lattice QCD calculations are more challenging for bottomonium. The large bottom quark mass requires extremely fine lattices ($a_s\ll 1/m_b$) to suppress discretization artifacts. As a consequence, lattice studies of bottomonium at finite temperature rely on NRQCD for efficient discretization. In addition, the lattice NRQCD scheme avoids several difficulties associated with relativistic quarks at finite temperature, such as dealing with temperature-dependent integration kernels and the transport peak contribution to the spectral function~\cite{Aarts:2002cc,Petreczky:2005nh}. 

\begin{figure}[!ht] 
\centering
    \includegraphics[height=5cm]{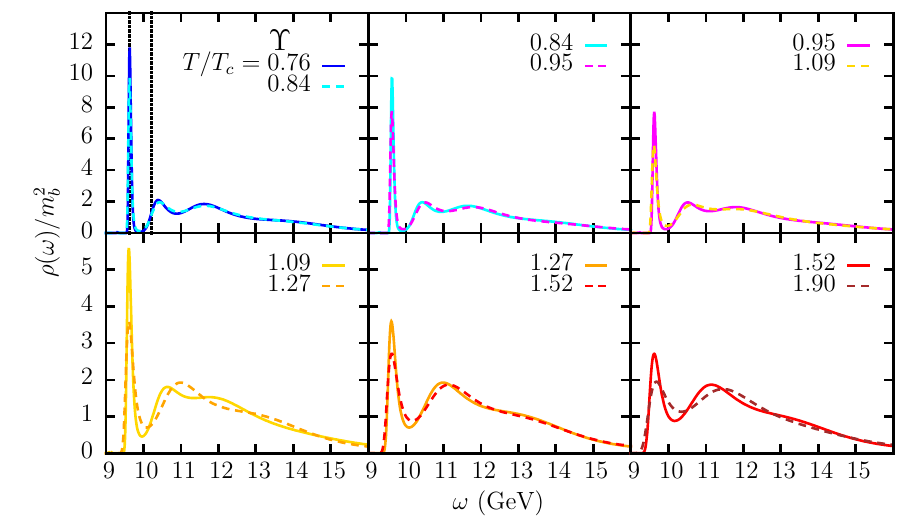}
    \includegraphics[height=5cm]{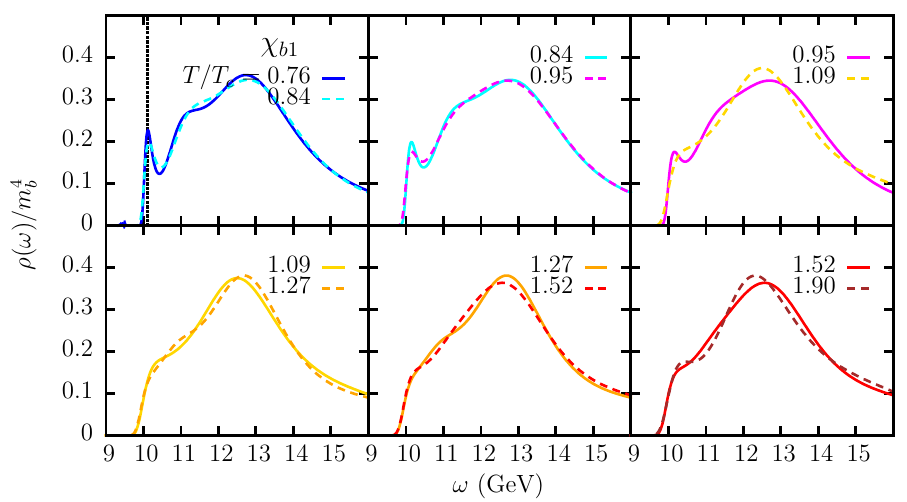}
\caption{S-wave (left) and P-wave (right) bottomonium spectral functions reconstructed from $N_f=2+1$ lattice NRQCD Euclidean correlators at various temperatures. Figures obtained from Ref.~\cite{Aarts:2014cda}.}
\label{fig:AartsBottomonia}
\end{figure}

A comprehensive effort to study bottomonium at finite temperature using lattice NRQCD was initiated by the FASTSUM collaboration in a series of publications~\cite{Aarts:2010ek,Aarts:2011sm,Aarts:2012ka,Aarts:2013kaa}. In this initial framework, the heavy $b$ quarks propagate nonrelativistically through a medium containing $N_f=2$ dynamical light quark flavors, simulated on highly anisotropic lattices ($a_s/a_\tau=6$) with $m_\pi\approx 400\mev$. This framework was later extended to include $N_f=2+1$ light flavors with finer spatial lattices and anisotropy $a_s/a_\tau=3.5$~\cite{Aarts:2014cda}. By reconstructing the bottomonium spectral functions using the MEM, these studies indicated that the S-wave ground states ($\eta_b$ and $\Upsilon$) survive up to the highest temperatures studied, $T\simeq2T_c$, while the first excited states and the P-wave states ($\chi_{b}$) appear to dissolve at temperatures near $T_c$, as seen in Fig.~\ref{fig:AartsBottomonia}.

\begin{figure}[!ht] 
\centering
    \includegraphics[width=0.8\linewidth]{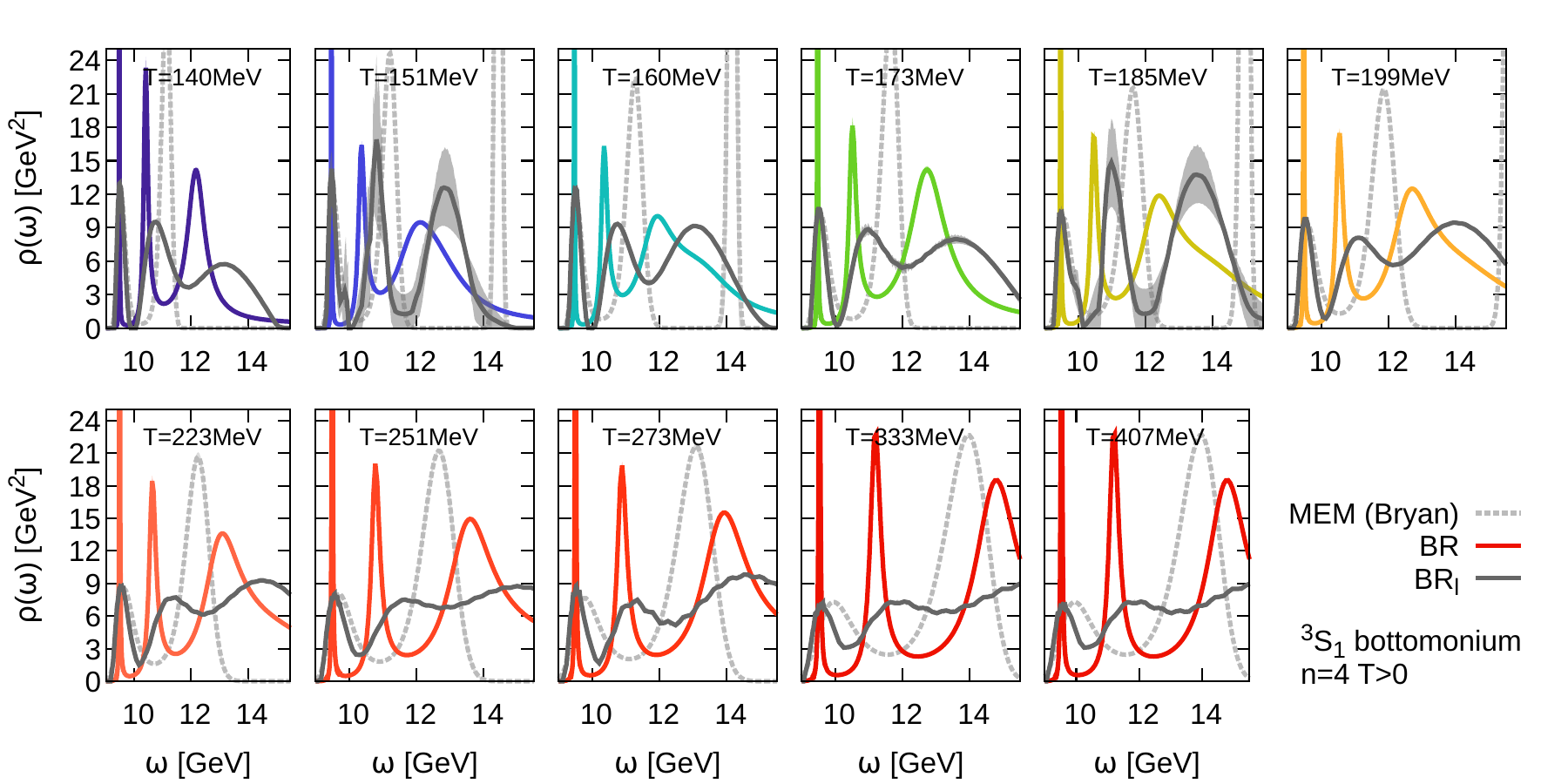}
\caption{Spectral functions reconstructed for S-wave bottomonium obtained using the MEM (gray dashed), standard BR (colored solid), and the smooth BR (dark gray solid) methods. Figure obtained from Ref.~\cite{Kim:2018yhk}.}
\label{fig:KimBottomonia}
\end{figure}

\begin{figure}[!ht] 
\centering
    \includegraphics[width=0.4\linewidth]{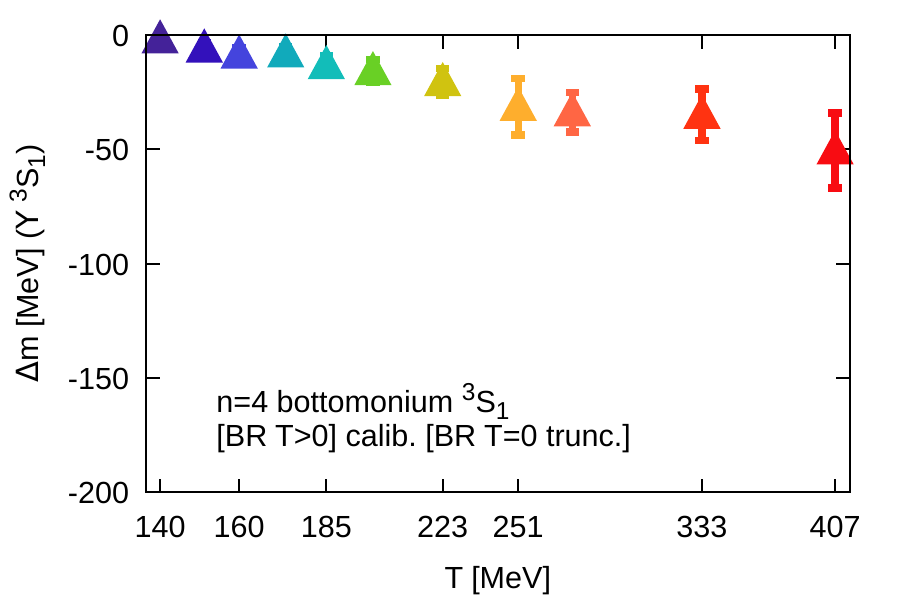}
    \includegraphics[width=0.4\linewidth]{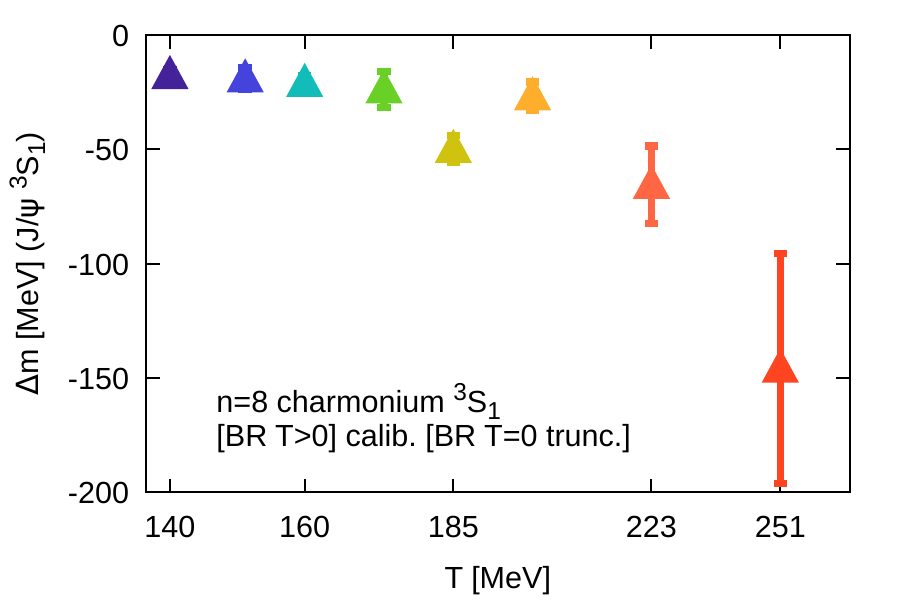}
    \includegraphics[width=0.4\linewidth]{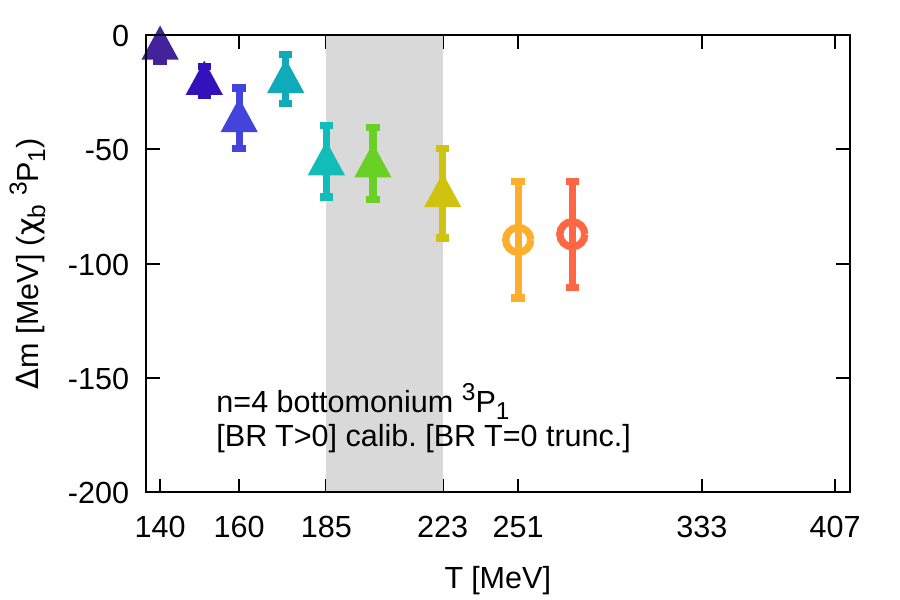}
    \includegraphics[width=0.4\linewidth]{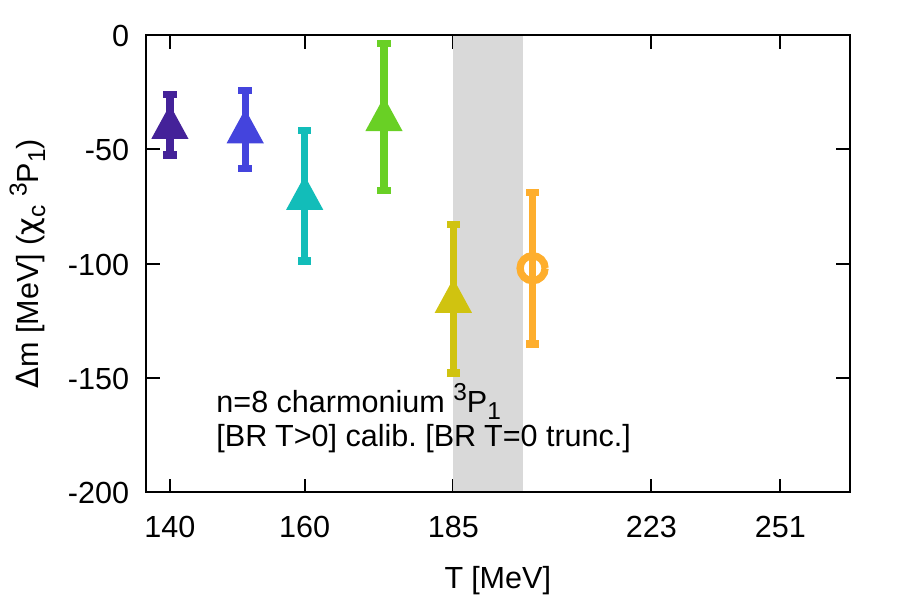}
\caption{Temperature dependence of the in-medium mass shift for bottomonium (left) and charmonium (right) S-wave (top) and P-wave (bottom) ground states. Figures obtained from Ref.~\cite{Kim:2018yhk}.}
\label{fig:KimMassShift}
\end{figure}

In parallel, an independent lattice NRQCD approach to study the spectral properties of quarkonia using $N_f=2+1$ configurations from the HotQCD collaboration with near-physical pion mass ($m_\pi\approx 161\mev$) was developed in Refs.~\cite{Kim:2014iga,Kim:2018yhk,Larsen:2019bwy,Larsen:2019zqv,Ding:2025fvo}.
Within this program, a comprehensive analysis by Kim et al.~\cite{Kim:2018yhk} significantly improved the understanding of the artifacts associated with various Bayesian reconstruction methods (MEM, standard BR, and smooth BR), demonstrating that previously reported discrepancies in melting temperatures were largely due to underestimated reconstruction uncertainties. All three methods consistently support a picture in which the S-wave bottomonium ground states survive up to very high temperatures ($T \sim 2.6 \, T_c$), as shown in Fig.~\ref{fig:KimBottomonia}, whereas the P-wave states are much more strongly modified and dissolve at considerably lower temperatures. The analysis covered both bottomonium and charmonium, with the temperature modification of charmonium states found to be substantially stronger than that of their bottomonium counterparts, consistent with the expectation of sequential melting, whereby more loosely bound states dissolve at lower temperatures. Furthermore, by comparing finite-temperature spectral functions with $T=0$ baselines that were carefully reconstructed from $T=0$ correlators truncated at the same time extent as the thermal correlator, the authors extracted negative in-medium mass shifts, shown in Fig.~\ref{fig:KimMassShift}. 

\begin{figure}[!ht] 
\centering
    \includegraphics[width=0.45\linewidth]{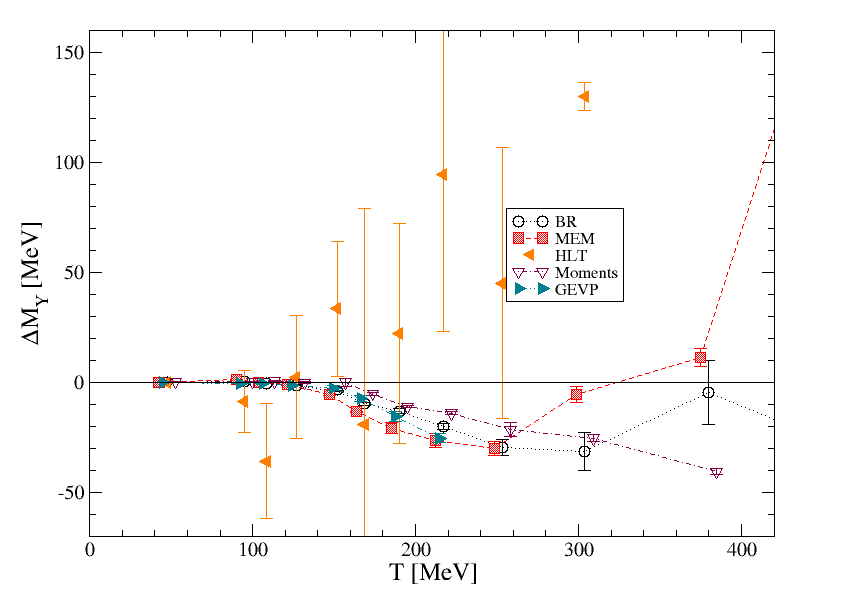}
    \includegraphics[width=0.45\linewidth]{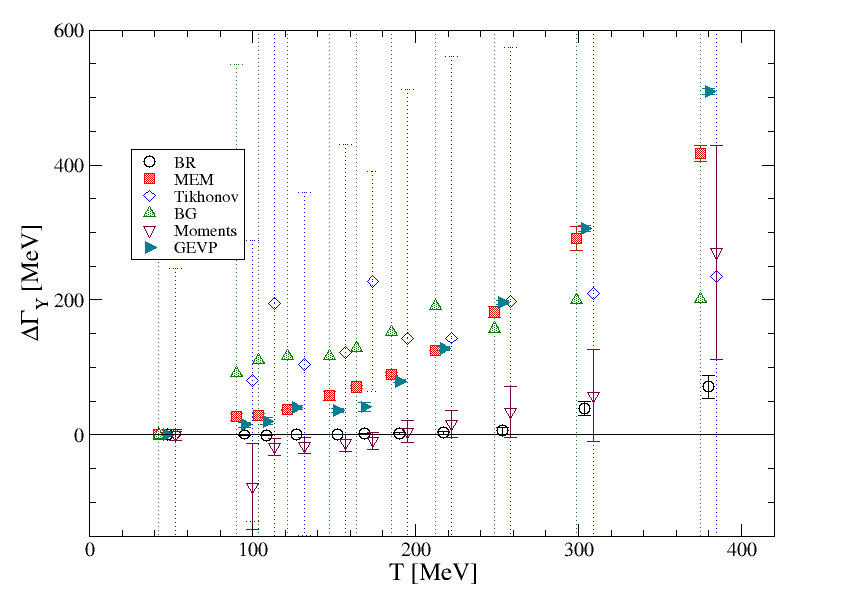}
    \includegraphics[width=0.45\linewidth]{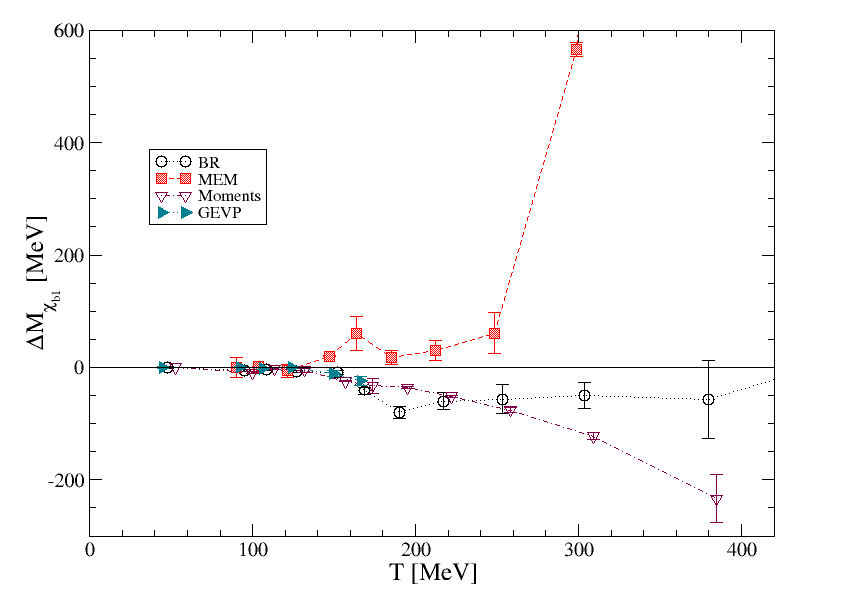}
    \includegraphics[width=0.45\linewidth]{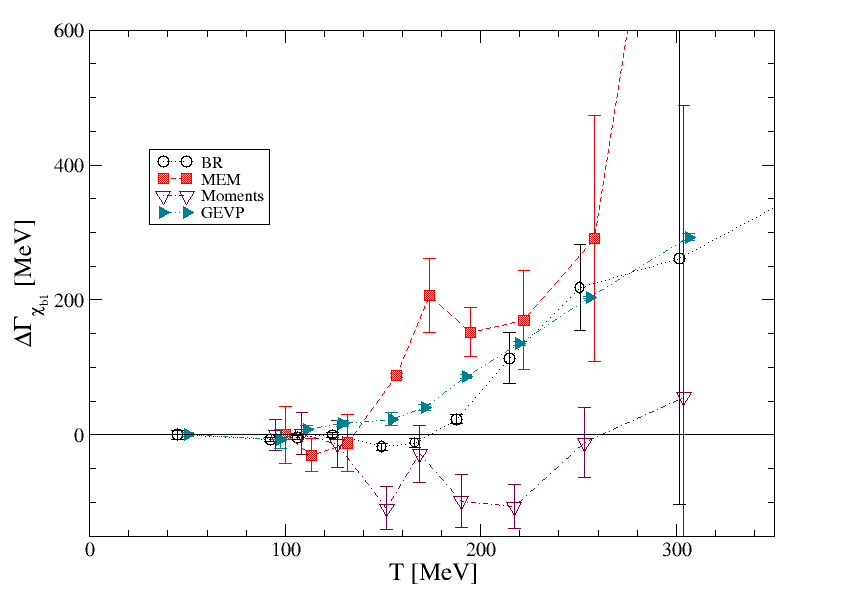}
\caption{Thermal mass shift (left) and width (right) of the $\Upsilon(1S)$ (top) and $\chi_{b1}(1P)$ (bottom) extracted from lattice NRQCD correlators with different methods. Figures obtained from Ref.~\cite{Skullerud:2025iqt}.}
\label{fig:Skullerud_comparison}
\end{figure}

A comprehensive study comparing a range of methods for extracting thermal mass shifts and widths from lattice NRQCD correlators in the bottomonium sector has been presented in Ref.~\cite{Skullerud:2025iqt}.
The analysis includes direct approaches based on the correlators, such as multi-exponential fits, or the calculation of time-derivative moments, which provide access to the location and width of the spectral peaks. It also considers spectral reconstruction techniques, including Bayesian methods (MEM~\cite{Asakawa:2000tr} and BR~\cite{Burnier:2013nla}) and linear approaches (Backus--Gilbert~\cite{Backus:1968svk}, Tikhonov~\cite{Tikhonov1943OnTS}, and Hansen--Lupo--Tantalo~\cite{Hansen:2019idp}). The latter regularizes the inverse problem by avoiding a point-by-point reconstruction and instead yields a ``smeared'' spectral function. Their results for the $\Upsilon$ and $\chi_{b1}$, shown in Fig.~\ref{fig:Skullerud_comparison}, indicate that a controlled extraction of thermal spectral properties is achievable when systematic uncertainties are assessed through a comparison of different methods.

\begin{figure}[!ht] 
\centering
    \includegraphics[width=0.4\linewidth]{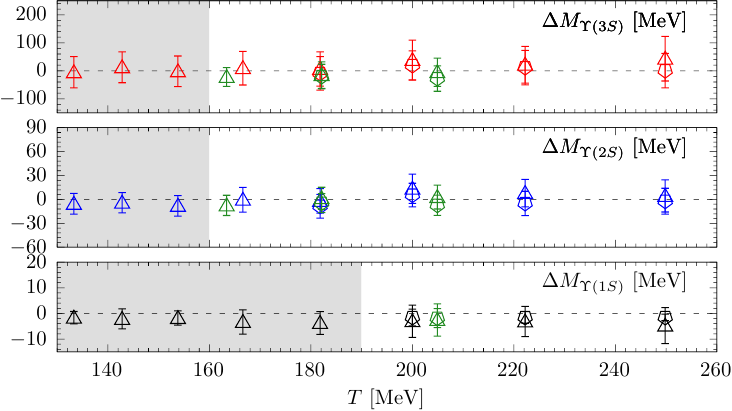}
    \includegraphics[width=0.4\linewidth]{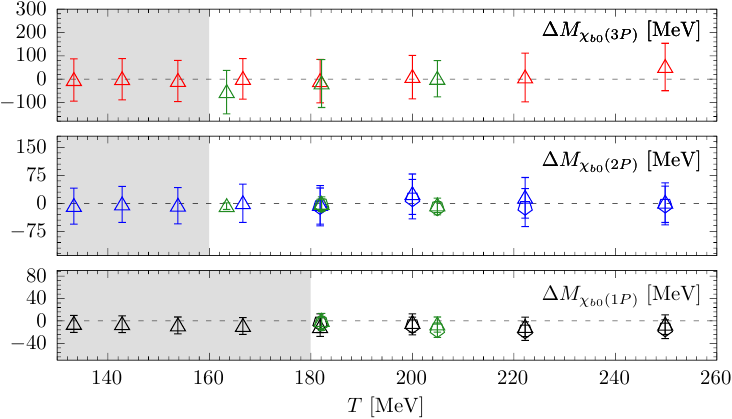}
\caption{Temperature dependence of the in-medium mass shift for $\Upsilon$ (left) and $\chi_{b0}$ (right) states from lattice NRQCD with extended operators. Different symbols correspond to various parametrization fits. Figures obtained from Ref.~\cite{Ding:2025fvo}.}
\label{fig:DingMassShift}
\end{figure}

\begin{figure}[!ht] 
\centering
    \includegraphics[width=0.3\linewidth]{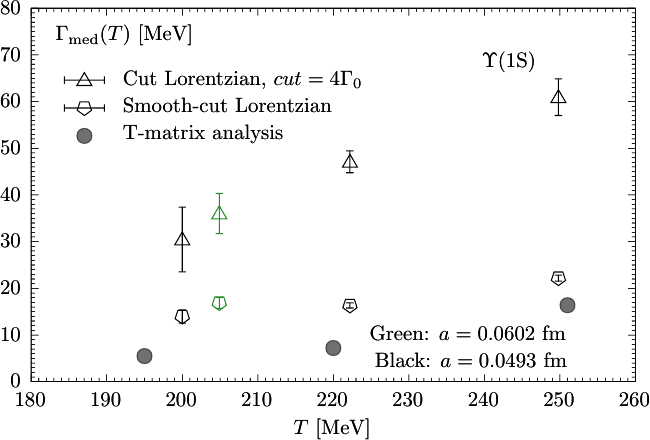}
    \includegraphics[width=0.3\linewidth]{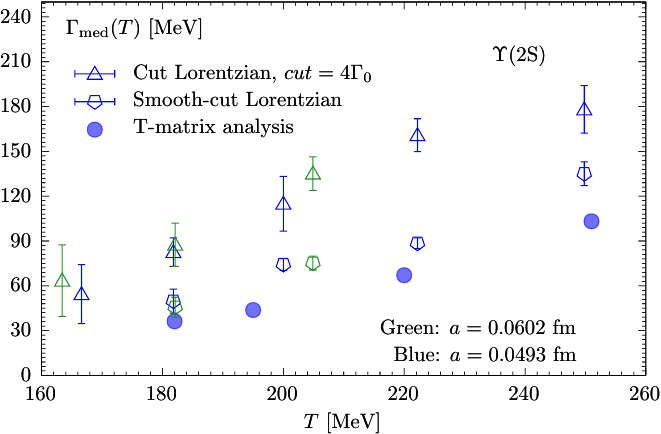}
    \includegraphics[width=0.3\linewidth]{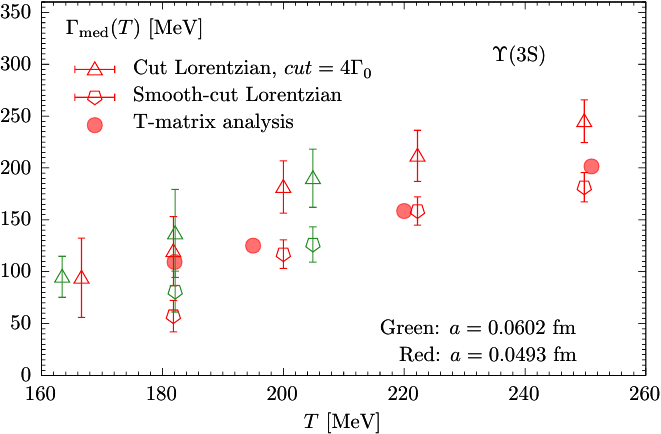}
    \includegraphics[width=0.3\linewidth]{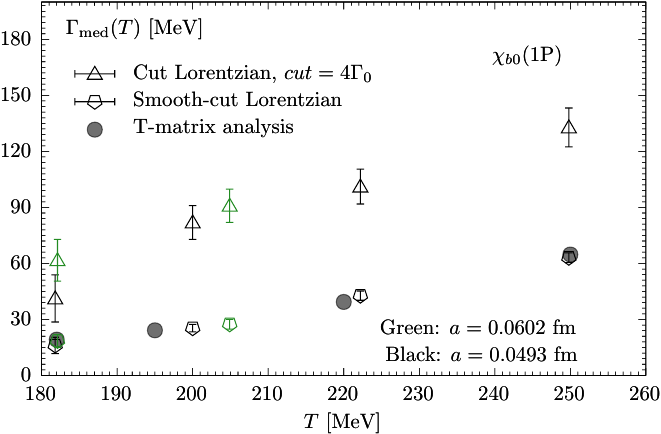}
    \includegraphics[width=0.3\linewidth]{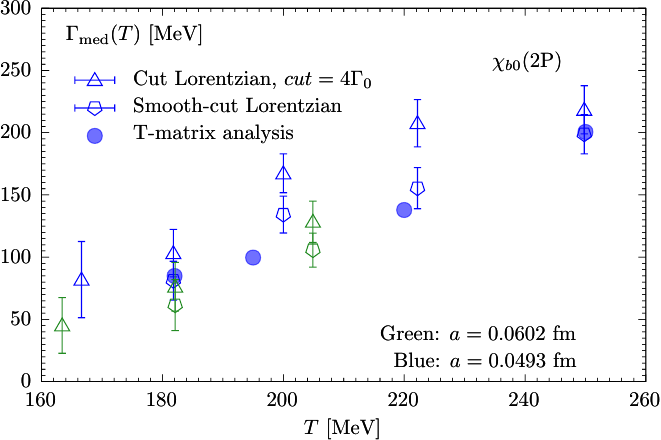}
    \includegraphics[width=0.3\linewidth]{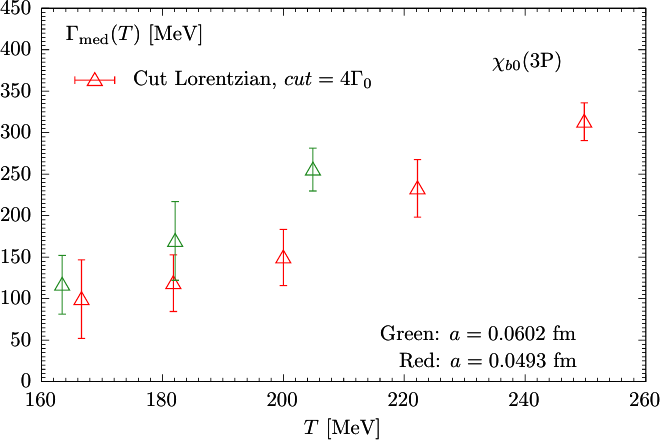}
\caption{Temperature dependence of the in-medium widths for $\Upsilon$ (top panels) and $\chi_{b0}$ (bottom panels) states from lattice NRQCD with extended operators. Different symbols correspond to various parametrization fits. Figures obtained from Ref.~\cite{Ding:2025fvo}.}
\label{fig:DingWidth}
\end{figure}

Recently, the lattice NRQCD program has been expanded to include extended meson operators for improved sensitivity to in-medium properties of quarkonium states~\cite{Larsen:2019bwy,Larsen:2019zqv,Ding:2025fvo}. These studies find nonzero thermal widths for various bottomonium states that increase with temperature, while no significant mass shifts are observed. Figure~\ref{fig:DingMassShift} displays the mass shift for the $\Upsilon$ and $\chi_{b0}$ ground states and their lowest radial excitations from Ref.~\cite{Ding:2025fvo}, while the corresponding widths are shown in Fig.~\ref{fig:DingWidth}, where the various symbols correspond to different parametrizations of the spectral function of the quasiparticle peak employed to fit the Euclidean correlators. The filled points correspond to a thermodynamic $T$-matrix analysis of the bottomonium lattice correlators from Ref.~\cite{Tang:2024dkz} (see Section~\ref{sec:QQ-other}). 

A complementary approach to obtaining the quarkonium spectral properties from lattice QCD that avoids the direct reconstruction of spectral functions from meson correlators is to combine lattice extractions of the complex thermal $Q\bar{Q}$ potential with the solution of the Schr\"odinger equation within the pNRQCD framework~\cite{Burnier:2017bod,Ali:2025iux}:
\be
\left(2m_Q-\frac{\nabla^2}{m_Q}+V(r)\right)\mathcal{C}^>(r,r',t)=\ii\frac{\partial \mathcal{C}^>(r,r',t)}{\partial t} \ ,
\ee
where $\mathcal{C}^>(r,r',t)$ is the point-split correlator, evaluated at separation $r'$ at the source and $r$ at the sink. The complex in-medium potential can be extracted from lattice simulations via a spectral decomposition of the thermal Wilson loop~\cite{Rothkopf:2011db}.  The spectral function is then obtained from the Fourier transform,
\be
\rho(\omega)=\lim_{r,r'\to 0}\frac{1}{2}\int_{-\infty}^\infty
\diff t \,e^{\ii\omega t}\mathcal{C}^>(r,r',t) \ .
\ee
The spectral functions obtained this way avoid the direct reconstruction of the meson spectral function from Euclidean correlators (although the extraction of the potential is itself an inverse problem, see below), but are valid only near the threshold region $\omega\sim 2m_Q$. Well above threshold, the spectral function is taken from vacuum perturbation theory.

\begin{figure}[!ht] 
\centering
\begin{subfigure}{0.48\textwidth}
    \includegraphics[height=3.3cm]{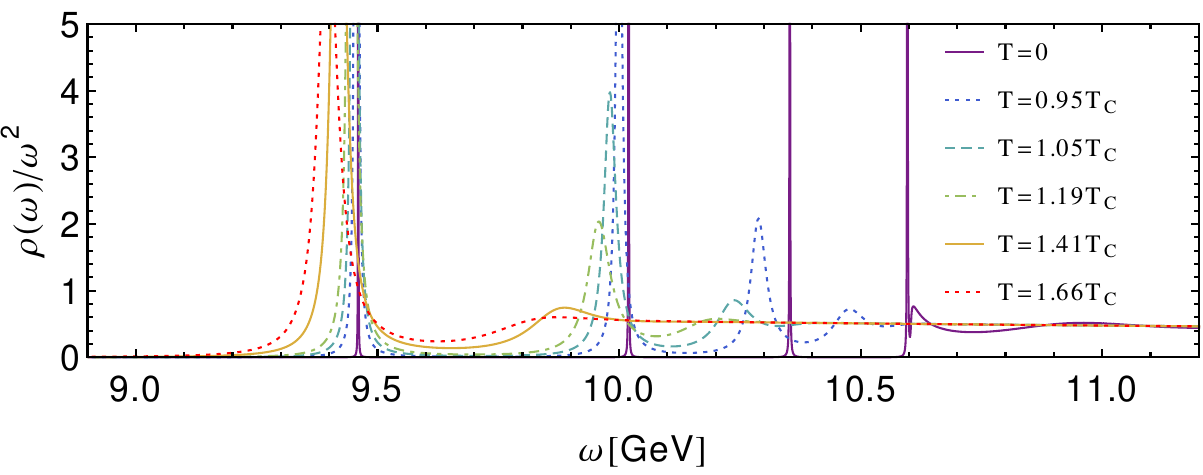}
    \caption{$\Upsilon$}\vspace{0.3cm}
\end{subfigure}
\begin{subfigure}{0.48\textwidth}
    \includegraphics[height=3.3cm]{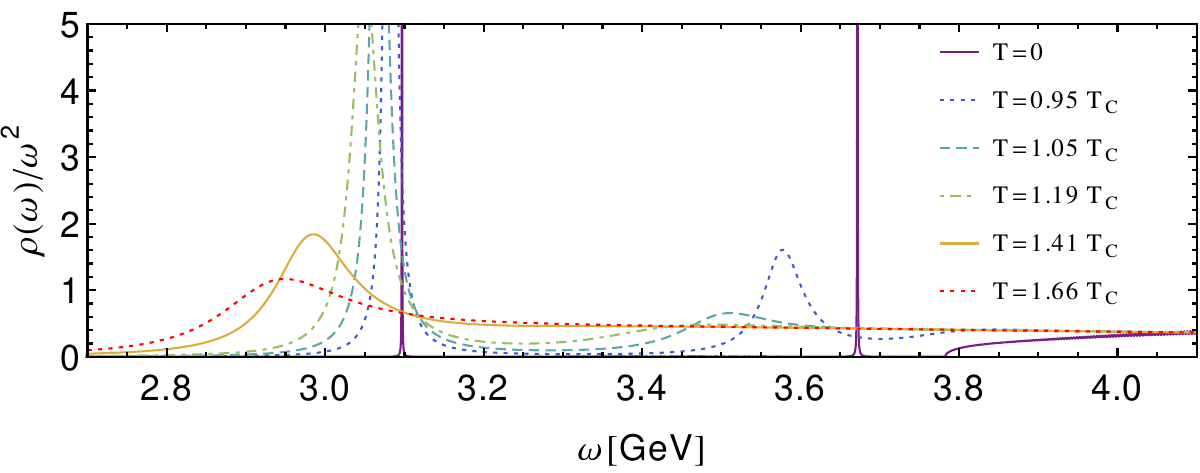}
    \caption{$J/\psi$}\vspace{0.3cm}
\end{subfigure}

\begin{subfigure}{0.48\textwidth}
    \includegraphics[height=3cm]{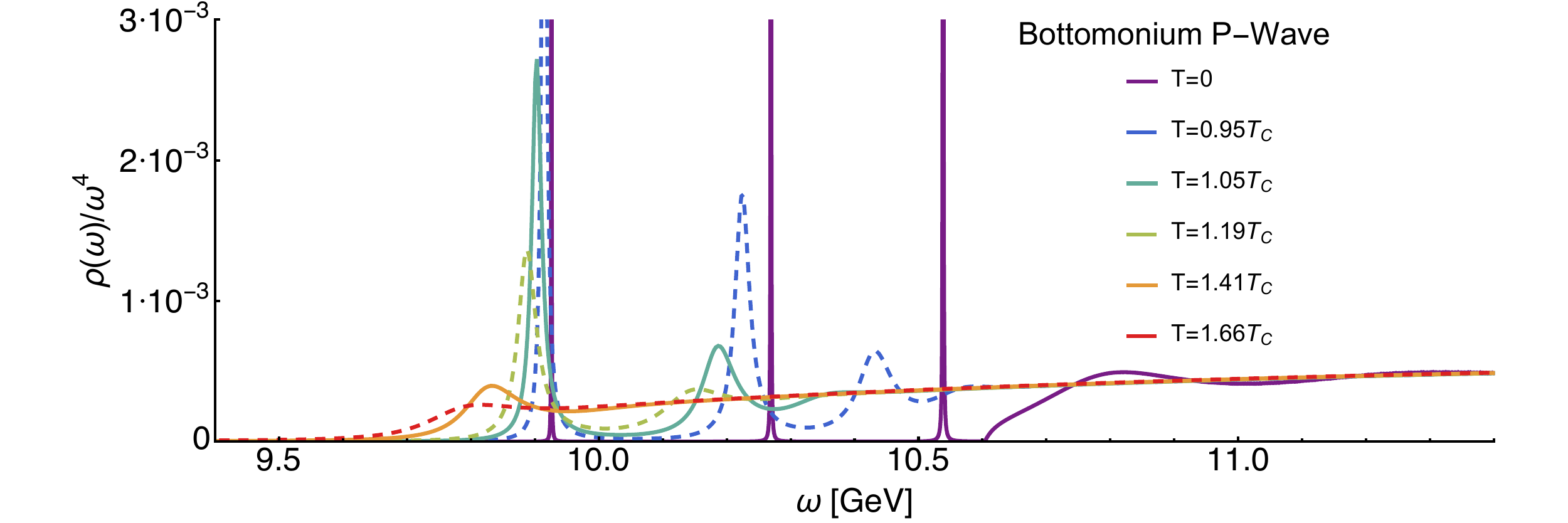}
    \caption{$\chi_{b0}$}
\end{subfigure}
\begin{subfigure}{0.48\textwidth}
    \includegraphics[height=3cm]{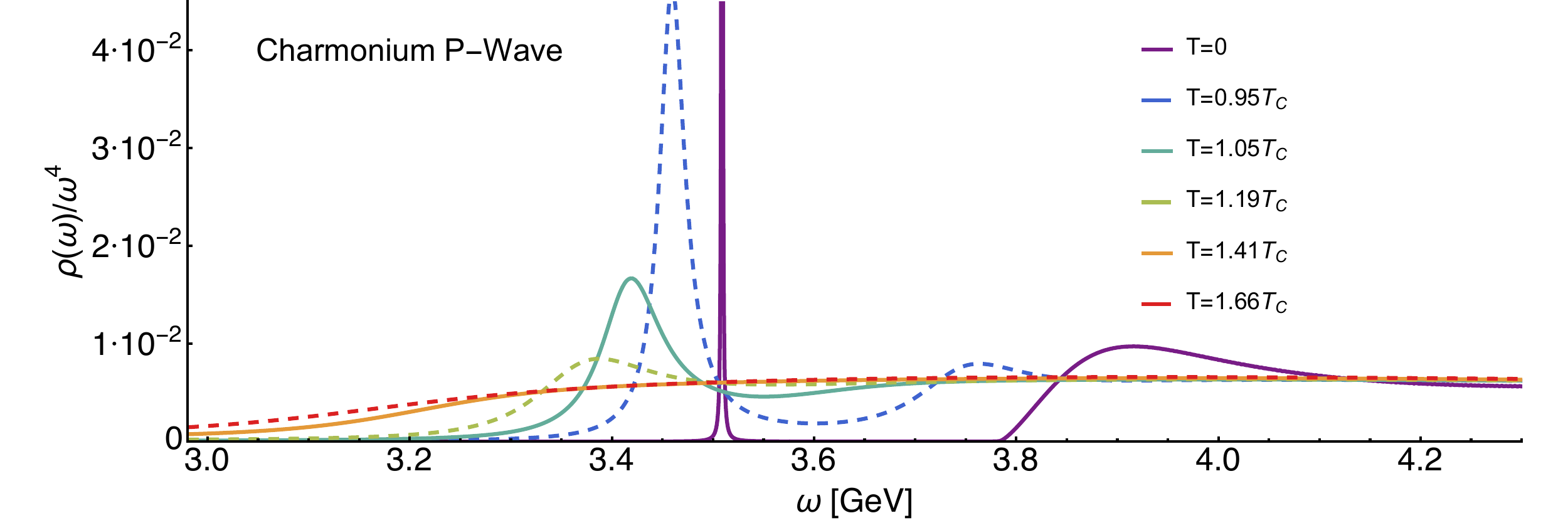}
    \caption{$\chi_{c0}$}
\end{subfigure}
\caption{Spectral functions from the lattice pNRQCD approach for bottomonium (left) and charmonium (right) S-wave (top) and P-wave (bottom) channels. Figures obtained from Refs.~\cite{Burnier:2015tda,Burnier:2016kqm}.}
\label{fig:BurnierSpectral}
\end{figure}

\begin{figure}[!ht] 
\centering
    \includegraphics[width=0.3\linewidth]{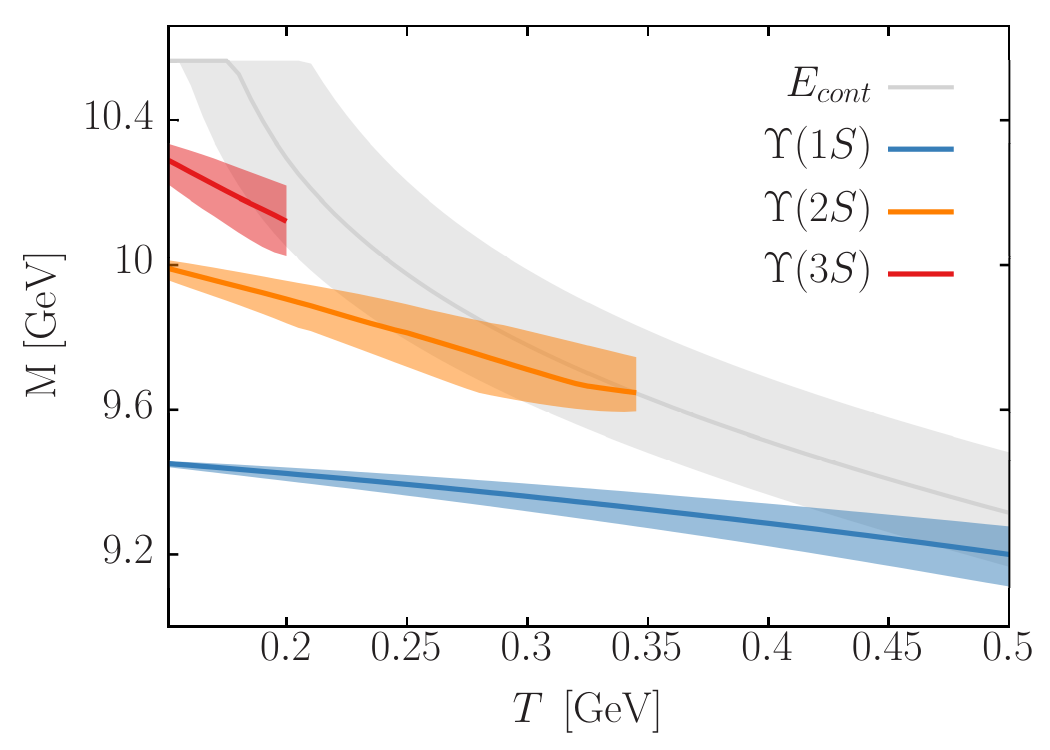}
    \includegraphics[width=0.3\linewidth]{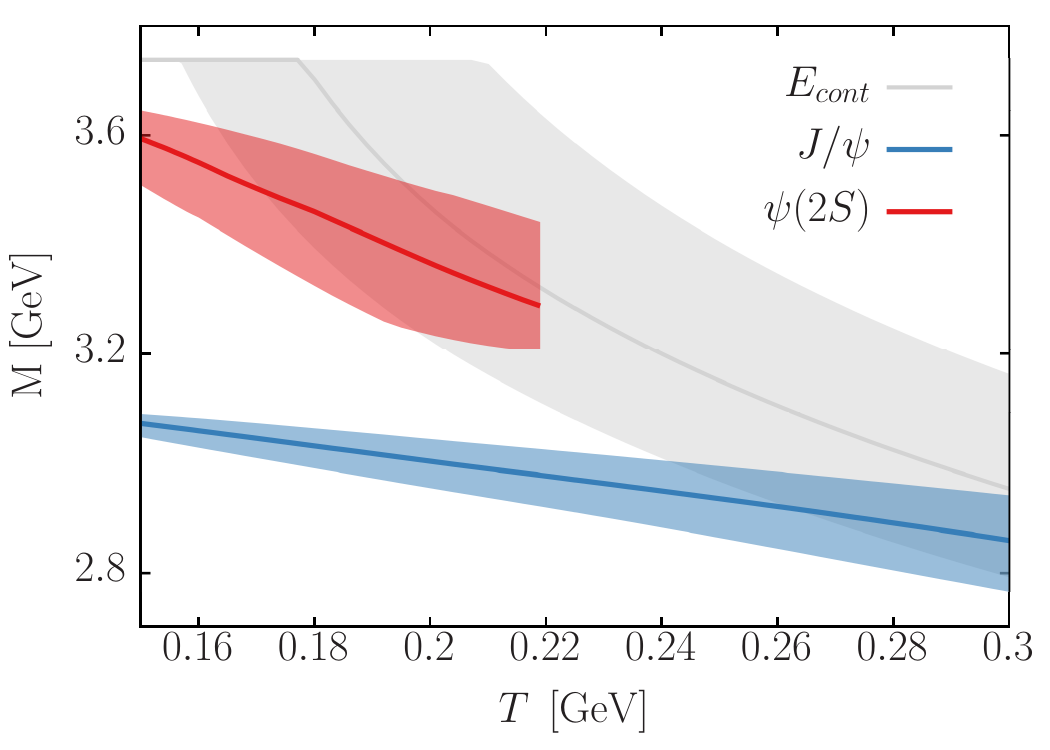}\\
    \includegraphics[width=0.3\linewidth]{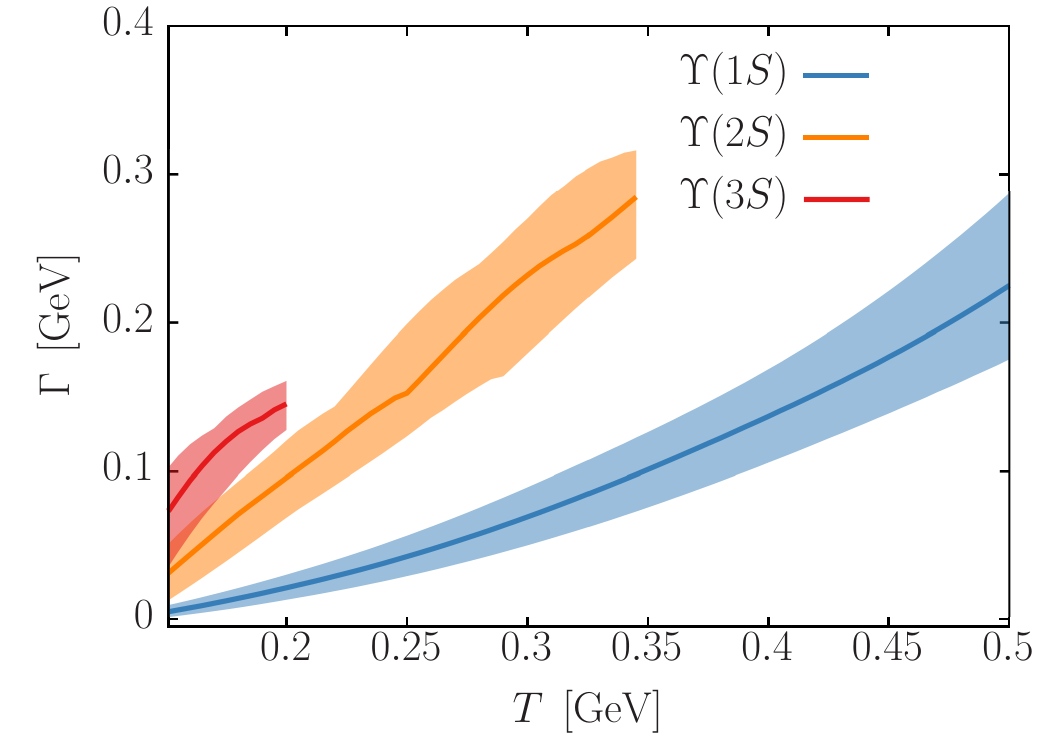}
    \includegraphics[width=0.3\linewidth]{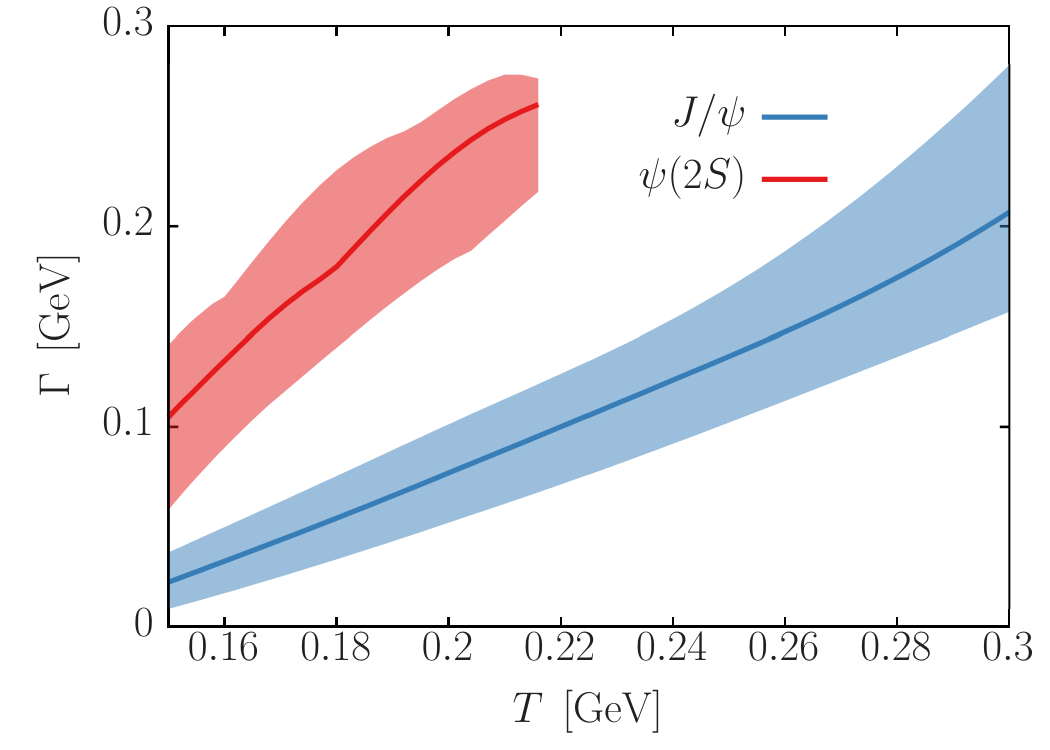}
\caption{Temperature dependence of the mass (top) and width (bottom) of the bottomonium (left) and charmonium (right) states from the lattice pNRQCD approach. Figures obtained from Ref.~\cite{Lafferty:2019jpr}.}
\label{fig:Lafferty}
\end{figure}

The first application of this approach to realistic unquenched lattice QCD configurations was presented in Refs.~\cite{Burnier:2015tda,Burnier:2016kqm,Burnier:2017bod}, where the complex potential was parametrized using a generalized Gauss law Ansatz and extracted from $N_f=2+1$ HotQCD ensembles. The single temperature-dependent parameter, the Debye mass, was tuned to reproduce the lattice values of $\re V$, and the resulting imaginary part $\im V$, which encodes thermal broadening via Landau damping, was found to be consistent with the lattice extraction. Solving the Schr\"odinger equation with this lattice-QCD-based complex potential for both charmonium and bottomonium vector~\cite{Burnier:2015tda} and scalar channels~\cite{Burnier:2016kqm}, the authors obtained spectral functions that exhibit broadening and a downward mass shift with temperature, as shown in Fig.~\ref{fig:BurnierSpectral}. This framework was improved and extended in Ref.~\cite{Lafferty:2019jpr}, which found qualitatively similar spectral functions. The values of the resulting in-medium masses and widths extracted via Breit--Wigner fits are shown in Fig.~\ref{fig:Lafferty}.

A similar approach has been used in Ref.~\cite{Ali:2025iux} to reconstruct the spectral functions in the quarkonium pseudoscalar channel using $N_f=2+1$ lattice ensembles, with a pion mass in the range $250\text{--}320\mev$, depending on the temperature. The reconstructed spectral functions exhibit significant in-medium modification: all excited states are found to melt already at $T_c$, and the ground states persist but undergo substantial thermal broadening.

The extraction of the complex potential from Euclidean Wilson line correlators is, however, itself an ill-posed inverse problem, and the screening of its real part depends on how it is addressed. While HTL-motivated parametrizations yield a screened $\re V$~\cite{Bala:2021fkm,Ali:2025iux}, other spectral analyses of HotQCD correlators find it essentially unscreened up to $T\simeq350\mev$, with the thermal effects absorbed into a large imaginary part~\cite{Bala:2021fkm,Bazavov:2023dci}. Since the in-medium potential is the central input of both potential-based calculations and the $T$-matrix approach discussed in Section~\ref{sec:QQ-other}, this question directly affects their predictions for binding energies and melting temperatures.

\subsection{Other approaches for quarkonia}\label{sec:QQ-other}
Having covered direct lattice QCD approaches above, we now briefly note other methods that have advanced our understanding of finite-temperature quarkonium properties. 

The in-medium modification of the mass and width of the $J/\psi$ and $\eta_c$ across $T_c$ was investigated in the QCD sum rule formalism in Refs.~\cite{Morita:2007pt,Morita:2007hv}. These studies were based on finite-temperature gluon condensates extracted from quenched lattice QCD data and employed a Breit--Wigner parametrization to model the quasiparticle peak of the spectral function. Although mass and width could not both be extracted simultaneously in this early approach, a significant mass decrease and thermal broadening were suggested as the temperature approached $T_c$.

\begin{figure}[!ht] 
\centering
    \includegraphics[width=0.3\linewidth]{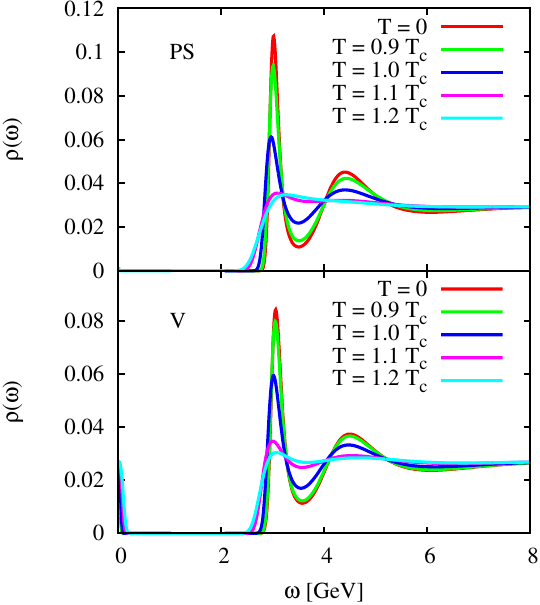}
    \includegraphics[width=0.3\linewidth]{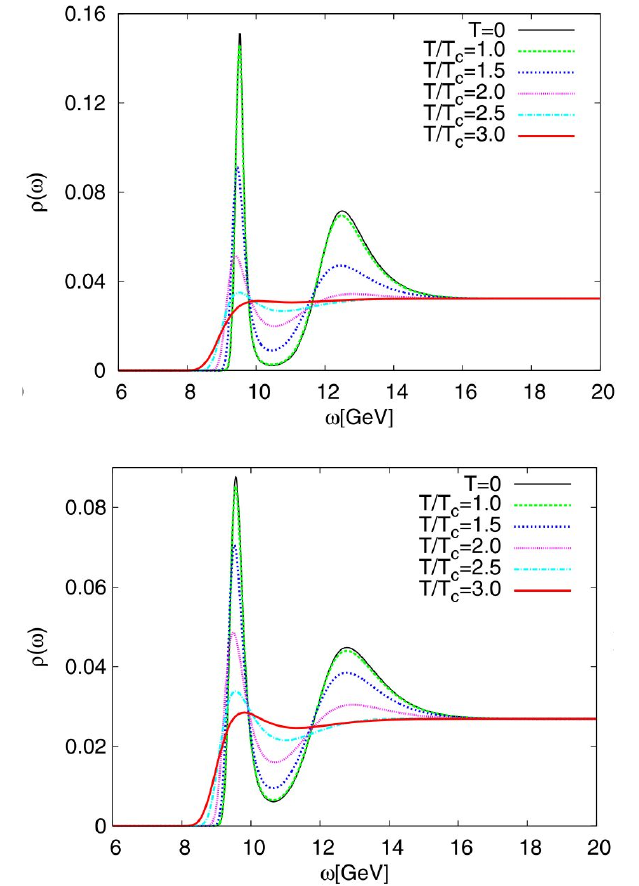}
\caption{Charmonium (left) and bottomonium (right) spectral functions in the pseudoscalar (top) and vector (bottom) channels from MEM analysis of QCD sum rules. Figures obtained from Refs.~\cite{Gubler:2011ua, Suzuki:2012ze}.}
\label{fig:Gubler_charmonia}
\end{figure}

An alternative analysis strategy of QCD sum rules based on MEM was carried out to study charmonium at finite temperature in Ref.~\cite{Gubler:2011ua}, and later extended to bottomonium in Ref.~\cite{Suzuki:2012ze}. This approach is particularly advantageous, as it allows for the direct extraction of the spectral function from the sum rules without assuming a specific functional form (such as Breit--Wigner or Gaussian). The results, shown in Fig.~\ref{fig:Gubler_charmonia}, indicate a rapid melting of the charmonium ground-state peaks right above $T_c$, while the bottomonium ground states survive above $2T_c$.

This MEM-based framework was subsequently extended in Ref.~\cite{Araki:2017ebb} through the use of complex Borel sum rules, which allowed for the first extraction of both ground-state ($J/\psi$, $\eta_c$) and excited-state ($\psi'$, $\eta_c'$) 
charmonium peaks from QCD sum rules at finite temperature. The analysis revealed an almost simultaneous melting of the ground and excited states in the vicinity of $T_c$.

The thermodynamic $T$-matrix formalism~\cite{Cabrera:2006wh,Liu:2017qah} provides a complementary framework to compute quarkonium thermal properties by studying the interaction of the $Q\bar{Q}$ pair with the partons of the QGP medium. This is achieved by solving the three-dimensional reduction of the Bethe--Salpeter equation, which gives the quarkonium $T$-matrix,
\begin{equation}
    T_{Q\bar{Q}}(z,{\bm p},{\bm p}';T)=V_{Q\bar{Q}}({\bm p},{\bm p}';T)+\int \frac{\diff ^3{\bm k}}{(2\pi)^3}V_{Q\bar{Q}}({\bm p},{\bm k};T)G_{Q\bar{Q}}(z,{\bm k};T)T_{Q\bar{Q}}(z,{\bm k},{\bm p}';T) \ ,
\end{equation} 
where $z=\ii E$ is the two-body Matsubara frequency, and $T$ in the arguments refers to the temperature. The in-medium $Q\bar{Q}$ propagator is given by
\begin{equation}
    G_{Q\bar{Q}}(z,{\bm k};T)=\int_{-\infty}^\infty \diff \omega_1 \diff \omega_2\frac{\rho_Q(\omega_1,{\bm k})\rho_{\bar{Q}}(\omega_2,{\bm k})}{z-\omega_1-\omega_2}(1- n_{\textrm{F},Q}(\omega_1)- n_{\textrm{F},\bar{Q}}(\omega_2)) \ ,
\end{equation}
with $n_{\textrm{F},Q} (\omega_1)=(\exp(\omega_1/T)+1)^{-1}$ the Fermi--Dirac distribution for the heavy quark $Q$. The single-quark spectral functions $\rho_i$ are modified by the medium via the one-quark self-energies calculated self-consistently by closing the light-parton line in the corresponding heavy-light $T$-matrix, $T_{Qi}$.

A central input is the finite-temperature potential $V_{Q\bar{Q}}(\bm{p},\bm{p}';T)$, which is generally complex-valued at $T>0$ and reduces to the standard Cornell potential in vacuum. The imaginary part accounts for the Landau damping of exchanged gluons and gluo-dissociation. This potential is typically constrained using lattice-QCD data for the heavy-quark free energy or the internal energy, or alternatively, from the quarkonium Euclidean correlator.

\begin{figure}[!ht] 
\centering
    \includegraphics[width=0.75\linewidth]{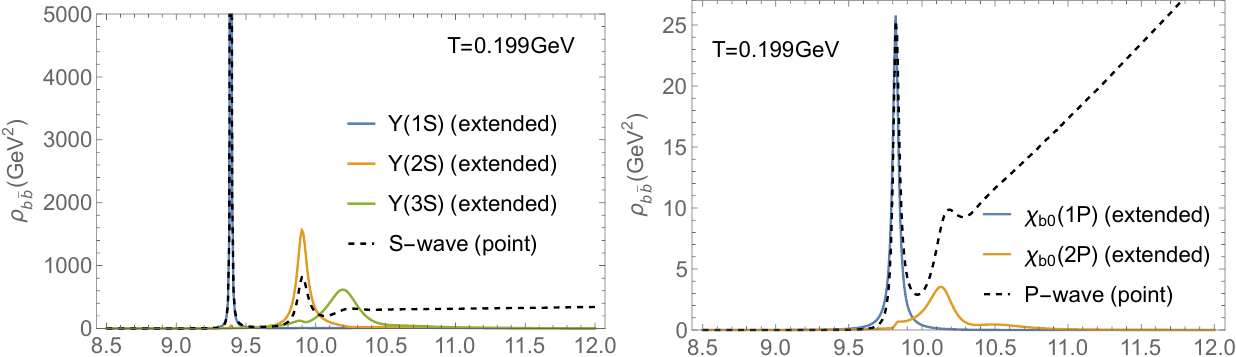}
    \includegraphics[width=0.75\linewidth]{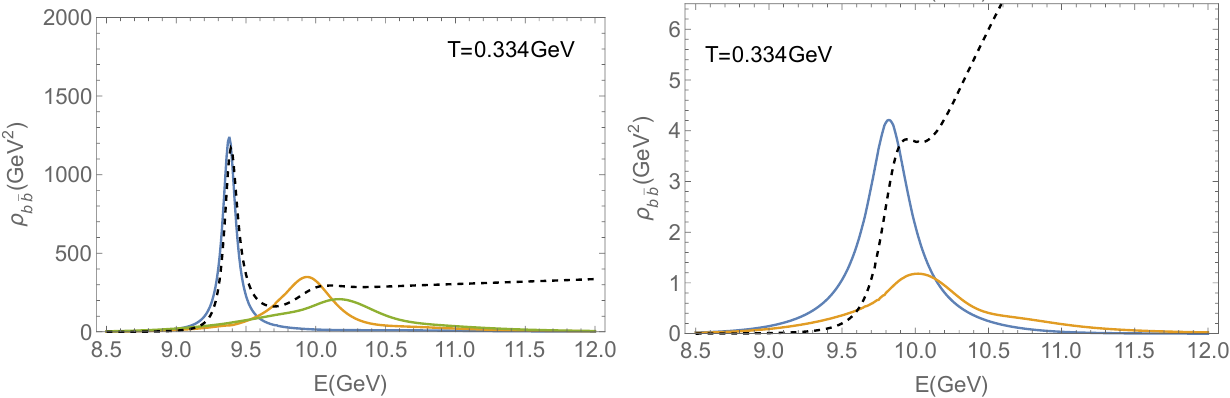}
\caption{Spectral functions for bottomonium states according to the calculation of Ref.~\cite{Tang:2024dkz}. The $L=0$ ($L=1$) states are given in the left (right) panel. Upper panels are for a temperature of $T=199\mev$, while lower panels are for $T=334\mev$. Figure adapted from Ref.~\cite{Tang:2024dkz}.}
\label{fig:ZhangduoBottom}
\end{figure}

In Ref.~\cite{Tang:2023tkm}, using the $T$-matrix approach at finite temperature, the authors computed the so-called Wilson line correlators, which can be matched to thermal lattice-QCD calculations with $N_f=2+1$ flavors, in order to better calibrate the model. This connection with lattice-QCD calculations was then exploited in the subsequent work~\cite{Tang:2024dkz} to compute constrained in-medium potentials for bottomonia (see also Ref.~\cite{Wu:2025hlf} for an application to charmonia). Finally, the thermal properties of bottomonium can be computed from its correlation functions. Examples are the spectral functions at different temperatures, a few of which are shown in Fig.~\ref{fig:ZhangduoBottom}. 
In the left panels of Fig.~\ref{fig:ZhangduoBottom} we show the spectral functions of the $L=0$ states at rest, namely, $\Upsilon(1S),\Upsilon(2S)$ and $\Upsilon(3S)$, while the right panels show the $L=1$ states ($\chi_{b0} (1P)$ and $\chi_{b0} (2P)$). The upper panels are for $T=199\mev$ and the lower panels for $T=334\mev$. 
One can clearly see the melting of the states due to the broadening of their spectral peaks, while the thermal masses decrease very slightly.
However, the authors of Ref.~\cite{Tang:2024dkz} point out that not all peaks in the spectral functions can be interpreted as bound states. From the poles of the $T-$matrix they find the $3S$, $2P$, $2S$, and $1P$ bottomonium states to melt sequentially at temperatures from $163\mev$ to $293\mev$, while the $1S$ state survives up to the highest temperature considered, $334\mev$.

\begin{figure}[!ht] 
\centering
    \includegraphics[width=0.9\linewidth]{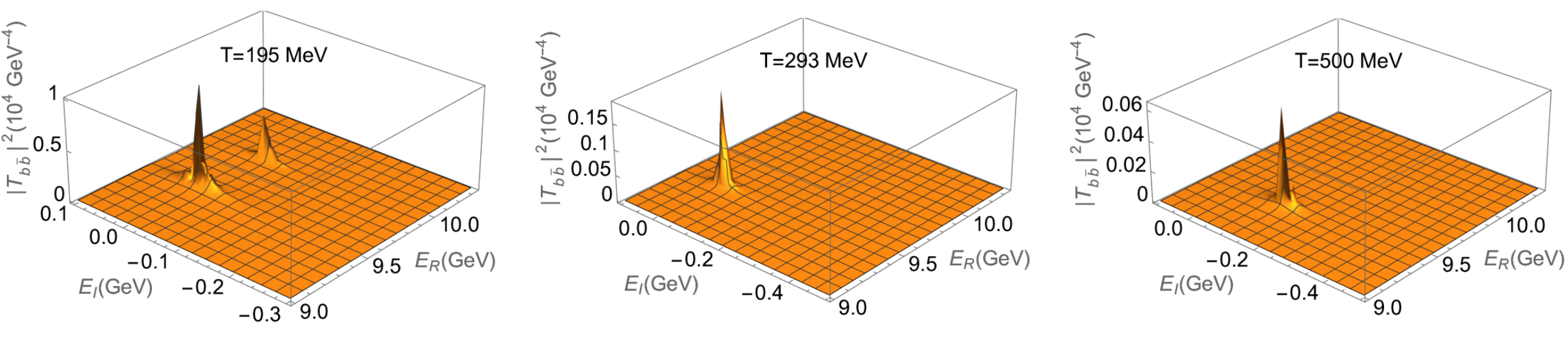}
\caption{The S-wave bottomonium $T$-matrices in the complex-energy plane, evaluated at zero center-of-mass momentum for different temperatures. Figures obtained from Ref.~\cite{Tang:2025ypa}.}
\label{fig:ZhangduoComplexPlane}
\end{figure}

A recent calculation in Ref.~\cite{Tang:2025ypa} extends the thermal $T$-matrix calculation into the complex-energy plane to search for poles on the second Riemann sheet. In Fig.~\ref{fig:ZhangduoComplexPlane} we reproduce the value of the scattering amplitude squared $|T|^2$ in the bottomonium sector, as a function of the real and imaginary parts of the center-of-mass energy.
The temperature increases from left to right, and several poles are clearly visible. These poles move in the complex-energy plane with temperature, reflecting the temperature dependence of their masses and widths. Some of the poles eventually disappear, meaning that the states have completely melted in the medium. 

\begin{figure}[!ht] 
\centering
    \includegraphics[width=0.4\linewidth]{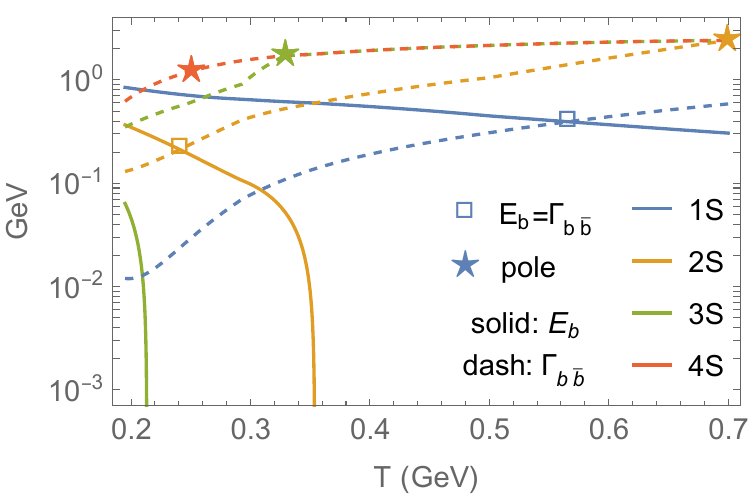}
\caption{Comparison of various dissociation criteria for S-wave bottomonia: open squares denote the temperatures at which the width (dashed lines) becomes equal to the binding energy (solid lines), while stars mark the disappearance of the pole in the complex-energy plane. Figure obtained from Ref.~\cite{Tang:2025ypa}.}
\label{fig:ZhangduoComplexPlane2}
\end{figure}

In Fig.~\ref{fig:ZhangduoComplexPlane2} we show the summary plot of Ref.~\cite{Tang:2025ypa}.
In this figure, the binding energies, defined as the difference between the thermal $2m_b$ value and the pole mass of the bound state, are depicted in solid lines, while thermal decay widths are plotted in dashed lines. All of them are extracted from the pole position for the four $\Upsilon$ states. The squares mark the point where the two are equal, providing a possible definition of the melting temperature. The stars mark the temperature at which the pole itself vanishes in the complex plane, providing another definition of the melting temperature, which is systematically higher than the previous one.

\subsection{Quarkonia: summary and future perspectives}\label{sec:hiddenHFsummary}
Unlike open heavy flavor, whose thermal properties have mostly been studied in the hadronic phase, quarkonia are probed mainly in the QGP. The approaches used here are closely interconnected. pNRQCD identifies the relevant scales and defines the in-medium heavy-quark potential, lattice QCD constrains this potential nonperturbatively and provides the Euclidean correlators against which all descriptions are ultimately tested, and potential-based, $T$-matrix, and sum-rule analyses turn this information into spectral functions. Each step has its own limitations. Weak-coupling scale hierarchies become questionable close to $T_c$. Relativistic lattice simulations, mostly restricted to charmonium, are affected by the transport peak and the temperature-dependent kernel, while lattice NRQCD makes bottomonium accessible at the cost of simulating a truncated EFT, and both require solving the inverse problem for spectral function reconstruction. Potential-based spectral functions avoid the direct reconstruction of meson spectral functions~\cite{Burnier:2015tda,Burnier:2016kqm,Lafferty:2019jpr,Ali:2025iux} but rely on a potential whose extraction is itself an inverse problem, and are reliable only near the $Q\bar Q$ threshold. QCD sum rules depend on the input for the thermal condensates, although their MEM-based analyses~\cite{Gubler:2011ua,Suzuki:2012ze,Araki:2017ebb} require no Ansatz for the spectral function. The $T$-matrix approach~\cite{Tang:2024dkz,Tang:2025ypa} explicitly accounts for the scattering of heavy quarks with the partons of the QGP using lattice-constrained potentials, and gives access to the pole structure of the bound states in the complex-energy plane.

With these caveats in mind, the following features are supported by most of the frameworks reviewed here:
\begin{itemize}
    \item \textbf{A complex in-medium potential.} pNRQCD predicts that the heavy-quark potential acquires an imaginary part at finite temperature, arising from Landau damping and singlet-octet transitions~\cite{Brambilla:2008cx,Brambilla:2010vq,Brambilla:2011sg}. Nonperturbative extractions from thermal Wilson loops support this prediction~\cite{Burnier:2015tda,Burnier:2016kqm}, although the size of the thermal effects in the real and imaginary parts depends on the extraction method, and lattice-constrained complex potentials are now routinely used as input for potential-based and $T$-matrix descriptions~\cite{Lafferty:2019jpr,Tang:2024dkz}.
    
    \item \textbf{Dissolution through broadening.} Lattice NRQCD~\cite{Larsen:2019bwy,Larsen:2019zqv,Skullerud:2025iqt,Ding:2025fvo}, potential-based~\cite{Burnier:2015tda,Burnier:2016kqm,Lafferty:2019jpr}, and $T$-matrix~\cite{Tang:2024dkz,Tang:2025ypa} studies all find quarkonium widths that are nonzero and grow with temperature. Relativistic lattice reconstructions~\cite{Asakawa:2003re,Ding:2012sp,Ikeda:2016czj,Kelly:2018hsi} and QCD sum rules~\cite{Morita:2007pt,Morita:2007hv,Gubler:2011ua,Suzuki:2012ze} are broadly consistent with this picture, although the temperatures at which the spectral peaks disappear differ among studies. Bound states are thus washed out dynamically by collisions with the medium and dissolve before static color screening alone would remove them.
    
    \item \textbf{Sequential suppression pattern.} The more tightly bound a state is, the longer it survives. In the bottomonium sector, the $\Upsilon(1S)$ and $\eta_b$ remain well defined far above $T_c$~\cite{Suzuki:2012ze,Aarts:2014cda,Kim:2018yhk,Tang:2024dkz}, while P-wave and radially excited states dissolve at considerably lower temperatures~\cite{Aarts:2014cda,Kim:2018yhk,Tang:2024dkz}. The same ordering is found for charmonium, where the $\chi_{c0}$ and $\chi_{c1}$ are much more strongly affected around $T_c$ than the $J/\psi$ and $\eta_c$~\cite{Datta:2003ww,Aarts:2007pk,Kelly:2018hsi,Kim:2018yhk}, and excited states melt before the ground states~\cite{Ali:2025iux}. QCD sum rules instead find sizable modifications of the ground states~\cite{Morita:2007pt,Morita:2007hv} and an almost simultaneous melting of ground and excited states near $T_c$~\cite{Gubler:2011ua,Araki:2017ebb}. Overall, charmonia are always more strongly affected than the corresponding bottomonia~\cite{Kim:2018yhk,Gubler:2011ua,Suzuki:2012ze}.
\end{itemize}

Beyond this qualitative agreement, most quantitative statements are still open. 
The thermal potential itself is not settled: while its real part is usually found to be screened~\cite{Burnier:2015tda,Burnier:2016kqm}, some analyses of lattice Wilson line correlators suggest that it may remain essentially unscreened, with the thermal effects absorbed into a large imaginary part~\cite{Bala:2021fkm,Bazavov:2023dci}. This uncertainty propagates directly into the predicted binding energies and widths. The size of the thermal widths still depends on how the spectral peak is parametrized and extracted~\cite{Skullerud:2025iqt,Ding:2025fvo}, and there is no agreement on the mass shifts, which range from clearly negative~\cite{Kim:2018yhk,Lafferty:2019jpr} to very small~\cite{Tang:2024dkz} or compatible with zero~\cite{Ding:2025fvo}. Melting temperatures are also uncertain. For the charmonium ground states, quenched lattice studies disagree on whether the peaks survive at $T\simeq1.5\,T_c$~\cite{Ding:2012sp,Ikeda:2016czj}, and QCD sum rules find their dissolution already close to $T_c$~\cite{Gubler:2011ua,Araki:2017ebb}. In addition, the melting temperature depends on its definition: requiring the width to equal the binding energy gives lower values than the disappearance of the pole from the complex-energy plane~\cite{Tang:2025ypa}.

A quantitative picture will require strengthening the links between these approaches. On the lattice, the strategies recently developed for bottomonium, namely the systematic cross-check of different reconstruction techniques~\cite{Skullerud:2025iqt} and the use of extended operators~\cite{Ding:2025fvo}, should be extended to charmonium, where the results remain least conclusive, together with relativistic lattice QCD calculations with physical quark masses and on finer lattices. For potential-based descriptions, settling the screening of the real part of the potential is particularly relevant, while going beyond the static limit would give access to the in-medium fine structure of quarkonia. QCD sum rules would benefit from improved lattice input for the thermal condensates, and the continued calibration of $T$-matrix approaches against lattice  correlators~\cite{Tang:2023tkm,Tang:2024dkz,Wu:2025hlf} would test the consistency of the overall picture. Below $T_c$, quarkonium physics connects directly with the open heavy-flavor sector (Section~\ref{sec:openHF}). The thermal modification of $D$ and $B$ mesons shifts the open-flavor thresholds and controls the dissociation of quarkonia by comoving hadrons, so both sectors should be described within a common framework.

\subsection{Doubly heavy exotic mesons}~\label{sec:exotics}

Experimental discoveries over the past two decades have revealed a rich spectrum of exotic heavy-quark states whose properties cannot be accommodated within conventional $Q\bar Q$ quarkonia. These so-called $XY\!Z$ states exhibit unexpected quantum numbers, unusual decay patterns, or proximity to open-flavor thresholds, which suggest more complex internal structures, such as compact tetraquarks arranged in diquark–antidiquark pairs, hadronic molecules formed from weakly bound mesons, or admixtures of both, as well as hybrid mesons with excited gluonic degrees of freedom. Understanding the nature of these states has become a central focus of heavy-quark spectroscopy in vacuum, as summarized in many reviews~\cite{Swanson:2006st,Chen:2016qju,Lebed:2016hpi,Esposito:2016noz,Guo:2017jvc,Liu:2019zoy,Brambilla:2019esw,Chen:2022asf,Meng:2022ozq}.

Among the many exotic candidates, the $X(3872)$ and the doubly charmed $T_{cc}^+(3875)$ stand as the clearest and best-measured examples of near-threshold heavy-quark exotics\footnote{Although the $T_{cc}^+$ is an open-charm state rather than a hidden-charm one, we include it in this section because its structure and phenomenology are dominated by the same near-threshold dynamics that govern hidden-charm exotics such as the $X(3872)$.}. Both lie extremely close to heavy-meson thresholds and exhibit very narrow widths. 

The $X(3872)$ was first reported by Belle in 2003~\cite{Belle:2003nnu} and later confirmed by multiple other experiments.
%~\cite{CDF:2003cab,D0:2004zmu,BaBar:2004oro,LHCb:2013kgk,BESIII:2013fnz}
Since its experimental discovery, its structure has been extensively debated, with still no consensus. Given the proximity of the observed mass ($m_{X(3872)}=3871.64\pm0.06\mev$~\cite{ParticleDataGroup:2024cfk}) to the $D^0\bar{D}^{*0}$ ($\bar D^0{D}^{*0}$) threshold ($m_{D^0}+m_{\bar D^{*0}}=3871.69\pm0.07\mev$), a natural interpretation is that of a shallow $D^0\bar{D}^{*0}$ molecular bound state~\cite{Wong:2003xk,Tornqvist:2004qy,Thomas:2008ja}
with a very narrow width ($\Gamma_{X(3872)}=1.19\pm0.21\mev$~\cite{ParticleDataGroup:2024cfk}). Alternative explanations include a $[cq][\bar{c}\bar{q}]$ diquark-antidiquark compact configuration~\cite{Maiani:2004vq,Ebert:2005nc,Matheus:2006xi}, and a mixture of a molecule and an excited $c\bar c$ charmonium state~\cite{Matheus:2009vq,Ortega:2009hj,Cincioglu:2016fkm}. These interpretations are typically constrained by comparisons to the charmonium spectrum and the analyses of two- and three-body decay branching fractions. 

More recently, the production of exotic hadrons in $pp$ collisions and HICs has emerged as a complementary probe of their underlying structure.
In particular, the evidence for the $X(3872)$ in Pb+Pb collisions by the CMS collaboration~\cite{CMS:2021znk} opened the possibility of studying this and other exotic candidates in the QCD medium. The behavior of the $X(3872)$ in a hot or dense environment is expected to depend strongly on whether it is a compact tetraquark or a loosely bound molecule. 

At finite temperature, molecular states, which typically have binding energies ranging from a few tens of keV to a few MeV and spatial extensions exceeding several fm, are particularly susceptible to thermal effects. Such states are expected to dissolve already at moderate temperatures in the hadronic medium through collisions with light mesons, leading to thermal broadening and the loss of well-defined hadronic constituents. Compact multiquark states, by contrast, are bound by short-range color interactions at the scale of hundreds of MeV and are expected to survive to temperatures around or above $T_c$, much like deeply bound quarkonia. These contrasting expectations motivate studying the in-medium behavior of heavy exotic states within appropriate theoretical frameworks.

As described in Section~\ref{sec:unitarizedEFT} for the open heavy-flavor sector, unitarized hadronic EFTs based on heavy-quark symmetries provide a systematic framework for describing heavy-meson molecules. HQSS and HQFS impose relations among the interactions of heavy mesons and organize near-threshold states into multiplets.
Based on these symmetries, such EFTs treat doubly heavy tetraquark states as bound states of two open heavy-flavor mesons interacting through contact terms and meson exchange. This approach has proven highly successful in reproducing the properties of many observed near-threshold states in vacuum and naturally predicts entire families of partner states consistent with HQSFS.

Building on the unitarized EFT framework, Refs.~\cite{Cleven:2019cre,Montana:2022inz} investigated the thermal properties of the $X(3872)$, interpreted as a $D\bar{D}^*+\text{c.c.}$ hadronic molecule, in the hadronic medium below $T_c$. The interaction is constructed from a coupled-channel $\sufour$ effective Lagrangian describing the $D\bar{D}^*$ and $D_s\bar{D}_s^*$ channels, including appropriate $\sufour$-symmetry breaking effects that suppress the exchange of heavy mesons~\cite{Gamermann:2006nm,Gamermann:2007fi}, and unitarized via the Bethe--Salpeter equation (cf.~Eq.~\eqref{eq:bs}). 

\begin{figure}[htbp!] 
\centering
\includegraphics[width=0.4\linewidth]{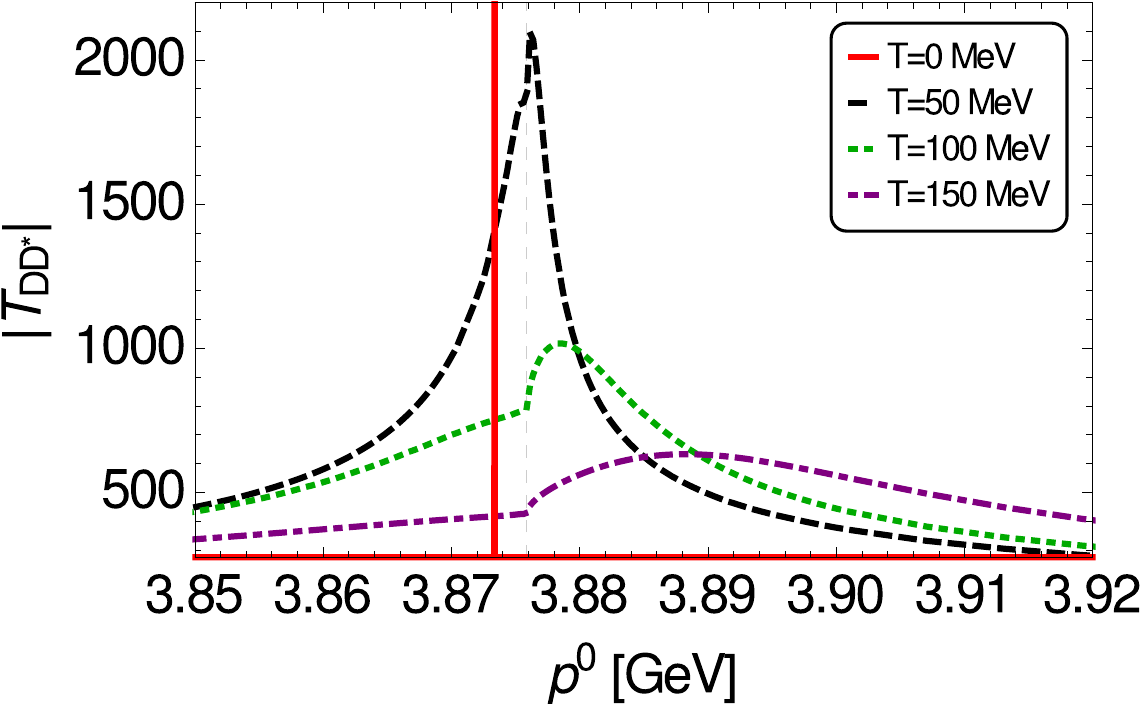}
\includegraphics[width=0.4\linewidth]{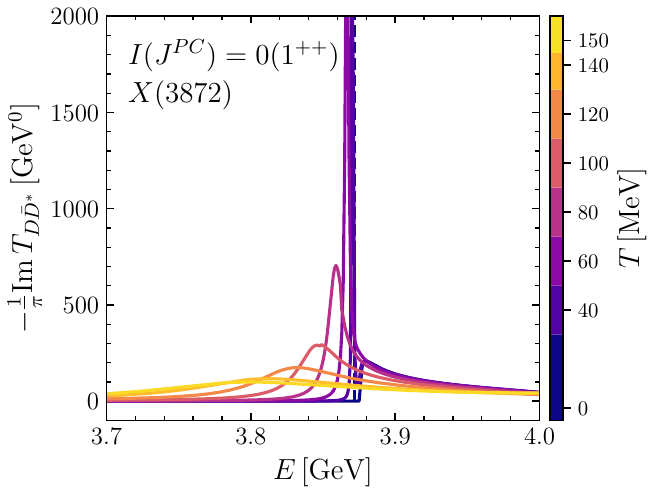}
\caption{Unitarized $D\bar{D}^*$ scattering amplitude illustrating the thermal behavior of the $X(3872)$ lineshape at several temperatures from two similar approaches that differ in their interaction kernels and implementation. Left: results from Ref.~\cite{Cleven:2019cre}. Right: results from Ref.~\cite{Montana:2022inz}.}
\label{fig:Cleven_vs_Montana_X3872}
\end{figure}

Medium effects enter through the thermal two-meson propagator loop function Eq.~\eqref{eq:thermalloop}, which incorporates the open-charm spectral functions derived self-consistently in previous works, i.e., in Refs.~\cite{Cleven:2017fun} and~\cite{Montana:2020lfi,Montana:2020vjg}. Depending on the implementation, the self-consistent approach leads either to a broadening only~\cite{Cleven:2017fun} or to a combined broadening and mass drop~\cite{Montana:2020lfi,Montana:2020vjg} of the $D$ and $D^*$ mesons. As a result, both studies find that the $X(3872)$ rapidly acquires a large width with increasing temperature, leading to the eventual dissolution of the molecular state well below $T_c$, as illustrated in Fig.~\ref{fig:Cleven_vs_Montana_X3872}.
In Ref.~\cite{Montana:2022inz}, the peak position also shifts to lower energies, following the temperature dependence of the $D\bar{D}^*$ threshold, as dictated by the input thermal spectral functions of the $D$ and $D^*$ mesons.

\begin{figure}[htbp!] 
\centering
\includegraphics[width=0.7\linewidth]{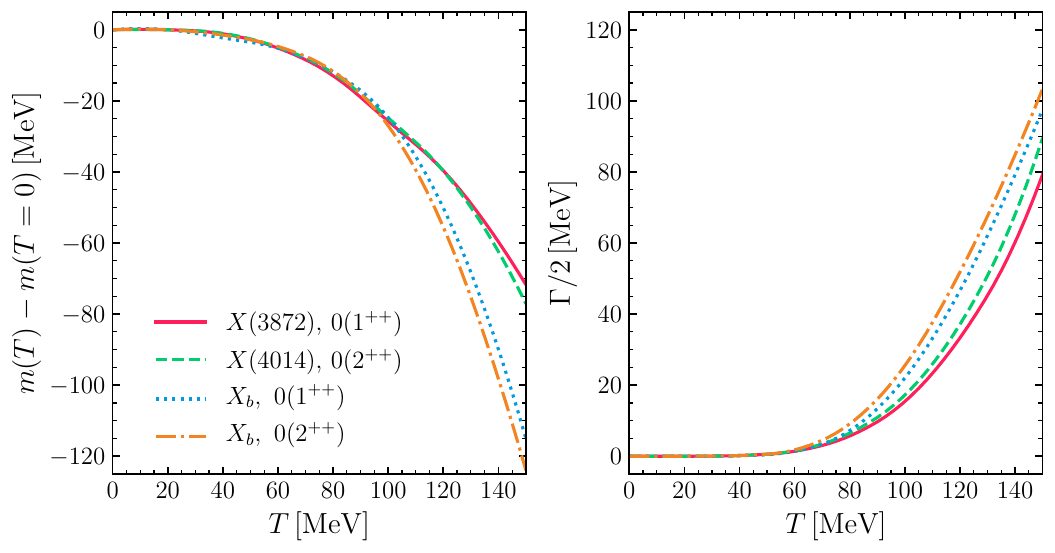}
\caption{Temperature dependence of the mass shift and half-width of the $X(3872)$ and its HQSFS partners obtained within a unitarized $D^{(*)}\bar D^{(*)}$ and $B^{(*)}\bar B^{(*)}$ framework. Figure taken from Ref.~\cite{Montana:2022inz}.}
\label{fig:Montana_X3872}
\end{figure}

The finite-temperature analysis of exotic molecular states in Ref.~\cite{Montana:2022inz} was extended to the spin partner of the $X(3872)$, the $X(4014)$, interpreted as a $D^*\bar{D}^*$ molecule, as well as to the bottom partners ($X_b$) generated from the $B^{(*)}\bar{B}^*$ interactions, with qualitatively similar in-medium behavior. The mass shifts and half-widths of the four states as functions of temperature are shown in Fig.~\ref{fig:Montana_X3872}.

The $T_{cc}^+(3875)$, discovered by the LHCb collaboration~\cite{LHCb:2021vvq,LHCb:2021auc}, is the first doubly heavy tetraquark candidate
established experimentally. It lies only $0.36\pm0.04\mev$ below the $D^0D^{*+}$ threshold and has an exceptionally small width of $48^{+2}_{-14}~\textrm{keV}$ and a mass of $3874.74\pm0.10\mev$, naturally suggesting its interpretation as a weakly bound hadronic molecule. Its extremely small binding energy implies a large spatial extent and makes it particularly sensitive to medium effects, suggesting that a molecular $T_{cc}$ should dissolve at relatively low temperatures.

\begin{figure}[htbp!] 
\centering
\includegraphics[width=0.4\linewidth]{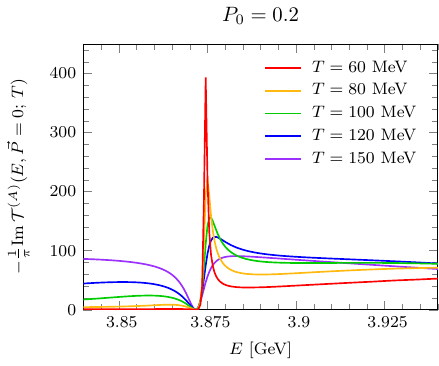}
\includegraphics[width=0.4\linewidth]{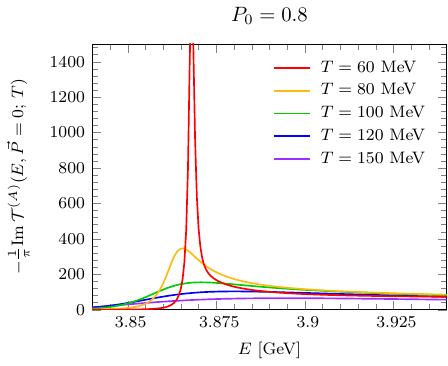}
\includegraphics[width=0.4\linewidth]{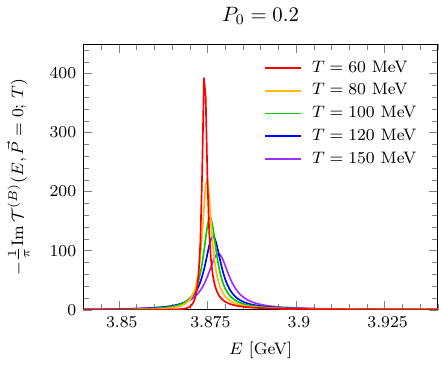}
\includegraphics[width=0.4\linewidth]{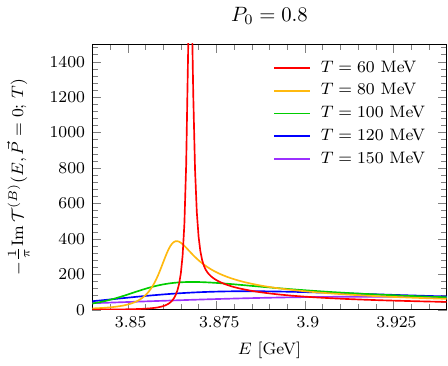}
\caption{Spectral functions of the $T_{cc}^+(3875)$ obtained with two different interaction kernels, $V_A$ (top panels) and $V_B$ (bottom panels), at several temperatures and for two values of the molecular probability (left and right columns). Figures taken from Ref.~\cite{Montesinos:2025mfx}.}
\label{fig:Montesinos_Tcc}
\end{figure}

The thermal behavior of the $T_{cc}$ has recently been investigated in Ref.~\cite{Montesinos:2025mfx} within a unitarized effective field theory framework analogous to that used for the $X(3872)$. The interaction in the $DD^*$ channel is modeled using different contact interaction potentials ($V_A$ and $V_B$), allowing one to assess the sensitivity of the results to the underlying dynamics. The medium effects are incorporated through the two-meson loop function, employing the thermal $D$ and $D^*$ spectral functions from Ref.~\cite{Montana:2020vjg}. The resulting spectral functions are displayed in Fig.~\ref{fig:Montesinos_Tcc}. The top and bottom panels correspond to the two different interaction potentials, while the left and right columns illustrate two scenarios with different assumed values of the molecular probability $P_0$, defined from the residue of the pole via the derivative of the loop function in vacuum. Thermal effects are included only in the molecular component, as medium modifications of a possible compact component are expected to be small below $T_c$. In all cases, the near-threshold peak of the $T$-matrix associated with the $T_{cc}$ broadens rapidly with increasing temperature, eventually dissolving into the continuum. For large molecular probabilities, this leads to the melting of the state already at temperatures of order $100\mev$. The study also included the spin partner $T_{cc}^*(4016)$.

\begin{figure}[htbp!] 
\centering
\includegraphics[width=0.32\linewidth]{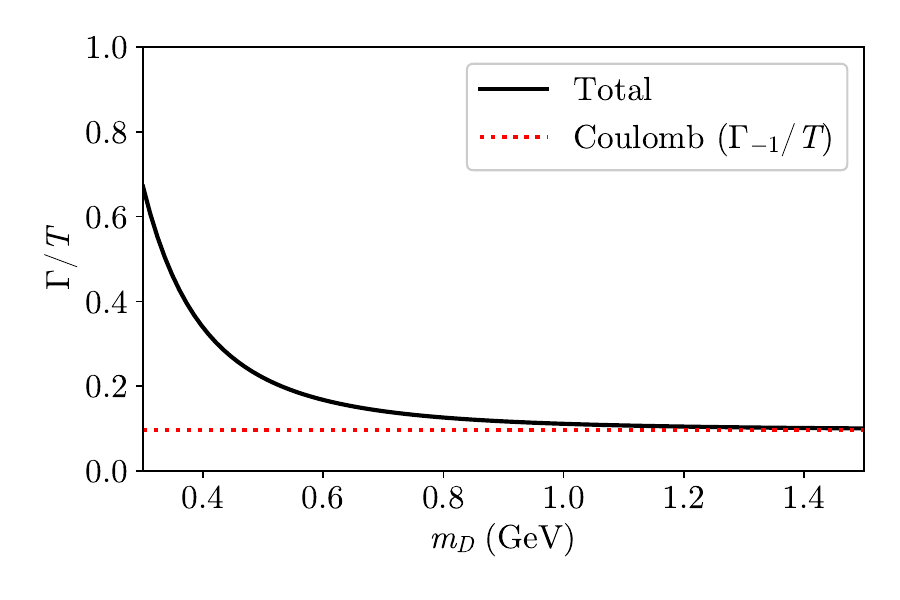}\hspace{0.2cm}
\includegraphics[width=0.32\linewidth]{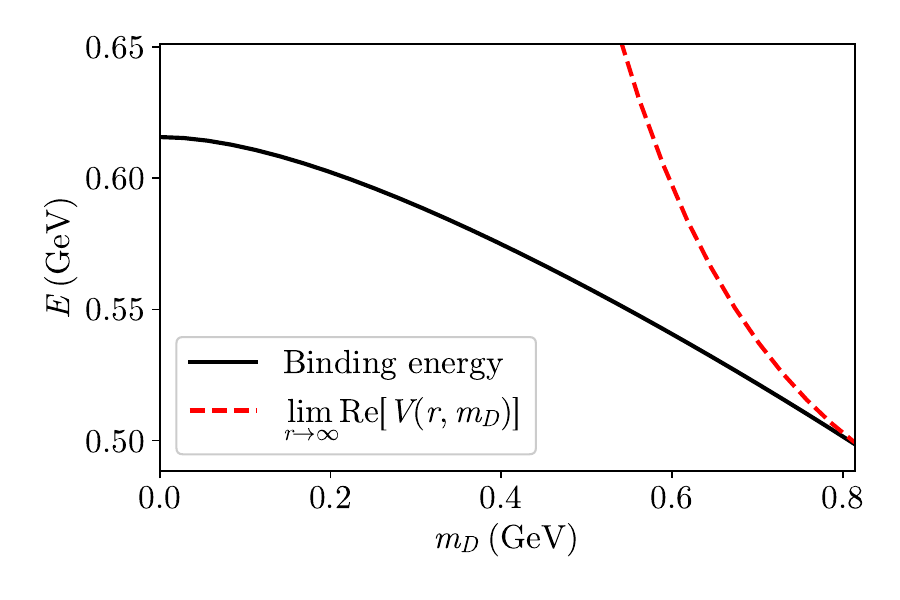}\hspace{0.2cm}\\
\includegraphics[width=0.32\linewidth]{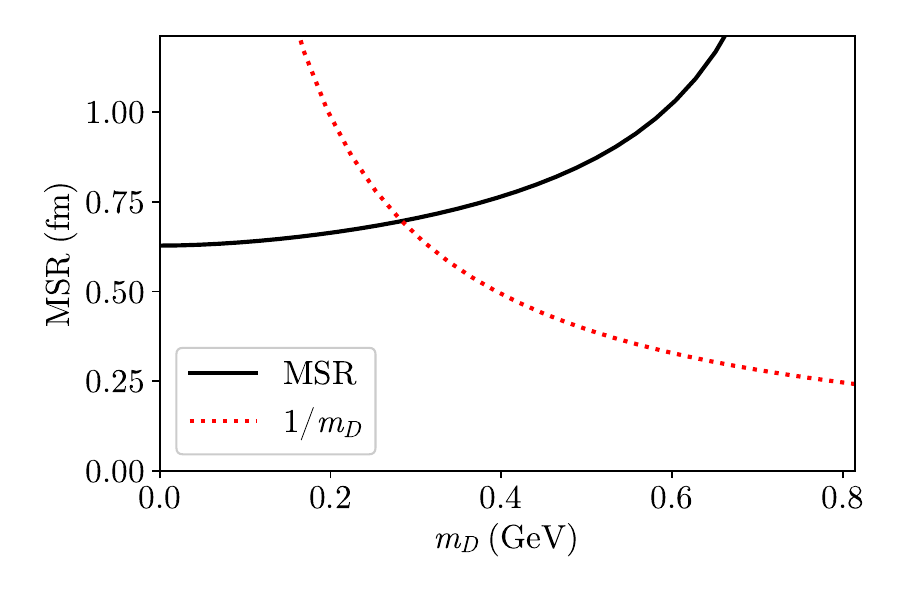}\hspace{0.2cm}
\includegraphics[width=0.32\linewidth]{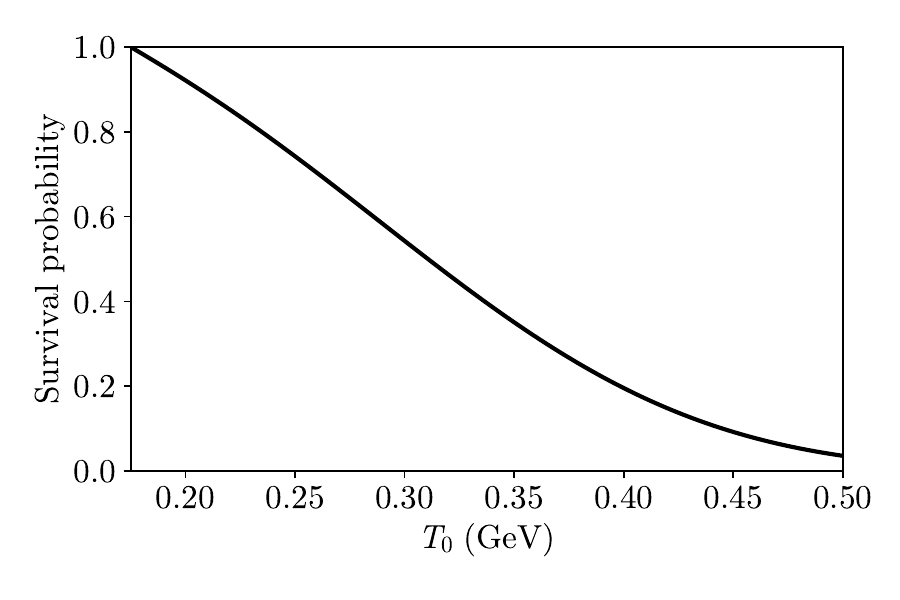}
\caption{Finite-temperature properties of the $X(3872)$ from the screened heavy-quark potential framework of Ref.~\cite{Armesto:2024zad}. Top left: Decay width scaled by the temperature as a function of the Debye mass. Top right: Binding energy and real part of the potential at large heavy-quark separations as functions of the Debye mass. Bottom left: Root-mean-square radius, $\sqrt{\langle r^2\rangle}$, and screening length, $1/m_D$, versus Debye mass. Bottom right: Survival probability as a function of the initial temperature. Figures taken from Ref.~\cite{Armesto:2024zad}.}
\label{fig:Armesto}
\end{figure}

A different approach to the in-medium behavior of the $X(3872)$ has been explored in Ref.~\cite{Armesto:2024zad}, where the state is described within a screened heavy-quark potential approach, using the Born--Oppenheimer approximation. In this picture, medium effects are encoded in the Debye screening mass, which is directly related to the temperature and modifies the vacuum interaction potential that is obtained from lattice QCD calculations. Since lattice QCD results for tetraquark potentials are not available, the authors rely on potentials extracted for heavy hybrid mesons, motivated by the structural similarity of these systems, which both involve a heavy quark-antiquark pair coupled to light degrees of freedom. The in-medium properties of the $X(3872)$ are then obtained by solving a Schr\"odinger equation. The results are shown in Fig.~\ref{fig:Armesto}.
The top panels show the width, which arises from the imaginary part of the in-medium potential, and the binding energy, which is defined as the difference between the eigenvalue of the bound state and the asymptotic value of the potential at large separations, both as functions of the Debye mass. The imaginary part of the Coulomb term and the asymptotic value of the real part of the potential are also shown in the left and right panels, respectively. As $m_D$ (or equivalently, $T$) increases, the binding energy decreases and eventually vanishes, signaling dissociation. The bottom-left panel compares the root-mean-square radius of the bound state, computed from the corresponding wave function, with the screening length $1/m_D$, illustrating that dissociation occurs when the bound-state size becomes comparable to the screening scale. Finally, the bottom-right panel displays the survival probability as a function of the initial temperature, indicating a gradual suppression of the $X(3872)$ with increasing temperature.

Overall, the thermal study of exotic heavy hadrons is still at an early stage, with only a few studies available, most of which assume a particular internal structure. Since the nature of the $X(3872)$ and the $T_{cc}^+(3875)$ is still debated even in vacuum, the most robust conclusion is that their thermal behavior strongly depends on their internal structure: weakly bound molecular configurations are expected to dissolve early in the hadronic phase, whereas compact ones should be more resilient. It would therefore be interesting to extend to finite temperature the frameworks that describe both components on the same footing in vacuum, such as coupled-channel approaches with an explicit $c\bar{c}$ component~\cite{Ortega:2009hj,Cincioglu:2016fkm} or the Born--Oppenheimer EFT, which incorporates tetraquark and quarkonium configurations together with the meson-meson thresholds~\cite{Brambilla:2024imu}, and to establish a closer connection to the production of exotic states in HICs.

\section{Phenomenological applications and experimental observables}\label{sec:applications}

Building upon the theoretical frameworks and lattice-QCD results detailed in the previous sections, we now explore how these in-medium modifications might translate into concrete experimental signatures. 
In HICs, the properties of hadrons are not static but evolve as the fireball cools toward the chiral transition and the subsequent freeze-out stages. While the vacuum expectations provide a baseline, the thermal shifts of masses and widths are essential for a consistent interpretation of the data collected in HICs.

In this section, we discuss how the thermal broadening of spectral functions and the shift in hadronic masses might manifest themselves in experimental observables, most notably in the production of electromagnetic probes (dileptons and photons) and the transport properties of the hadronic gas. Furthermore, we discuss how these medium-modified properties can influence the yields of hadrons, particularly exotic states, and the dynamics of the late-stage expansion of the fireball, e.g., from the perspective of hadron correlation (femtoscopy) measurements.

\subsection{Fireball evolution in the hadronic phase}\label{sec:app-hics}

The hadronic phase is the last stage of HICs, spanning the interval between hadronization and kinetic freeze-out. In typical hybrid simulations, this stage starts with a transition from a hydrodynamic expansion to a microscopic particle evolution in a more dilute medium (particlization). In the dilute regime, hadrons are then evolved following Boltzmann transport equations---or their quantum analogs---in a microscopic realization, generically known as hadronic cascades, such as IQMD~\cite{Aichelin:1991xy,Hartnack:1997ez}, HSD~\cite{Ehehalt:1996uq,Cassing:1999es}, UrQMD~\cite{Bass:1998ca}, JAM~\cite{Nara:1999dz}, AMPT~\cite{Lin:2004en}, GiBUU~\cite{Buss:2011mx}, PHSD~\cite{Bratkovskaya:2011wp}, and SMASH~\cite{SMASH:2016zqf}, among others. For heavy probes, such as $D$ or $B$ mesons, the Langevin equation is frequently used to describe their diffusion through the thermal medium, mostly at the heavy-quark level in the QGP~\cite{Svetitsky:1987gq,GolamMustafa:1997id,Moore:2004tg,vanHees:2004gq,vanHees:2005wb,Cao:2013ita,Ozvenchuk:2014rpa}, although the Boltzmann transport equation has also been used~\cite{Gossiaux:2008jv,Uphoff:2012gb,Das:2013kea}.

The time evolution of the space-momentum coordinates in these models depends directly on the particle mass. At temperatures $T=120$--$150\mev$, right below the chiral transition, but still above the kinetic freeze-out, thermal masses should, in principle, be utilized to maintain consistency with the underlying EFT or lattice-QCD results. Furthermore, the presence of a finite thermal decay width implies that the quasiparticle evolution is inherently off-shell. Consequently, an extension of the standard (semi-)classical evolution is required. Only a few of the transport models mentioned, such as GiBUU~\cite{Buss:2011mx} and HSD~\cite{Cassing:1999es} (and its extensions), incorporate these in-medium modifications~\cite{Leupold:2009kz} with genuine off-shell transport.  

Such off-shell propagation can be implemented through the Kadanoff-Baym equations~\cite{Baym:1961zz}. At finite temperature, the thermal modifications of quasiparticles were incorporated in the PHSD (Parton-Hadron-String Dynamics) transport approach to describe HICs. In particular, by combining it with the Dynamical Quasiparticle Model~\cite{Cassing:2007nb}, which treats quarks and gluons as thermal quasiparticles, the dynamical masses and widths of these species have been addressed. 

\begin{figure}[bp!] 
\centering
\includegraphics[width=0.4\linewidth]{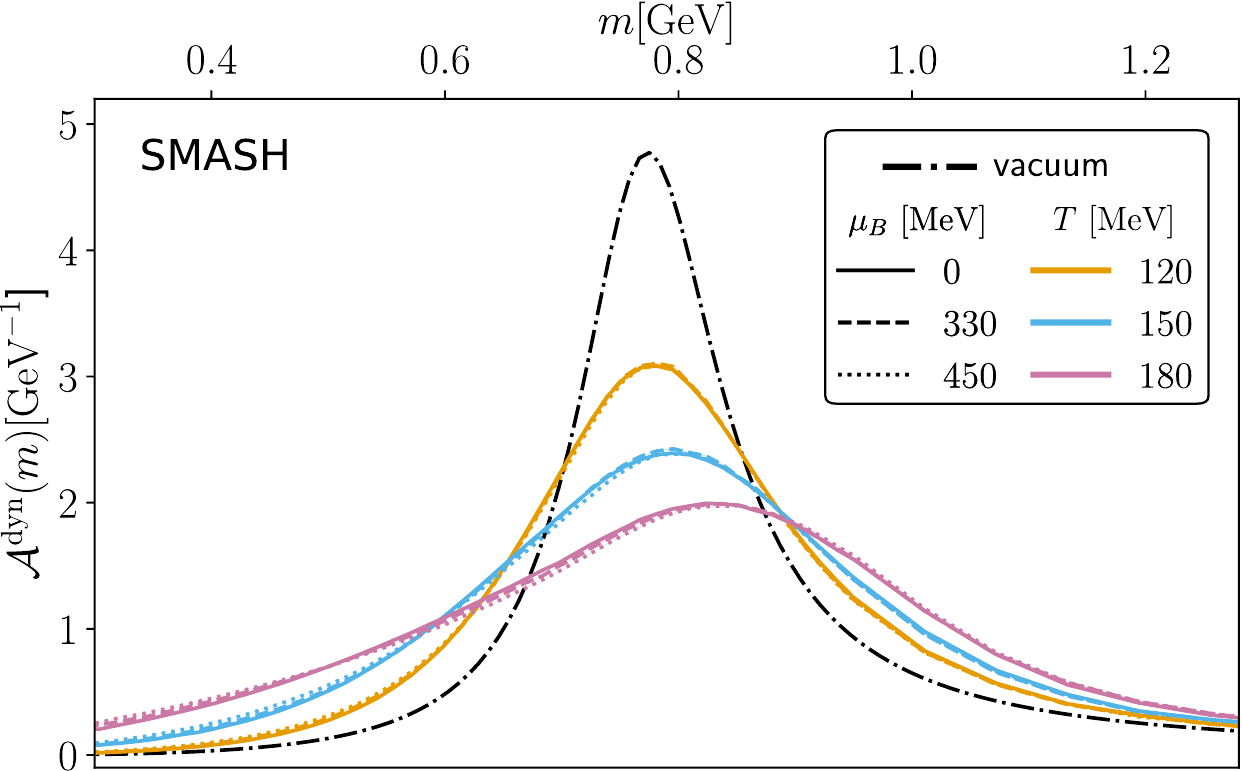}
\caption{$\rho$ meson spectral function at different temperatures and chemical potentials used by SMASH simulations in Ref.~\cite{Hirayama:2022rur}.}
\label{fig:SMASHrhomedium}
\end{figure}

Beyond these transport codes, off-shell propagation with thermal masses and widths for hadrons has not yet been widely incorporated, but it could lead to new effects stemming from the thermal modification of the hadronic mixture. Examples of the existing implementations addressing the collisional broadening of vector mesons are found in HSD~\cite{Bratkovskaya:2007jk}, GiBUU~\cite{Larionov:2020fnu}, and SMASH~\cite{Hirayama:2022rur}. In Fig.~\ref{fig:SMASHrhomedium} we reproduce the medium modification of the $\rho$ meson spectral function with temperature and baryochemical potential. This modification was used in SMASH simulations in Ref.~\cite{Hirayama:2022rur}.

\begin{figure}[htbp!] 
\centering
\includegraphics[width=0.4\linewidth]{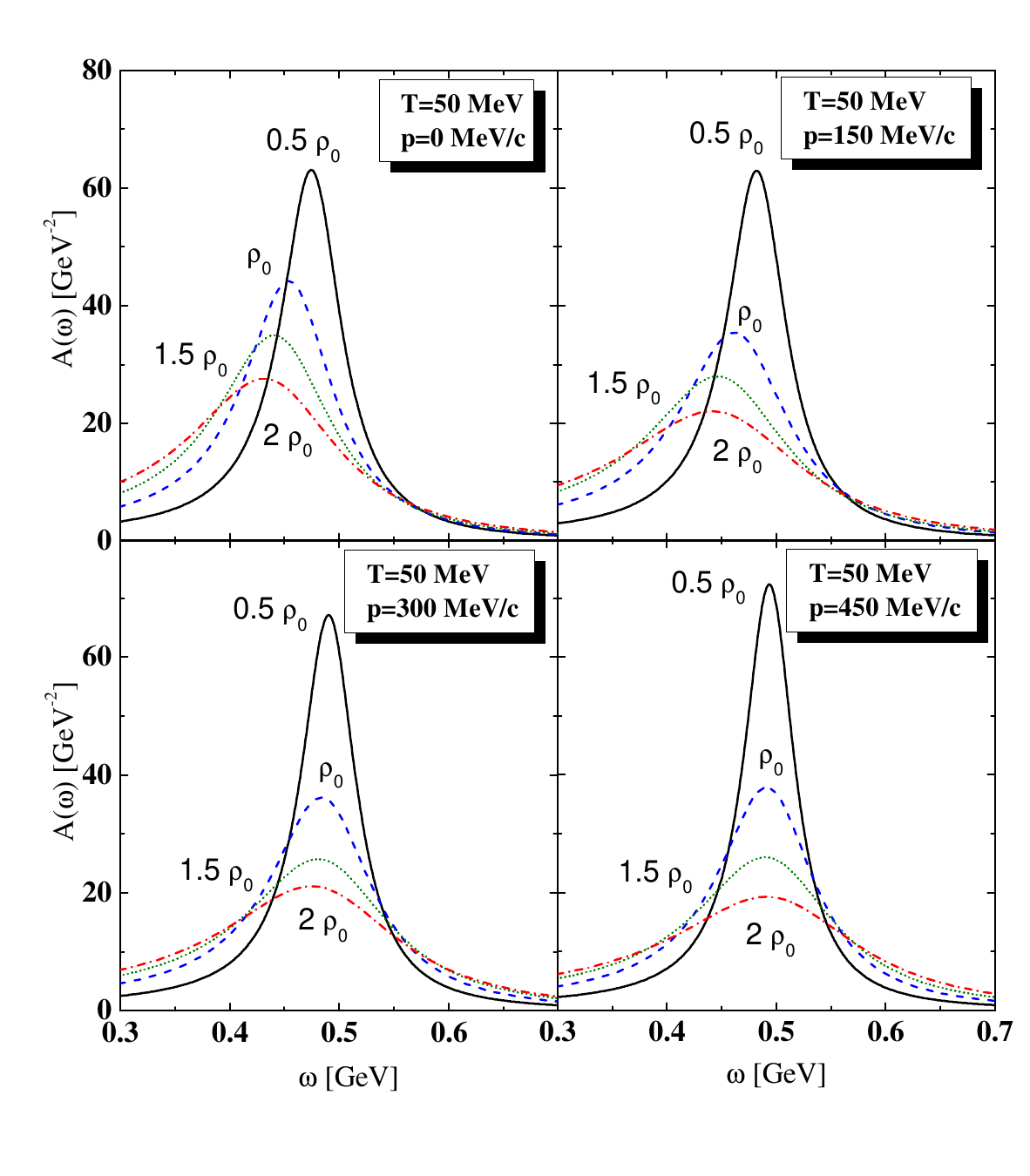}
\includegraphics[width=0.4\linewidth]{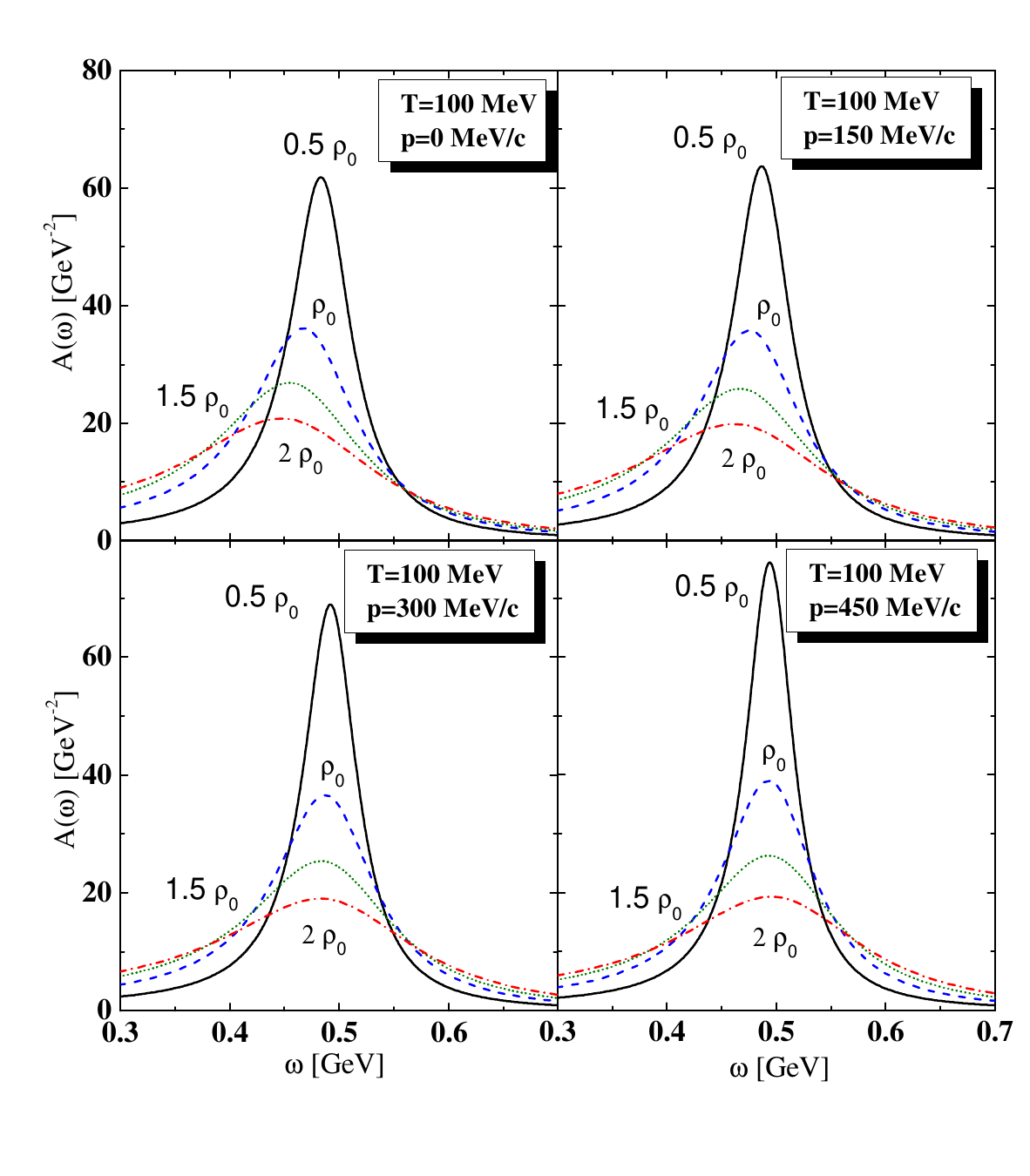}
\caption{$\bar{K}$ meson spectral function at different temperatures and chemical potentials used by PHSD simulations in Ref.~\cite{Song:2020clw}.}
\label{fig:PHSDantikaonmedium}
\end{figure}

In Fig.~\ref{fig:PHSDantikaonmedium} we show the temperature and density modifications of the $\bar{K}$ spectral function incorporated into the PHSD formalism of Ref.~\cite{Song:2020clw}, where strangeness production in low-energy HICs is analyzed.

We note that both temperature and density effects are accounted for. Although this review focuses on purely thermal effects, a high baryon density turns out to be more important for the spectral broadening of states. In the absence of net baryon density ($\mu_B=0$), the interaction with baryons and antibaryons, which appear in equal amounts, is still important for the modification of the spectral function. This is seen in the broadening of the $\rho$ meson in high-energy HICs where the net baryochemical potential is small. Because the effect of the collisions is determined not by the difference but by the sum of baryon and antibaryon densities, the broadening is expected to survive in these collisions with $\mu_B \simeq 0$. Since the thermal population of (anti)baryons is Boltzmann suppressed at the temperatures considered, the need for such broadening for the description of dielectron spectra and other observables imposes constraints on the strength of the meson-baryon interaction.

Implementing such modifications for a representative set of hadrons would be computationally challenging, especially when the time evolution of the system is involved. While one might argue that such implementations are unnecessary due to the modest medium effects for many light hadrons at temperatures below $T_c$, they become crucial for specific states---particularly resonances close to thresholds or loosely bound exotic states---where thermal effects can significantly alter their nature or open new decay channels, as is the case for the exotic $X(3872)$ state just below the open-charm threshold, discussed in Section~\ref{sec:exotics}. This may lead to visible consequences in experimental observables.  

\subsection{Dilepton emission and chiral symmetry restoration}\label{sec:app-dileptons}

Probably the most direct experimental window into these in-medium hadronic modifications is provided by the emission of electromagnetic probes~\cite{Rapp:2004zh,Hohler:2013eba}. Unlike hadrons, dileptons ($l^+l^-=\{e^+e^-,\mu^+\mu^- \}$) do not interact strongly and thus escape the hot and dense fireball without further strong interactions, carrying information about the specific stage of the collision in which they were produced (see the reviews on the topic~\cite{Rapp:2009yu,Leupold:2009kz,Geurts:2022xmk}).

The thermal dilepton emission rate per unit volume and momentum is related to the retarded electromagnetic current-current correlation function, $\Pi_{\text{EM}}^{\mu\nu}$~\cite{Rapp:2009yu},
\begin{equation}
\frac{\diff N_{l^+l^-}}{\diff^4x \diff^4q} = -\frac{\alpha_{\text{EM}}^2}{\pi^3 M^2} \ n_{\text{B}}(q_0; T) \frac13g_{\mu \nu} \ \im \Pi_{\text{EM}}^{\mu\nu, \text{R}}(M, q; T,\mu_B) \ ,
\end{equation}
where $M$ is the dilepton invariant mass, $\alpha_{\text{EM}}$ is the fine-structure constant and $n_{\text{B}}(q_0; T) = 1/(e^{q_0/T}-1)$ is the Bose--Einstein distribution function. 

At low invariant masses ($M \le 1\gev$), this rate is dominated by the spectral function of the light vector mesons $\rho, \omega, \phi$,  through the Vector Dominance Model (VDM)~\cite{Gale:1990pn,Rapp:2009yu},
\be j^\mu_{\textrm{EM}} = \frac{m^2_\rho}{g_\rho} \rho^\mu + 
\frac{m^2_\omega}{g_\omega} \omega^\mu +\frac{m^2_\phi}{g_\phi} \phi^\mu  \ ,
\ee 
so that,
\be 
\im \Pi_{\text{EM}}^{\mu\nu, \text{R}} (M,q; T,\mu_B)  = \sum_{V=\{ \rho,\omega,\phi\}} \left( \frac{m^2_V}{g_V} \right)^2 \im {\cal D}^{\mu\nu,\text{R}}_V (M,q;T,\mu_B) \ ,
\ee 
where the main contribution comes from the $\rho$ meson.

\begin{figure}[htbp!] 
\centering
\includegraphics[width=0.38\linewidth]{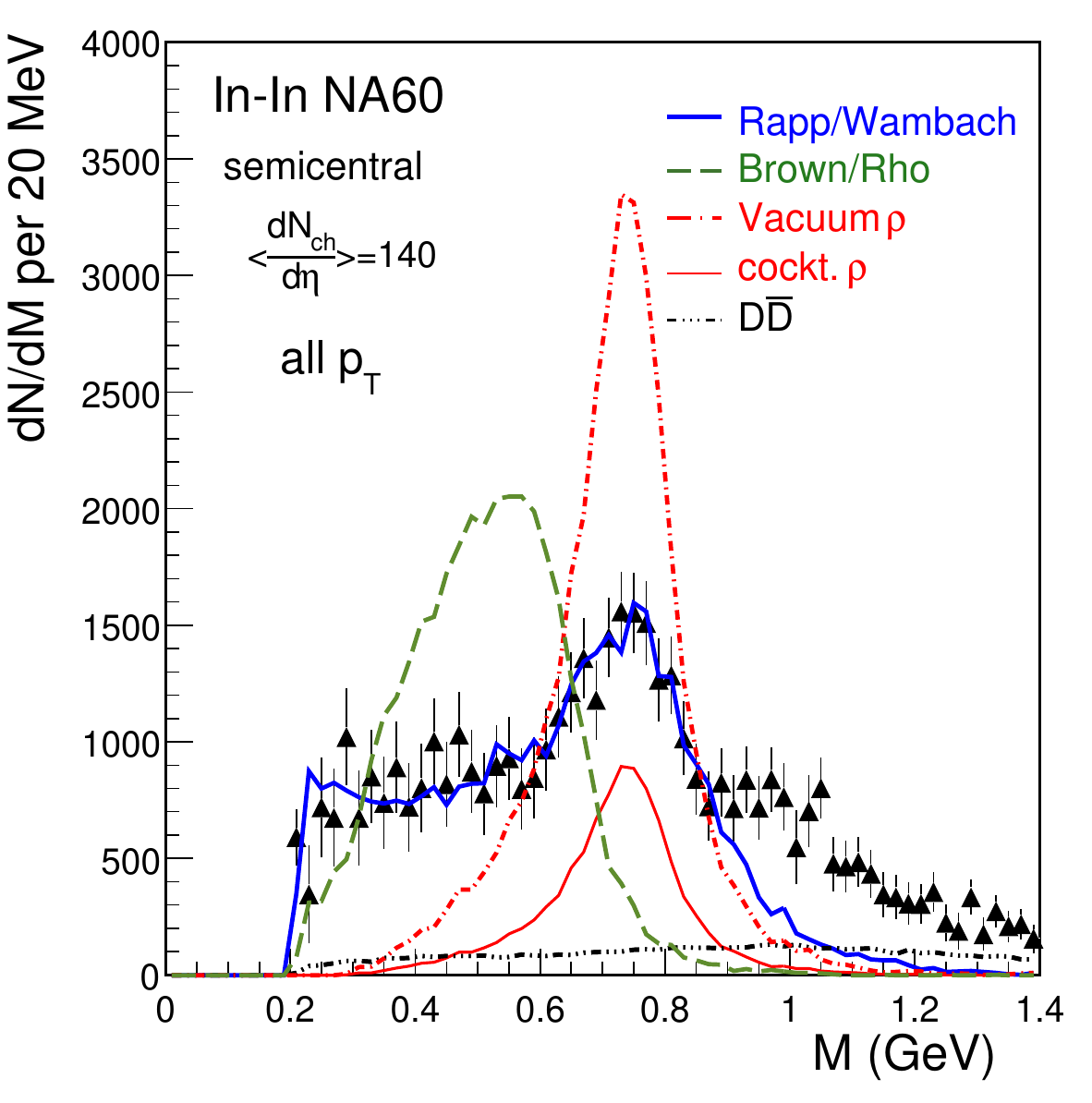}
\includegraphics[width=0.45\linewidth]{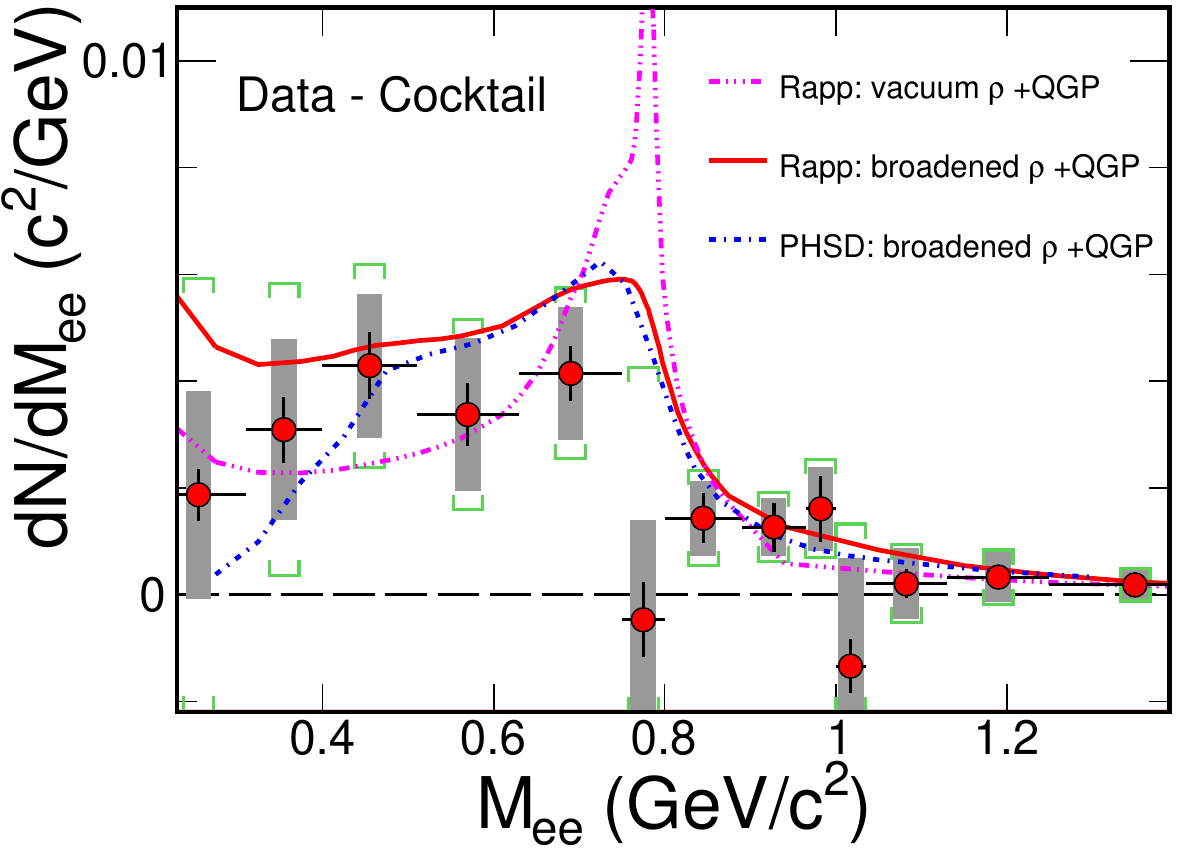}
\caption{Left panel: Experimental dimuon spectra in semicentral In+In collisions at $158A\gev$, measured by the NA60 collaboration at SPS~\cite{NA60:2008ctj}. Theoretical predictions include a vacuum $\rho$ meson spectral function (red dot-dashed line), a $\rho$ meson spectral function with a shifted mass (green dashed line, ``Brown/Rho''); and a $\rho$ meson spectral function including collisional broadening but no mass shift (solid blue line, ``Rapp/Wambach''). Right panel: Experimental dielectron excess in minimum-bias Au+Au collisions at $\sqrt{s_{NN}}=200\gev$, measured by the STAR collaboration at the Relativistic Heavy-Ion Collider (RHIC)~\cite{STAR:2015tnn}. Again, theoretical predictions show a preference for a $\rho$ meson with a broad spectral function and no mass shift, rather than the vacuum spectral function.}
\label{fig:Expdilepton}
\end{figure}

As the system approaches the chiral restoration temperature $T_c$, the broadening of the $\rho$ meson leads to a measurable enhancement of the dilepton yield in the low-mass region ($0.2 \gev < M < 1.1\gev$). This enhancement was observed at SPS energies at CERN by the CERES collaboration~\cite{CERES:1995vll} and CERES/NA45~\cite{CERESNA45:1997tgc}. A thermal reduction of the $\rho$ mass (motivated by the Brown--Rho scaling~\cite{Brown:1991kk}) was eventually disfavored, in favor of the broadening of the $\rho$ spectral function, by subsequent experimental data from the NA60 collaboration~\cite{NA60:2006ymb} and the CERES collaboration~\cite{CERES:2006wcq} at SPS energies. This result can be seen in the left panel of Fig.~\ref{fig:Expdilepton} taken from the NA60 collaboration~\cite{NA60:2006ymb,NA60:2008ctj}.

This panel shows the excess mass spectra of dimuons in In+In collisions measured at $158A\gev$. In this case, the vacuum $\rho$ spectral function (dot-dashed red line) cannot explain the data, and a thermal shift to lower masses (green dashed line) also disagrees with it. Among the scenarios shown, only the theoretical model of Ref.~\cite{Rapp:1999ej}, incorporating the thermal broadening of the state, describes both the region around the nominal mass of the resonance and the low-$M$ shoulder of the dimuon spectrum~\cite{vanHees:2007th}. At RHIC, both the PHENIX~\cite{PHENIX:2009gyd} and STAR~\cite{STAR:2015tnn} experiments have also observed the dielectron enhancement in the low-$M$ region in Au+Au collisions at $\sqrt{s_{NN}}=200\gev$. While the large enhancement initially reported by PHENIX could not
be explained by theoretical models, the later PHENIX measurement~\cite{PHENIX:2015vek} and the STAR data are consistent with vector-meson broadening, as implemented in different theoretical calculations~\cite{Cassing:2009vt,Bratkovskaya:2011wp, Rapp:2013ema}. In the right panel of Fig.~\ref{fig:Expdilepton}, we present the result from Ref.~\cite{STAR:2015tnn} where the mass spectrum excess of dielectrons in the low-$M$ region is compared with three model calculations. First, the calculation of Ref.~\cite{Rapp:2013ema} with a vacuum $\rho$ spectral function does not describe the data around the $\rho$ pole mass. Then, the same model with a $\rho$ meson undergoing collisional broadening can describe the data well. Finally, a transport model based on off-shell propagation~\cite{Cassing:2009vt,Bratkovskaya:2011wp} also shows good agreement. 
The HADES collaboration has also observed the dielectron excess in Ar+KCl collisions at $1.756A\gev$~\cite{HADES:2011nqx}.

While dilepton measurements serve as continuous probes of the light vector sector throughout the fireball's evolution, the final yields of vector and axial-vector mesons offer a complementary picture of a possible chiral symmetry restoration close to the chemical freeze-out. 

In the strangeness sector, the $K^*(892)$ and the $K_1(1270)$ mesons can be seen as chiral partners with a mass difference of about $360\mev$ and relatively narrow widths in vacuum. At temperatures close to the chiral transition, they are expected to become partly degenerate, implying deviations from vacuum-based expectations, such as those of the Statistical Hadronization Model~\cite{Braun-Munzinger:2003pwq,Andronic:2017pug}. In Ref.~\cite{Sung:2021myr} this effect is used to predict the ratio $K_1/K^*$ by modeling the interactions in thermal equilibrium of the two states, e.g., the $K_1$ is dissolved in the medium by reactions like $K_1 \pi \rightarrow K \pi, K^* \rho$ or $K_1 \rho \rightarrow K^* \pi, K \rho$ and their inverse reactions, which maintain detailed balance. In the symmetry-restored scenario, the production yields of $K^*$ and $K_1$ are the same at chemical freeze-out. For peripheral HICs, where the time evolution is shorter and the kinetic freeze-out temperature is higher, the effect is larger. The ratio $K_1/K^*$ is, in these conditions, a factor of 6 larger than the expectation from the Statistical Hadronization Model.

\begin{figure}[htbp!] 
\centering
\includegraphics[width=0.4\linewidth]{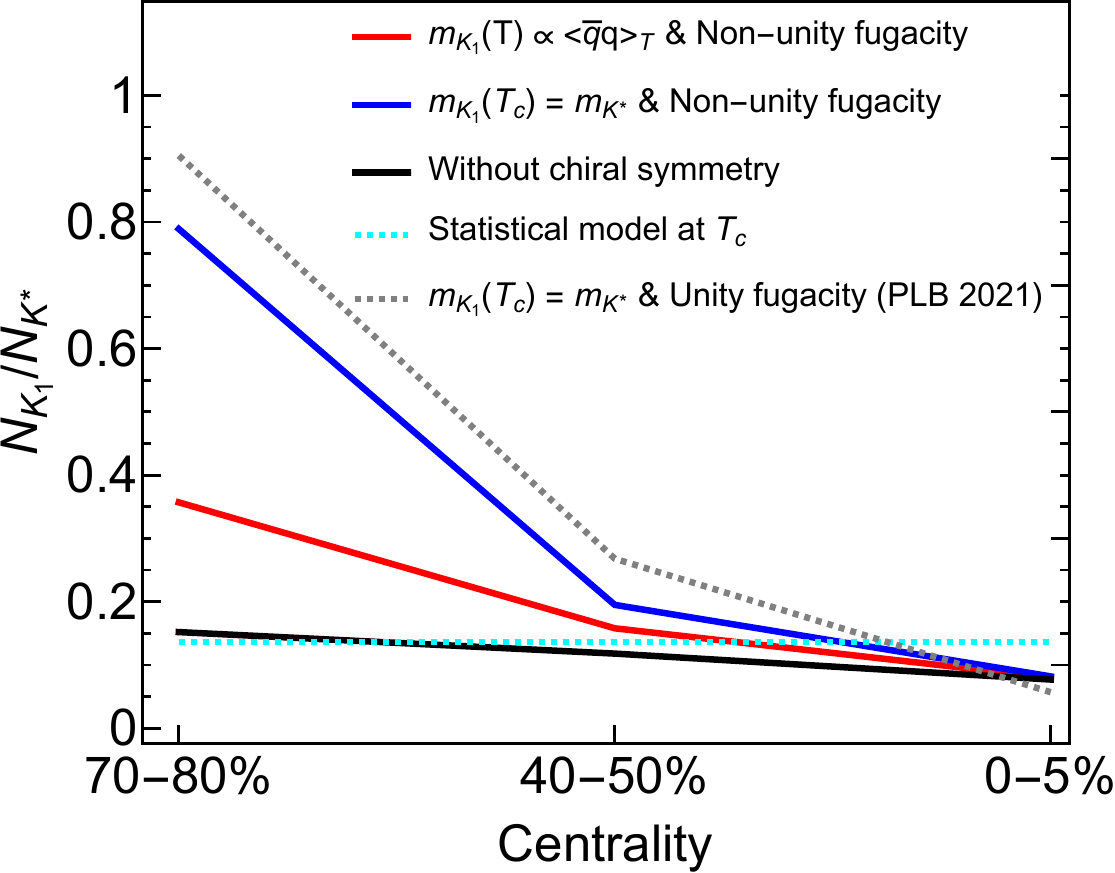}
\includegraphics[width=0.4\linewidth]{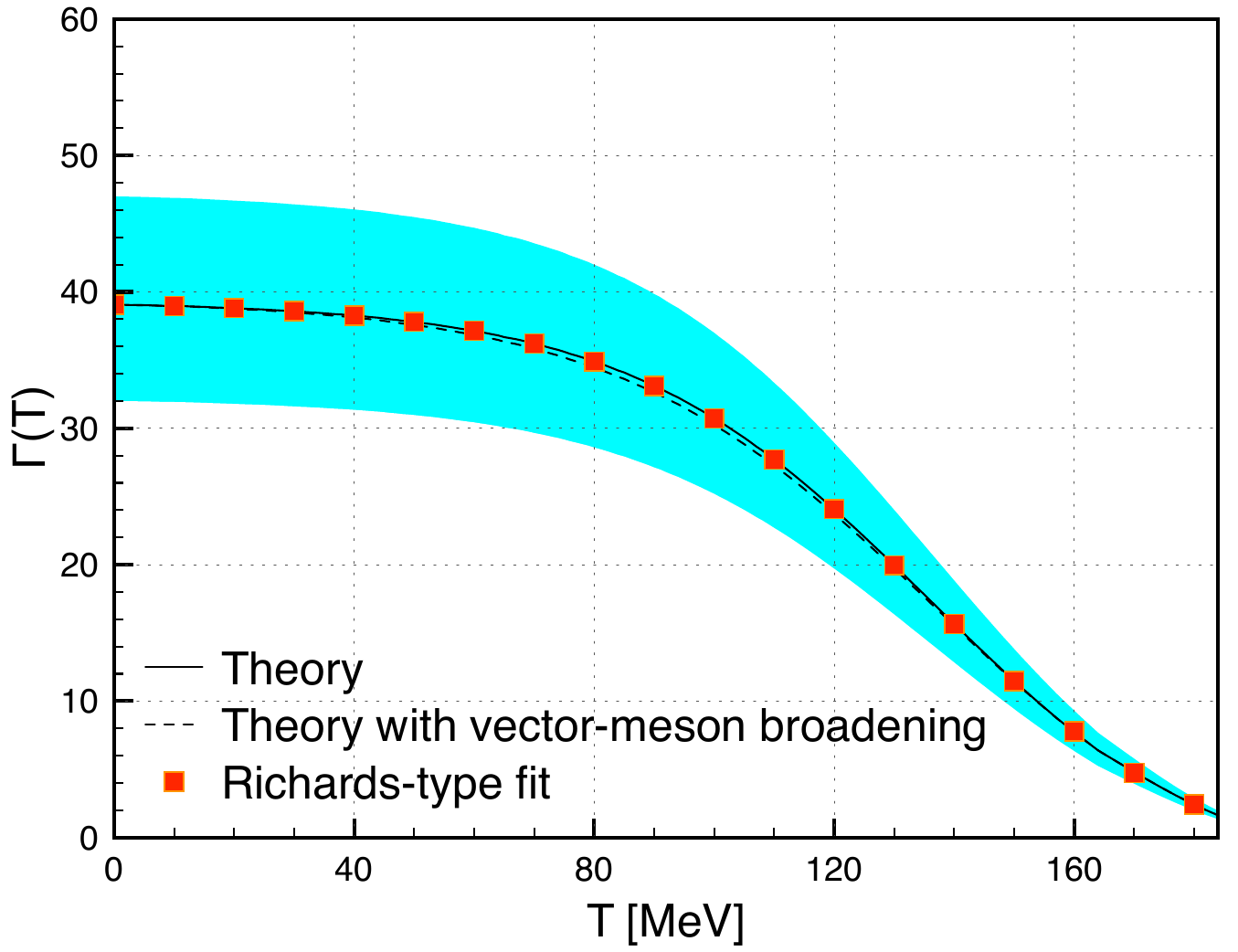}
\caption{Left panel: Ratio of $K_1$ over $K^*$ yields in Pb+Pb collisions at $\sqrt{s_{NN}}=5.02$ TeV at different centralities. From Ref.~\cite{Sung:2023oks}. Right panel: Temperature dependence of the integrated three-body decay width of $K_1^+ (1270) \rightarrow \pi^+ \pi^- K^+$. Figure adapted from Ref.~\cite{Nam:2026mrg}.}
\label{fig:kstark1}
\end{figure}

In Ref.~\cite{Sung:2023oks}, a thermal shift for the $K_1$ mass from the QCD sum-rule expectations~\cite{Lee:2023ofg} was also included, together with light-meson fugacities in the time evolution equation. In this case, the enhancement over the Statistical Hadronization Model is only a factor of 2.4 for the most peripheral collisions. The two scenarios are shown as gray dotted lines and red solid lines, respectively, in the left panel of Fig.~\ref{fig:kstark1}.

More recently, Ref.~\cite{Nam:2026mrg} studied the temperature modification of the three-body decay process $K_1^+(1270) \rightarrow \pi^+ \pi^- K^+$ using the in-medium $K_1$ mass obtained from Weinberg sum rules, finding a remarkable suppression of the decay width due to the reduction of the available final-state phase space (right panel of Fig.~\ref{fig:kstark1}).

\subsection{Determination of transport coefficients}\label{sec:app-transport}

The transport coefficients of an interacting system---like the shear viscosity $\eta$, the bulk viscosity $\zeta$, and the thermal and electrical conductivities $\kappa, \sigma$, or diffusion coefficients $D$---are macroscopic manifestations of the underlying microscopic interactions among the components of the system. In the context of HICs, these coefficients govern the dissipative evolution of the fireball and its transition from a strongly interacting system of quarks and gluons to a dilute gas of hadrons. 

For the hadronic gas at finite temperature, while traditionally transport models have used vacuum cross sections (interaction widths) and constant masses, the temperature-dependent modifications to hadron properties discussed in the previous sections necessitate a more sophisticated treatment. This becomes particularly important as the system approaches the transition temperature $T_c$, where the melting of the quark condensate and the corresponding shifts in spectral functions can significantly alter the scattering amplitudes and the relaxation times of hadronic species.

This becomes evident in those calculations in which both phases of a phase transition can be computed explicitly, like the chiral transition in the L$\sigma$M. In Ref.~\cite{Chakraborty:2010fr}, Chakraborty and Kapusta used this model with thermal masses for the pions and the massive mode to maintain thermodynamic consistency. The shear and bulk viscosities were computed by solving the linearized version of the Boltzmann equation in the relaxation-time approximation. The Chapman--Enskog expansion was also applied for $\eta$ (this method was later also applied to the bulk viscosity in the same model in Ref.~\cite{Dobado:2012zf}).

\begin{figure}[htbp!] 
\centering
\includegraphics[width=0.4\linewidth]{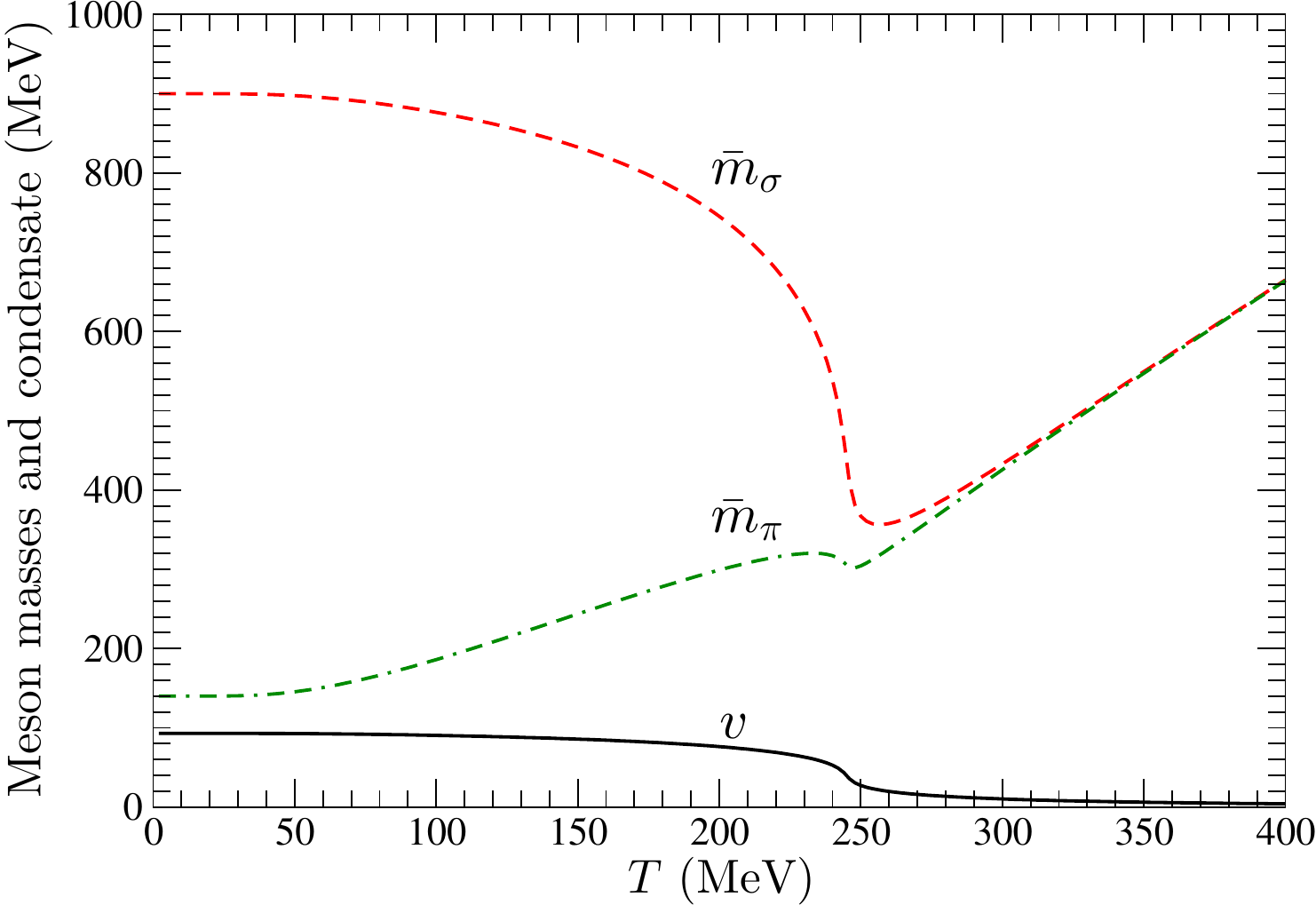}
\includegraphics[width=0.4\linewidth]{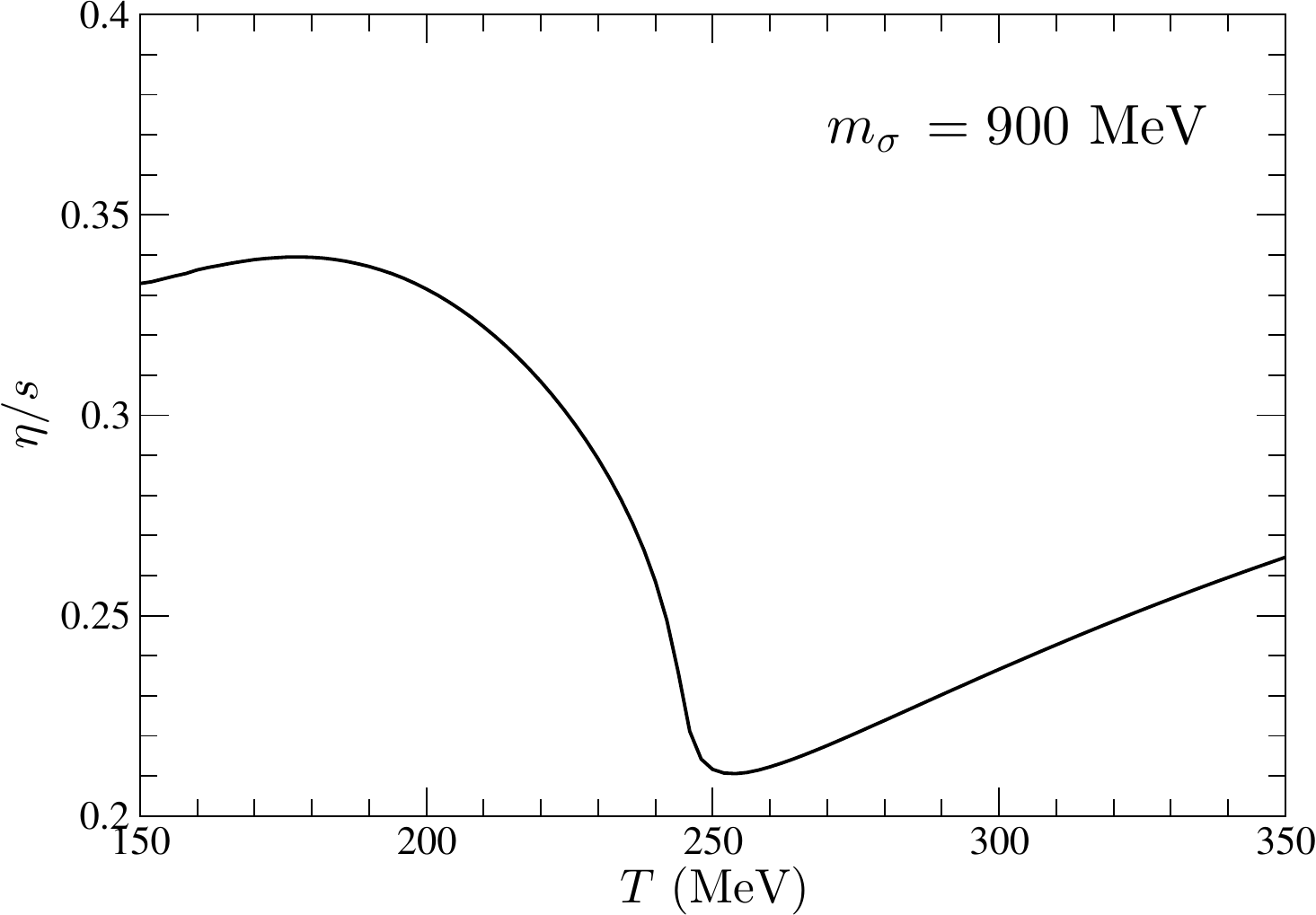}
\caption{Left panel: Pion and $\sigma$ meson masses and condensate in the L$\sigma$M as functions of the temperature as calculated in Ref.~\cite{Chakraborty:2010fr}. Right panel: Corresponding value of the shear viscosity over entropy density calculated in the same model.}
\label{fig:chakrabortykapusta}
\end{figure}

In the left panel of Fig.~\ref{fig:chakrabortykapusta} we show the pion and $\sigma$ masses as functions of the temperature, together with the thermal expectation value of the $\sigma$ field, $v=\langle \sigma\rangle$, which coincides with the pion decay constant in vacuum. The figure corresponds to the physical vacuum pion mass, for which a crossover transition is obtained around $T_c=250\mev$ (a much higher temperature than the one found in lattice QCD). The right panel of the same figure shows the shear viscosity over entropy density as a function of the temperature, which presents a minimum at $T_c$.

Similar results have been obtained for the same model under different approximations in Refs.~\cite{Petropoulos:2004bt,Dobado:2009ek,Dobado:2012zf,Torres-Rincon:2012sda}. For the bulk viscosity, it is interesting to note the existence of a maximum, instead of a minimum, at the transition temperature. In a second-order transition, like the one shown in Ref.~\cite{Dobado:2012zf} in the chiral limit, the correlation length diverges at the critical temperature, but conformal invariance is strongly broken: the trace anomaly $(\epsilon-3P)$ and the deviation of the speed of sound from its conformal value, $1/3-c_s^2$, peak near $T_c$. Since the bulk viscosity determines how fast the system returns to equilibrium under a uniform expansion or compression, and it vanishes for a conformal system,
it is expected to peak there as well. Capturing this behavior requires a consistent thermal evolution of the masses of the medium constituents.

In the QGP phase, it is worth mentioning analogous calculations in the (P)NJL model and the Dynamical Quasiparticle Model~\cite{Marty:2013ita,Berrehrah:2013mua,Puglisi:2014sha}, where the thermal (dynamical) quark and gluon masses and thermal widths directly affect the relevant transport coefficients. A similar effect can also be expected in the hadronic phase.

\begin{figure}[htbp!] 
\centering
\includegraphics[width=0.4\linewidth]{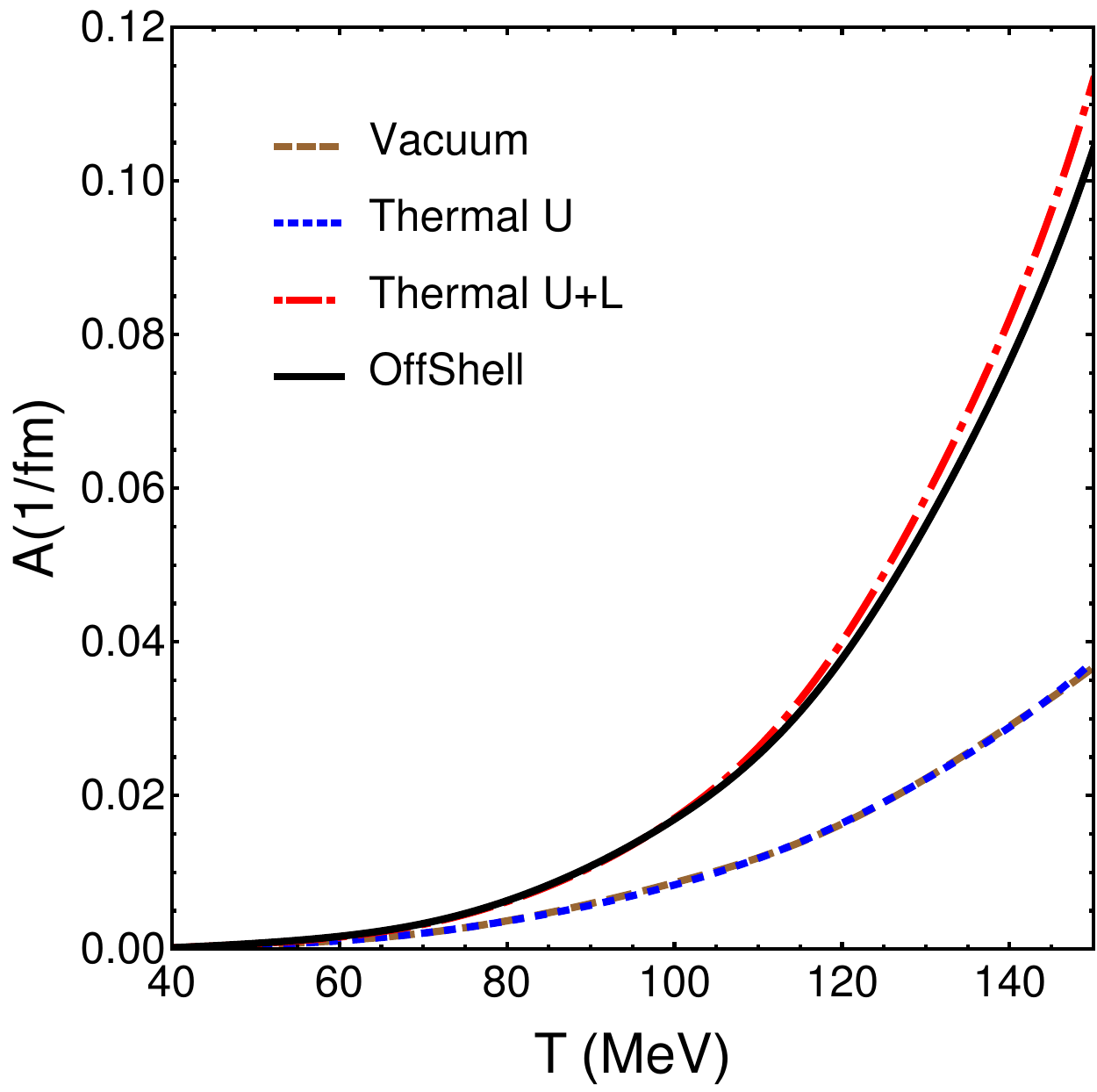}
\hspace{5mm}
\includegraphics[width=0.4\linewidth]{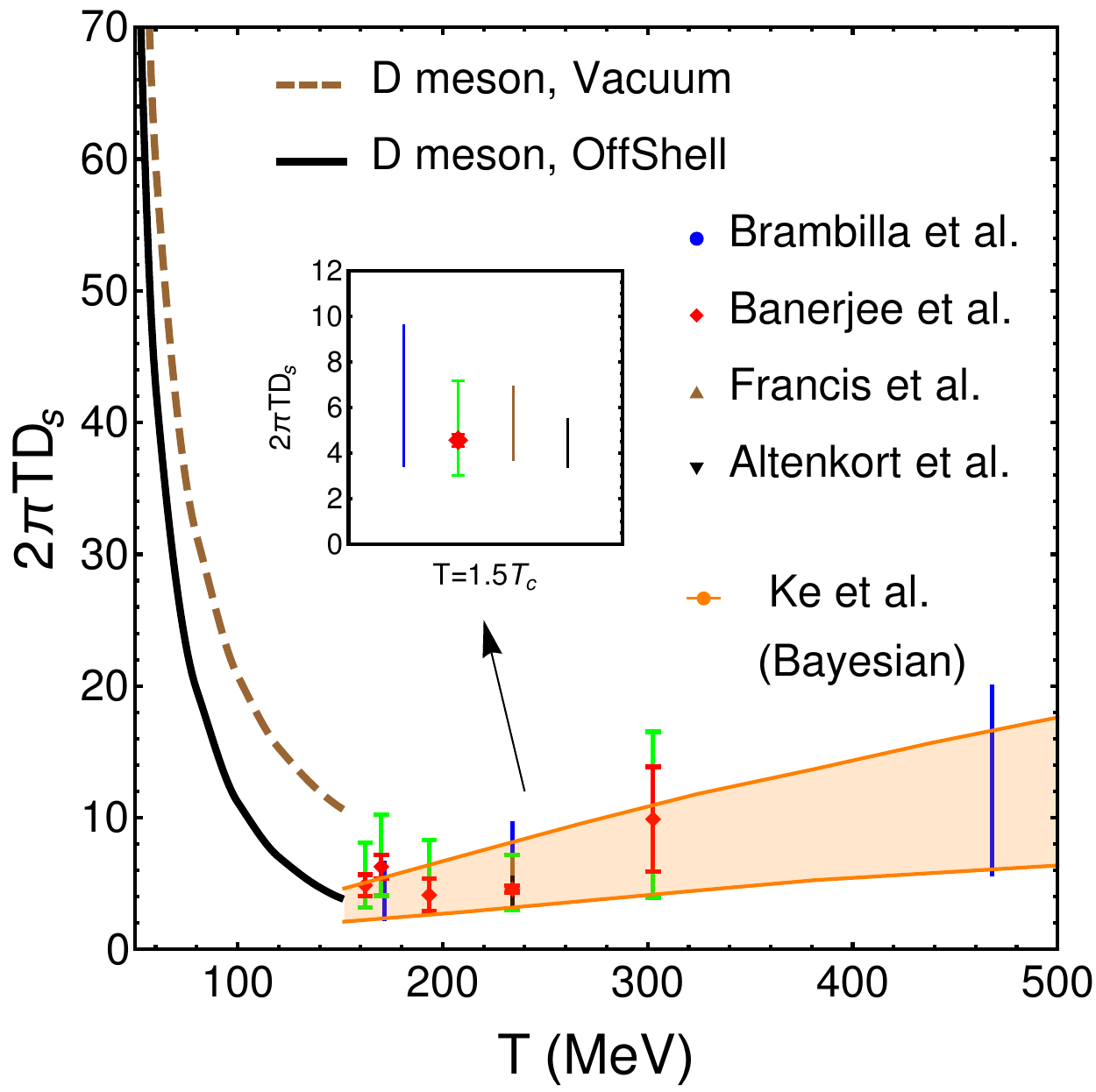}
\caption{Drag force coefficient (left panel) and spatial diffusion coefficient (right panel) of the $D$ meson as calculated within the unitarized EFT~\cite{Torres-Rincon:2021yga}. The panels show the differences between using vacuum masses and interactions and using consistent thermal ones.}
\label{fig:DmesonAandDs}
\end{figure}

In the heavy-flavor sector, a similar situation occurs with the transport coefficients that govern the heavy-particle dynamics, the drag and diffusion coefficients in momentum space, and the diffusion coefficient in coordinate space $D_s$~\cite{Das:2024vac}. The heavy-meson masses and in-medium interactions with light mesons obtained in Refs.~\cite{Montana:2020lfi,Montana:2020vjg} have been incorporated in a microscopic calculation of the off-shell version of the Fokker--Planck equations, starting from the Kadanoff--Baym transport equations, where the transport coefficients received medium-dependent corrections~\cite{Torres-Rincon:2021yga}. The additional scattering processes that can happen at finite temperature but not in vacuum---the so-called Landau scattering---effectively increase the interaction rates of the heavy meson with thermal particles of the medium, even allowing for particle-number-changing processes ($1 \leftrightarrow 3$ scattering) otherwise not possible without thermal spectral broadening of hadrons. The drag coefficient of $D$ mesons, $A(\bm{k})$, measures the relaxation rate of a heavy meson due to collisions with the medium, as it is clear from the deterministic part of the Langevin equation,
\be 
\left\langle \frac{ \diff\bm{k}}{\diff t} \right\rangle = - A(\bm{k}) \bm{k} \ , 
\ee
where the brackets on the left-hand side denote an average over realizations of the fluctuations (that is, over thermal noise). This coefficient is found to be enhanced when thermal effects are incorporated in the calculation, especially at temperatures around the transition temperature, as can be seen in the left panel of Fig.~\ref{fig:DmesonAandDs}.

On the other hand, the spatial diffusion coefficient is defined as the coefficient relating the nonequilibrium heavy-particle current to the local gradient in the heavy-particle density that generates it. In nonrelativistic Brownian motion, it sets the growth rate of the mean-square displacement with time,  
\be 
\langle (\bm{r}(t)- \bm{r}(t=0) )^2\rangle \simeq 6 D_s t \ . 
\ee
The temperature dependence of $D_s$ after incorporating medium effects in the microscopic calculation of Ref.~\cite{Torres-Rincon:2021yga} shows a reduction of this coefficient with respect to the calculation with vacuum masses and $T=0$ interactions. We reproduce the quantitative results in the right panel of Fig.~\ref{fig:DmesonAandDs}, where the $D$-meson calculation is the one appearing at low temperatures. After incorporating thermal effects, the result goes from the dashed brown line to the black solid one, with an important correction at temperatures close to the crossover temperature.

\subsection{Hadron femtoscopy}\label{sec:app-femtoscopy}

Hadron correlation studies in nuclear collisions are a powerful tool to probe the spatial characteristics of the particle-emitting source~\cite{Pratt:1995,Wiedemann:1999qn,Lisa:2005dd}. Typically, the most abundant particles, identically charged pions, are used, exploiting their Bose--Einstein correlations and the negligible role of their strong final-state interaction. In general, the two-particle correlation function at small relative momentum can also reflect the integrated effect of the strong interaction between the studied pair. For this reason, femtoscopy techniques have been used to learn about the interaction between pairs of hadrons for which scattering experiments cannot be performed~\cite{Gyulassy:1979yi,Lednicky:1981su,Heinz:1999rw,Fabbietti:2020bfg}. 

For a pair of hadrons, labeled 1 and 2, the femtoscopic correlation function $C(\bm{p}_1, \bm{p}_2)$ is defined in terms of their momenta $\bm{p}_i$ and energies $E_i=\sqrt{\bm{p}^2_i+m^2_i}$ as

\be
C(\bm{p}_1, \bm{p}_2) =
\frac{E_1 E_2 \, \frac{\diff^6 N}{\diff^3p_1 \diff^3p_2}}
{\left(E_1 \, \frac{\diff^3 N}{\diff^3 p_1}\right)\left(E_2 \, \frac{\diff^3 N}{\diff^3p_2}\right)}
= \frac{N(\bm{p}_1, \bm{p}_2)}{N(\bm{p}_1)\, N(\bm{p}_2)} \ .
\ee
Here, $N(\bm{p}_1, \bm{p}_2)$ denotes the number of hadron pairs with momenta $\bm{p}_1$ and $\bm{p}_2$, while $N(\bm{p}_i)$ represents the number of single hadrons of type $i$ with momentum $\bm{p}_i$. This function quantifies deviations from the uncorrelated scenario $N(\bm{p}_1, \bm{p}_2) = N(\bm{p}_1) N(\bm{p}_2)$, which arise due to quantum effects or final-state interactions between the two hadrons.

In the pair center-of-mass frame (also called the pair-rest frame, indicated by an asterisk $*$), the correlation function can be expressed as a function of the pair’s relative momentum $k^*$:
\be
C(k^*) = \xi(k^*) \, \frac{N_{\text{same}}(k^*)}{N_{\text{mixed}}(k^*)} \ ,
\ee
where $N_{\text{same}}(k^*)$ is the number of particle pairs with relative momentum $k^*$ measured within the same collision event, whereas $N_{\text{mixed}}(k^*)$ corresponds to pairs formed from different (mixed) events, which ensures that they contain no correlations due to mutual interactions or quantum effects. In practice, a normalization factor $\xi$ is introduced to account for differences in conditions and normalizations between the same-event and mixed-event samples.

Theoretically, the correlation function can be computed with the Koonin--Pratt formula~\cite{Koonin:1977fh,Pratt:1990zq},
\be 
C(k^*)= \int \diff ^3r^* \ \sum_i w_i S_i(\bm{r}^*) |\Psi_{i} (\bm{k}^*,\bm{r}^*)|^2 \ , 
\ee
where $w_i$ are the weights for a given initial two-body channel $i$ that can be connected to the observed one, $S_i(\bm{r}^*)$ is the source function containing information on the distribution of the distance between the particles in the pair $i$, and $\Psi_i (\bm{k}^*,\bm{r}^*)$ is the two-body wave function connecting the initial state $i$ with the observed one.

Temperature effects are implicitly contained in $w_i$, which are usually calculated from a thermal model for hadron production, and, to a lesser degree, in $S_i$ through details of the fireball geometry. The type (proton-proton, proton-nucleus, nucleus-nucleus) and centrality of the HIC will determine the details of the source function. All dynamical information about the interaction of particles is contained in the wave function, which is assumed not to depend on temperature. However, at freeze-out temperatures, even for high-multiplicity $pp$ collisions, the masses and interactions might also be modified by the medium, and the wave function can reflect these effects for some systems. In-medium femtoscopy is not yet a developed field, but it should be noted that the initial state $i$ is not an asymptotic one, but a correlated system created at a high temperature. Therefore, the few-body dynamics may need to be modified with respect to vacuum expectations.

A recent example is provided by the proton-pion and deuteron-pion correlation functions measured by the ALICE collaboration in high-multiplicity $pp$ collisions at $\sqrt{s}=13$ TeV~\cite{ALICE:2025aur,ALICE:2025byl}. For $p\pi^+$ and $d\pi^+$, the correlation functions reflect an overall repulsive interaction due to the Coulomb force at small relative momentum, as indicated by values of the correlation function smaller than one. In addition, a broad peak in the correlation function with values larger than one is clearly visible, associated with the $\Delta$ resonance (from $\Delta^{++} \rightarrow p\pi^+$ decays, and from $\pi d$ scattering through $\Delta$ excitation for $d\pi^+$). However, the position of the peak is clearly shifted to lower momentum compared to the expected maximum given by the nominal $\Delta$ mass.

\begin{figure}[htbp!] 
\centering
\includegraphics[width=0.9\linewidth]{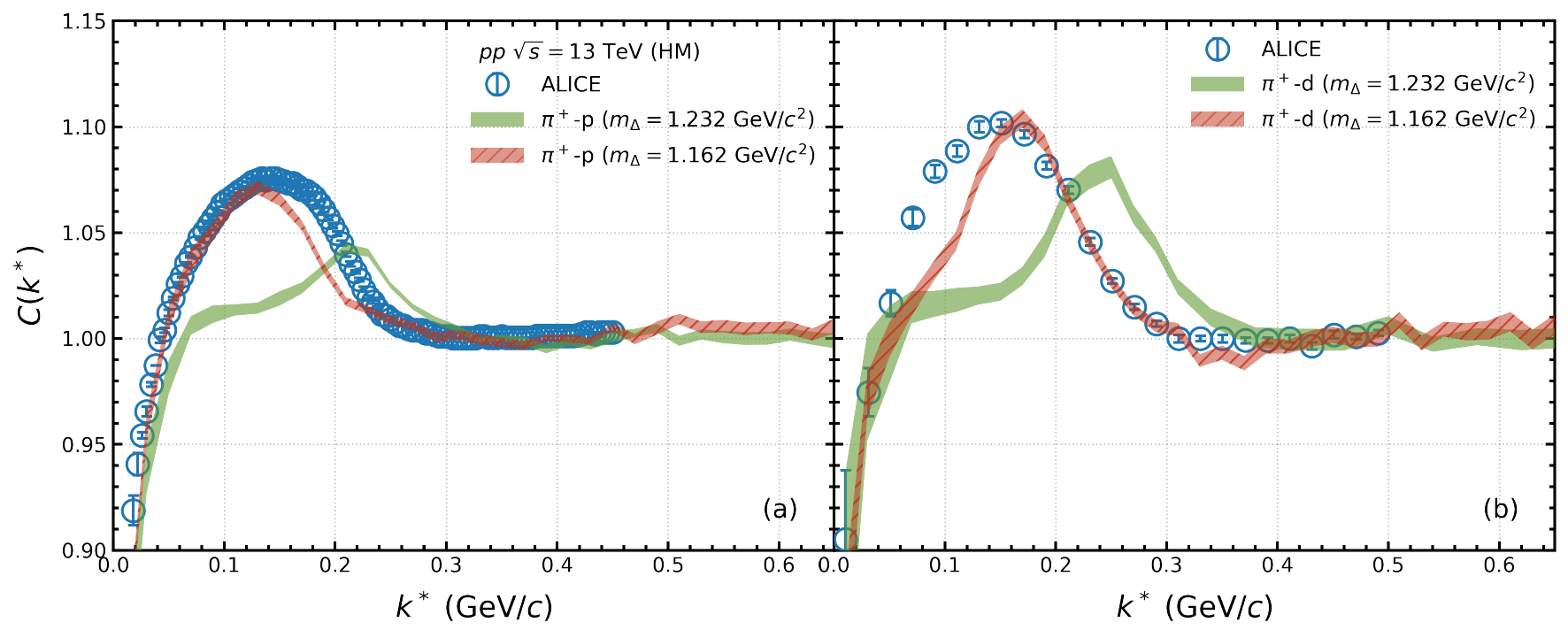}
\caption{Pion-proton (left panel) and pion-deuteron (right panel) femtoscopy correlation functions measured in high-multiplicity $pp$ collisions at $\sqrt{s}=13$ TeV by the ALICE collaboration~\cite{ALICE:2025aur, ALICE:2025byl}. Theoretical calculations of Ref.~\cite{Zhang:2025tfd} are shown using the nominal (green band) and thermally shifted (orange band) $\Delta$ baryon mass.}
\label{fig:Zhang_Ko}
\end{figure}

This effect was explained in Ref.~\cite{ALICE:2025aur} by accounting for the rescattering effect and a ``$\Delta$ spectral temperature'' that is used to convolute the vacuum resonance distribution. In contrast, Ref.~\cite{Zhang:2025tfd} suggests that the effect can be explained by the reduction of the thermal mass of the $\Delta$ baryon at temperatures around the freeze-out temperature. In Fig.~\ref{fig:Zhang_Ko}, reproduced from Ref.~\cite{Zhang:2025tfd}, one can see that a $\Delta$ with its vacuum mass of $m_\Delta = 1232\mev$ does not fit the ALICE data, but a reduction of $\simeq 70\mev$ can fit not only the $\pi^+p$ correlation function (left panel), but also the $\pi^+-d$ correlation function (right panel). Such a reduction is expected from theoretical calculations using thermal QCD sum rules~\cite{Xu:2015jxa,Azizi:2016ddw}, and also the (P)NJL model~\cite{Blanquier:2011zz, Torres-Rincon:2015rma}.

\subsection{Model dependence of in-medium observables}\label{sec:th_uncertainties}

The different microscopic approaches that consider finite temperature predict quantitatively different thermal masses, widths, and spectral functions. The observables discussed in this section differ in how much they depend on the framework used. The main sources of model dependence are the following: EFT results, like chiral approaches, are controlled only well below the crossover temperature, $T_c$. Unitarized extensions can go further, but only as long as new degrees of freedom (in particular, the deconfined ones) can be ignored. In addition, they typically depend on the regularization technique, since the unitarization introduces some model dependence. 
Chiral models can reproduce the transition, but only qualitatively, since the crossover temperature could be as high as $T_c \simeq 250\mev$, as in the L$\sigma$M~\cite{Chakraborty:2010fr,Torres-Rincon:2021wyd}. Widths and mass shifts also depend on the particle content of the medium. In the case of the (P)NJL model, mesons above the quark-antiquark threshold acquire unphysical decay widths into the quark-antiquark continuum already in vacuum, and the lighter ones above their Mott temperature. On the lattice, screening masses and susceptibilities are computed directly, but pole masses, widths, and spectral functions require an ill-posed inversion from Euclidean correlators, in addition to discretization, volume, and quark-mass effects. Finally, converting spectral functions into measurable observables requires a model of the collision dynamics (fireball profile, freeze-out condition, off-shell transport, feed-down corrections, etc.), which introduces uncertainties unrelated to the microscopic calculation.
From the examples presented, the most robust, largely model-independent information comes from the low-mass dilepton excess, which points to a strong broadening of the $\rho$ meson without a sizable mass shift.
The effect has been tested by different experiments (NA60~\cite{NA60:2006ymb,NA60:2008ctj}, CERES~\cite{CERES:1995vll,CERESNA45:1997tgc,CERES:2006wcq}, STAR~\cite{STAR:2015tnn}, PHENIX~\cite{PHENIX:2009gyd,PHENIX:2015vek}, and HADES~\cite{HADES:2011nqx}), and the broadening conclusion can be accommodated by different theoretical frameworks (hadronic many-body theory and off-shell transport). This can be considered a robust result since the emission rate is directly given through vector dominance by the in-medium electromagnetic spectral function, whose vacuum limit is constrained by $e^+ e^- \rightarrow$ hadrons data, and by lattice-QCD and perturbative input. The disfavoring of a dropping $\rho$-mass scenario also follows the same reasoning. Some lattice-QCD results can also be considered rather robust, like the computation of thermal quantities (screening masses and susceptibilities), the degeneracy of chiral partners across the crossover, and the ordering of sequential quarkonium suppression, always within the inherent lattice-QCD limitations (finite-size effects, discretization, etc.).
Other observables are partially model dependent. For example, the detailed $\rho$ line shape and its connection to chiral symmetry restoration require a model of the axial-vector channel. Since the in-medium $a_1$ spectral function is not directly observable~\cite{Hohler:2013eba}, several scenarios remain compatible with the data. 
Heavy-flavor transport coefficients (the drag and spatial-diffusion coefficients) close to $T_c$ are sensitive to the widths and scattering rates of heavy hadrons, which differ among hadronic approaches and models. However, the values obtained from these theoretical calculations, lattice QCD and heavy-ion phenomenology lie in the same ballpark near $T_c$.
Finally, several observables remain strongly model dependent. Quarkonium melting temperatures and thermal widths depend on the choice of potential, on the lattice spectral reconstruction, and on the mechanism considered (color screening or collisional dissociation). The size of the mass shifts and widths of open-heavy-flavor mesons below $T_c$ differs among approaches, and lattice-QCD results in the hadronic phase, obtained with heavier than-physical pion masses, are not yet precise enough to discriminate between them. The $K_1/K^*$ resonance ratio is also highly sensitive to the inputs: the enhancement over the statistical-hadronization expectation changes from a factor $\sim 6$ to $\sim 2.4$ when a thermal $K_1$ mass shift and light-meson fugacities are included. In femtoscopy, the shift of the $\Delta$ peak in $p \pi^+$ and $d \pi^+$ correlations can be explained either by a thermal mass reduction or by rescattering and a modified line shape, so the observable mixes medium effects with source and production modeling.
Reducing this model dependence requires comparing approaches with the same medium composition, varying regularization scales to estimate EFT truncation and regularization uncertainties, propagating lattice-QCD reconstruction uncertainties, and using a common fireball evolution for different spectral functions.

\section{Conclusions and future outlook}\label{sec:summary}

In this review, we present a comprehensive overview of the thermal modifications of hadron properties, with a specific focus on the hadron masses, decay widths, and spectral functions. Through the interplay of finite-temperature quantum field theory, EFTs describing confined matter, and lattice-QCD approaches, a consistent picture emerges detailing how hadrons---from pions and nucleons to the bottomonia---behave when they are immersed in a hot thermal bath resembling the conditions of the early universe or the final stages of HICs.

Our discussion started by establishing the theoretical foundations: the imaginary-time formalism, which accounts for temperature effects in a consistent framework, combined with the EFT techniques that describe the details of the dynamics of different systems. This combination can then be used to extract equilibrium properties such as pole masses, screening masses, thermal decay widths, spectral functions, and transport coefficients.  

In the light-flavor sector, the physics is largely dominated by the spontaneous breaking of chiral symmetry and the emergence of (pseudo-)Goldstone bosons. We have reviewed how ChPT, either in its original formulation or with corrections from imposing exact unitarity on the scattering amplitudes, captures the temperature dependence of the properties of light mesons. Other extensions and models based on chiral symmetry breaking, such as the (N)L$\sigma$M or the (P)NJL models, can also reflect the thermal behavior of light mesons and baryons. A rather general outcome from these models is the systematic modification of the spectral function with channel-dependent mass shifts as the temperature starts to increase, and a broadening of the spectral function due to collisional effects and the opening of new possible decay channels at finite temperature. Moreover, the restoration of chiral symmetry at high temperature (partial in ChPT, or full in other models that can describe both sides of the transition) leads to the predicted degeneracy of chiral partners when the temperature exceeds $T_c$. 

The heavy-flavor sector also presents an interesting phenomenology when these states are put into a thermal bath. The heavy-quark mass scale appears in the problem, and it generates a wide variety of possibilities. For open-heavy-flavor mesons, their interaction with light degrees of freedom that compose the medium produces moderate shifts in their masses (several tens of MeV at $T\simeq 150\mev$) and a sizable increase of their decay widths. Calculations in EFTs show a decrease of heavy-meson masses and a broadening of their widths at temperatures below $T_c$. On the lattice, the screening masses increase above $T_c$, and those of parity partners also eventually become degenerate. For the hidden-heavy sector, the suppression and sequential melting of quarkonia have been extensively studied with lattice-QCD approaches (including NRQCD formulations on the lattice) and also self-consistent $T$-matrix calculations. The sequential melting of these states provides, as long predicted, a qualitative thermometer of the produced QGP, although the melting temperatures remain model dependent. Most frameworks indicate that the states dissolve through collisional broadening before static color screening alone would remove them, while the screening of the real part of the potential is still under discussion. The medium effects can also be extended to the newly discovered exotic states, such as the $X(3872)$ and the $T_{cc}^+(3875)$, for which finite temperature opens a new way to access their internal structure, helping to distinguish between a compact, multi-quark configuration and a more extended, loosely bound molecule. Incorporating these states into simulations of HICs that study their survival probability in the hot medium can serve to test the different hypotheses about their internal structure.

Finally, the thermal effects on hadron properties can have direct phenomenological consequences for relativistic HICs. We have discussed how in-medium modifications need to be self-consistently incorporated into the calculation of transport coefficients, such as the shear and bulk viscosities of the hadronic gas, as well as the drag and diffusion coefficients for the heavy systems. Furthermore, thermal spectral functions are intimately linked to dilepton emission rates, providing clear signatures of medium modifications that have already been observed for light vector mesons, especially the $\rho$ meson. A rather different sector in which these effects can produce observable consequences is hadron femtoscopy. Since correlation functions are used to extract valuable information about hadron-hadron interactions, their in-medium corrections could, in principle, also be tested. As a particular example, the thermal mass shift predicted for the $\Delta$ baryon has been used to improve the description of pion-proton and pion-deuteron correlation functions compared with the vacuum pole-mass expectations.

\vspace{0.5cm}
\noindent\textbf{Outlook}
\vspace{0.2cm}

We have reviewed a substantial body of theoretical results regarding the medium modifications of hadrons and the potential implications for HIC experiments. Looking forward, several key research avenues remain open:

\begin{itemize}
    \item \textbf{Advances in lattice-QCD methods and spectral reconstruction:} Extracting transport coefficients and spectral functions from Euclidean correlators at finite temperature remains a formidable challenge. While significant progress has been made using the MEM and other Bayesian reconstruction techniques, developing entirely model-independent methods for extracting real-time quantities is still an active area of research. Utilizing finer lattices with physical quark masses at temperatures near and below the crossover temperature will be essential to resolve the behavior of hadron properties close to the chiral transition. While this review has focused on finite temperature, the study of medium modifications due to a finite baryochemical potential $\mu_B$ is also an important avenue. This requires overcoming the lattice-QCD ``sign problem'' to establish first-principles constraints on the QCD phase diagram.

    \item \textbf{Internal structure of exotic states in the medium:} The nature of exotics is currently a subject of intense debate. Compact configurations (tetraquark, pentaquark) compete with extended molecular structures to describe near-threshold states, such as the well-known $X(3872)$. Quite interestingly, it is now possible to reconstruct these states in relativistic HICs. Although such measurements are still limited, thermal corrections due to the surrounding medium can provide additional tests for these models. Even if thermal modifications are subtle, the fact that many of these states reside just below mass thresholds---a characteristic that supports the hypothesis of molecular structures---means that thermal effects can significantly influence them by shifting thresholds and opening new decay channels. Here, we have reported theoretical works studying this situation for the $X(3872)$ and the $T_{cc}^+(3875)$, but many other exotic states exist. In addition, numerical simulations of real-time dynamics of these states affected by thermal potentials are still lacking, but these could enormously help to constrain their inner structure and interactions.

    \item \textbf{Simulations and nonequilibrium dynamics:} The incorporation of medium modifications of specific hadrons has already improved the description of dilepton spectra and other sensitive observables. However, there is a need for new observables directly linked to thermal modifications, particularly those reflecting chiral symmetry restoration. This could manifest itself as modified decay rates or corrections to the yields predicted by the Statistical Hadronization Model. While small mass shifts may not always result in detectable variations, specific systems---particularly those near thresholds or sensitive to chiral symmetry restoration---may show significant changes. In addition, the internal structure can be a reliable indicator. For example, scalar mesons typically exhibit greater temperature dependence than the pseudoscalars; the thermal reduction in their mass could lead to a measurable enhancement in their production yields or a modification of their decay products.
    
    \item \textbf{Hadron femtoscopy:} In recent years, there has been a wealth of new data from hadron femtoscopy, especially from the ALICE (Large Hadron Collider, LHC) and STAR (RHIC) collaborations. These collisions, whether nucleus-nucleus or high-multiplicity $pp$, create a hadronic system that may be close to thermal equilibrium, in which strong and Coulomb interactions shape the two-particle correlation functions of different pairs. These interactions are inherently subject to medium effects. Although such modifications are often subtle or even unobservable, specific systems may produce visible signals in correlation functions. For example, thermal resonances---whose mass or width is modified by the medium---decaying into hadrons could be observed, as seen in the case of the $\Delta$ baryon in proton-pion and deuteron-pion femtoscopy. Upcoming data from the ALICE Run 3 will provide unprecedented precision, offering an excellent opportunity to identify systems where these medium effects are most prominent.

     \item \textbf{Future experimental opportunities:} The theoretical predictions reviewed here can be confronted with a new generation of measurements. At the LHC, the upgraded ALICE detector~\cite{ALICE:2023udb} has collected large samples of Pb+Pb collisions during Run 3 (and will continue to do so in the high-luminosity phase, Run 4, after the Long Shutdown 3), which can improve the precision of low-mass dielectron spectra, of resonance yields such as the $K_1/K^*$ ratio discussed in Section~\ref{sec:applications}, of the ratios of excited to ground-state quarkonia, and of femtoscopic correlations, including those involving charm hadrons. In the longer term, ALICE 3~\cite{ALICE:2022wwr} is designed to measure dielectrons with very low combinatorial background. This will give access to the intermediate-mass region ($1 \lesssim M \lesssim 1.5\gev$), where the mixing of the vector and axial-vector spectral functions, a direct signal of chiral symmetry restoration~\cite{Hohler:2013eba}, is expected, and where the slope of the thermal dilepton spectrum provides a direct thermometer of the medium. In ALICE 3, the early-time temperature is expected to be extracted with a statistical uncertainty of about $1.5 \%$~\cite{ALICE:2022wwr}. ALICE 3 will also allow systematic studies of multi-charm hadrons and exotic states, whose production in HICs has already been observed for the $X(3872)$~\cite{CMS:2021znk}. In particular, ALICE 3 will be able to reconstruct $X(3872)$ decays down to $p_T \simeq 5\gev$, which is the optimal region to probe the thermal effects in the production yield of this exotic state~\cite{ALICE:2022wwr}. At lower collision energies, where the baryochemical potential is larger, the beam energy scan at RHIC~\cite{STAR:2023wta}, the HADES experiment~\cite{HADES:2011nqx}, the future CBM experiment at FAIR~\cite{CBM:2016kpk}, and the proposed NA60+ experiment at the CERN SPS~\cite{NA60:2022sze} will measure dilepton spectra with high statistics as a function of the collision energy. These data will constrain the in-medium $\rho$ spectral function at high baryon density, as well as the temperature of the fireball, testing the extension of the finite-temperature results reviewed here to finite $\mu_B$.
\end{itemize}

In conclusion, the study of hadron properties at finite temperature remains a remarkable field that links the microscopic QCD dynamics and the macroscopic observables of high-energy nuclear collisions. Continued synergy between theoretical advances, first-principles calculations, and phenomenological modeling will be essential to determine the remaining unknowns of the QCD phase diagram in the confined phase. 

\section*{Acknowledgements}

This work has been supported by the projects \mbox{CEX2024-001451-M} (Unidad de Excelencia “Mar\'ia de Maeztu”) and  \mbox{PID2023-147112NB-C21}, financed by \mbox{MICIU/AEI/10.13039/501100011033/} and FEDER, UE.
JMT-R also thanks the Contract 2021 SGR 171 by the Generalitat de
Catalunya and Grant No. 402942/2024-8 by
the Brazilian CNPq (National Council for Scientific and
Technological Development).
GM was supported by the Beatriu de Pin\'os program by AGAUR, Grant \mbox{No. BP 2024 00189}, and also thanks U.S. Department of Energy contract \mbox{DE-AC05-06OR23177}, under which Jefferson
Science Associates, LLC operated Jefferson Lab.

\section*{Authors' contributions}
Both authors made equal contributions to the conceptualization, literature review, drafting, and final revision of this manuscript.

\bibliography{MesonProp}
%Please use Bib\TeX\ to generate your bibliography and include DOIs whenever available. Example of bib file: 

%%%%%%%%%%%%%%%%%%%%%%%%%%%%%%%%%%%%%%%%%%%%%%%%%%%%%%%%%%%%%%%%%%%
% Encoding: ISO-8859-1

%@Article{Eichmann:2016yit,
  %author        = {Eichmann, Gernot and Sanchis-Alepuz, Helios and Williams, Richard and Alkofer, Reinhard and Fischer, Christian S.},
  %title         = {{Baryons as relativistic three-quark bound states}},
  %journal       = {Prog. Part. Nucl. Phys.},
  %year          = {2016},
  %volume        = {91},
  %pages         = {1-100},
  %archiveprefix = {arXiv},
  %doi           = {10.1016/j.ppnp.2016.07.001},
  %eprint        = {1606.09602},
  %owner         = {chfi},
  %primaryclass  = {hep-ph},
  %slaccitation  = {%%CITATION = ARXIV:1606.09602;%%},
  %timestamp     = {2018.08.02},
%}

%@Comment{jabref-meta: databaseType:bibtex;}
%%%%%%%%%%%%%%%%%%%%%%%%%%%%%%%%%%%%%%%%%%%%%%%%%%%%%%%%%%%%%%%%%%%

\appendix
\renewcommand*{\thesection}{\Alph{section}}

% \section{Appendices, if necessary}\label{appendix}

\end{document}